\documentclass[usenatbib]{mn2e}
\bibpunct{(}{)}{;}{a}{}{,}
\usepackage{psfig}

\begin{document}

\title
[The runaway instability of thick discs around black holes. II.]
{The runaway instability of thick discs around black holes. \\ 
II. Non constant angular momentum discs}

\author[Daigne \& Font]
{Fr\'{e}d\'{e}ric Daigne$^{1}$ and Jos\'{e} A. Font$^{2}$\\
\\
$^{1}$ Institut d'Astrophysique de Paris, 98 bis boulevard Arago,
75014 Paris, France\\
$^{2}$ Departamento de Astronom\'{\i}a y Astrof\'{\i}sica,
Universidad de Valencia, Dr. Moliner, 50, 46100 Burjassot (Valencia), Spain\\
}
\maketitle

\begin{abstract}
We present results from a comprehensive number of relativistic, time-dependent,
axisymmetric simulations of the runaway instability of {\it non-constant} angular 
momentum thick discs around black holes. This second paper in the series extends 
earlier results where only constant angular momentum discs were considered. 
All relevant aspects of the theory of stationary thick discs around rotating
black holes, necessary to build the equilibrium initial data used in our simulations,
are presented in great detail. The angular momentum of the evolved discs is assumed 
to increase outwards with the radial distance according to a power law, 
$l=K r^{\alpha}$, where $K>0$ corresponds to prograde discs (with respect to the
black hole rotation) and $K<0$ to retrograde discs. The main simplifying assumptions 
of our approach are not to include magnetic fields and self-gravity in the discs 
(test-fluid approximation). Furthermore, the dynamics of the spacetime is accounted 
for by computing the transfer of mass and angular momentum from the disc to the black 
hole through the event horizon. In this approximation the evolution of the central 
black hole, which initially is non-rotating, is assumed to follow a sequence of Kerr 
black holes of increasing mass and spin. All discs we build slightly overflow the 
potential barrier at the cusp, departing from equilibrium, so that accretion is 
possible. In agreement with previous results based on stationary models we find that 
by allowing the mass and the spin of the black hole to grow, constant angular momentum 
discs rapidly become unstable on a dynamical timescale (a few orbital periods). The 
comparison with the results of our first paper shows that the effect of the angular 
momentum transfer from the tori to the black hole is to make constant angular momentum 
discs less unstable, increasing the timescale for the runaway instability to grow. 
However, we find that non-constant angular momentum discs are dramatically stabilized 
for very small values of the angular momentum slope $\alpha$, much smaller than the 
Keplerian value $\alpha=1/2$. Our fully relativistic and time-dependent simulations 
confirm, thus, the predictions of stationary studies concerning the stabilizing 
effect of non-constant angular momentum distributions. For the various disc-to-hole 
mass ratios considered we systematically find that the critical values of $\alpha$ 
below which the runaway instabilty can exist are slightly smaller than those reported 
previously in the literature based on stationary studies.
\end{abstract}

\begin{keywords}
accretion, accretion discs -- black hole physics -- hydrodynamics -- 
instabilities -- gamma rays: bursts.
\end{keywords}

\section{Introduction}

In a recent paper \citep{font:02a} (Paper I hereafter) we presented the first time-dependent, 
hydrodynamical simulations in general relativity of the so-called runaway instability of  
geometrically thick discs (or tori) around a Schwarzschild black hole. The origin of this 
instability, discovered by \citet{abramowicz:83}, is a dynamical process by which the 
cusp of a disc which initially is filling its Roche lobe penetrates inside the disc 
due to mass transfer from the disc to the accreting black hole. This process leads to the 
complete destruction of the disc on a dynamical timescale. \citet{abramowicz:83} found that 
the runaway instability occurs for a large range of parameters such as the disc-to-hole mass 
ratio and the location of the inner radius of the disc. More detailed studies followed (see, 
e.g. Paper I for an up-to-date list), most of which were based on stationary models in which 
a fraction of the mass and angular momentum of the initial disc is instantaneously transferred 
to the black hole. The new gravitational field allows to compute the new position of the cusp, 
which controls the occurrence of the runaway instability. The self-gravity of the tori favours 
the instability, as shown from both, studies based on a pseudo-potential for the black hole 
\citep{khanna:92,masuda:98} and from fully relativistic (stationary) calculations \citep{nishida:96a}. 
However, there are some parameters in the models which have stabilizing effects. One of this is 
the rotation of the black hole \citep{wilson:84,abramowicz:98}, but the most important one seems 
to be the radial distribution of the specific angular momentum, increasing with the radial 
distance as a power law \citep{daigne:97,abramowicz:98}. Despite the abundant literature on the 
subject, it is commonly accepted that the very existence of the instability remains still 
uncertain.

In Paper I we initiated a program aimed at investigating the runaway instability through 
time-dependent hydrodynamical simulations of increasing complexity in general relativity. 
The starting point was simple. In the simulations reported in Paper I the distribution 
of the angular momentum in the disc was assumed to be constant. We also neglected the 
self-gravity of the disc as well as the presence of magnetic fields. Our model, however, 
accounted for the dynamics of the gravitational field of the black hole plus torus system 
in an approximate yet satisfactory manner. The evolution of the central black hole 
was assumed to follow a sequence of Schwarzschild black holes of increasing mass, whose 
increase rate was controlled by computing the mass flux from the disc to the black hole 
through the innermost point of the numerical grid. We notice that the ability to take 
into account such dynamics is of paramount importance to investigate the runaway instability 
at all.

Our study showed that by allowing the mass of the black hole to grow the runaway instability 
appears on a dynamical timescale, in agreement with previous estimates from stationary 
models. Since our study was mainly motivated by understanding the hydrodynamical evolution of 
the central engine of gamma-ray bursts (GRBs hereafter), we only considered the case of stellar 
mass black holes, leaving aside thick discs around supermassive black holes which can also be 
present in active galactic nuclei. For a black hole of $2.5\mathrm{M_{\odot}}$ and disc-to-hole 
mass ratios in the range 0.05 to 1, we found in Paper I that the timescale of the instability 
never exceeds $1\ \mathrm{s}$ for a large range of mass fluxes and it is typically a few 10 ms. 
We note in passing that the runaway instability seems to be a robust feature of non-self-gravitating, 
constant angular momentum discs irrespective of the way accretion is induced, as \citet{zanotti:02} 
have recently reported. 

Whatever the progenitor system which collapses to form a stellar mass black hole surrounded by a 
thick disc is, it is very unlikely that the initial distribution of the angular momentum in the 
disc is constant. As numerical simulations show this is not the case in discs formed after the 
coalescence of a close compact binary (consisting either of two neutron stars or of a black 
hole and a neutron star) \citep{davies:94,ruffert:96,kluzniak:98,ruffert:99,lee:00,shibata2:03}, 
in the collapsar model of {\it failed} supernova \citep{macfadyen:99}, or, at different scales, in the 
collapse of a rotating supermassive star leading to a rotating supermassive black hole 
\citep{shibata:02,shapiro:02}. For instance, in the Newtonian SPH simulations of \citet{davies:94} 
the final configuration consists of a core of mass $2.44$M$_{\odot}$ in uniform rotation 
surrounded by a differentially rotating thick disc of mass $0.36$M$_{\odot}$. The distribution 
of the angular momentum in the disc is close to a power law of index 0.2. Similarly, in the most 
recent three-dimensional SPH coalescence simulations of \citet{lee:00} the corresponding angular 
momentum distribution follows a power law of index 0.45, quite close to Keplerian. We note in 
particular that recent relativistic simulations of unequal mass neutron star coalescence 
\citep{shibata2:03} yield a black hole surrounded by a thick disc, following tidal disruption 
of the less massive star. \citet{shibata2:03} found that for a neutron star rest-mass ratio
$\sim 0.85$ the final mass of the disc may be several percents of the total mass of the system
if the stars are not very compact. Furthermore, the angular momentum distribution in the disc 
that forms is not constant (M.~Shibata, private communication). In addition to the evidence 
provided by the results of numerical simulations, as the dynamical timescale on which the runaway 
instability occurs is much shorter than the viscous timescale necessary to achieve angular momentum 
transport, it seems much more realistic to assume a \textit{non-constant} angular momentum 
distribution in the disc, increasing outwards from the cusp to larger radii. 

Therefore, in the present paper we extend the study initiated in Paper I to such most interesting 
case of non-constant angular momentum discs. The main motivation of our work is to check through 
time-dependent simulations in general relativity whether such distributions have indeed the 
stabilizing effect previously found in non self-gravitating stationary models 
\citep{daigne:97,abramowicz:98,lu:00}. We note that preliminary results of this investigation were 
already presented in a previous letter \citep{font:02b} (Paper II hereafter). In the current paper 
we extend those results in two main directions: firstly, we present in great detail all relevant 
aspects of the theory of stationary, non-constant angular momentum thick discs around rotating
black holes, which is necessary to build the equilibrium initial data we use in our simulations.
Secondly, we report on the results of a comprehensive number of simulations which allows us
to explore the dependence of the absence/presence of the runaway instability on various parameters 
of the model such as the disc-to-hole mass ratio, the potential barrier at the cusp (or in an equivalent 
manner, the initial mass accretion rate), or the efficiency of the angular momentum transfer from 
the disc to the black hole. The extended number of models considered here permits to draw some 
conclusive answers on the likelihood of the runaway instability of non-self-gravitating discs, as 
well as to pin-down the critical value of the exponent of the angular momentum distribution separating 
stable and unstable models.

As we showed in Paper II, and we further clarify in the present investigation, the runaway instability 
is most likely avoided in non rigidly rotating discs as those formed in the coalescence of a binary neutron 
star system or in the gravitational collapse of a massive star. This result is in agreement with 
recent perturbative analysis of vertically integrated discs performed by \citet{rezzolla:03} who found 
that the innermost regions of discs tend to behave as regions of {\it evanescent-wave propagation}, 
reducing the mass flux and, as a result, suppressing the runaway instability. If the growth of 
non-axisymmetric modes in the disc by the Papaloizou-Pringle instability is suppressed by the accretion 
process itself, as suggested by \citet{blaes:87}, systems consisting of a Kerr black hole surrounded by 
a high density torus may then be long lived. The lifetime is probably controlled by the viscous 
timescale (a few seconds) rather than by the dynamical one. This may provide enough time for any 
plausible magneto-hydrodynamical process to efficiently transfer part of the energy reservoir of 
the system to a relativistic outflow. Therefore, the most favoured current models of GRBs (see 
\citet{woosley:01}) can indeed be based on such central engines. This is especially important in the 
context of the so-called internal shock model for the prompt $\gamma$-ray emission \citep{rees:94}, where 
the observed lightcurve reflects the activity of the central engine, with no modification of the timescales 
other than the time dilation due to the redshift. In particular, the central engine has to survive for 
a duration at least comparable with the observed duration of the GRB.

The paper is organized as follows: Section~\ref{sec:initialmodel} describes in detail the 
relevant aspects of the theory of relativistic, stationary, non-constant angular momentum
thick discs necessary to build the initial state for our simulations. Part of this
information is presented in three appendices at the end of the paper. Section~\ref{sec:hydro} 
deals with a brief description of the numerical framework we use. Section~\ref{sec:tests} 
shows specific tests for the Kerr spacetime that the code has successfully passed. The 
description of the simulations and the discussion of the results are presented in 
Section~\ref{sec:simulations}. Finally Section~\ref{sec:conclusions} summarizes our 
conclusions and presents possible directions for future work. As we did in Papers I and II we 
use geometrized units ($G=c=1$) unless explicitely stated. Usual cgs units are obtained by 
using the gravitational radius of the black hole $r_\mathrm{g}=G M_\mathrm{BH}/c^{2} = 
1.5\times 10^{5}\ (M_{\mathrm{BH}}/M_{\odot})\ \mathrm{cm}$ as unit of length. Greek (Latin) 
indices run from 0 to 3 (1 to 3), and we use a spacelike signature for the metric (- + + +).

\section{Equilibrium initial tori}
\label{sec:initialmodel}


The theory of stationary, relativistic thick discs (or tori) with a baryotropic equation 
of state (EoS) was derived by \citet{abramowicz:78} (see also \citet{fishbone:76}). We 
included in Paper I all relevant aspects of this theory necessary to construct the initial 
state for our simulations. In particular we described in detail the procedure to build 
equilibrium configurations of the system in the case of a Schwarzschild (non-rotating) 
black hole and a constant angular momentum disc. Here we extend this procedure to the more 
general case of a Kerr black hole and a non-constant angular momentum disc which follows a 
power-law distribution with the radial distance. The notations and conventions we use are 
the same than in Paper I.

\subsection{Gravitational field}

As mentioned before the self gravity of the disc is neglected. In Boyer-Lindquist 
$(t,r,\theta,\phi)$ coordinates, the Kerr line element, $ds^2=g_{\mu\nu} dx^{\mu} dx^{\nu}$, 
reads
\begin{eqnarray}
ds^2 &=& - { {\Delta - a^2\sin^2\theta} \over {\varrho^2} } dt^2 -
       2a { {2M_{\mathrm{BH}}r\sin^2\theta} \over {\varrho^2} } dt d\phi
\nonumber \\ &+&
       { {\varrho^2} \over {\Delta} } dr^2                     +
       \varrho^2 d\theta^2 +
       { {\Sigma} \over {\varrho^2} }
       \sin^2\theta d\phi^2,
\label{blform}
\end{eqnarray}
\noindent
with the usual definitions:
\begin{eqnarray}
\Delta &\equiv& r^2 - 2M_{\mathrm{BH}}r + a^2,
\\
\varrho^2 &\equiv& r^2 + a^2\cos^2\theta,
\label{metrho}
\\
\Sigma &\equiv& (r^2+a^2)^2 - a^2\Delta\sin^2\theta,
\end{eqnarray}
\noindent
where $M_{\mathrm{BH}}$ is the mass of the black hole and $a$ is the black hole angular momentum 
per unit mass ($J_{\mathrm{BH}}/M_{\mathrm{BH}}$). We assume here that $0 \le a \le M_{\mathrm{BH}}$ 
so that prograde (respectively, retrograde) orbits correspond to positive (respectively, negative) 
values of the angular momentum of the disc $l$ (see below). The above metric, Eq.~(\ref{blform}), 
describes the spacetime exterior to a rotating and non-charged black hole. The metric has a 
coordinate singularity at the roots of the equation $\Delta=0$, which correspond to the horizons 
of a rotating black hole, $r=r_{\pm}=M_{\mathrm{BH}}\pm(M_{\mathrm{BH}}^2-a^2)^{1/2}$. In the 
following, we use the notation $r_\mathrm{h}=r_{+}$ and we call it ``horizon of the black hole''. 
Kerr black holes are also characterized by the presence of the ergosphere, a region inside which 
no static observers exist, whose boundary is given by
\begin{equation}
r_\mathrm{e}(\theta) = M_\mathrm{BH}+\sqrt{M_{\mathrm{BH}}^2-a^{2}\cos^{2}{\theta}}\ .
\end{equation}
Furthermore, the ``distance to the rotation axis'' is defined by 
\begin{equation}
\varpi^{2} = g_{t\phi}^{2}-g_{tt} g_{\phi\phi} = \Delta\sin^2\theta.
\end{equation}

The 3+1 decomposition (see, e.g., \citet{misner:73}) of this form of the metric 
leads to a spatial 3-metric $\gamma_{ij}$ with non-zero elements given by 
$\gamma_{rr} = \varrho^{2}/\Delta, \gamma_{\theta\theta} = \varrho^2, 
\gamma_{\phi\phi} = {\Sigma} / {\varrho^2}  \sin^2\theta$. In addition,
the azimuthal {\em shift vector} $\beta_{\phi}\equiv g_{t\phi}$ is given by
\begin{equation}
\beta_{\phi} = -\frac{2 a M_\mathrm{BH} r\sin^2\theta}{\varrho^{2}},
\end{equation}
and the {\em lapse function} is given by
\begin{equation}
\alpha = \left( \frac{\varrho^{2} \Delta}{\Sigma} \right)^{1/2}.
\end{equation}

For convenience we assume in the following that the initial mass of the black hole 
equals unity (therefore $0 \le a \le 1$). 

\subsection{Angular momentum}

The angular momentum per unit inertial mass (hereafter ``angular momentum'') of a fluid element 
is related to the four-velocity by
\begin{equation}
l = -\frac{u_{\phi}}{u_{t}}\ .
\label{eq:defl}
\end{equation}
The corresponding angular velocity $\Omega$ is
\begin{equation}
\Omega = \frac{u^{\phi}}{u^{t}} = -\frac{g_{t\phi}+g_{tt} l}{g_{\phi\phi}+g_{t\phi} l}\ .
\label{eq:defomega}
\end{equation}
The energy per unit inertial mass of a fluid element $-u_{t}$ can be obtained from the condition 
$u^{2}\equiv u_{\mu}u^{\mu}=-1$:
\begin{equation}
-u_{t} = \sqrt{\frac{\varpi^{2}}{g_{tt} l^{2} + 2 g_{t\phi} l + g_{\phi\phi}}}\ .
\label{eq:defut}
\end{equation}
This imposes a condition on the angular momentum:
\begin{equation}
g_{tt} l^{2} + 2 g_{t\phi} l + g_{\phi\phi} > 0\ ,
\end{equation}
which is equivalent to
\begin{equation}
\left\lbrace
\begin{array}{rcl}
l < l_\mathrm{cr}^{+}(r,\theta)\ \mathrm{or}\  l > l_\mathrm{cr}^{-}(r,\theta) & \mathrm{for} 
& r\le r_\mathrm{e}(\theta)\\
l_\mathrm{cr}^{-}(r,\theta) < l < l_\mathrm{cr}^{+}(r,\theta) & \mathrm{for} & r > r_\mathrm{e}(\theta)
\end{array}
\right.
\label{eq:critical}
\end{equation}
with
\begin{equation}
l_\mathrm{cr}^{\pm}(r,\theta) = \frac{g_{t\phi}\pm\varpi}{-g_{tt}}\ .
\end{equation}

A fluid element will be bound if $-u_{t} < 1$, which is equivalent to
\begin{equation}
\left\lbrace
\begin{array}{rcl}
l_\mathrm{b}^{+}(r,\theta) < l\ \mathrm{or}\  l > l_\mathrm{b}^{-}(r,\theta) & 
\mathrm{for} & r\le r_\mathrm{e}(\theta)\\
l_\mathrm{b}^{-}(r,\theta) < l < l_\mathrm{b}^{+}(r,\theta) & \mathrm{for} & 
r > r_\mathrm{e}(\theta)
\end{array}
\right.
\label{eq:bound}
\end{equation}
with
\begin{equation}
l_\mathrm{b}^{\pm}(r,\theta) = \frac{g_{t\phi}\pm\varpi\sqrt{1+g_{tt}}}{-g_{tt}}.
\end{equation}

Notice that inside the ergosphere ($r < r_\mathrm{e}(\theta)$), we have $l_\mathrm{b}^{+} 
< l_\mathrm{cr}^{+} < l_\mathrm{cr}^{-} < l_\mathrm{b}^{-}$ and outside the ergosphere 
($r > r_\mathrm{e}(\theta)$), we have $l_\mathrm{cr}^{-} < l_\mathrm{b}^{-} < l_\mathrm{b}^{+} 
< l_\mathrm{cr}^{+}$. On the surface of the ergosphere ($r=r_\mathrm{e}(\theta)$), 
$l_\mathrm{b}^{-}=l_\mathrm{cr}^{-}=-\infty$, $l_\mathrm{b}^{+}=(r_\mathrm{e}^{2}+a^{2})/2a$ 
and $l_\mathrm{cr}^{+}=(r_\mathrm{e}^{2}-r_\mathrm{e}+a^{2})/a$. Then the angular momentum 
corresponding to the marginally bound case fullfils condition~(\ref{eq:critical}). At the 
horizon ($r=r_\mathrm{h}$), we have $l_\mathrm{b}^{\pm}=l_\mathrm{cr}^{\pm}=2 r_\mathrm{h}/a$.

\begin{figure}
\centerline{\psfig{file=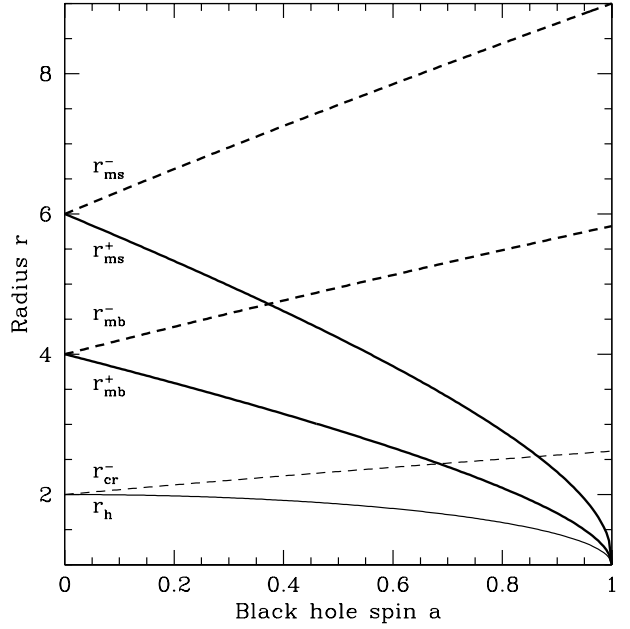,width=9.0cm}}
\caption{\textbf{Characteristic radii in the equatorial plane:} the radius of the horizon 
$r_\mathrm{h}$, of the last marginally bound Keplerian orbit $r_\mathrm{mb}^{\pm}$ and of the 
last marginally stable Keplerian orbit $r_\mathrm{ms}^{\pm}$ are plotted as a function of the 
spin $a$ of the black hole. The radius $r_\mathrm{cr}^{-}$ where the Keplerian angular momentum 
of a particle with a retrograde orbit becomes infinite is also plotted. 
The schwarzschild limit ($a=0$) corresponds to $r_\mathrm{h}=r_\mathrm{cr}^{-}=2$, 
$r_\mathrm{mb}^{\pm}=4$ and $r_\mathrm{ms}^{\pm}=6$. The maximum black hole rotation limit ($a=1$) 
corresponds to $r_\mathrm{h}=r_\mathrm{mb}^{+}=r_\mathrm{ms}^{+}=1$, $r_\mathrm{mb}^{-}=
3+2\sqrt{2}\simeq 5.83$, $r_\mathrm{ms}^{-}=9$ and $r_\mathrm{cr}^{-}=(3+\sqrt{5})/2\simeq 2.62$.
}
\label{fig:radii}
\end{figure}

\subsection{Keplerian orbits}

In the particular case of a Keplerian orbit in the equatorial plane, the angular momentum is given by
\begin{equation}
l_\mathrm{K,eq}^{\pm}   = \pm \frac{r^{2} \mp 2 a \sqrt{r} + a^{2}}{(r-2)\sqrt{r} \pm a}.
\label{eq:lk}
\end{equation}
Throughout the paper the subscript `eq' means that the considered quantity is evaluated at the
equatorial plane $\theta=\pi/2$. In Eq.~(\ref{eq:lk}) $l^{\pm}_\mathrm{K,eq}$ stands for a prograde 
(+) or a retrograde (-) orbit.  A Keplerian orbit will be stable for $d|l_\mathrm{K,eq}|/dr > 0$. 
The radius of the last (marginally) stable orbit then corresponds to the case where $dl_\mathrm{K,eq}/dr 
=0$, i.e. the minimum (respectively, maximum) of $l_\mathrm{K,eq}$ for a prograde orbit (respectively, 
retrograde orbit). This radius is given by 
\begin{equation}
r_\mathrm{ms}^{\pm} = 3+Z_{2}\mp\sqrt{(3-Z_{1})(3+Z_{1}+2 Z_{2})}\ ,
\end{equation}
where $Z_{1}=1+(1-a^{2})^{1/3} \left[ (1+a)^{1/3}+(1-a)^{1/3} \right]$ and $Z_{2}=\sqrt{ 3 a^{2} + 
Z_{1}^{2}}$.  A Keplerian orbit will be bound if $l_\mathrm{K,eq}$ verifies equation~(\ref{eq:bound}). 
The radius of the last (marginally) bound orbit corresponds to the case where 
$l_\mathrm{K,eq}^{\pm}=l_\mathrm{b,eq}^{\pm}$. This radius is given by
\begin{equation}
r_\mathrm{mb}^{\pm} = 2 \mp a + 2\sqrt{1\mp a}\ .
\end{equation}

\begin{figure*}
\begin{center}
\begin{tabular}{ccc}
$a=0$ & $a=\sqrt{5}/3$ & $a=1$\\
\psfig{file=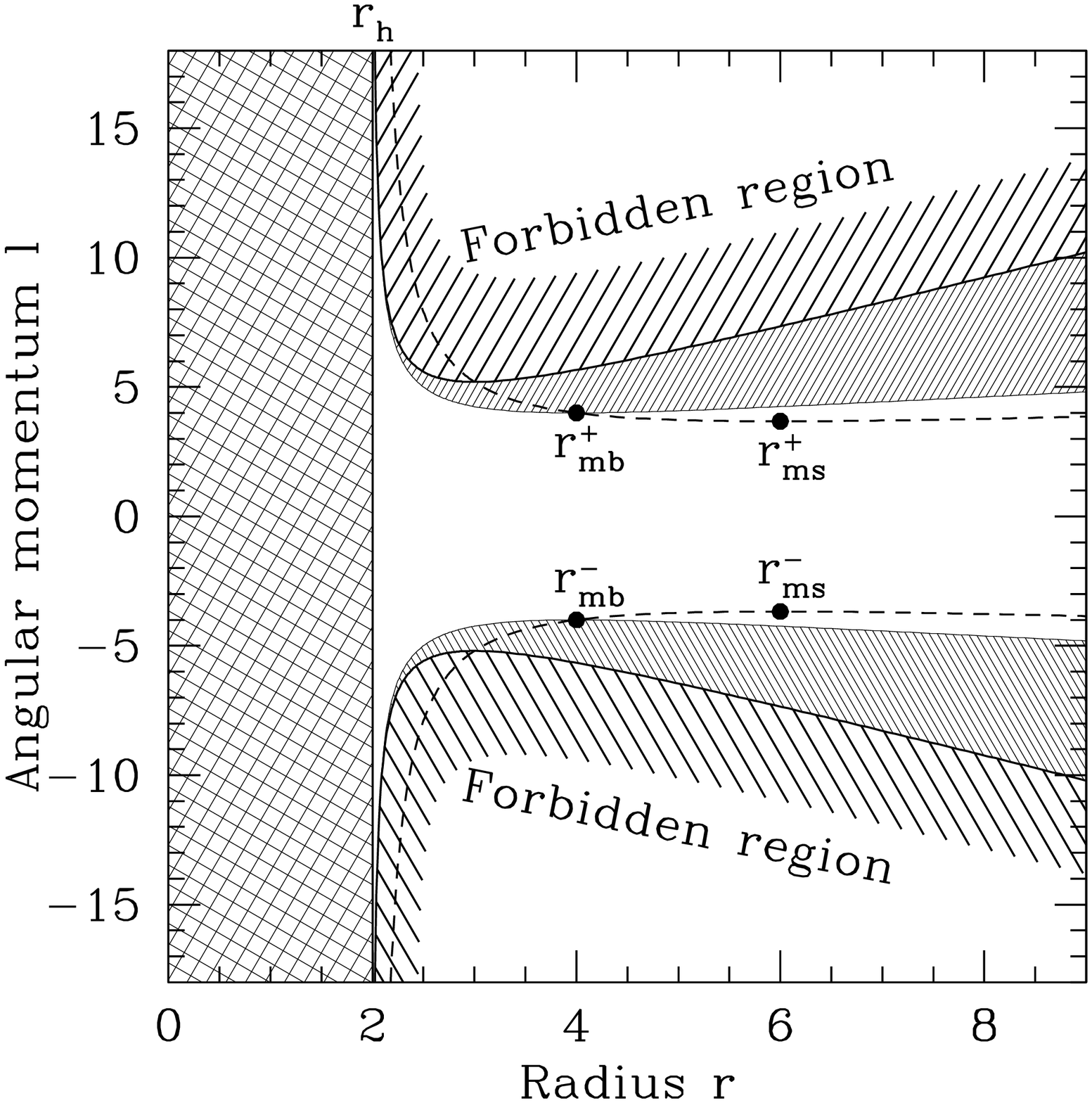,width=0.32\textwidth} &
\psfig{file=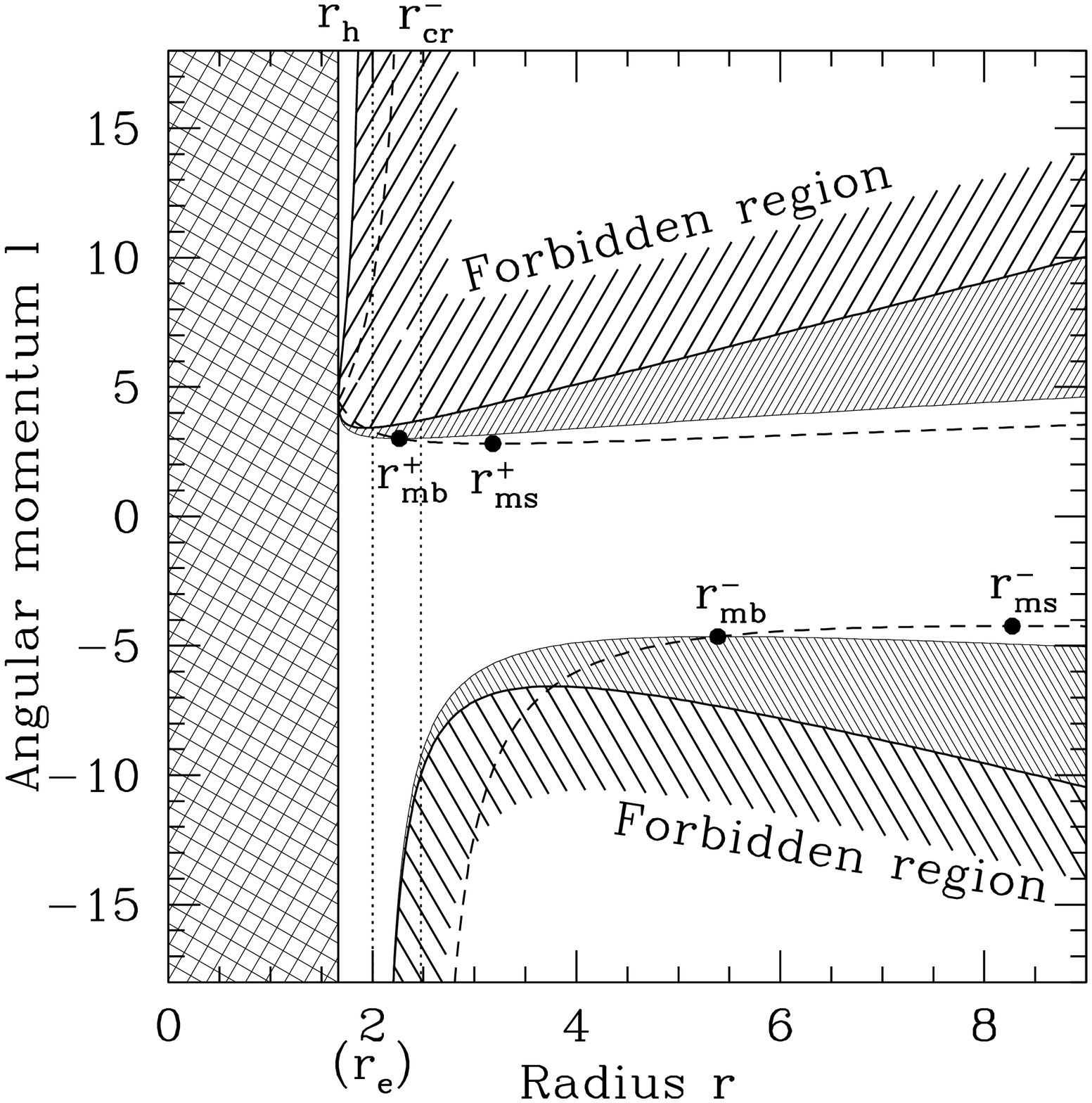,width=0.32\textwidth} &
\psfig{file=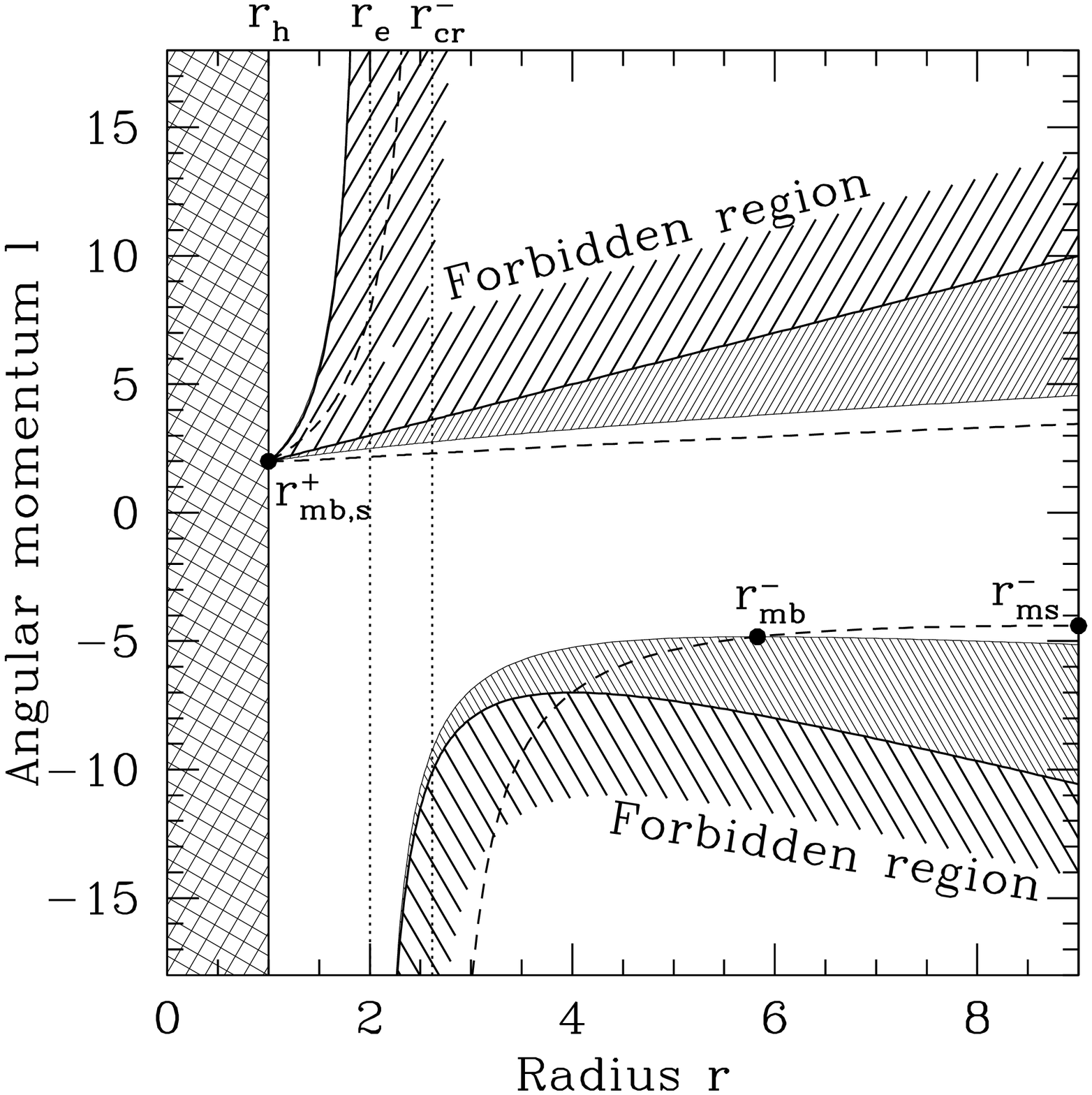,width=0.32\textwidth}
\end{tabular}
\end{center}
\caption{\textbf{Characteristic values of the angular momentum in the equatorial plane:} the critical 
value $l_\mathrm{cr,eq}^{\pm}$ for which the energy per unit inertial mass of a fluid element $-u_{t}$ 
becomes infinite is plotted as a function of $r$ in thick solid line and defines a forbidden region for 
the angular momentum which is indicated (grey area). The angular momentum $l_\mathrm{b,eq}^{\pm}$ for 
which the orbit is marginally bound is plotted in thin solid line. The region where orbits are not bound 
is also indicated (between thick and thin solid lines). The angular momentum of a Keplerian orbit 
$l_\mathrm{K,eq}^{\pm}$ is plotted in dashed line. The radius of the horizon $r_\mathrm{h}$, the 
ergosphere $r_\mathrm{e}$, the marginally bound Keplerian orbit $r_\mathrm{b}^{\pm}$ and the marginally 
stable Keplerian orbit $r_\mathrm{ms}^{\pm}$ are indicated, as well as the radius $r_\mathrm{cr}^{-}$ 
where the Keplerian angular momentum of a retrograde orbit becomes infinite. The interior of the black 
hole horizon has been shaded with crossing lines.}
\label{fig:angularmomentum}
\end{figure*}

Notice that outside the horizon, the Keplerian angular momentum becomes infinite for retrograde 
orbits at a radius $r_\mathrm{cr}^{-} \ge 2$ solution of $(r-2)\sqrt{r}-a=0$. The characteristic 
radii $r_\mathrm{h}$, $r_\mathrm{ms}^{\pm}$, $r_\mathrm{mb}^{\pm}$ and $r_\mathrm{cr}^{-}$ are 
plotted as a function of $a$ in Figure~\ref{fig:radii}. All radii are given in units of the 
black hole gravitational radius. Correspondingly, the characteristic values of 
the angular momentum in the equatorial plane $l_\mathrm{K,eq}^{\pm}$, $l_\mathrm{b,eq}^{\pm}$ and 
$l_\mathrm{cr,eq}^{\pm}$ are plotted as a function of $r$ in Figure~\ref{fig:angularmomentum} 
for $a=0$ (non rotating black hole), $a=\sqrt{5}/3$ and $a=1$ (maximally rotating black hole). 
The intermediate value has been chosen to have $r_\mathrm{h}=1+2/3$, where deviations from the
Schwarzschild case start being more prominent. Notice that at the horizon $l_\mathrm{K,eq}^{\pm}=
l_\mathrm{b,eq}^{\pm}=l_\mathrm{cr,eq}^{\pm}= 2 r_\mathrm{h}/a$.

\subsection{Rotation law in the disc}
\label{sec:rotation}

The distribution of angular momentum in the disc is assumed to follow a power-law in the equatorial plane:
\begin{equation}
l_\mathrm{eq}(r) = K r^{\alpha}\ ,
\label{eq:leq}
\end{equation}
with $\alpha \ge 0$. Prograde (respectively, retrograde) discs correspond to $K>0$ (respectively, $K<0$). 
From the rotation law in the equatorial plane, we can compute the value of the angular momentum at any 
location $(r,\theta)$ in the disc, by solving the equations of the constant $l$ surfaces, i.e. the 
so-called von Zeipel's cylinders: 
\begin{equation}
l(r,\theta)=l_\mathrm{eq}(r_{0}),
\label{eq:r0}
\end{equation}
where $r_{0}$ is the radius at which the von Zeipel's cylinder passing at $(r,\theta)$ intersects the 
equatorial plane. The structure of these cylinders is presented in detail in Appendix~\ref{sec:vonzeipel}. 
For practical purposes the value of $l(r,\theta) = l_\mathrm{eq}(r_{0})$ is obtained by solving numerically 
equation~(\ref{eq:cylinders}) to find $r_{0}$.

\subsection{Iso--pressure and iso--density surfaces}

The integral form of the relativistic Euler equation that governs the dynamics of the gas flow can be 
written as
\begin{equation}
W-W_\mathrm{in} = \ln(-u_{t})-\ln(-u_{t\ \mathrm{in}})-\int_{l_\mathrm{in}}^{l}\frac{\Omega dl}{1-\Omega l}.
\label{eq:euler}
\end{equation}
The subscript `$\mathrm{in}$' here refers to the inner edge of the disc in the equatorial plane. 
The ``potential'' $W$ is defined by
\begin{equation}
W - W_\mathrm{in} = -\int_{0}^{p}\frac{dp}{\rho h},
\end{equation}
where $p$, $\rho$ and $h$ are the pressure, the density and the specific enthalpy of the gas (comoving frame). 
We limit ourselves here to the case of an isentropic fluid, where
\begin{equation}
\int_{0}^{p}\frac{dp}{\rho h} = \ln{\frac{h}{h_{\mathrm{in}}}}.
\end{equation}
In this case, the equipotentials $W=\mathrm{const}$ coincide with the equi-pressure $p=\mathrm{const}$ and 
the equi-density $\rho=\mathrm{const}$ surfaces. Once the structure of the equipotentials $W(r,\theta)$ has 
been found, the thermodynamical quantities can be easily deduced through the condition 
$h=h_\mathrm{in}\exp{(W_\mathrm{in}-W)}$ and the equation of state (EoS). In this paper, we will 
consider the case of polytropic fluids where 
\begin{equation}
p=\kappa \rho^{\gamma},
\end{equation}
and 
\begin{equation}
h = 1+\frac{\gamma}{\gamma-1}\kappa \rho^{\gamma-1},
\label{eq:h}
\end{equation}
so that
\begin{eqnarray}
\rho & = & \left(\frac{\gamma-1}{\gamma}\frac{\left(h_\mathrm{in}\ e^{W_\mathrm{in}-W}-1\right)}
{\kappa}\right)^{\frac{1}{\gamma-1}},
\label{eq:rho}\\
p    & = & \left(\frac{\gamma-1}{\gamma}\frac{\left(h_\mathrm{in}\ e^{W_\mathrm{in}-W}-1\right)}
{\kappa^{\frac{1}{\gamma}}}\right)^{\frac{\gamma}{\gamma-1}}.
\label{eq:p}
\end{eqnarray}
These expressions of $h$, $p$ and $\rho$ are valid only in the interior of the disc, which is defined by 
$W\ge W_\mathrm{in}$. The total mass in the disc is given by the integral
\begin{eqnarray}
M_\mathrm{D} 
& = & 2\pi\int\!\!\!\!\int\limits_{\!\!\!\!\rho > 0} \frac{g_{\phi\phi}-g_{tt} l^{2}}
{g_{\phi\phi}+2 g_{t\phi} l +g_{tt} l^{2}}\left(\rho h + 2 P\right)\nonumber\\
& & \times\left(r^{2}+a^{2}\cos^{2}{\theta}\right)\sin{\theta}\ dr d\theta.
\label{eq:md}
\end{eqnarray}
In the models we consider in this paper, we have usually $\frac{P}{\rho} \ll 1$ so that 
$\rho h + 2P \simeq \rho$.

According to these expressions, the construction of the equilibrium configuration of a thick disc reduces 
to the calculation of the function $W(r,\theta)$. In the case of a constant angular momentum in the disc 
(see Paper I), the last term in Eq.~(\ref{eq:euler}) vanishes and $W(r,\theta)$ reduces to the first 
term, the estimation of which is trivial through Eq.~(\ref{eq:defut}) once the distribution $l(r,\theta)$ 
has been computed as described in the previous subsection. Here we consider more general distributions 
of $l$ for which all terms have to be taken into account. 

\subsection{Geometry of the equipotentials}

\begin{figure*}
\begin{center}
\begin{tabular}{ccc}
$a=0$ & $a=\sqrt{5}/3$ & $a=1$\\
\psfig{file=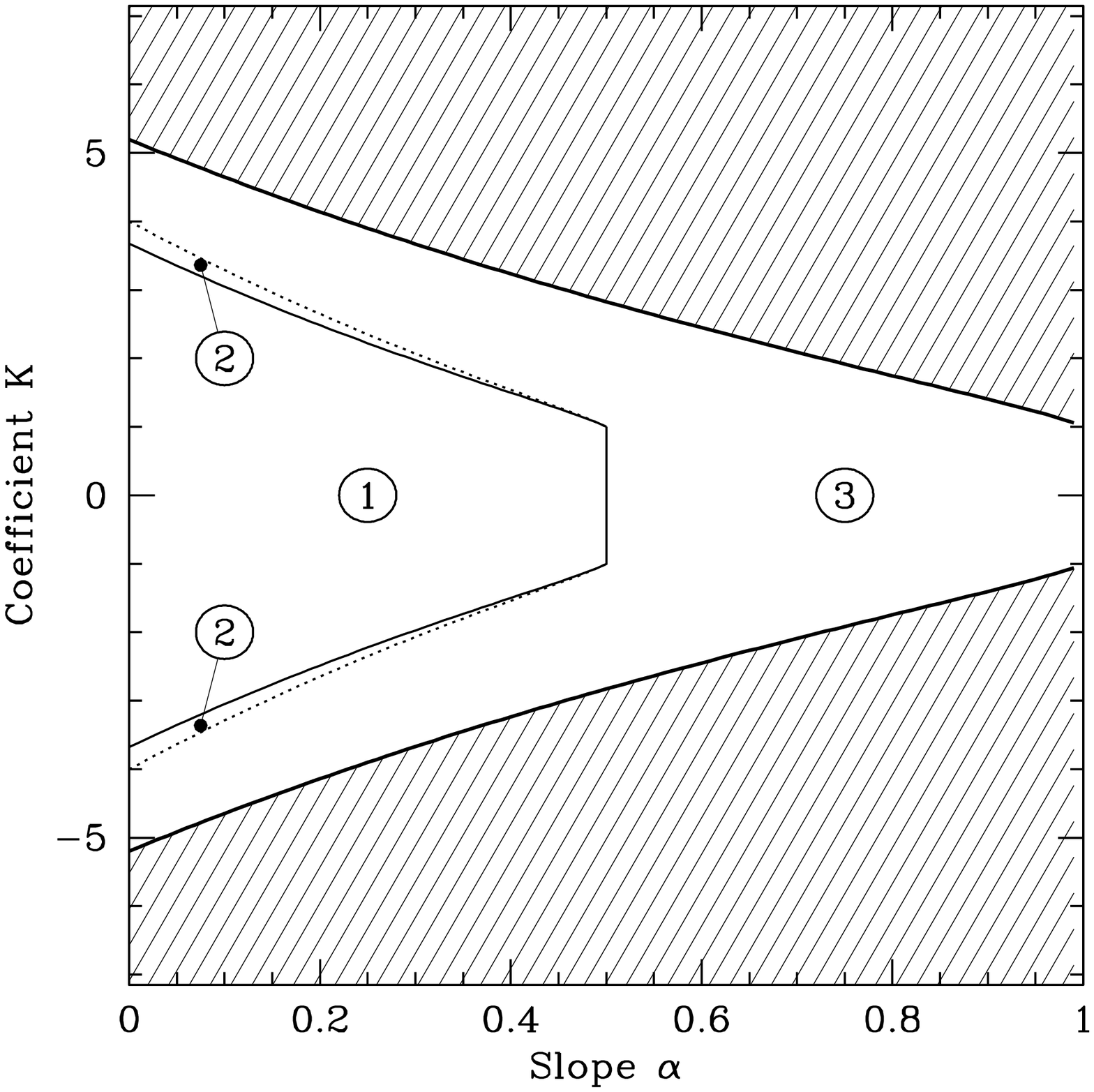,width=0.3\textwidth} &
\psfig{file=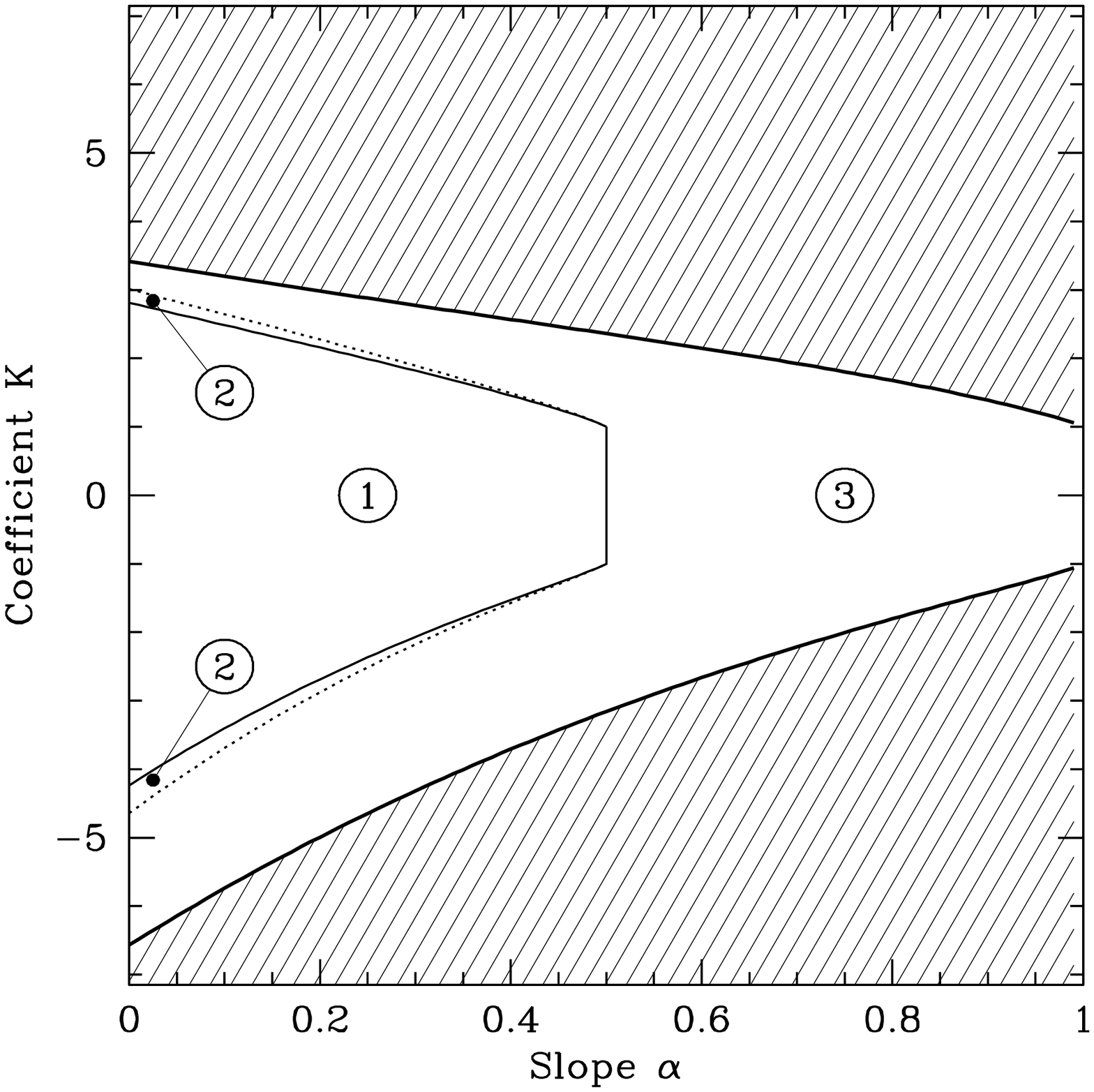,width=0.3\textwidth} &
\psfig{file=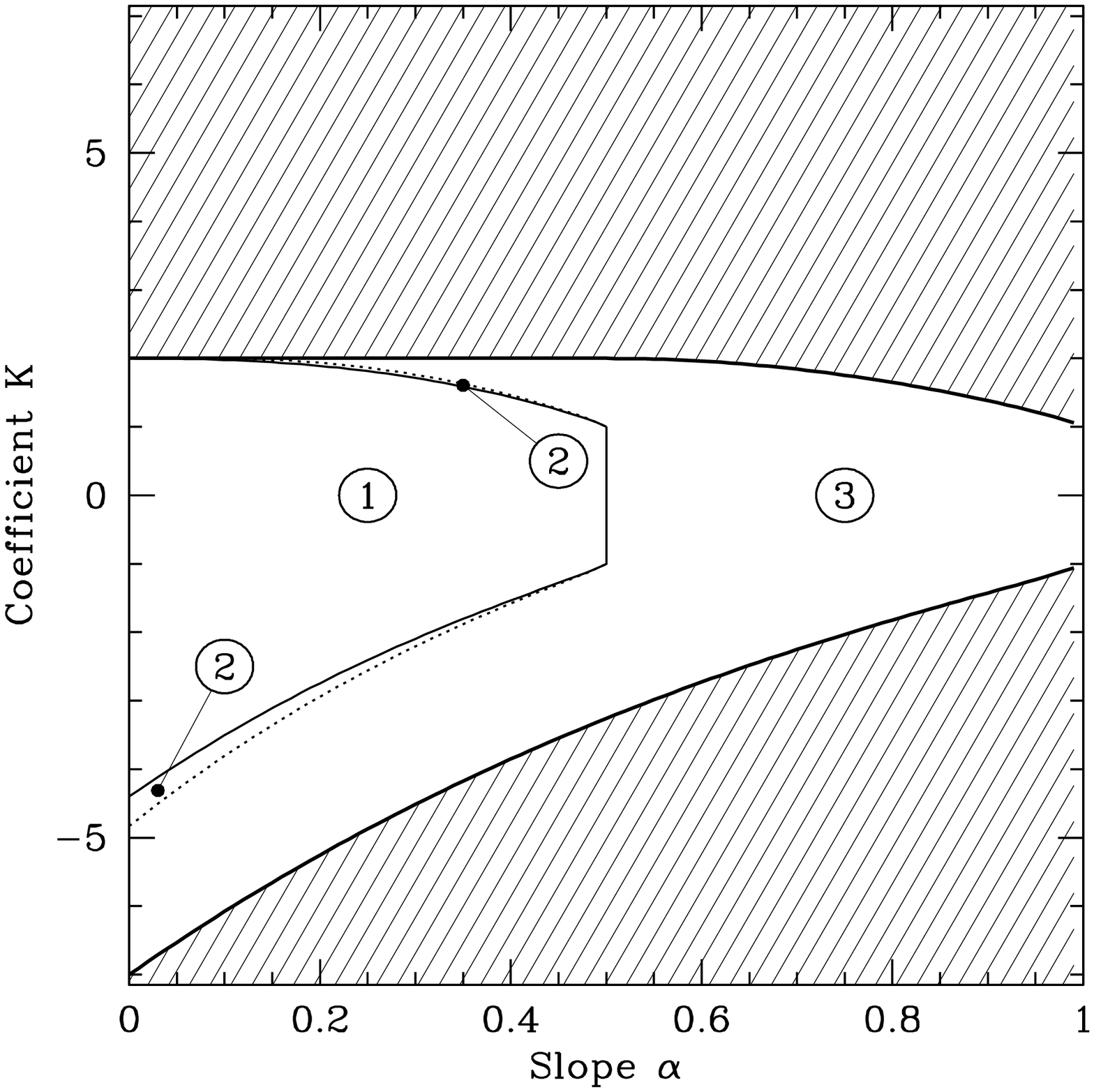,width=0.3\textwidth}\\
\end{tabular}
\end{center}
\caption{\textbf{Classification of the possible geometries of the equipotentials for three 
representative values of the black hole spin $a$.} In the parameter space $K$-$\alpha$ of 
the rotation law $l_\mathrm{eq}(r)=K r^{\alpha}$ in the equatorial plane, the three lines 
$K=K_\mathrm{cr}$ (thick solid line), $K=K_\mathrm{ms}$ (thin solid line) 
and $K=K_\mathrm{mb}$ (dotted line) define three regions. In region 1, there is 
neither a cusp nor a centre; in region 2, there are a cusp and a centre and the equipotential 
of the cusp is closed, which defines a disc; finally, in region 3, there are a cusp and a 
centre (the centre being rejected at $+\infty$ for $\alpha \ge 1/2$) but the equipotential 
of the cusp is not closed, so that the disc is infinite. The dashed region in the plots is 
not considered in this paper as the derivative of the potential in the equatorial plane
is not defined everywhere outside the horizon.}
\label{fig:alphaK}
\end{figure*}

It is interesting to describe the structure of the equipotentials in the general case where 
$\alpha \ne 0$ (the constant angular momentum case $\alpha=0$ has been fully described in Paper I). 
In the equatorial plane, the expression of $W_\mathrm{eq}(r)=W(r,\pi/2)$ can be written as
\begin{eqnarray}
W_\mathrm{eq}(r) & = & -\int_{r}^{+\infty} \left(\frac{\partial \ln({-u_{t,\mathrm{eq}}})}
{\partial r}(r)-\frac{\Omega_\mathrm{eq}(r) \frac{dl_\mathrm{eq}}{dr}}{1-\Omega_\mathrm{eq} 
l_\mathrm{eq}}\right)dr\nonumber\\
& = & -\int_{r}^{+\infty} w_\mathrm{eq}(r) dr
\label{eq:w}
\end{eqnarray}
where we have used the condition $W(r,\theta)\to 0$\,for\,$r\to+\infty$ to eliminate the radius 
$r_\mathrm{in}$ and the corresponding value $W_\mathrm{in}$. From the definition of the von Zeipel's 
cylinders, the potential at any given position is related to the potential in the equatorial plane by
\begin{equation}
W(r,\theta) = W_\mathrm{eq}(r_{0})+\ln{\left(\frac{-u_{t}(r,\theta)}{-u_{t}(r_{0},\pi/2)}\right)},
\label{eq:woutside}
\end{equation}
where $r_{0}$ is the radius where the von Zeipel's cylinder passing at $(r,\theta)$ intersects the 
equatorial plane ($r_{0}$ is solution of Eq.~(\ref{eq:r0})). The closed (respectively, open) equipotentials 
correspond to $W<0$ (respectively, $W>0$) and the marginal equipotential $W=0$ is closed at infinity. 
In order to understand the geometry of the equipotentials, it is useful to study the derivative 
$w_\mathrm{eq}(r)=dW_\mathrm{eq}/dr$ which is given from equations~(\ref{eq:w}), (\ref{eq:defut}) and 
(\ref{eq:defomega}) by
\begin{equation}
w_\mathrm{eq}(r) =  \frac{1}{2}\left[
 \frac{\partial\ln{\tilde{\varpi}^{2}}}{\partial r}
-\frac{\frac{\partial\tilde{g}_{tt}}{\partial r} l_\mathrm{eq}^{2} + 2 \frac{\partial\tilde{g}_{t\phi}}
{\partial r} l_\mathrm{eq} + \frac{\partial\tilde{g}_{\phi\phi}}{\partial r}}{\tilde{g}_{tt} 
l_\mathrm{eq}^{2} + 2 \tilde{g}_{t\phi} l_\mathrm{eq} + \tilde{g}_{\phi\phi}}
\right].
\label{eq:dwdr}
\end{equation}
Hereafter, the notation $\tilde{g}_{tt}$, $\tilde{g}_{t\phi}$, $\tilde{g}_{\phi\phi}$ and 
$\tilde{\varpi}$ means that these quantities are computed in the equatorial plane $\theta=\pi/2$. 
The function given by Eq.~(\ref{eq:dwdr}) is studied in Appendix~\ref{sec:potential}. The condition 
for it to vanish is
\begin{equation}
l_\mathrm{eq}(r) = l_\mathrm{K,eq}^{\pm}(r)\ ,
\label{eq:cuspcenter}
\end{equation}
which is not surprising: by the definition of a Keplerian orbit, when the angular momentum equals the 
Keplerian value, the gravitational forces are exactly compensated by the centrifugal forces and the 
pressure gradient vanishes; the pressure (and the density) are either minimum (surface of the disc) 
or maximum (centre of the disc). Therefore the geometry of the equipotentials is mainly fixed by the 
distribution $l_\mathrm{eq}(r)$. We now describe the geometrical structure of the equipotentials 
$W=\mathrm{const}$ for the particular choice of the distribution of angular momentum made in this paper 
(see Eq.~(\ref{eq:leq})). 

\subsubsection{Allowed values of the constant $K$}

We restrict this study to the case where the derivative of the potential $w_\mathrm{eq}(r)$ in the 
equatorial plane has only finite values outside the horizon. This corresponds to the fact that 
equation $l_\mathrm{eq}(r)=l_\mathrm{cr,eq}^{\pm}(r)$ has no solution (see Appendix~\ref{sec:potential}). 
This is equivalent to 
\begin{equation}
| K | \le |K_\mathrm{cr}|\ , 
\end{equation}
where the marginal case $K=K_\mathrm{cr}$ is given by 
\begin{equation}
K_\mathrm{cr}=\frac{l_\mathrm{cr,eq}^{\pm}(r_\mathrm{cr})}{r_\mathrm{cr}^{\alpha}}\ .
\end{equation}
The radius $r_\mathrm{cr}$ is the solution of 
\begin{equation}
\frac{d\ln{|l_\mathrm{cr,eq}^{\pm}|}}{d\ln{r}}(r_\mathrm{cr}) = \alpha\ .
\label{eq:Kcr}
\end{equation}
The slope of $|l_\mathrm{cr,eq}^{\pm}|$ is an increasing function of $r$. For prograde orbits, the slope 
at the horizon is
\begin{equation}
\lim_{r\to r_\mathrm{h}^{+}} \frac{d\ln{|l_\mathrm{cr,eq}^{+}(r)|}}{d\ln{r}} =\left\lbrace\begin{array}{cl}
-\infty     & 0 \le a < 1\\
\frac{1}{2} & a = 1
\end{array}\right.
\end{equation}
and for retrograde orbits, the slope at the ergosphere is
\begin{equation}
\lim_{r\to 2} \frac{d\ln{|l_\mathrm{cr,eq}^{-}(r)|}}{d\ln{r}} = -\infty\ .
\end{equation}
In both cases, the maximal slope equals
\begin{equation}
\lim_{r\to+\infty} \frac{d\ln{|l_\mathrm{cr,eq}^{\pm}(r)|}}{d\ln{r}}=1\ .
\end{equation}
Then, there are four cases: (i) for $0 \le \alpha < 1/2$ (sub-Keplerian case) and $0 \le a \le 1$, 
there is one unique root $r_\mathrm{cr}$ outside the horizon (respectively, outside the ergosphere) 
for prograde (respectively, retrograde) orbits except for $a=1$ and prograde orbits. In this last case, 
the root of equation~(\ref{eq:Kcr}) is inside the horizon and has to be replaced by 
$r_\mathrm{cr}=r_\mathrm{h}=1$ so that $K_\mathrm{cr}=2$; (ii) for $1/2 \le \alpha < 1$ 
(Keplerian and super-Keplerian case) and $0 \le a \le 1$, there is one unique root $r_\mathrm{cr}$ 
outside the horizon (respectively, outside the ergosphere) for prograde (respectively, retrograde) orbits; (iii) for 
$\alpha=1$, this unique root is rejected at infinity so that $r_\mathrm{cr}=+\infty$ and 
$K_\mathrm{cr}=1$ (respectively, $-1$) for prograde (respectively, retrograde) orbits; (iv) for $\alpha>1$, 
equation $l_\mathrm{eq}(r)=l_\mathrm{cr,eq}^{\pm}(r)$ has always a solution. For this reason, we will 
not consider the $\alpha>1$ case in this paper. Notice that the case presented in Paper I corresponds 
to $\alpha=0$ and $a=0$. In this case $r_\mathrm{cr} = 3$ and $K_\mathrm{cr}=
3\sqrt{3}\simeq 5.20$.

\subsubsection{Presence of a cusp and centre}

The geometry of the equipotentials is mainly determined by the roots of the equation $l_\mathrm{eq}(r) 
= l_\mathrm{K,eq}^{\pm}(r)$. The condition to have two roots $r_\mathrm{cusp}$ and $r_\mathrm{centre}$, 
corresponding respectively to the cusp and the centre of the disc, is
\begin{equation}
| K | \ge |K_\mathrm{ms}|\ ,
\end{equation}
where the marginal case $K=K_\mathrm{ms}$ is given by 
\begin{equation}
K_\mathrm{ms}=\frac{l_\mathrm{K,eq}^{\pm}(r_\mathrm{ms})}{r_\mathrm{ms}^{\alpha}}.
\label{eq:Kms}
\end{equation}
The radius $r_\mathrm{ms}$ is the solution of 
\begin{equation}
\frac{d\ln{|l_\mathrm{K,eq}^{\pm}|}}{d\ln{r}}(r_\mathrm{ms}) = \alpha.
\end{equation}
The slope of $|l_\mathrm{K,eq}^{\pm}|$ is an increasing function of $r$. It equals 0 at the radius of the 
last marginally stable orbit $r_\mathrm{ms}$. The maximal slope equals
\begin{equation}
\lim_{r\to+\infty} \frac{d\ln{|l_\mathrm{K,eq}^{\pm}(r)|}}{d\ln{r}}=\frac{1}{2},
\end{equation}
so that for $0 \le \alpha < 1/2$ (sub-Keplerian case) the root $r_\mathrm{ms}$ is unique and 
finite, for $\alpha=1/2$ (Keplerian case) the root $r_\mathrm{ms}$ goes towards infinity 
($K_\mathrm{ms}\to \pm 1$), and for $\alpha > 1/2$ (super-Keplerian case) there is no root: 
Eq.~(\ref{eq:cuspcenter}) has always only one root, the radius of the cusp $r_\mathrm{cusp}$. 
Notice that for $\alpha=0$ (constant angular momentum, see Paper I), the root $r_\mathrm{ms}$ 
coincides with the the radius of the last marginally stable orbit and $K_\mathrm{ms}=l_\mathrm{K,eq}
(r_\mathrm{ms}^{\pm})$ in this case.

\begin{table*}
\caption{All possible equipotential configurations for a Kerr black hole and a disc with an
angular momentum distribution in the equatorial plane following a power-law 
$l_\mathrm{eq}(r) = K r^{\alpha}$.}
\label{tab:geometry}
\begin{tabular}{lccccc}
\multicolumn{6}{c}{\textbf{\underline{Prograde rotation : $K>0$}}}\\
\\
\multicolumn{6}{c}{sub-Keplerian case $0 \le \alpha < 1/2$}\\
\hline
Angular momentum & $r_\mathrm{cusp}$ & $W_\mathrm{cusp}$ & $r_\mathrm{centre}$ & 
$W_\mathrm{centre}$ & Comments\\
\hline
$0 < K<K_\mathrm{ms}$ & & & & & No cusp. No centre. No closed disc.\\
$K=K_\mathrm{ms}$ & $r_\mathrm{cusp} =r_\mathrm{ms}$ & $<0$ & 
$r_\mathrm{centre}=r_\mathrm{cusp}$ & $<0$ & Cusp = centre. Disc reduced to one point.\\
$K_\mathrm{ms} < K < K_\mathrm{mb}$ & $r_\mathrm{mb} < 
r_\mathrm{cusp} < {r}_\mathrm{ms} 
$ & $<0$ & $r_\mathrm{centre} > {r}_\mathrm{ms}$ & $<0$ & Cusp. Centre. Disc closed.\\
$K={K}_\mathrm{mb}$ & $r_\mathrm{cusp}={r}_\mathrm{mb}$ & $0$ & 
$r_\mathrm{centre} > {r}_\mathrm{ms}$ & $<0$ & Cusp. Centre. Disc closed at infinity.\\
${K}_\mathrm{mb} < K < {K}_\mathrm{cr}$ & $r_\mathrm{cusp}<{r}_\mathrm{mb}$ & 
$>0$ & 
$r_\mathrm{centre} > {r}_\mathrm{ms}$ & $<0$ & Cusp. Centre. Disc infinite$^{*}$.\\
$K = {K}_\mathrm{cr}$ & & & 
$r_\mathrm{centre} > {r}_\mathrm{ms}$ & $<0$ & No cusp. Centre. Disc infinite$^{*}$.\\ 
\hline
\\
\multicolumn{6}{c}{Keplerian case $\alpha = 1/2$}\\
\hline
Angular momentum & $r_\mathrm{cusp}$ & $W_\mathrm{cusp}$ & $r_\mathrm{centre}$ & 
$W_\mathrm{centre}$ & Comments\\
\hline
$0 < K< 1$ & & & & & No cusp. No centre. No closed disc.\\
$K=1$ & $+\infty$ & 0 & $+\infty$ & 0 & Cusp and centre at infinity. No closed disc.\\
$1 < K < {K}_\mathrm{cr}$ & $r_\mathrm{cusp}<+\infty$ & $>0$ & $+\infty$ &0 & 
Cusp. Center at infinity. No closed disc.\\
$K={K}_\mathrm{cr}$ & & & $+\infty$ & 0 & No cusp. Center at infinity. No closed 
disc.\\
\hline
\\
\multicolumn{6}{c}{super-Keplerian case $1/2 < \alpha < 1$}\\
\hline
Angular momentum & $r_\mathrm{cusp}$ & $W_\mathrm{cusp}$ & $r_\mathrm{centre}$ & 
$W_\mathrm{centre}$ & Comments\\
\hline
$0< K < {K}_\mathrm{cr}$ & $r_\mathrm{cusp}<+\infty$ & $>0$ & $+\infty$ &0 & 
Cusp. Center at infinity. No closed disc.\\
$K={K}_\mathrm{cr}$ & & & $+\infty$ & 0 & No cusp. Center at infinity. No closed 
disc.\\
\hline
\\
\\
\multicolumn{6}{c}{\textbf{\underline{Retrograde rotation : $K<0$}}}\\
\\
\multicolumn{6}{c}{sub-Keplerian case $0 \le \alpha < 1/2$}\\
\hline
Angular momentum & $r_\mathrm{cusp}$ & $W_\mathrm{cusp}$ & $r_\mathrm{centre}$ & 
$W_\mathrm{centre}$ & Comments\\
\hline
${K}_\mathrm{ms} < K < 0$ & & & & & No cusp. No centre. No closed disc.\\
$K={K}_\mathrm{ms}$ & $r_\mathrm{cusp} ={r}_\mathrm{ms}$ & $<0$ & 
$r_\mathrm{centre}=r_\mathrm{cusp}$ & $<0$ & Cusp = centre. Disc reduced to one point.\\
${K}_\mathrm{mb} < K < {K}_\mathrm{ms}$ & ${r}_\mathrm{mb} < 
r_\mathrm{cusp} < {r}_\mathrm{ms} 
$ & $<0$ & $r_\mathrm{centre} > {r}_\mathrm{ms}$ & $<0$ & Cusp. Centre. Disc 
closed.\\
$K={K}_\mathrm{mb}$ & $r_\mathrm{cusp}={r}_\mathrm{mb}$ & $0$ & 
$r_\mathrm{centre} > {r}_\mathrm{ms}$ & $<0$ & Cusp. Centre. Disc closed at 
infinity.\\
${K}_\mathrm{cr} \le K < {K}_\mathrm{mb}$ & $r_\mathrm{cusp}<{r}_\mathrm{mb}$ 
& $>0$ & 
$r_\mathrm{centre} > {r}_\mathrm{ms}$ & $<0$ & Cusp. Centre. Disc infinite$^{*}$.\\
\hline
\\
\multicolumn{6}{c}{Keplerian case $\alpha = 1/2$}\\
\hline
Angular momentum & $r_\mathrm{cusp}$ & $W_\mathrm{cusp}$ & $r_\mathrm{centre}$ & 
$W_\mathrm{centre}$ & Comments\\
\hline
$-1 < K < 0$ & & & & & No cusp. No centre. No closed disc.\\
$K=-1$ & $+\infty$ & 0 & $+\infty$ & 0 & Cusp and centre at infinity. No closed 
disc.\\
$ {K}_\mathrm{cr} \le K < -1$ & $r_\mathrm{cusp}<+\infty$ & $>0$ & $+\infty$ &0 & 
Cusp. Center at infinity. No closed disc.\\
\hline
\\
\multicolumn{6}{c}{super-Keplerian case $1/2 < \alpha < 1$}\\
\hline
Angular momentum & $r_\mathrm{cusp}$ & $W_\mathrm{cusp}$ & $r_\mathrm{centre}$ & 
$W_\mathrm{centre}$ & Comments\\
\hline
${K}_\mathrm{cr} \le K < 0$ & $r_\mathrm{cusp}<+\infty$ & $>0$ & $+\infty$ &0 
& Cusp. Center at infinity. No closed disc.\\
\hline
\end{tabular}
\begin{flushleft}
$^{*}$ Some closed equipotentials are still present around the centre.\\
\end{flushleft}

\end{table*}

\subsubsection{Closed equipotential at the cusp}

In the case where there is a cusp at $r_\mathrm{cusp}$ the corresponding equipotential will be closed 
only if
\begin{equation}
W_\mathrm{eq}(r_\mathrm{cusp}) \le 0.
\end{equation} 
This is equivalent to 
\begin{equation}
-u_{t,\mathrm{eq}}(r_\mathrm{cusp}) \le e^{-F_\mathrm{eq}(r_\mathrm{cusp})}.
\end{equation}
In case of equality, the equipotential of the cusp is closed at infinity ($W_\mathrm{cusp}=0$). The 
quantity $F_\mathrm{eq}(r)$ is defined as
\begin{eqnarray}
F_\mathrm{eq}(r) & = & \int_{r}^{+\infty} \frac{\Omega_\mathrm{eq}}
{1-\Omega_\mathrm{eq}l_\mathrm{eq}}\frac{dl_\mathrm{eq}}{dr}dr\\
& = & -\int_{r}^{+\infty} \frac{g_{t\phi}+g_{tt}l_\mathrm{eq}}{g_{tt}l_\mathrm{eq}^{2}+
2g_{t\phi}l_\mathrm{eq}+g_{\phi\phi}}\frac{dl_\mathrm{eq}}{dr} dr.
\end{eqnarray}
As already mentioned, in the case where $\alpha=0$ the integral $F_\mathrm{eq}(r)$ vanishes 
and this condition is equivalent to the condition for a fluid element in orbit at $r_\mathrm{cusp}$ 
with angular momentum $l_\mathrm{eq}(r_\mathrm{cusp})$ to be bound (Eq.~(\ref{eq:bound})), 
i.e. $|K|<|l_\mathrm{K,eq}^{\pm}(r_\mathrm{mb}^{\pm})|$. In the more general case it can be shown 
that the condition is equivalent to
\begin{equation}
| K | \le | {K}_\mathrm{mb} |,
\end{equation}
where ${K}_\mathrm{mb}$ is the solution of 
\begin{equation}
W_\mathrm{eq}(r_\mathrm{cusp})=0\ \mathrm{for}\ K={K}_\mathrm{mb}.
\label{eq:Kmb}
\end{equation}
For $0\le\alpha\le 1/2$, we have $|{K}_\mathrm{cr}| > |{K}_\mathrm{mb}| \ge 
|{K}_\mathrm{ms}| \ge 1$ (with ${K}_\mathrm{mb}={K}_\mathrm{ms}=\pm 1$ for 
$\alpha=1/2$): the cusp and the centre are respectively a local maximum and minimum of the 
potential $W_\mathrm{eq}(r)$. For $\alpha>1/2$, the centre is rejected at infinity where 
$W_\mathrm{eq}(r)\to 0$ and the cusp is still the maximum of $W_\mathrm{eq}(r)$ and then 
must have a positive potential. This means that ${K}_\mathrm{mb}=0$ in this case.

Figure~\ref{fig:alphaK} shows the critical coefficients ${K}_\mathrm{cr}$, 
${K}_\mathrm{mb}$ and ${K}_\mathrm{ms}$ as a function of $\alpha$ for pro- 
and retrograde discs and for three different values of the spin of the black hole: $a=0$ 
(non-rotating), $a=\sqrt{5}/3$, and $a=1$ (extreme rotation). This allows to classify the 
possible geometries of the equipotentials, which is presented in Table~\ref{tab:geometry}. 
This table is the generalized version of Table 2 in Paper I. The lines appearing in 
Fig.~\ref{fig:alphaK} separate different equipotential geometries. In region 1, there is 
neither a cusp nor a centre. Correspondingly, in region 2, there are cusp and centre and 
the equipotential of the cusp is closed, which defines a disc. Finally, in region 3, 
there are cusp and centre but the equipotential of the cusp is not closed, so that the 
disc is infinite. The corresponding cases are illustrated in Figures~\ref{fig:equi1pro} 
(for $a=0$), \ref{fig:equi2pro} (for $a=\sqrt{5}/3$), and \ref{fig:equi3pro} (for $a=1$). 
We notice that in order to limit the number of figures to a reasonable amount we do not include 
the case of retrograde discs/rotating black holes. Figs.~\ref{fig:equi1pro}-\ref{fig:equi3pro} 
show the various equipotential geometries that may arise depending on the value of $K$, and 
this is done for the sub-Keplerian ($\alpha<1/2$), Keplerian ($\alpha=1/2$) and super-Keplerian 
($\alpha>1/2$) cases. In all these figures, the left panels show the radial distribution of
the angular momentum in the equatorial plane and the corresponding potential. In addition, 
the right panels show the equipotentials in the $x-y$ plane. In the case $a>0$ and a retrograde 
disc (i.e. $K<0$), the evolution of the geometry of the equipotentials with $|K|$ is essentially 
identical and we do not include it here. The main difference is that the cusp -- when it exists -- 
is at a larger distance from the horizon, as well as the centre of the disc in case (2).
The interesting case for the study of the runaway instability corresponds to the situation when 
there are both a cusp and a centre, and where the equipotential of the cusp is closed, which means 
$0 \le \alpha < 1/2$ and $|{K}_\mathrm{ms}| < |K| < |{K}_\mathrm{mb}|$. Notice that the size of
the disc evolves from zero ($K=K_\mathrm{ms}$) to infinite ($K=K_\mathrm{mb}$), which allows to 
fix its mass by adjusting the constant $K$. This case of interest is illustrated in the third 
row of the first page of Figs.~\ref{fig:equi1pro}-\ref{fig:equi3pro}. In this case, we define 
the potential barrier at the surface of the disc by
\begin{equation}
\Delta W_\mathrm{in} = W_\mathrm{in}-W_\mathrm{cusp}.
\end{equation}
There are three possible cases to consider: (i) for $\Delta W_\mathrm{in} < 0$, the disc is 
inside its Roche lobe (defined as the equipotential of the cusp) and there is no possible mass 
transfer from the disc to the black hole; (ii) for $\Delta W_\mathrm{in}>0$, the disc is overfilling 
its Roche lobe and mass transfer is possible through the cusp; (iii) the marginal case, 
$\Delta W_\mathrm{in}=0$, which corresponds to the case where the disc is exactly filling its 
Roche lobe.


\begin{figure*}
\begin{center}
\begin{tabular}{cc}
\multicolumn{2}{c}{\textbf{\textit{-- Schwarzschild black hole}} ($a=0$) --}\\
\\
\multicolumn{2}{c}{\underline{\textbf{Sub-Keplerian case (here with $\alpha=1/4$)}}}\\
\multicolumn{2}{c}{\textbf{Case (1): $K<K_\mathrm{ms}$} (here $K=\pm 1.5$ ; no cusp ; no centre)}\\
\psfig{file=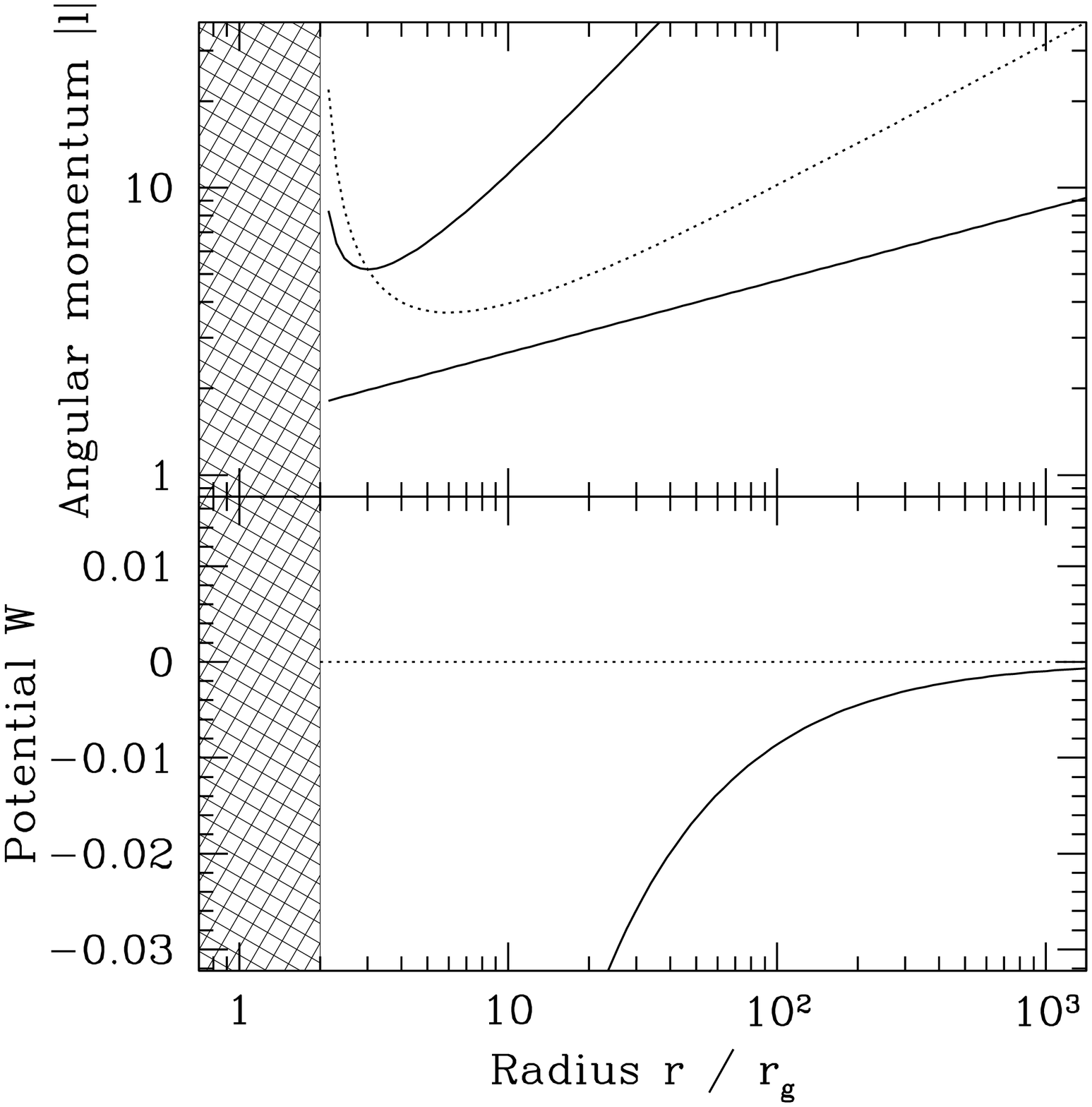,width=6.25cm} &
\psfig{file=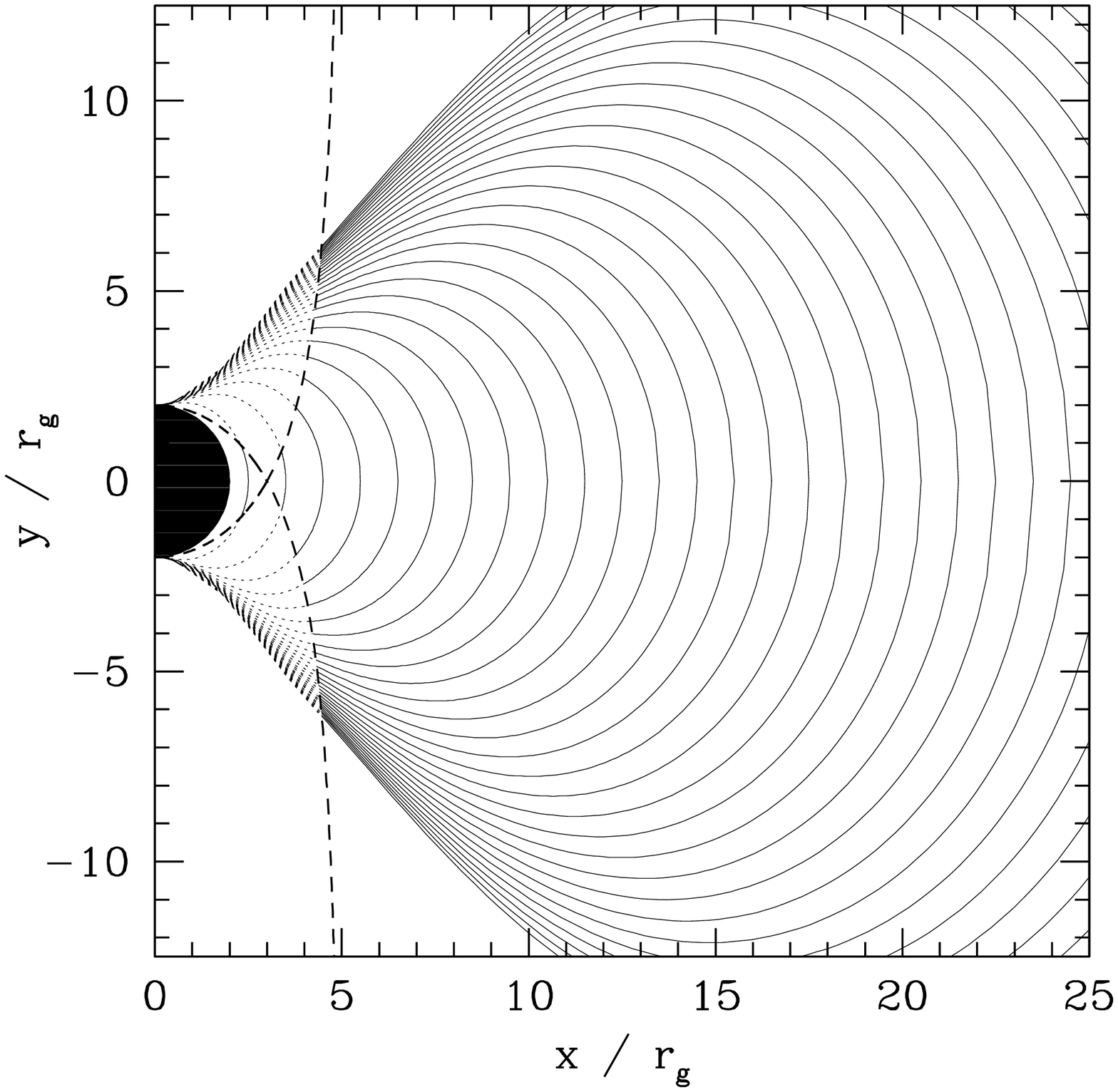,width=6.25cm}\\
 \multicolumn{2}{c}{\textbf{Critical case: $K=K_\mathrm{ms}$} (here $K_\mathrm{ms}\simeq \pm 2.22$; 
$r_\mathrm{cusp}=r_\mathrm{centre} = 10$)}\\
\psfig{file=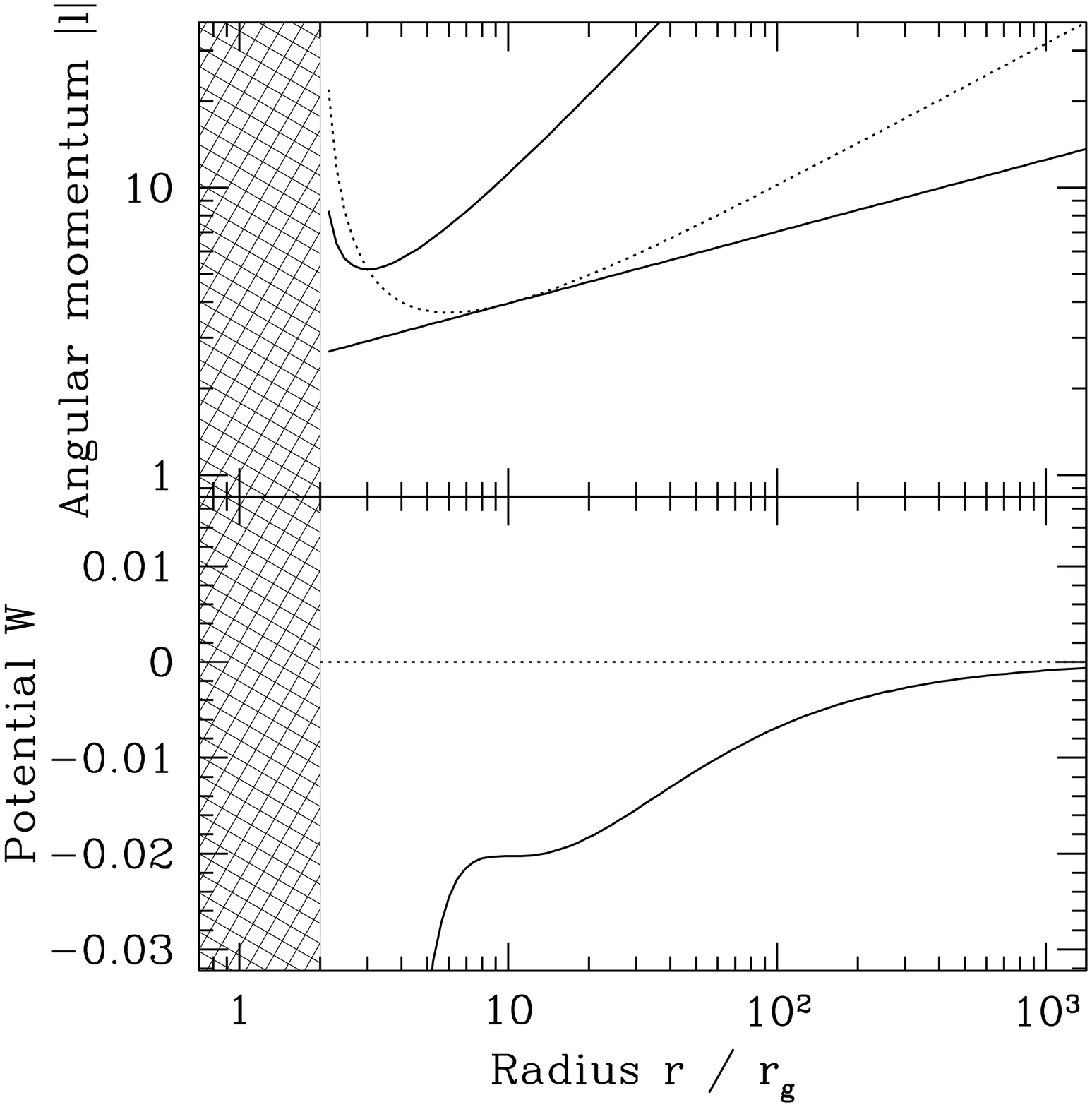,width=6.25cm} &
\psfig{file=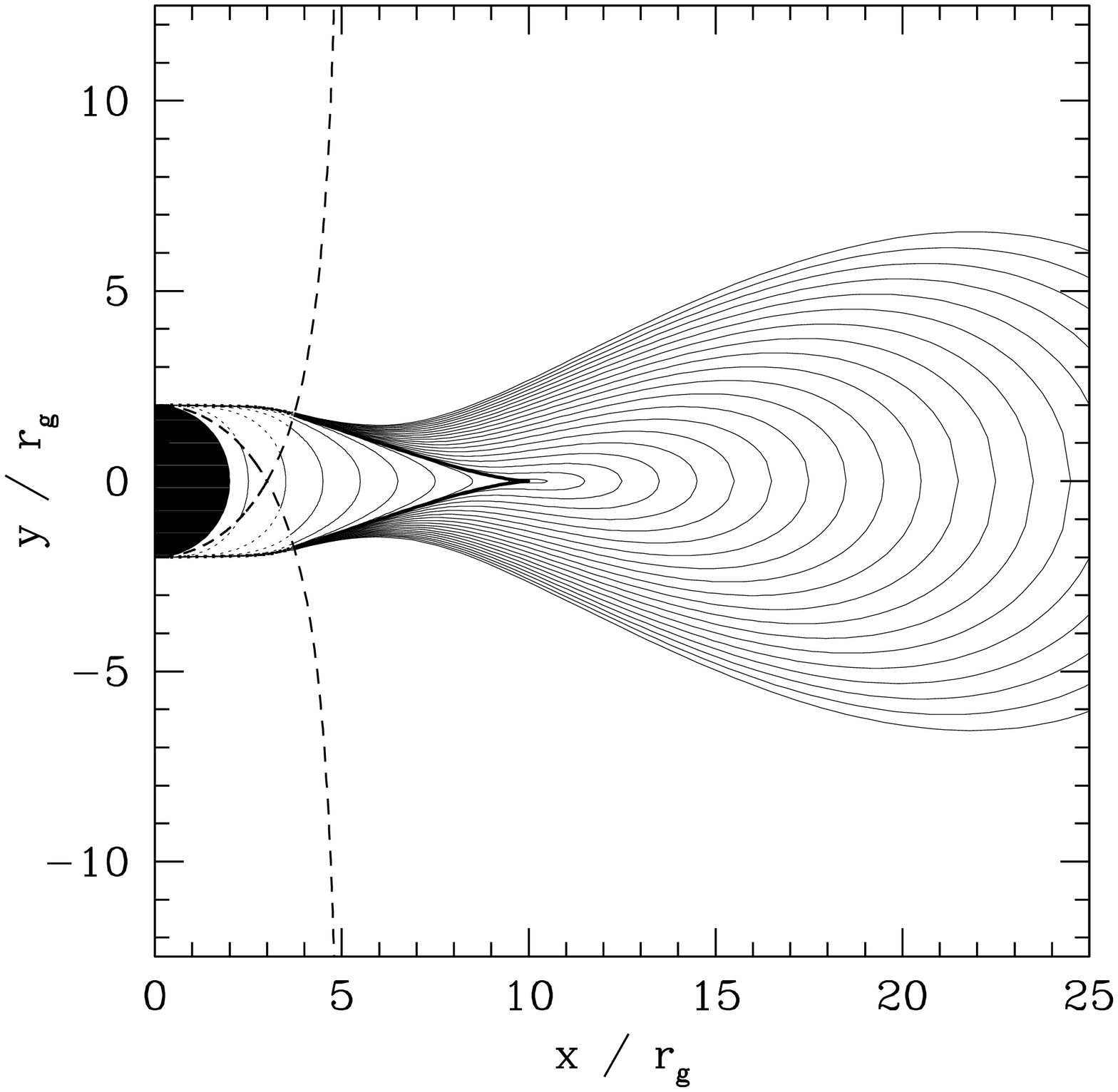,width=6.25cm}\\
\multicolumn{2}{c}{\textbf{Case (2): $K_\mathrm{ms}<K<K_\mathrm{mb}$} (here $K= \pm 2.25$; 
$r_\mathrm{cusp}\simeq 7.71$ ; $r_\mathrm{centre}\simeq 13.5$)}\\
\psfig{file=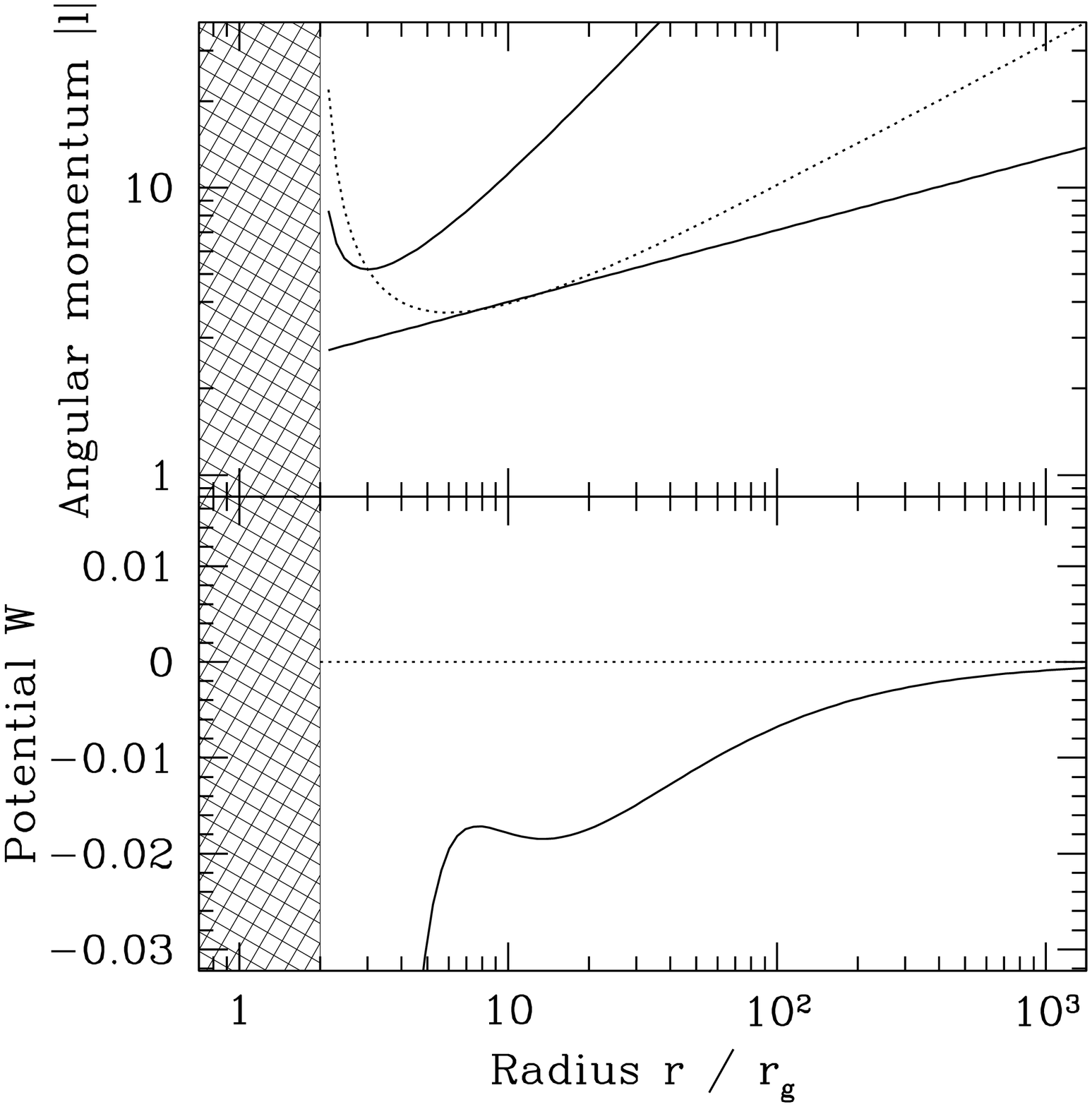,width=6.25cm} &
\psfig{file=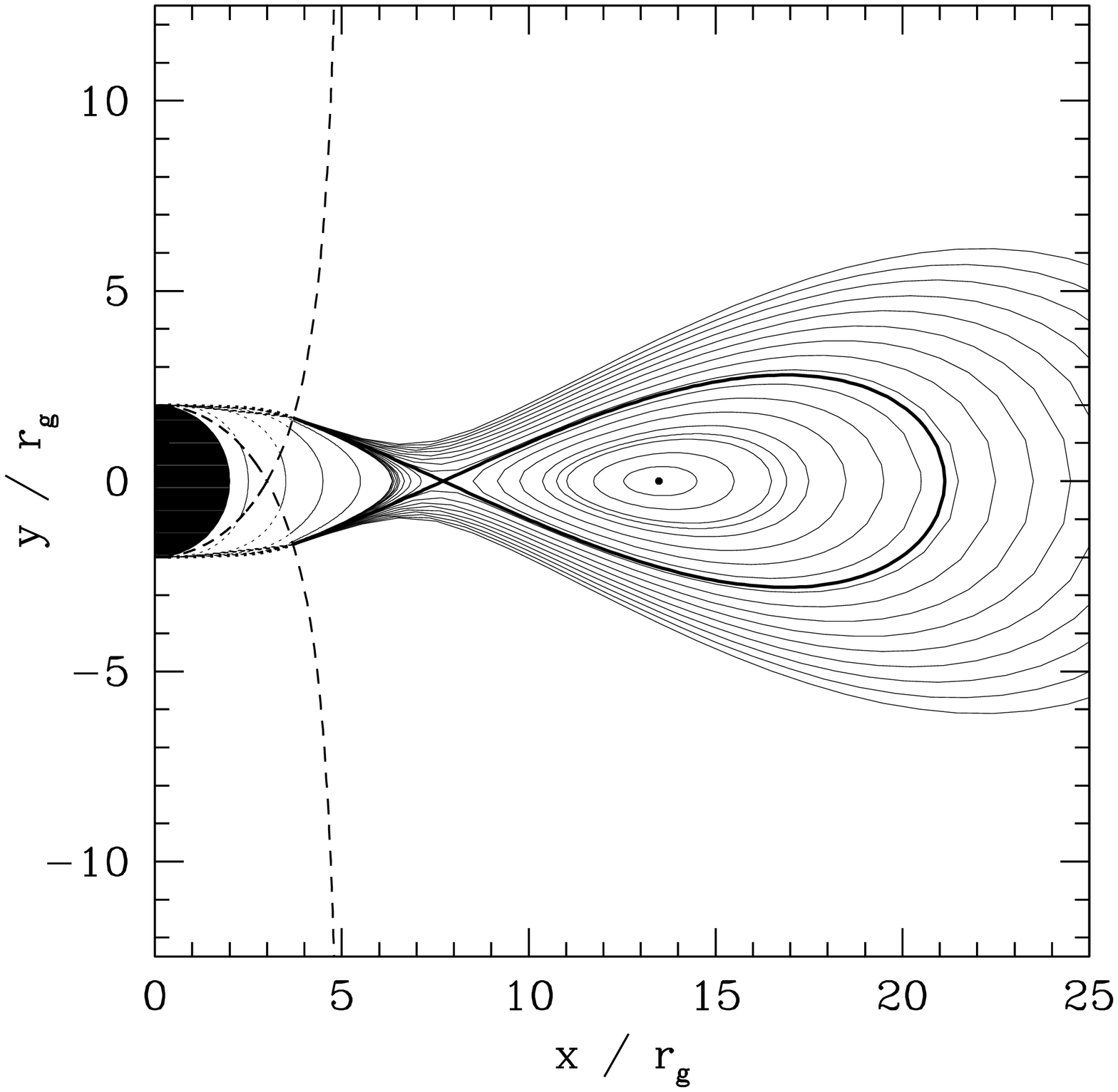,width=6.25cm}\\
\end{tabular}
\end{center}
\vspace*{-3.4ex}
\caption{\textbf{Geometry of the equipotentials: Schwarzschild black hole ($a=0$)  and pro- or 
retrograde disc. Sub-Keplerian case.} For each case, the angular momentum $l_\mathrm{eq}(r)$ and 
the potential $W_\mathrm{eq}(r)$ in the equatorial plane are plotted in the left panel, and the 
equipotentials in the $x=r\sin{\theta}$-$y=r\cos{\theta}$ plane in the right panel. \textit{Left:} 
the critical (respectively, Keplerian) angular momentum $l_\mathrm{cr,eq}^{\pm}(r)$ (respectively 
$l_\mathrm{K,eq}^{\pm}(r)$) is plotted in thin solid (respectively, dotted) line. The intersections 
of $l_\mathrm{eq}(r)$ and $l_\mathrm{K,eq}(r)$ (if any) correspond to the cusp or to the centre, 
which are respectively a local maximum or minimum of $W_\mathrm{eq}(r)$. \textit{Right:} 
the interior of the black hole horizon is filled in black. The critical von Zeipel cylinder is 
indicated with a dashed line. The equipotentials outside the critical cylinder are plotted in 
dotted line and are extrapolated using a constant angular momentum in this region, equal to the 
value on the critical cylinder. When present, the equipotential of the cusp is plotted in thick 
line and the centre is indicated with a big dot.}
\label{fig:equi1pro}
\end{figure*}

\begin{figure*}
\begin{center}
\begin{tabular}{cc}
\multicolumn{2}{c}{\textbf{\textit{-- Schwarzschild black hole}} ($a=0$) --}\\
\\
\multicolumn{2}{c}{\underline{\textbf{Sub-Keplerian case (here with $\alpha=1/4$)}}}\\
\multicolumn{2}{c}{\textbf{Critical case: $K=K_\mathrm{mb}$} (here $K_\mathrm{mb}\simeq \pm 2.35$; 
$r_\mathrm{cusp}\simeq 5.96$ ; $r_\mathrm{centre}\simeq 20.1$)}\\
\psfig{file=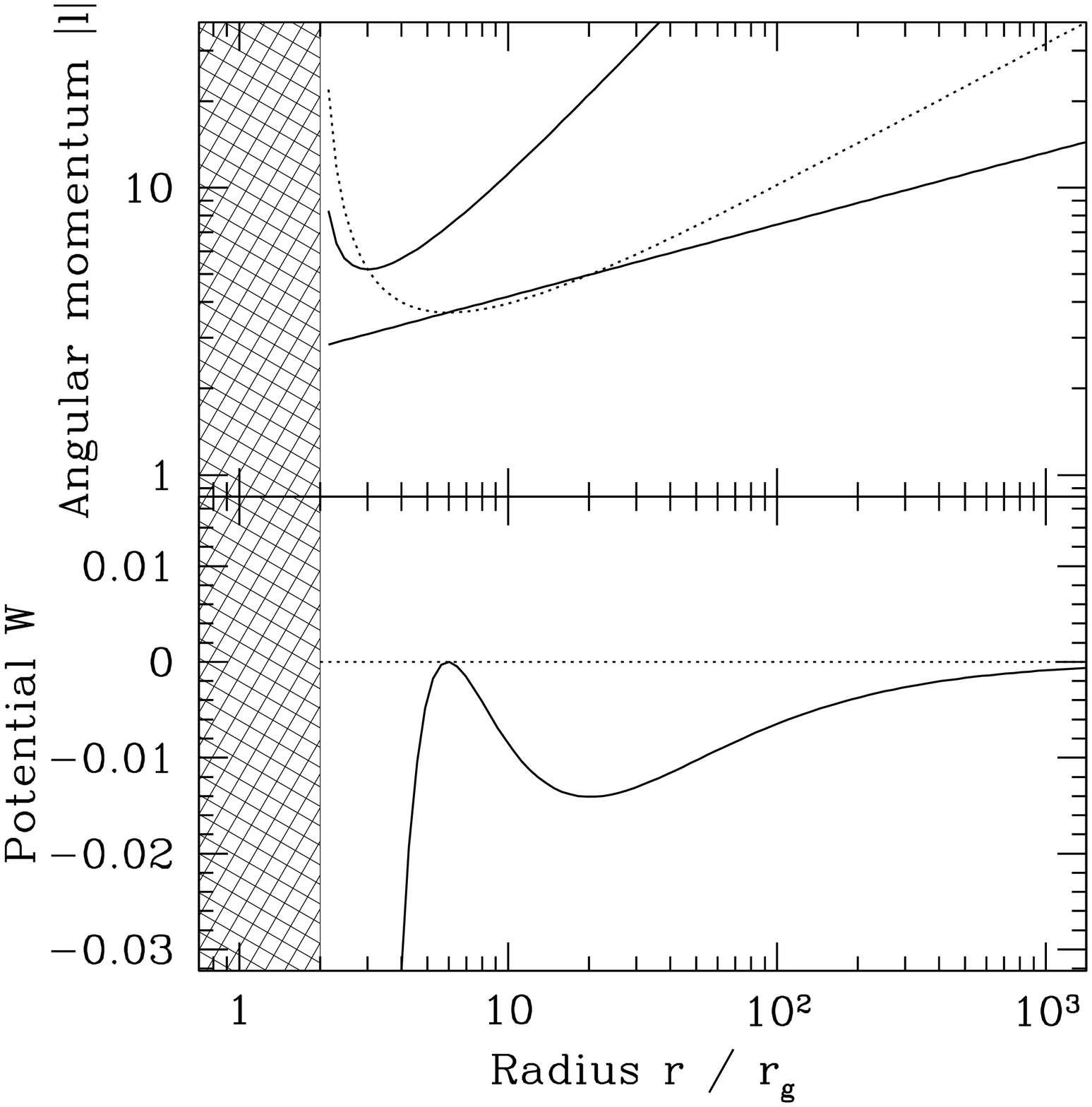,width=6.25cm} &
\psfig{file=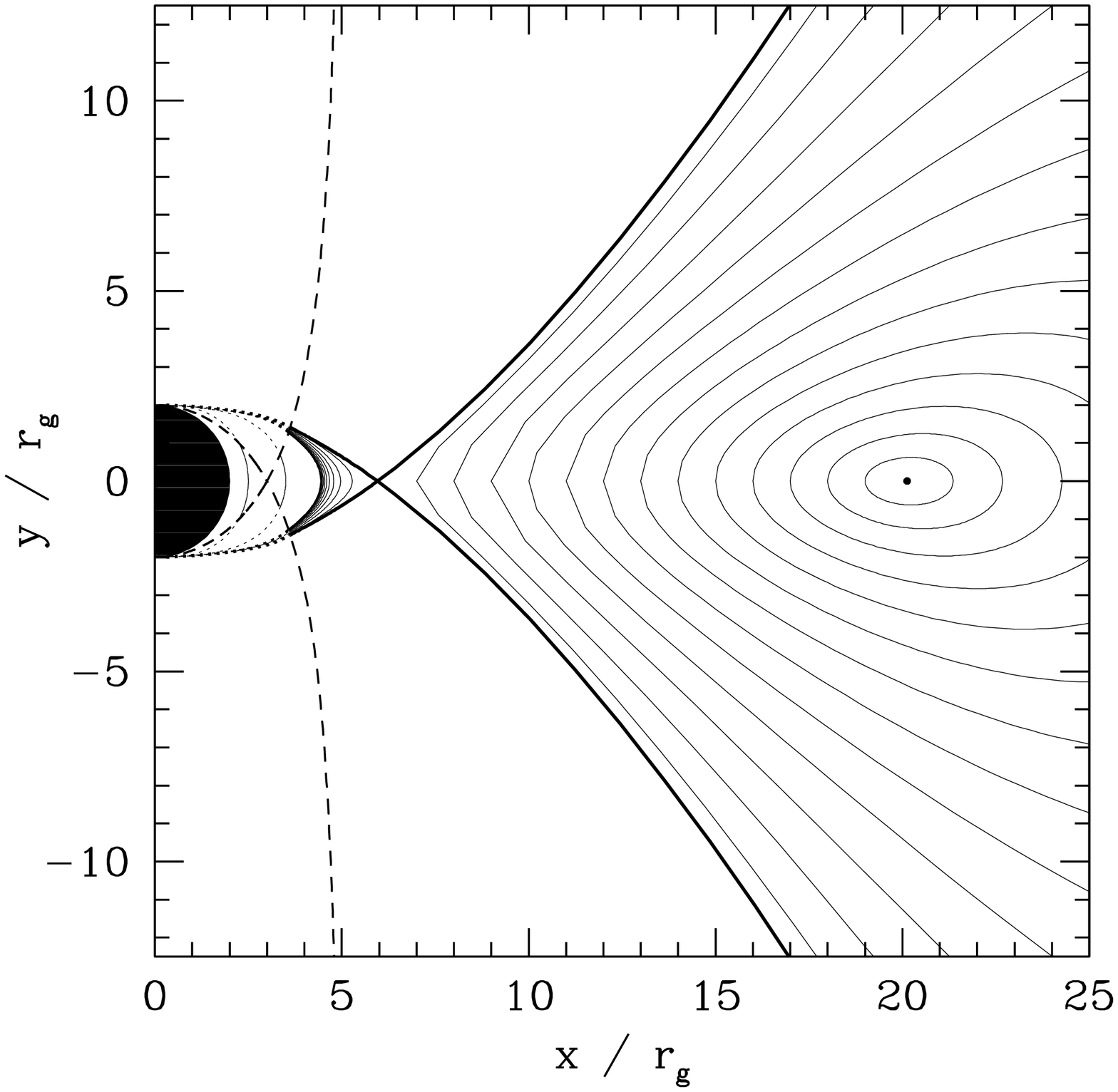,width=6.25cm}\\
\multicolumn{2}{c}{\textbf{Case (3): $K_\mathrm{mb} < K < K_\mathrm{max}$} (here $K= \pm 2.4$; 
$r_\mathrm{cusp}\simeq 5.55$ ; $r_\mathrm{centre}\simeq 23.1$)}\\
\psfig{file=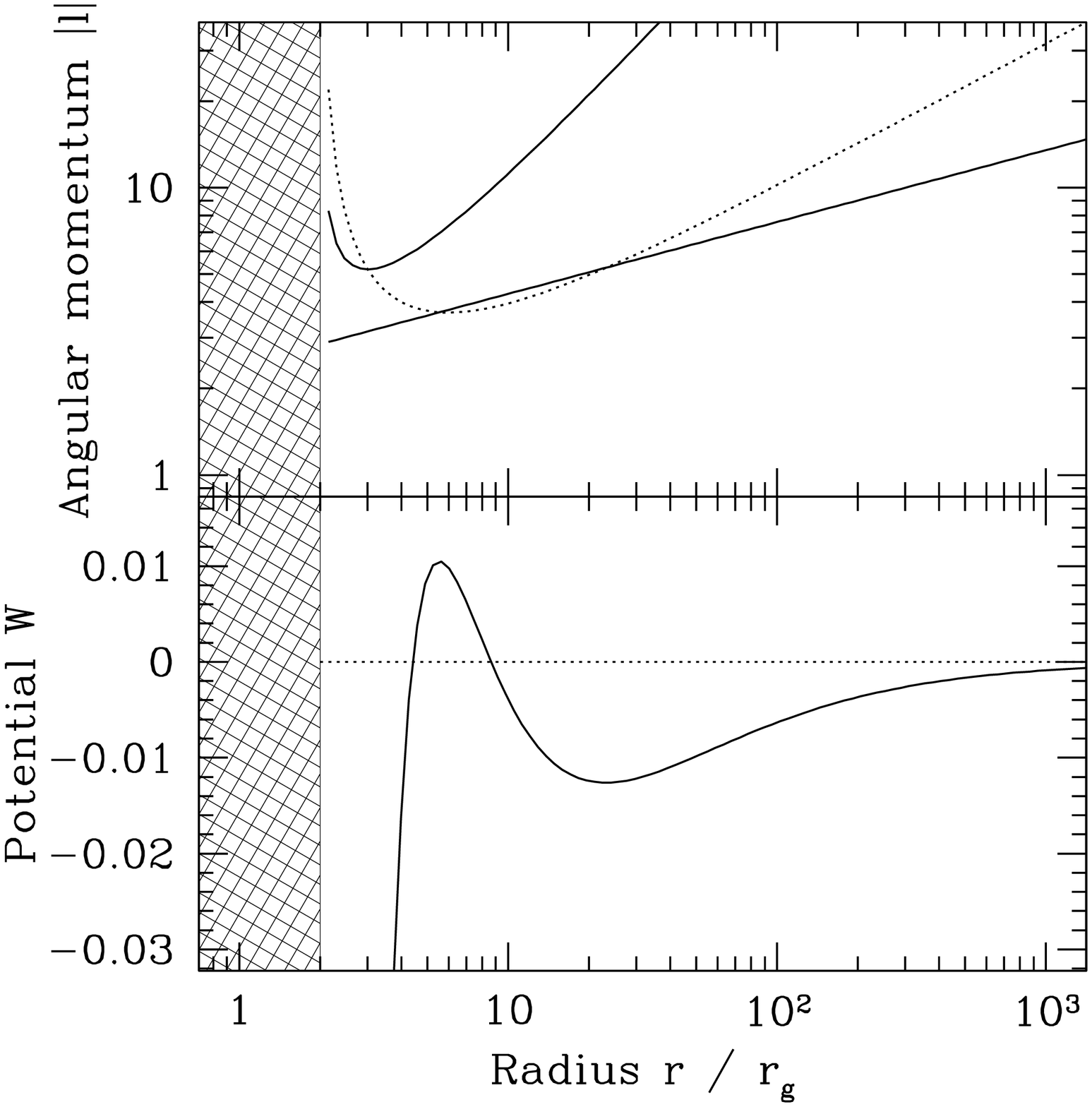,width=6.25cm} &
\psfig{file=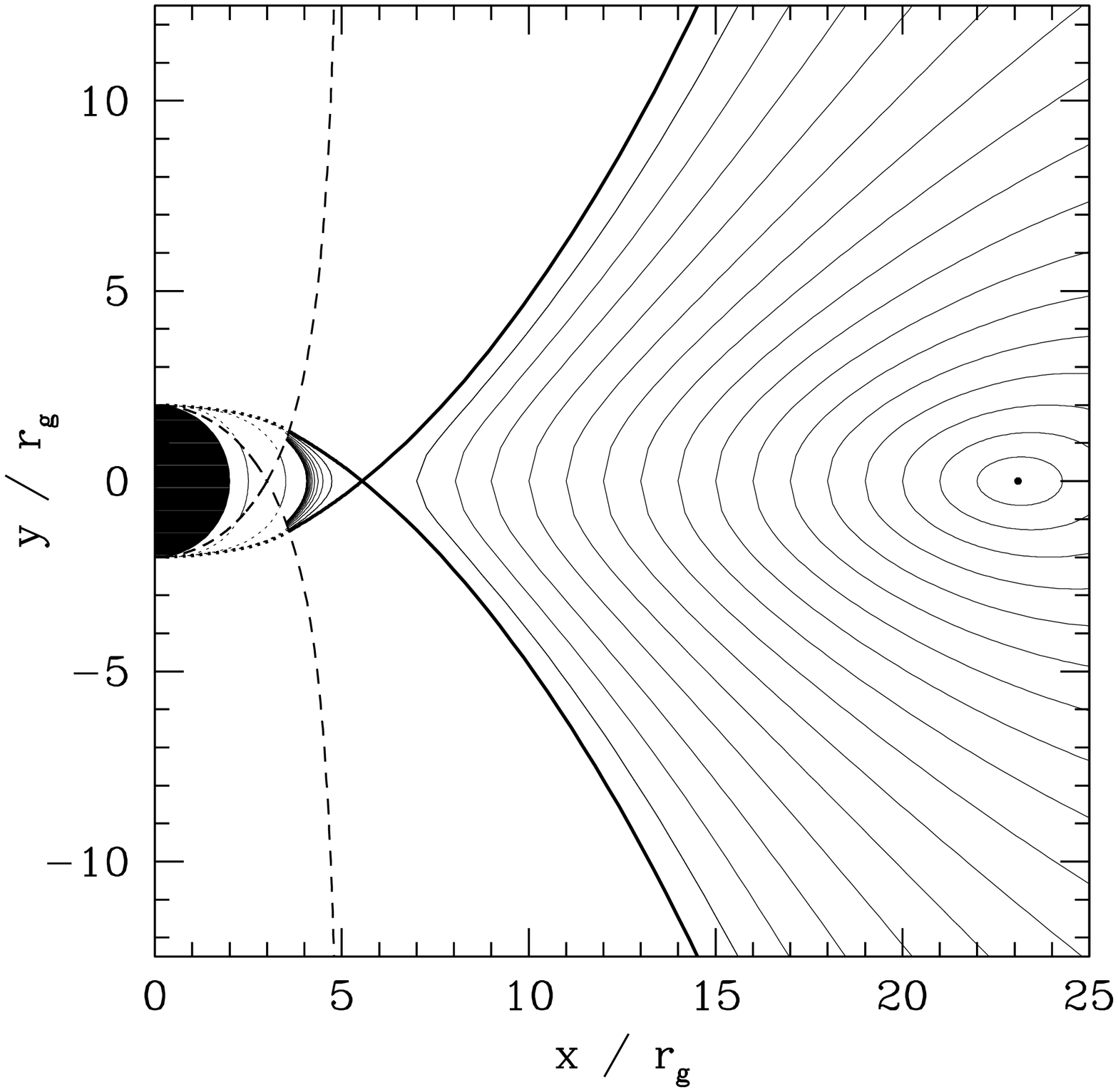,width=6.25cm}\\
\multicolumn{2}{c}{\textbf{Critical case: $K = K_\mathrm{max}$} (here $K_\mathrm{max}\simeq \pm 3.90$; 
no cusp ; $r_\mathrm{centre}\simeq 223$)}\\
\psfig{file=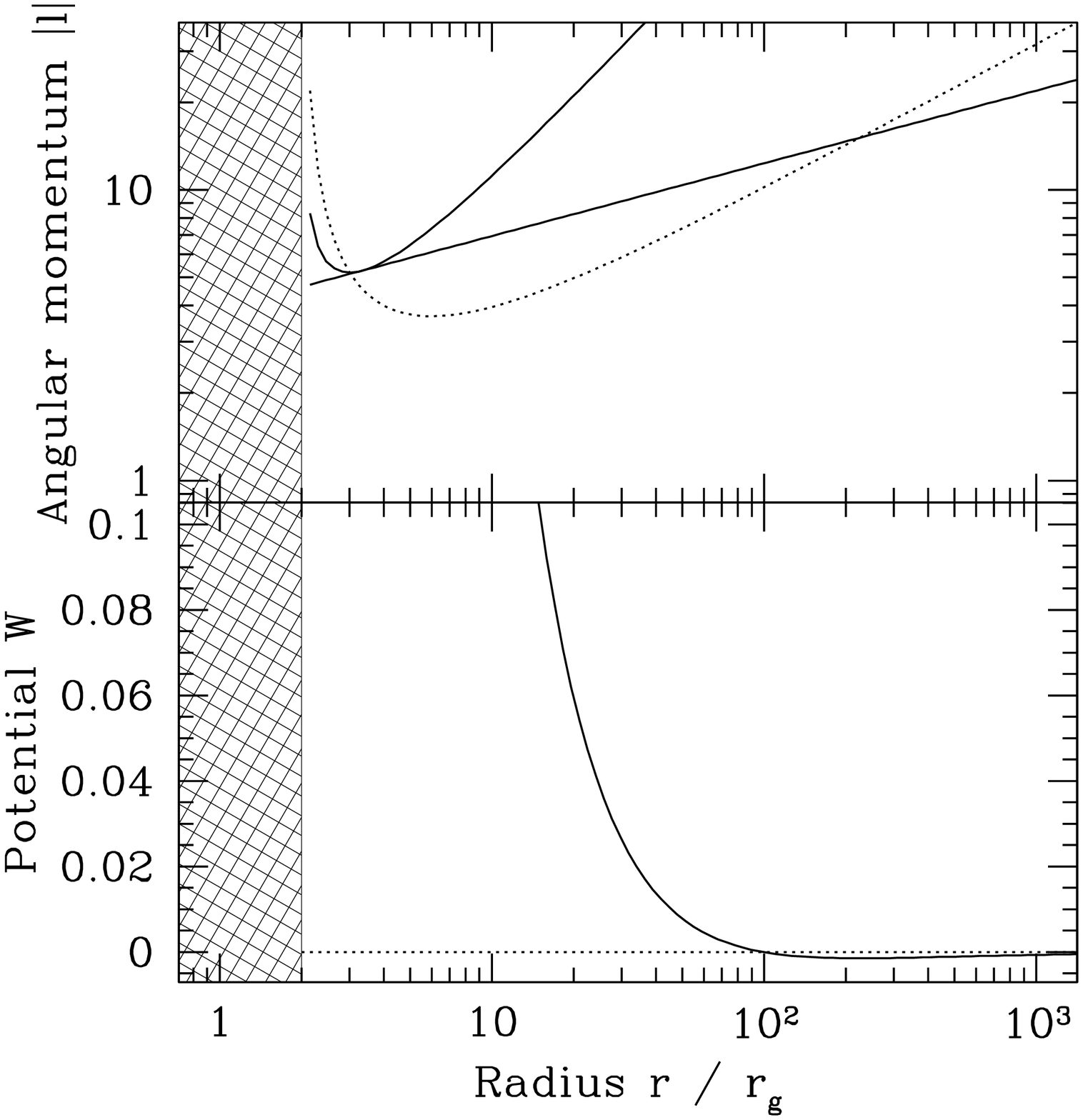,width=6.25cm} &
\psfig{file=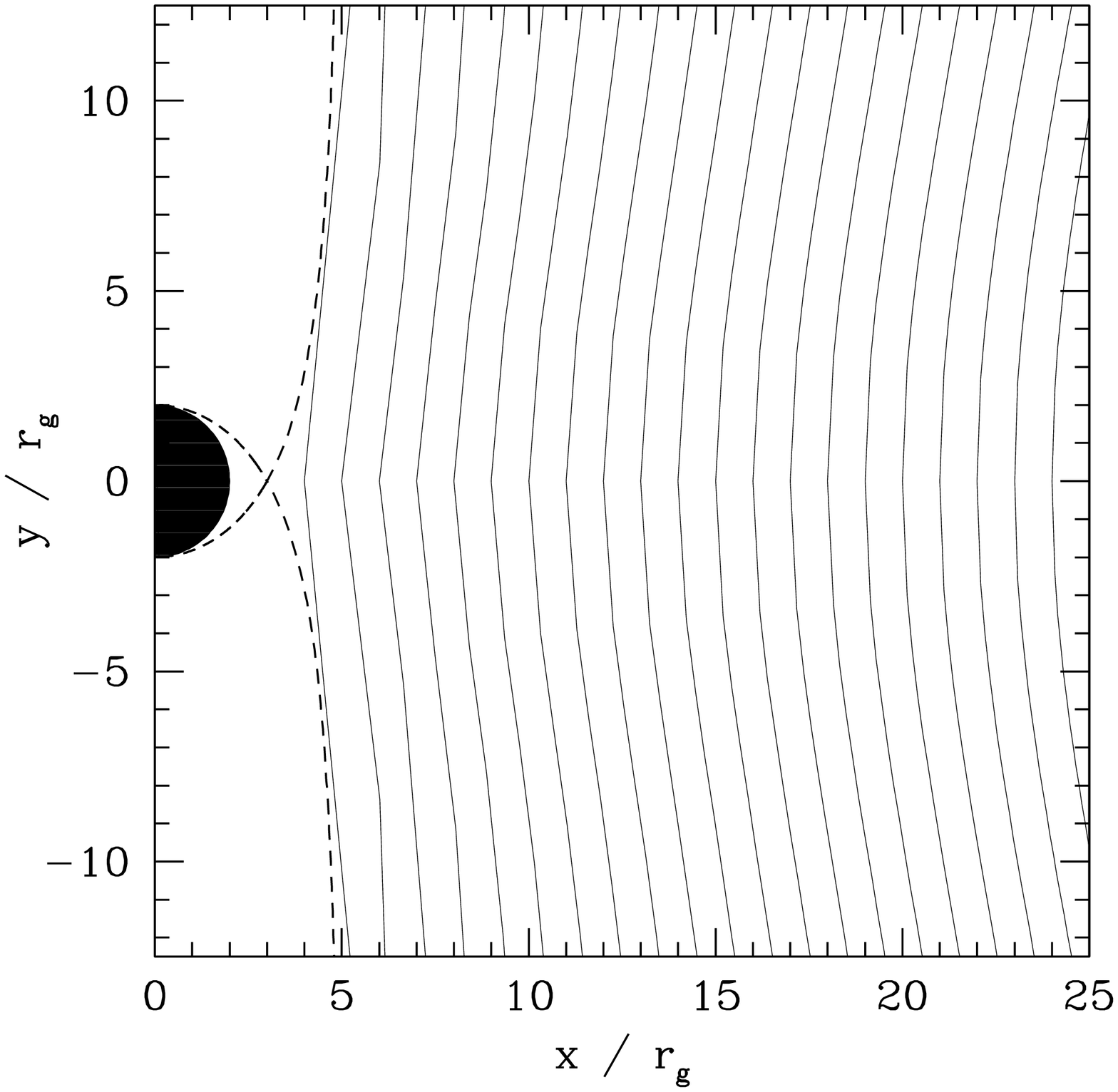,width=6.25cm}\\
\end{tabular}
\end{center}
\contcaption{\textbf{Geometry of the equipotentials: Schwarzschild black hole ($a=0$) and pro- or 
retrograde disc. Sub-Keplerian case.}
In the critical case 
$K=K_\mathrm{max}$ (third line), the thick line corresponds to the critical equipotential $W=+\infty$.
}
\end{figure*}

\begin{figure*}
\begin{center}
\begin{tabular}{cc}
\multicolumn{2}{c}{\textbf{\textit{-- Schwarzschild black hole}} ($a=0$) --}\\
\\
\multicolumn{2}{c}{\underline{\textbf{Keplerian case ($\alpha=1/2$)}}}\\
\multicolumn{2}{c}{\textbf{Case (1): $K<K_\mathrm{ms}=K_\mathrm{mb}=1$} (here $K= \pm 0.5$; 
no cusp ; no centre)}\\
\psfig{file=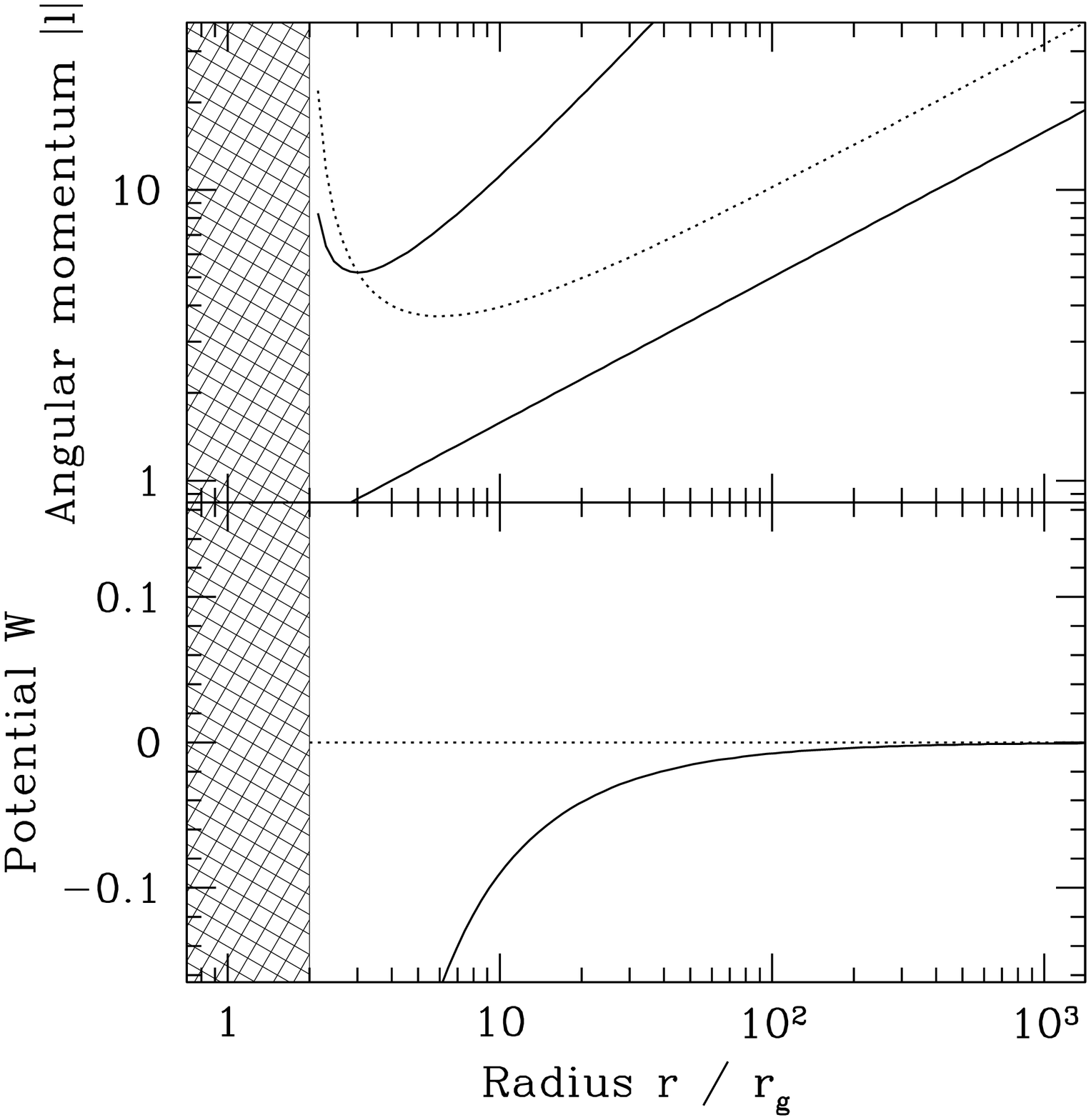,width=6.25cm} &
\psfig{file=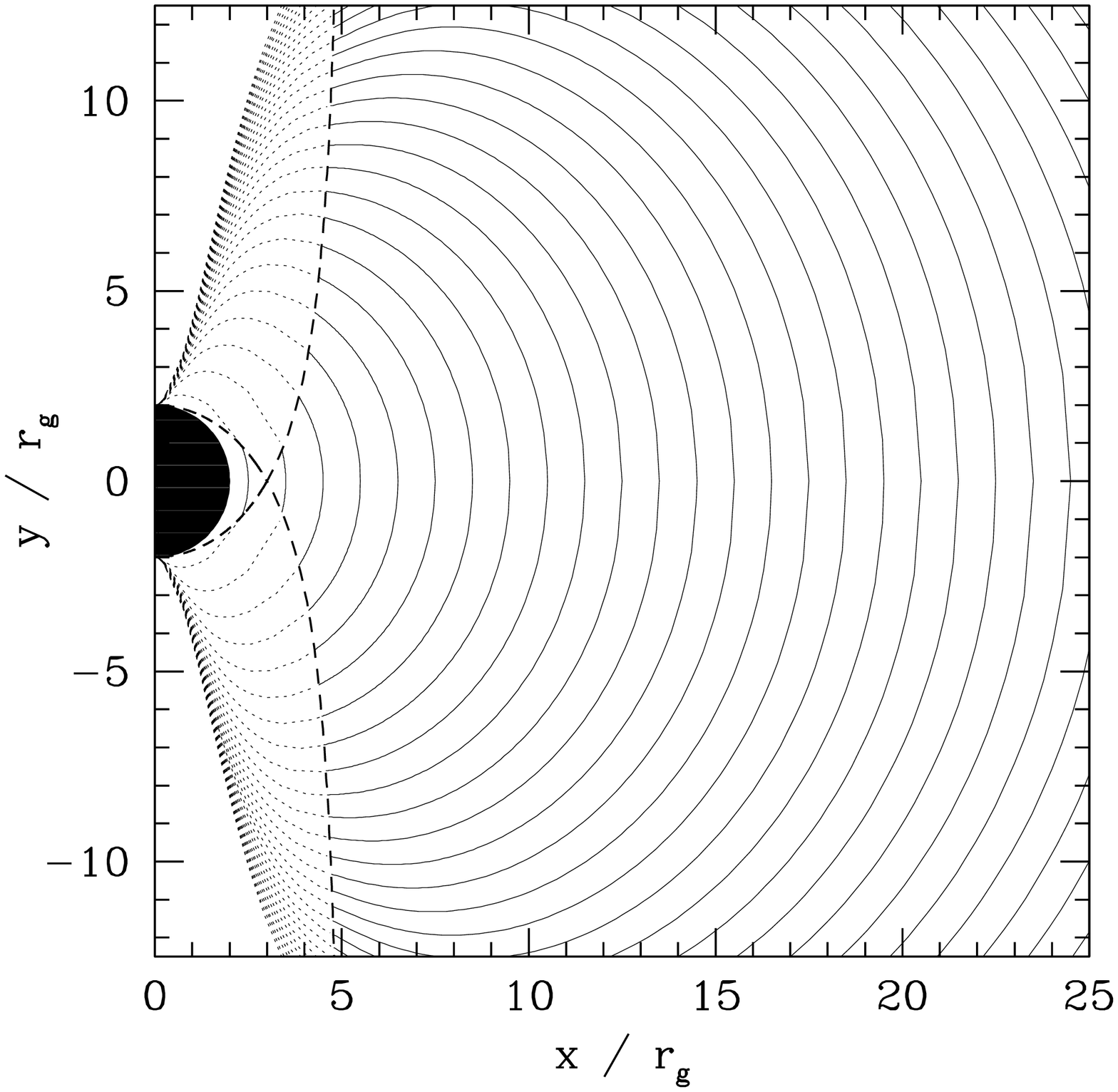,width=6.25cm}\\
\multicolumn{2}{c}{\textbf{Case (3): $1=K_\mathrm{ms}=K_\mathrm{mb} < K < K_\mathrm{max}$} 
(here $K=\pm 1.5$ ; $r_\mathrm{cusp} =6 $; centre at $\infty$)}\\
\psfig{file=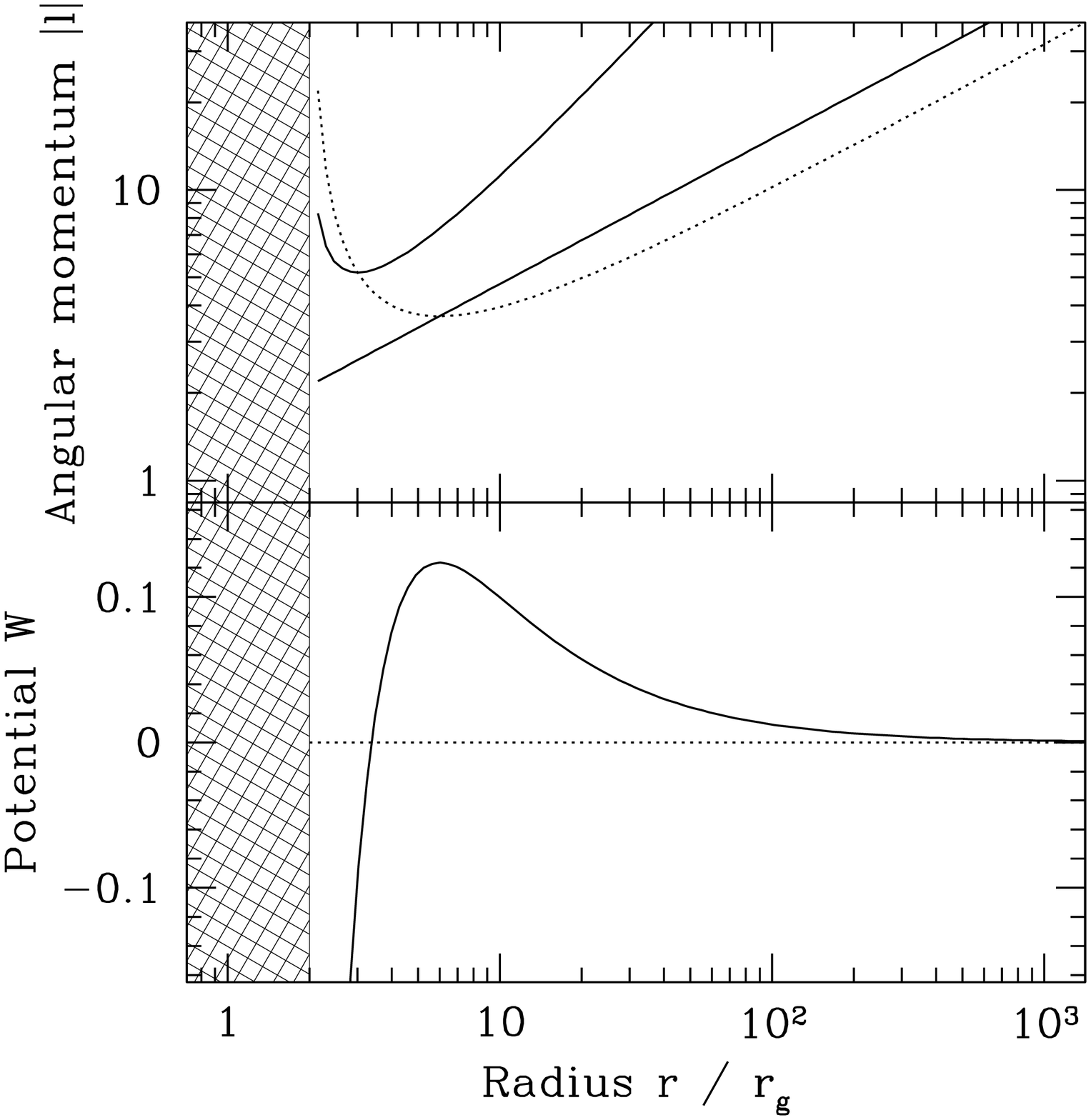,width=6.25cm} &
\psfig{file=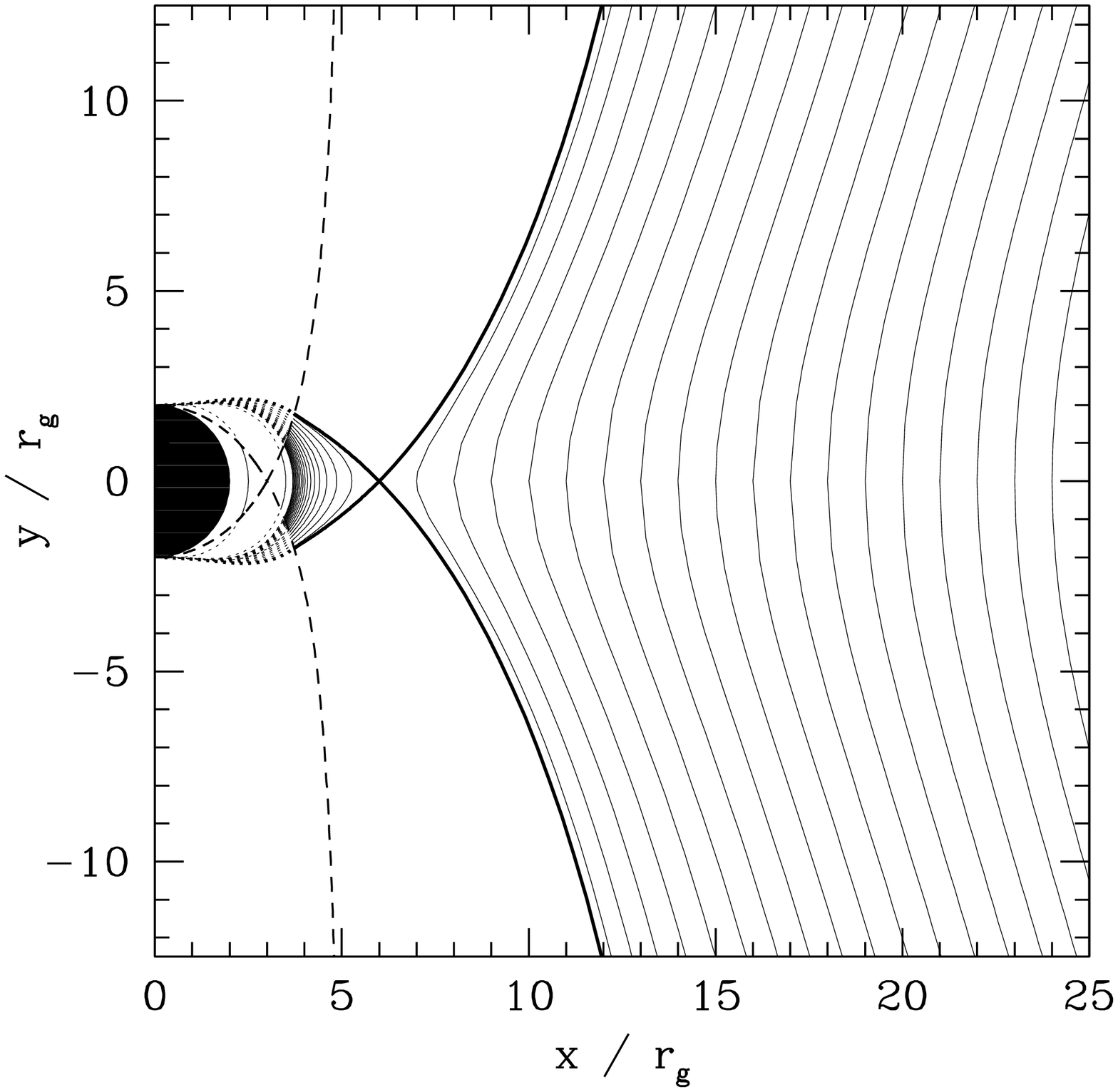,width=6.25cm}\\
\\
\multicolumn{2}{c}{\textbf{\underline{Super-Keplerian case (here with $\alpha=3/4$)}}}\\
\multicolumn{2}{c}{\textbf{Case (3): $K < K_\mathrm{max}$} (here $K=\pm 1$; 
$r_\mathrm{cusp}\simeq 5.68$; centre at $\infty$)}\\
\psfig{file=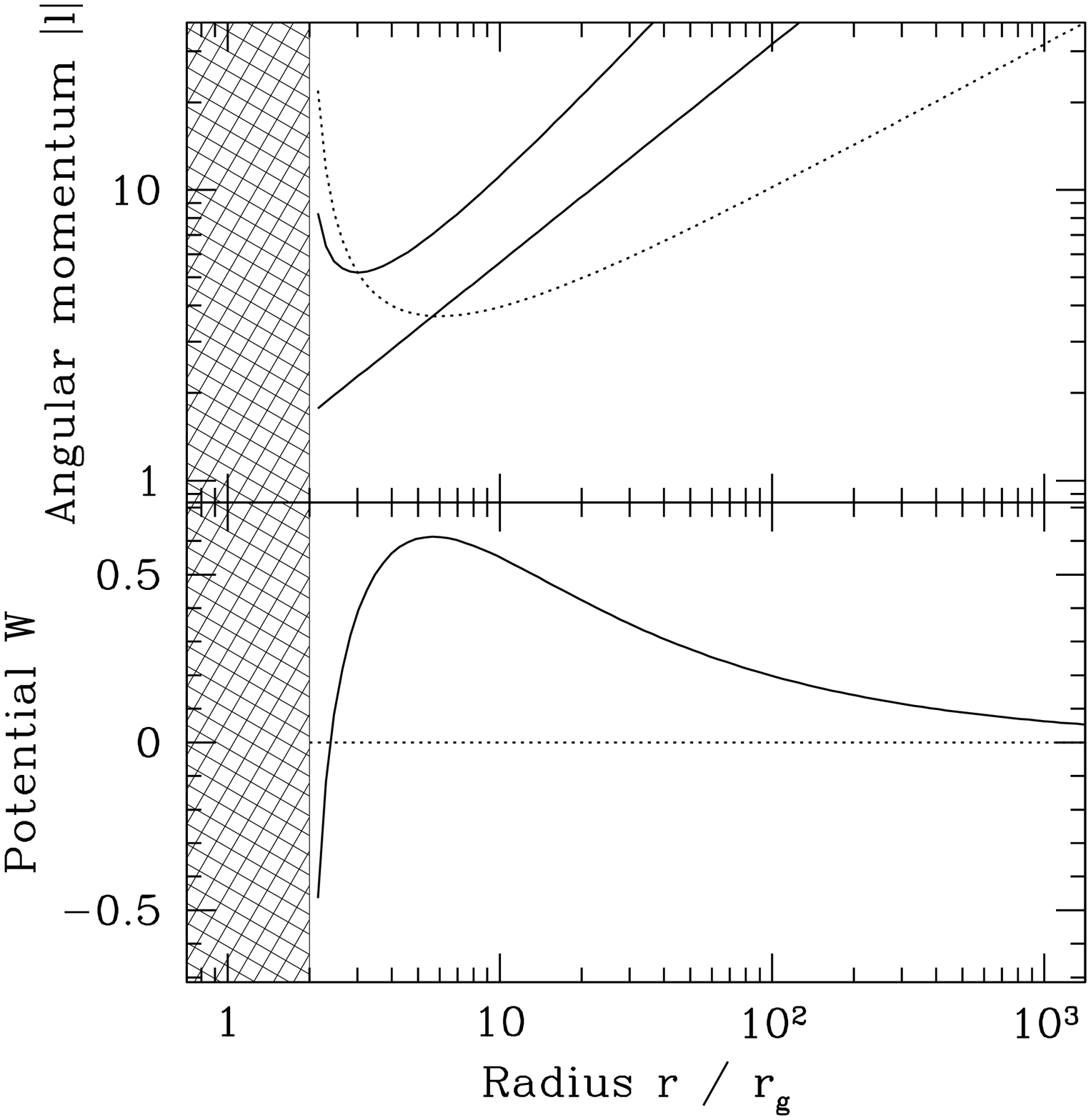,width=6.25cm} &
\psfig{file=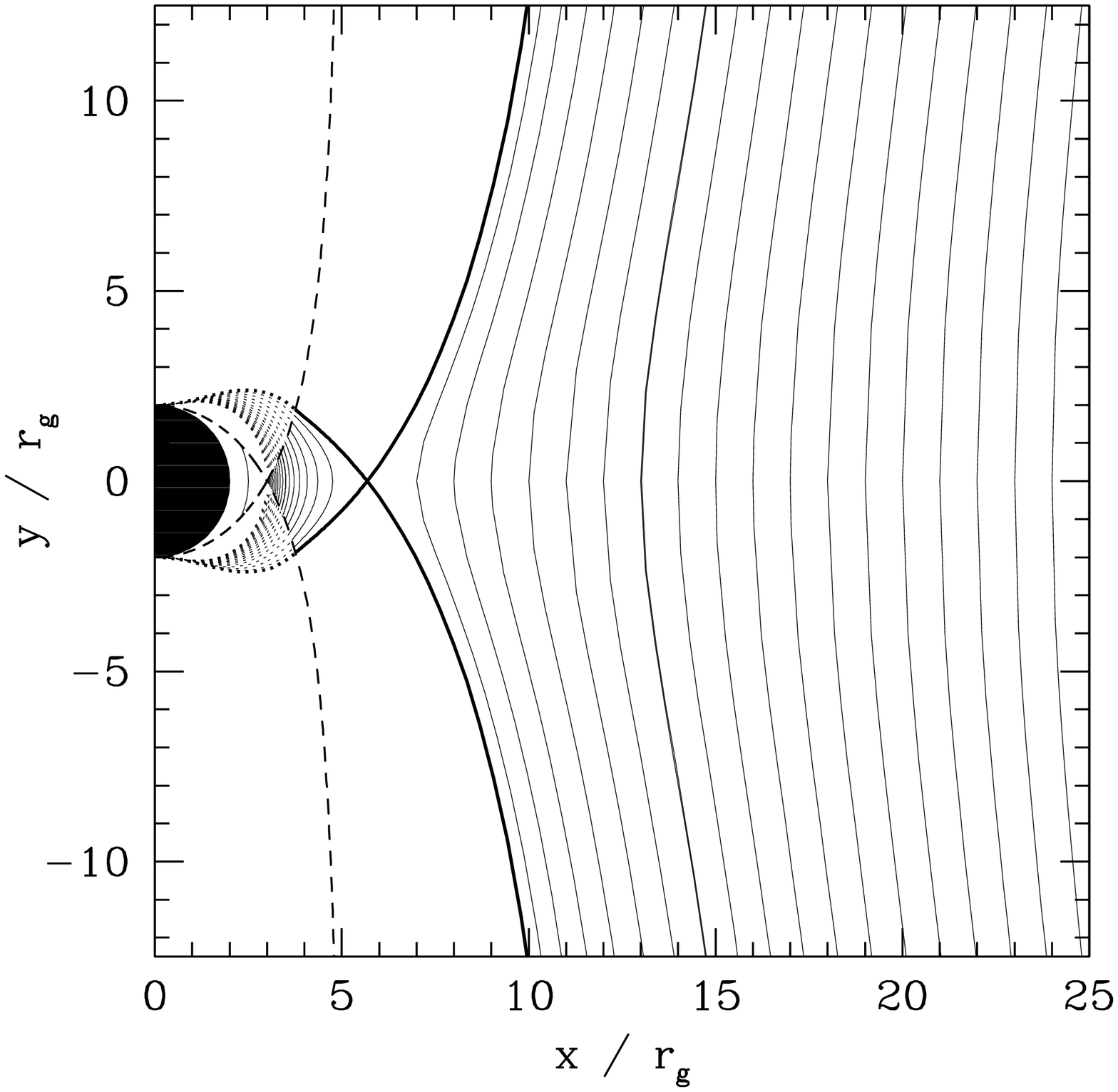,width=6.25cm}\\
\end{tabular}
\end{center}
\contcaption{\textbf{Geometry of the equipotentials: Schwarzschild black hole ($a=0$) and pro- or 
retrograde disc. Keplerian and super-Keplerian cases.}}
\end{figure*}



\begin{figure*}
\begin{center}
\begin{tabular}{cc}
\multicolumn{2}{c}{\textbf{\textit{-- Mildly rotating Kerr black hole}} ($a=\sqrt{5}/3$) --}\\
\\
\multicolumn{2}{c}{\underline{\textbf{Sub-Keplerian case (here with $\alpha=1/4$)}}}\\
\multicolumn{2}{c}{\textbf{Case (1): $K<K_\mathrm{ms}$} (here $K= 1.5$; no cusp; no centre)}\\
\psfig{file=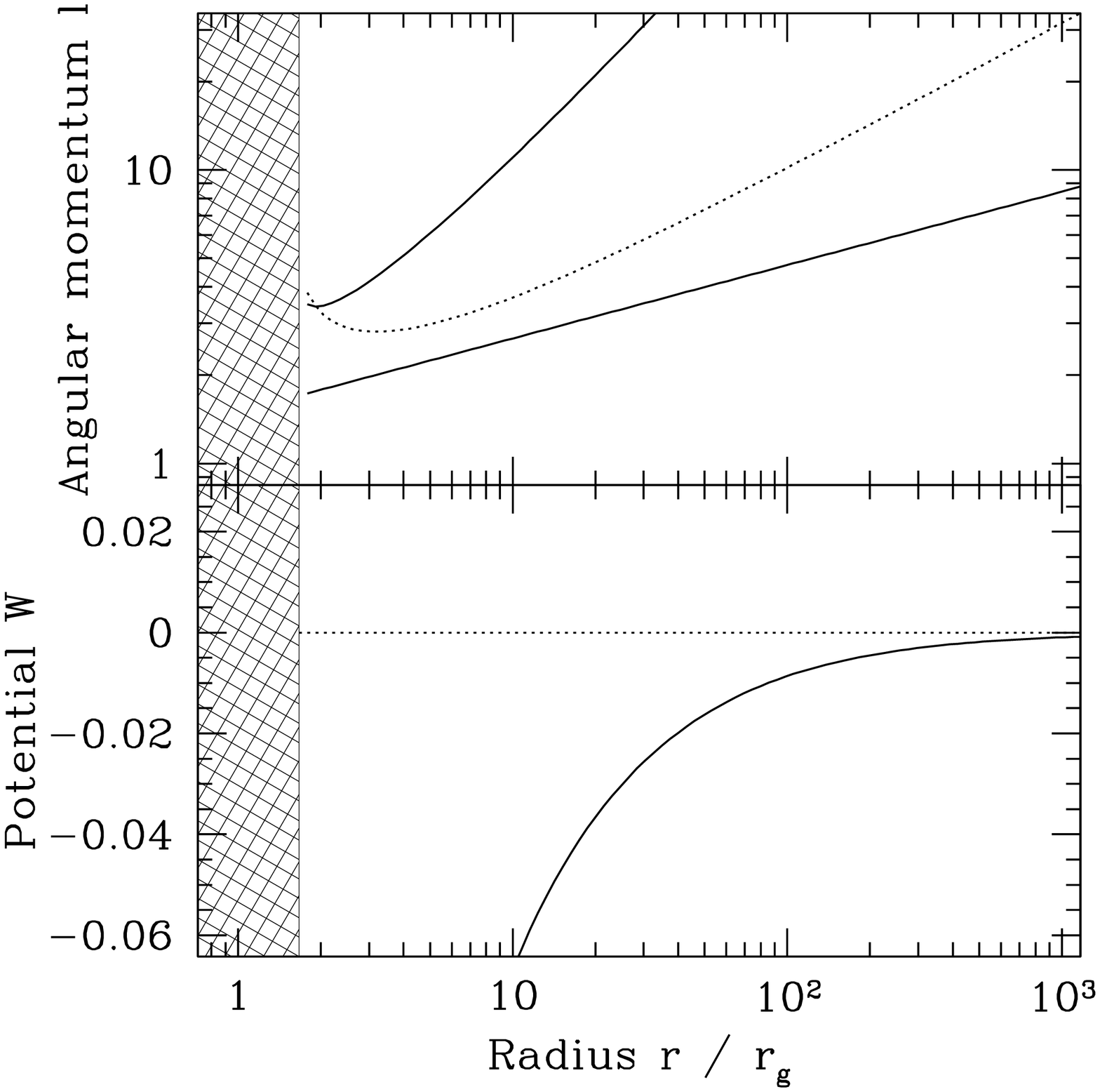,width=6.25cm} &
\psfig{file=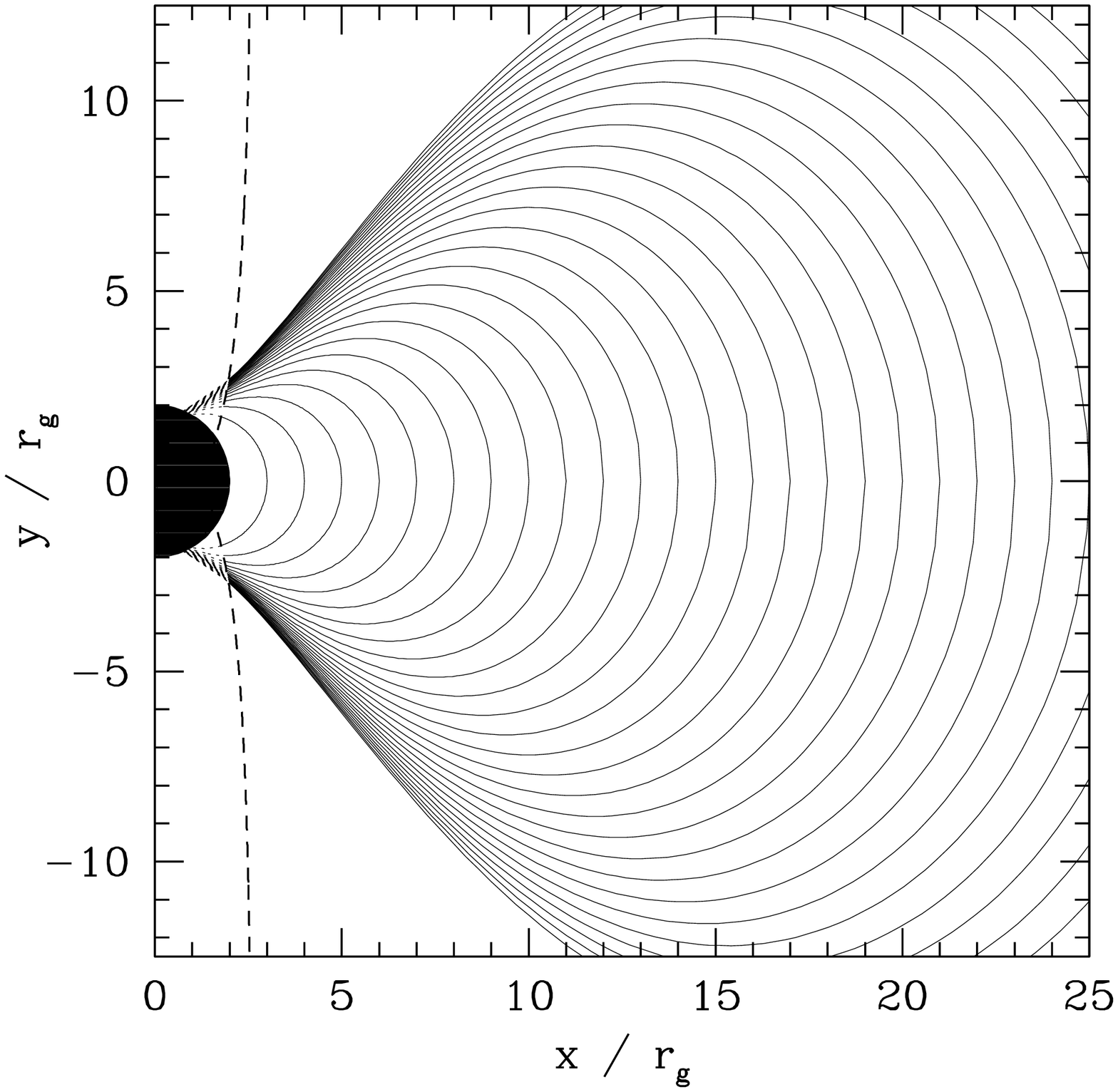,width=6.25cm}\\
\multicolumn{2}{c}{\textbf{Critical case: $K=K_\mathrm{ms}$} (here $K_\mathrm{ms}\simeq  1.99$; 
$r_\mathrm{cusp}=r_\mathrm{centre} \simeq 5.45$)}\\
\psfig{file=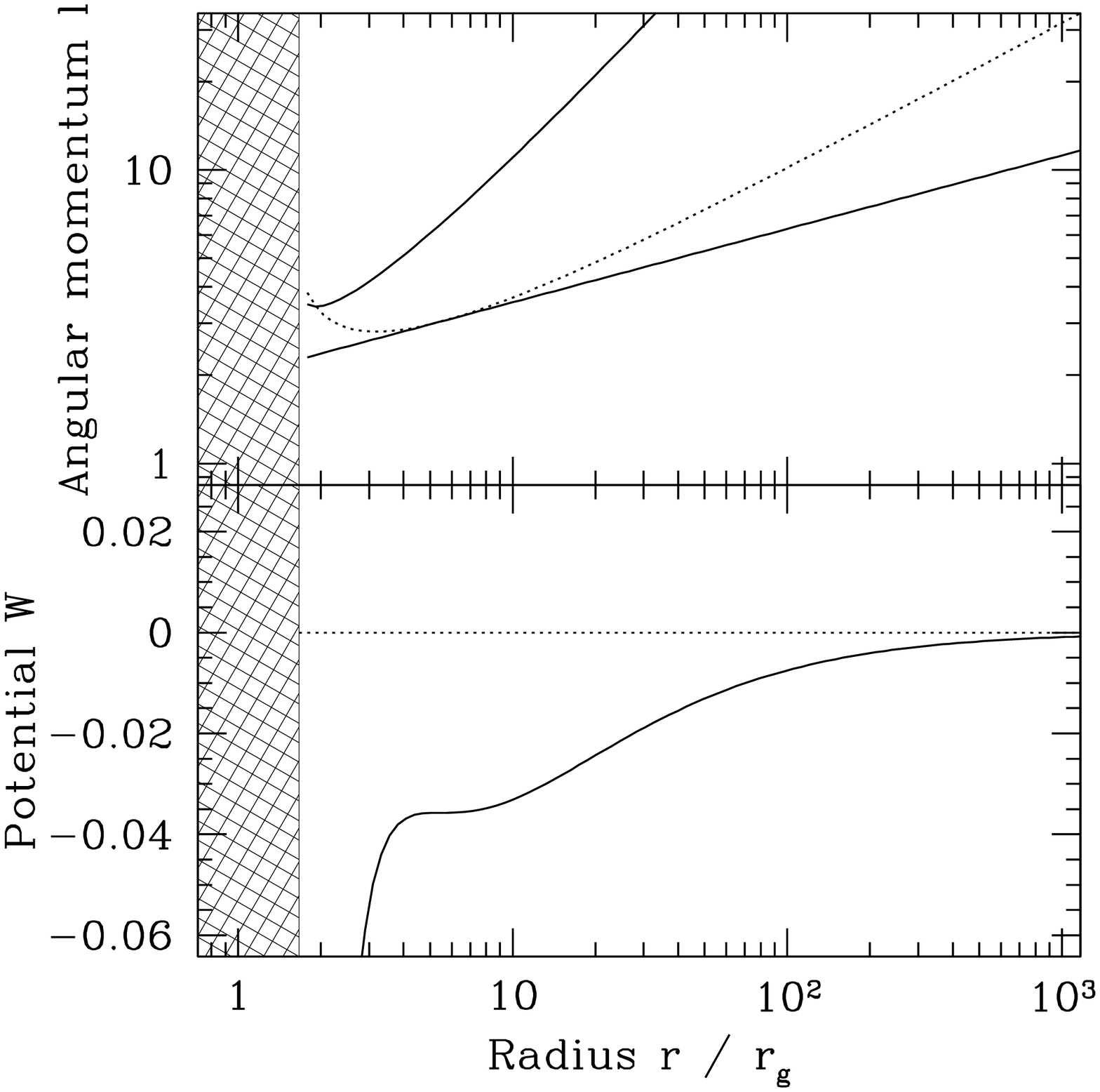,width=6.25cm} &
\psfig{file=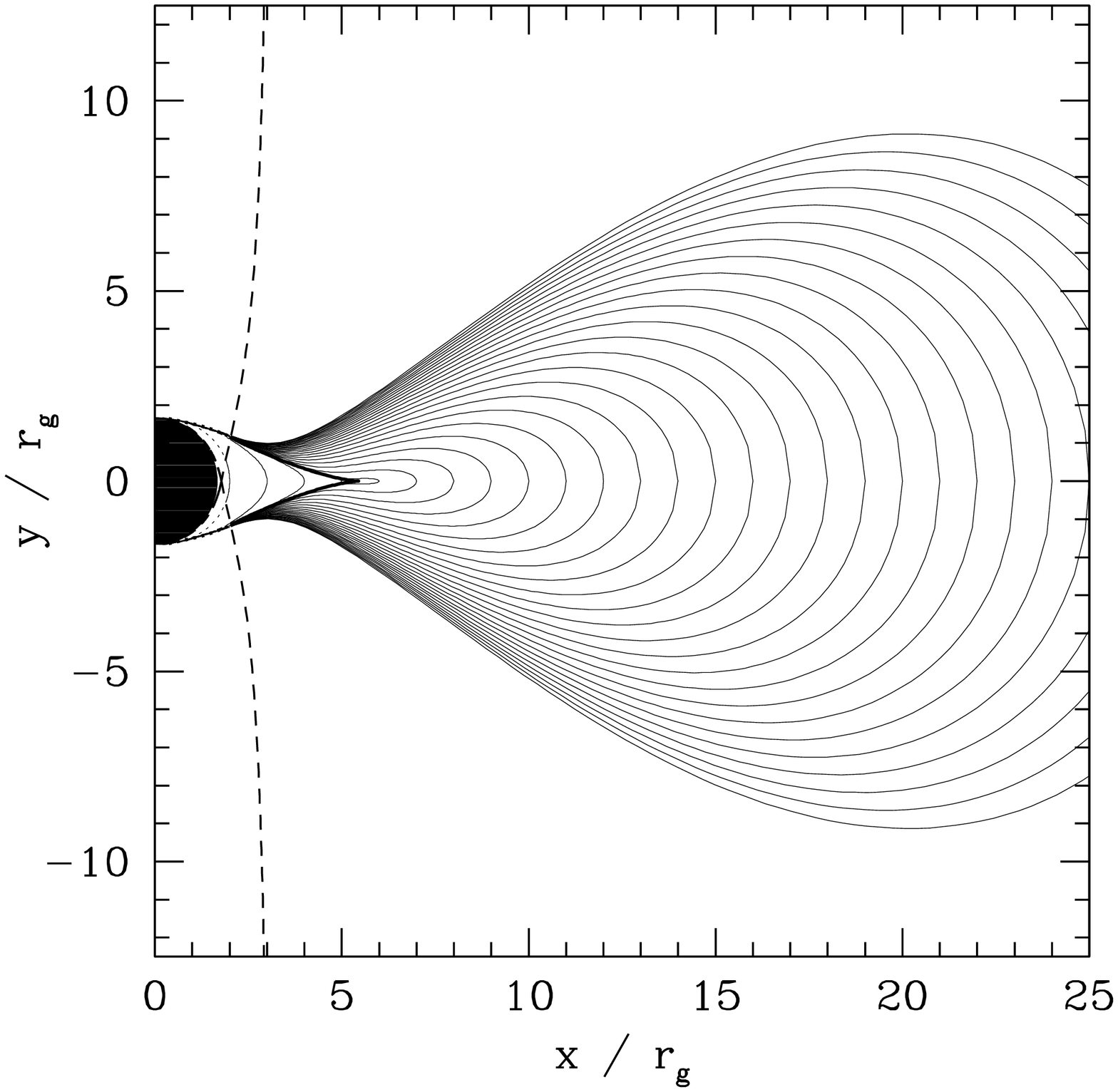,width=6.25cm}\\
\multicolumn{2}{c}{\textbf{Case (2): $K_\mathrm{ms}<K<K_\mathrm{mb}$} (here $K=  2.02$; 
$r_\mathrm{cusp}\simeq 4.05$; $r_\mathrm{centre}\simeq 7.74$)}\\
\psfig{file=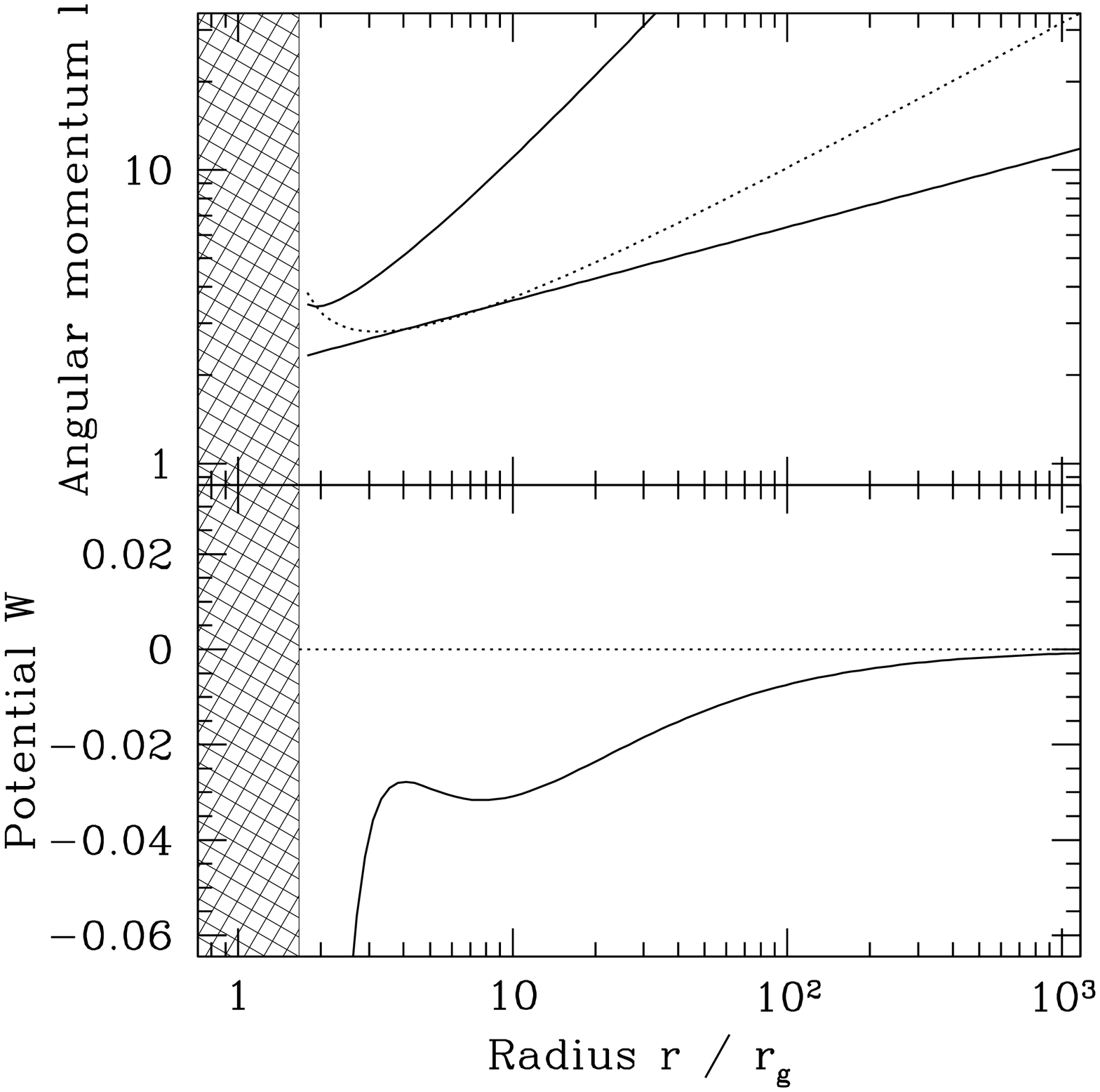,width=6.25cm} &
\psfig{file=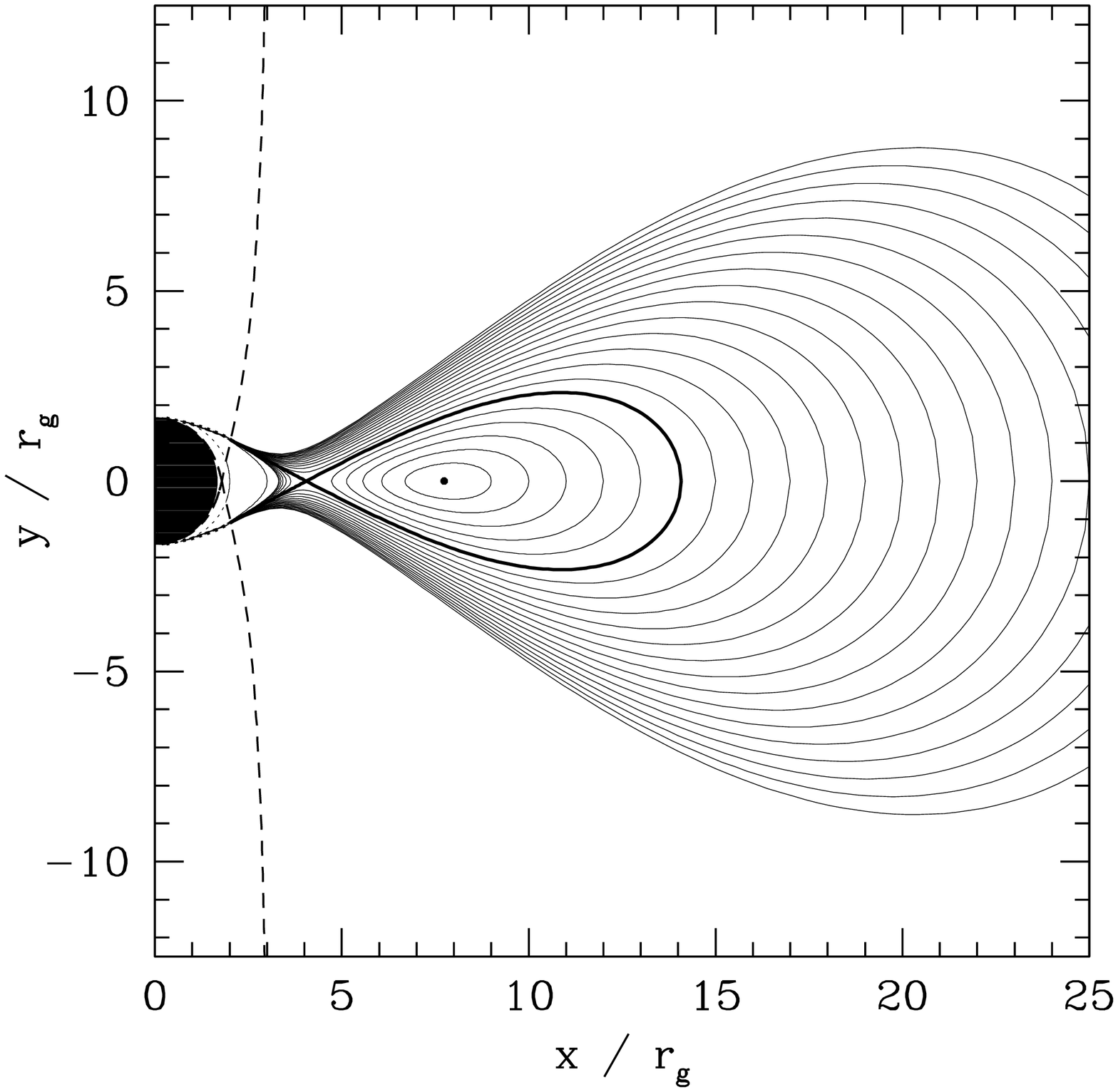,width=6.25cm}\\
\end{tabular}
\end{center}
\caption{\textbf{Geometry of the equipotentials: mildly rotating Kerr black hole ($a=\sqrt{5}/3$) 
and prograde disc. Sub-Keplerian case.} All line styles and notations are identical to those used
in Fig.~\ref{fig:equi1pro}.}
\label{fig:equi2pro}
\end{figure*}

\begin{figure*}
\begin{center}
\begin{tabular}{cc}
\multicolumn{2}{c}{\textbf{\textit{-- Mildly rotating Kerr black hole}} ($a=\sqrt{5}/3$) --}\\
\\
\multicolumn{2}{c}{\underline{\textbf{Sub-Keplerian case (here with $\alpha=1/4$)}}}\\
\multicolumn{2}{c}{\textbf{Critical case: $K=K_\mathrm{mb}$} (here $K_\mathrm{mb}\simeq  2.09$; 
$r_\mathrm{cusp}\simeq 3.29$ ; $r_\mathrm{centre}\simeq 10.8$)}\\
\psfig{file=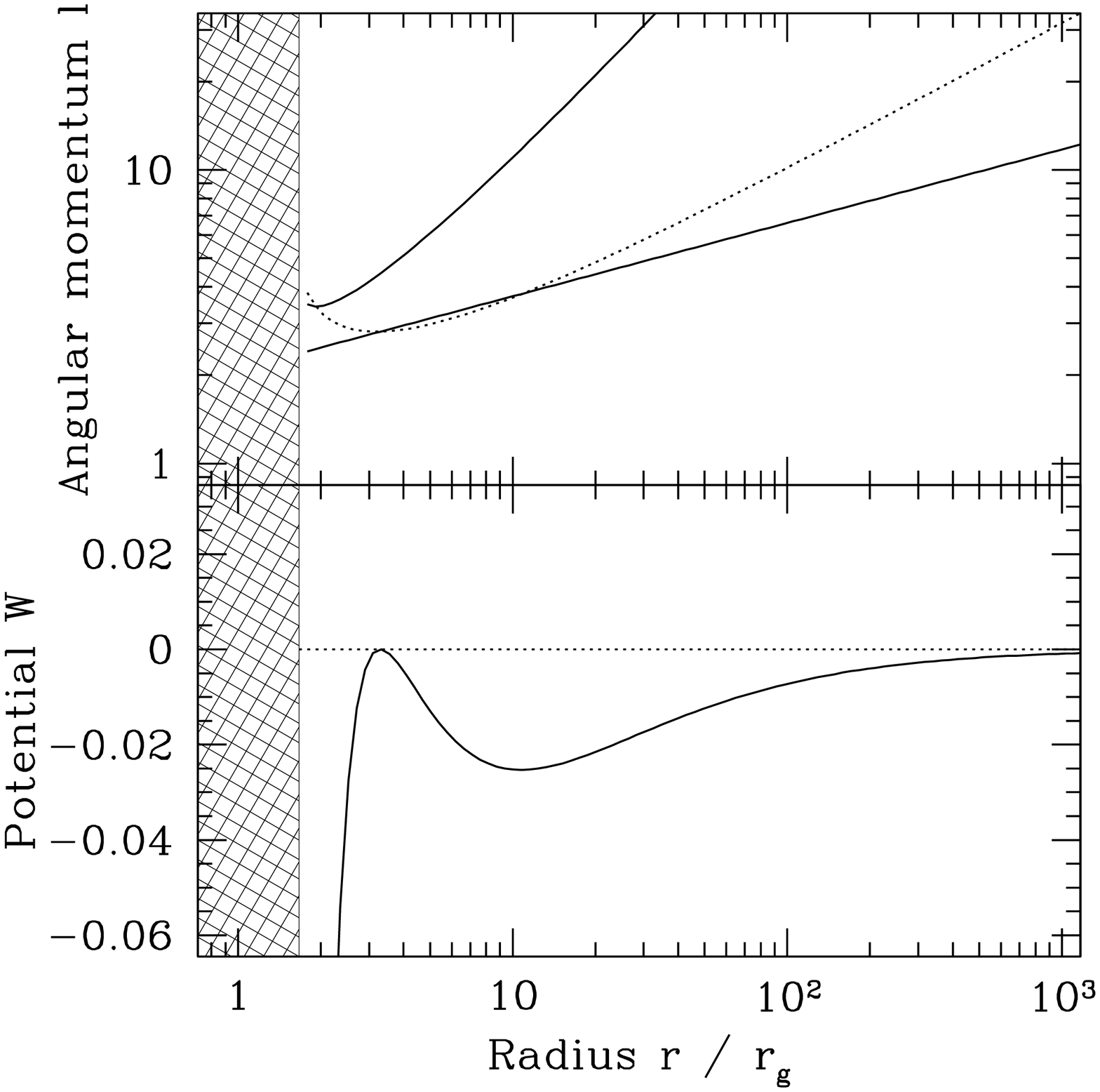,width=6.25cm} &
\psfig{file=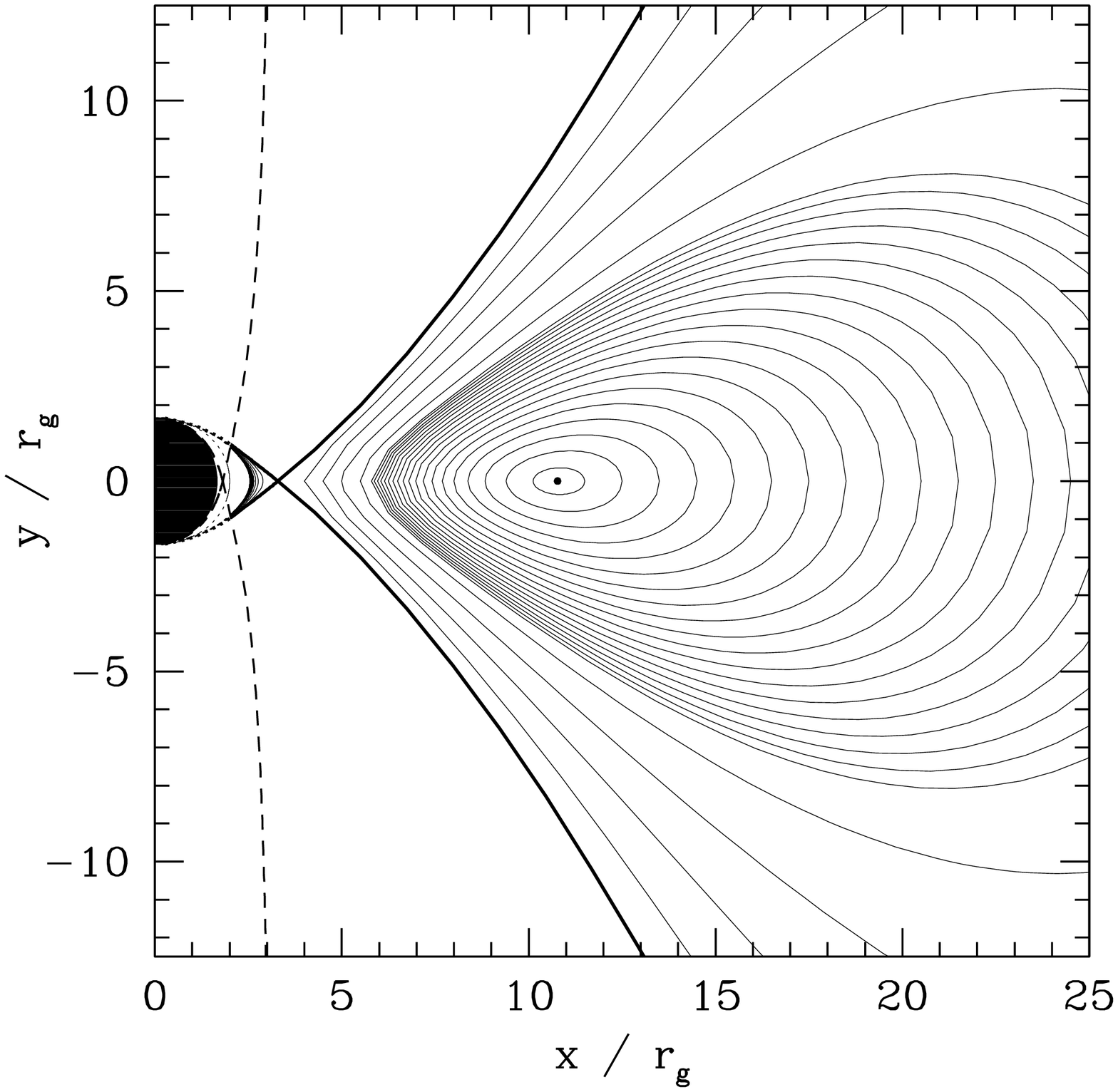,width=6.25cm}\\
\multicolumn{2}{c}{\textbf{Case (3): $K_\mathrm{mb} < K < K_\mathrm{max}$} (here $K=  2.2$; 
$r_\mathrm{cusp}\simeq 2.76$ ; $r_\mathrm{centre}\simeq 15.6$)}\\
\psfig{file=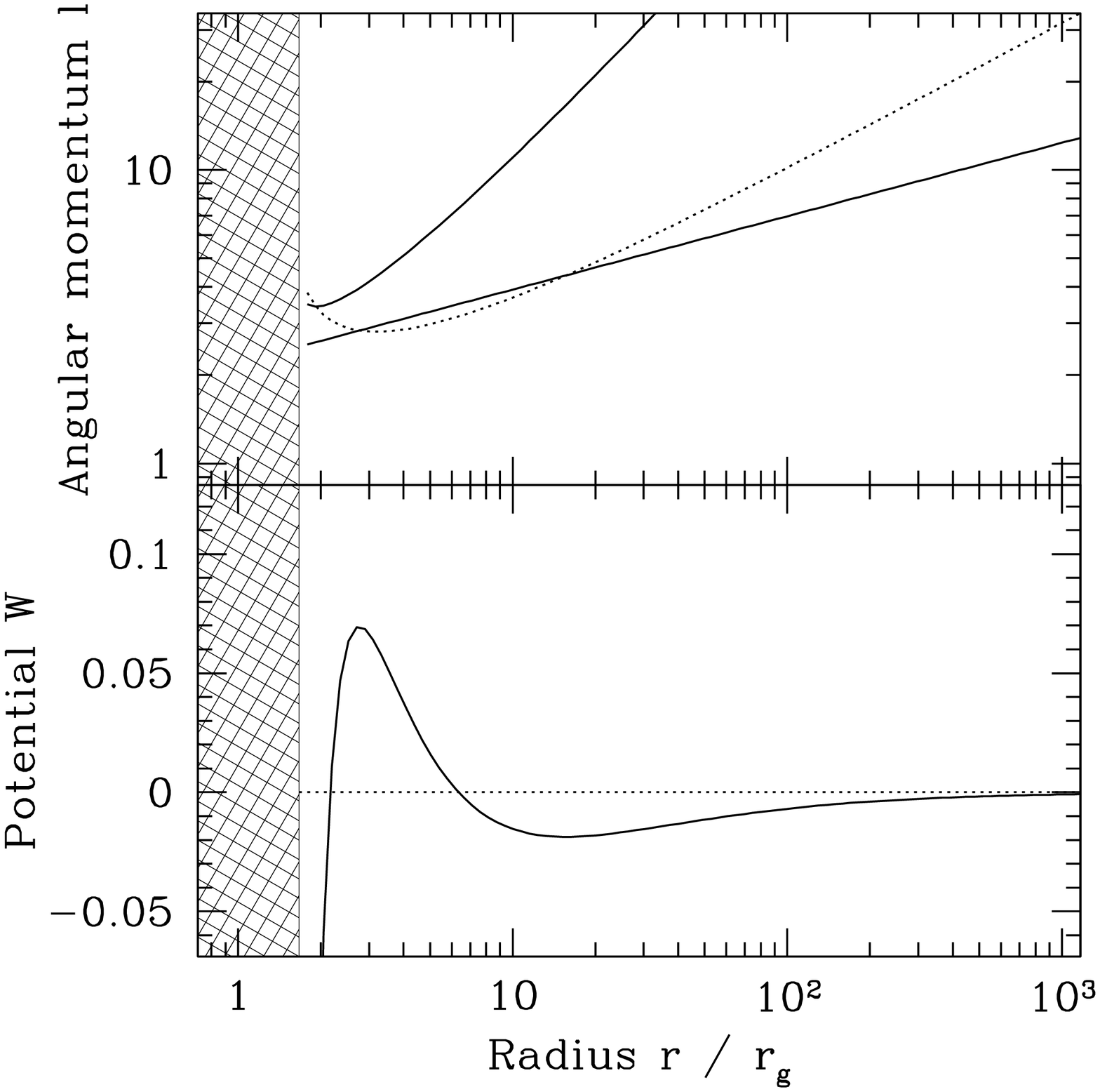,width=6.25cm} &
\psfig{file=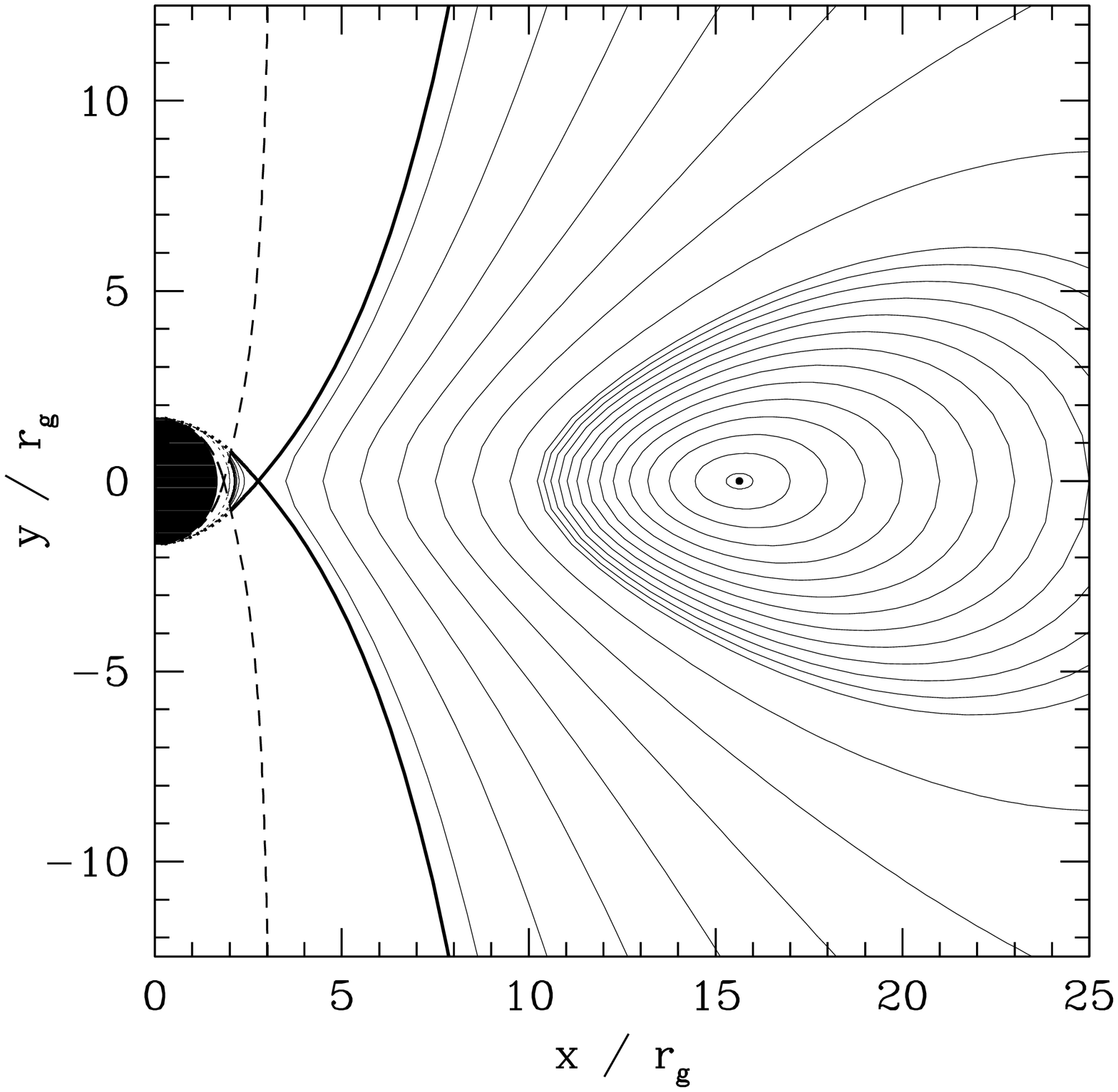,width=6.25cm}\\
\multicolumn{2}{c}{\textbf{Critical case: $K = K_\mathrm{max}$} (here $K_\mathrm{max}\simeq  2.88$; 
no cusp ; $r_\mathrm{centre}\simeq 61.3$)}\\
\psfig{file=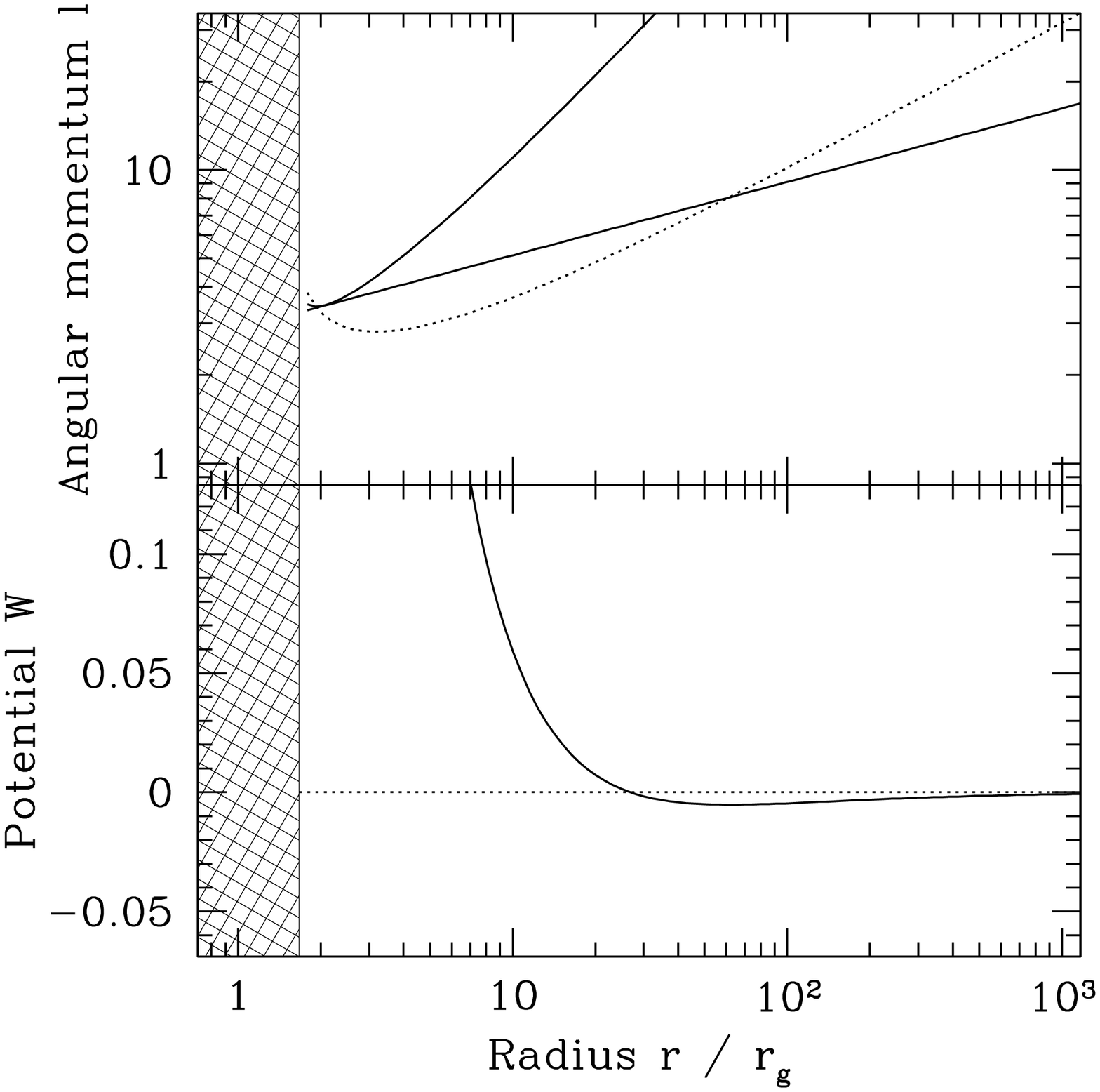,width=6.25cm} &
\psfig{file=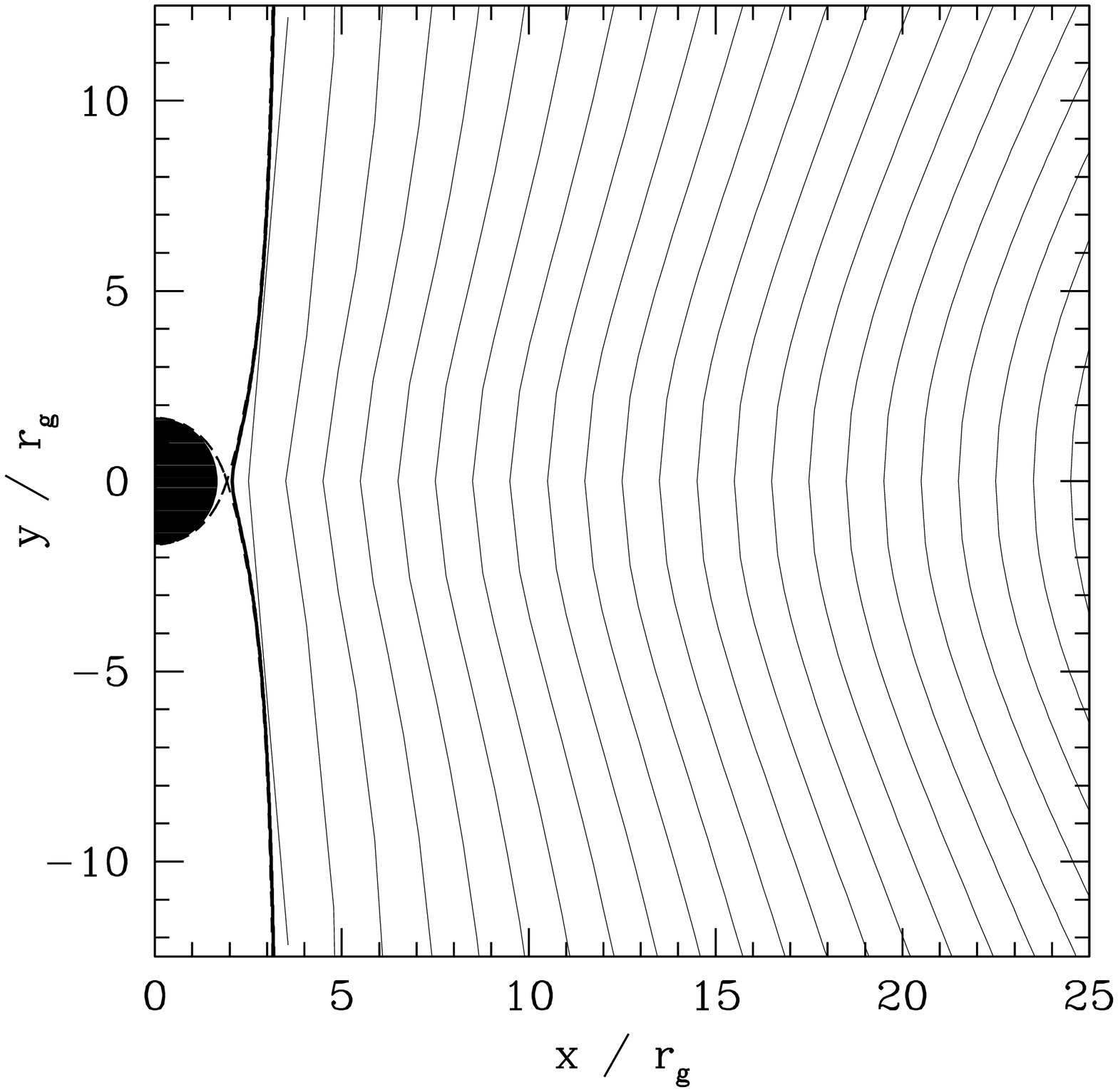,width=6.25cm}\\
\end{tabular}
\end{center}
\contcaption{\textbf{Geometry of the equipotentials: mildly rotating Kerr black hole ($a=\sqrt{5}/3$) 
and prograde disc. Sub-Keplerian case.}}
\end{figure*}

\begin{figure*}
\begin{center}
\begin{tabular}{cc}
\multicolumn{2}{c}{\textbf{\textit{-- Mildly rotating Kerr black hole}} ($a=\sqrt{5}/3$) --}\\
\\
\multicolumn{2}{c}{\underline{\textbf{Keplerian case ($\alpha=1/2$)}}}\\
\multicolumn{2}{c}{\textbf{Case (1): $K<K_\mathrm{ms}=K_\mathrm{mb}=1$} (here $K=0.5$; no cusp; 
no centre)}\\
\psfig{file=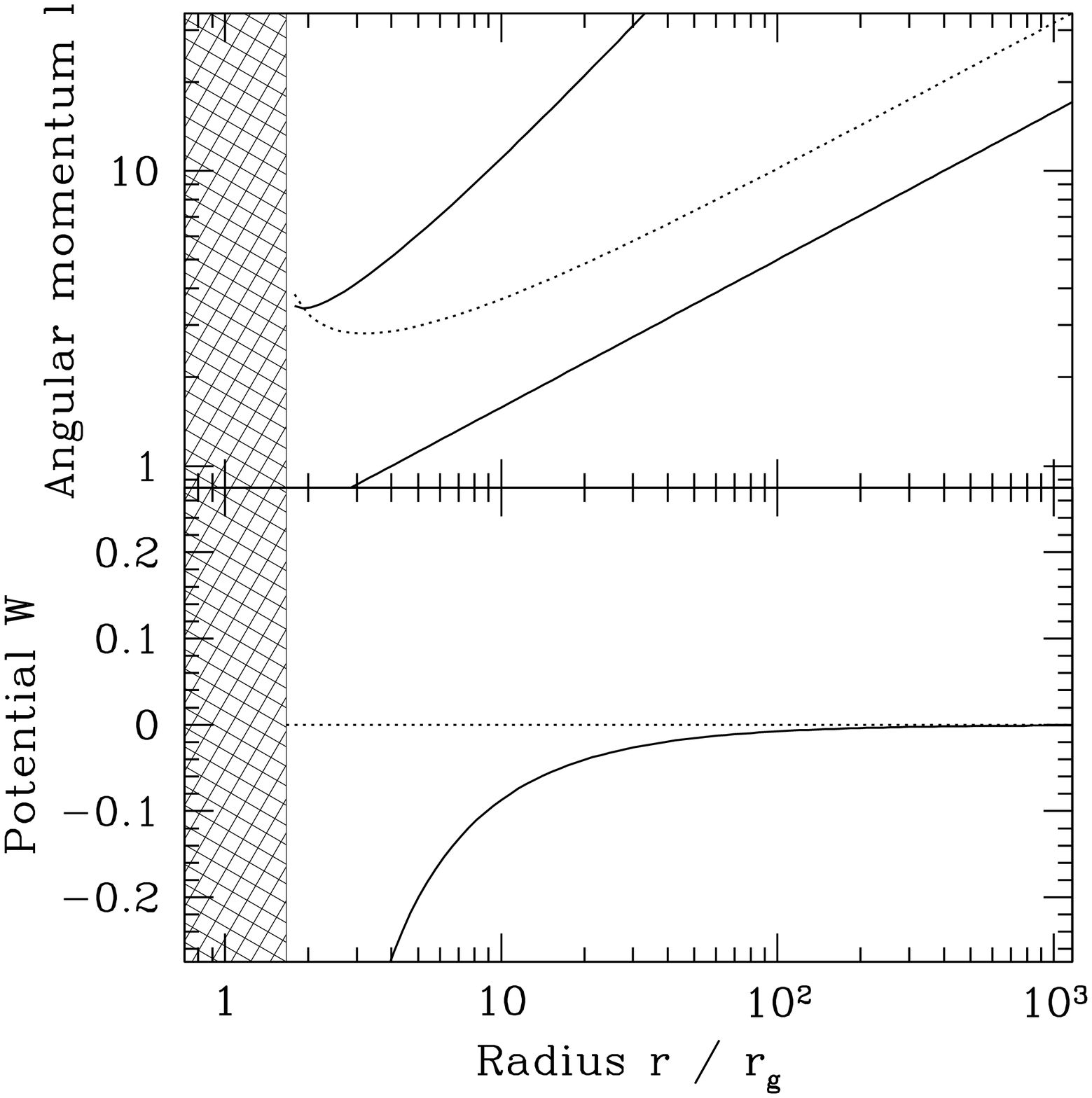,width=6.25cm} &
\psfig{file=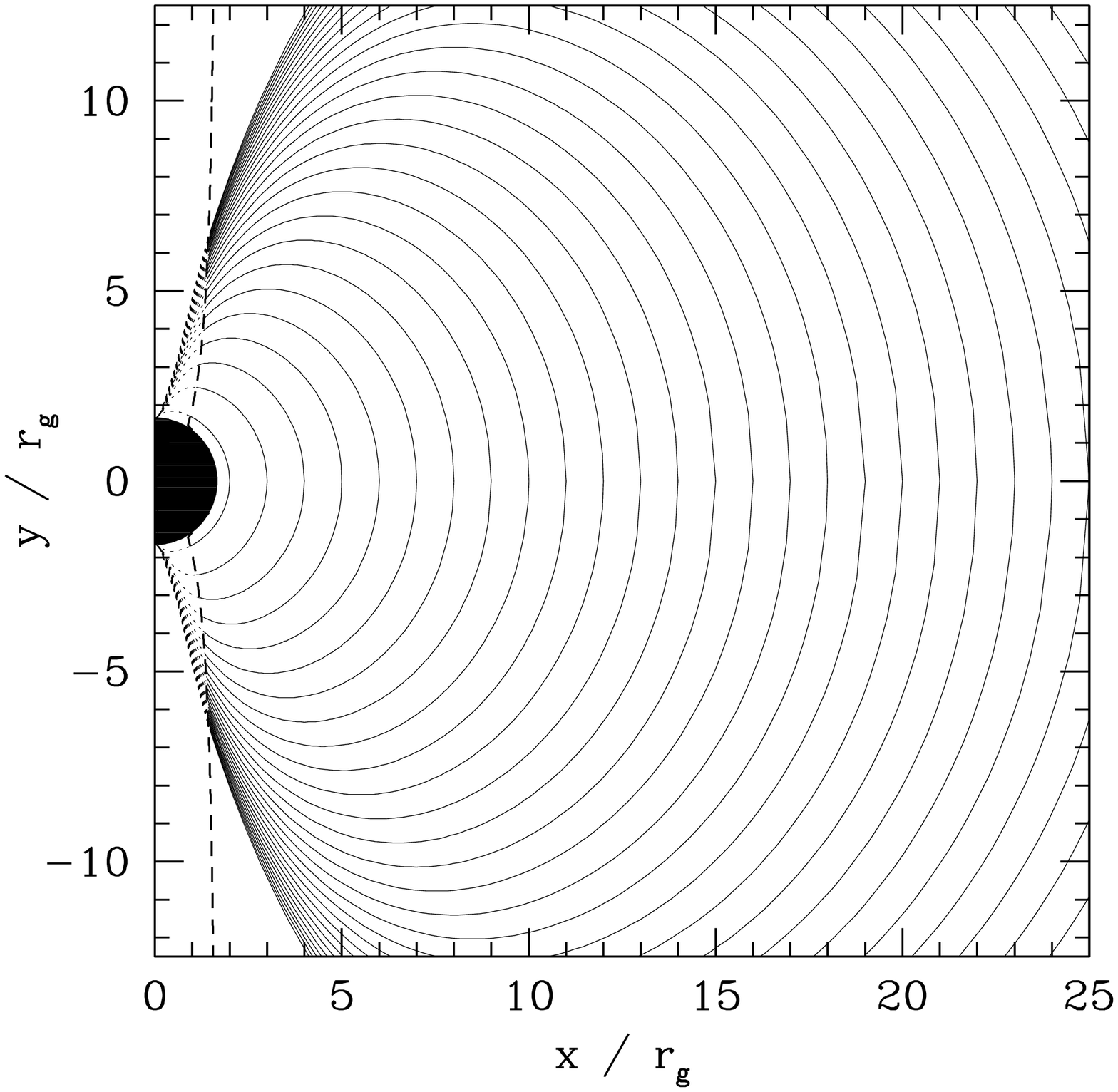,width=6.25cm}\\
\multicolumn{2}{c}{\textbf{Case (3): $1=K_\mathrm{ms}=K_\mathrm{mb} < K < K_\mathrm{max}$} 
(here $K= 3.54$; $r_\mathrm{cusp}=6 $; centre at $\infty$)}\\
\psfig{file=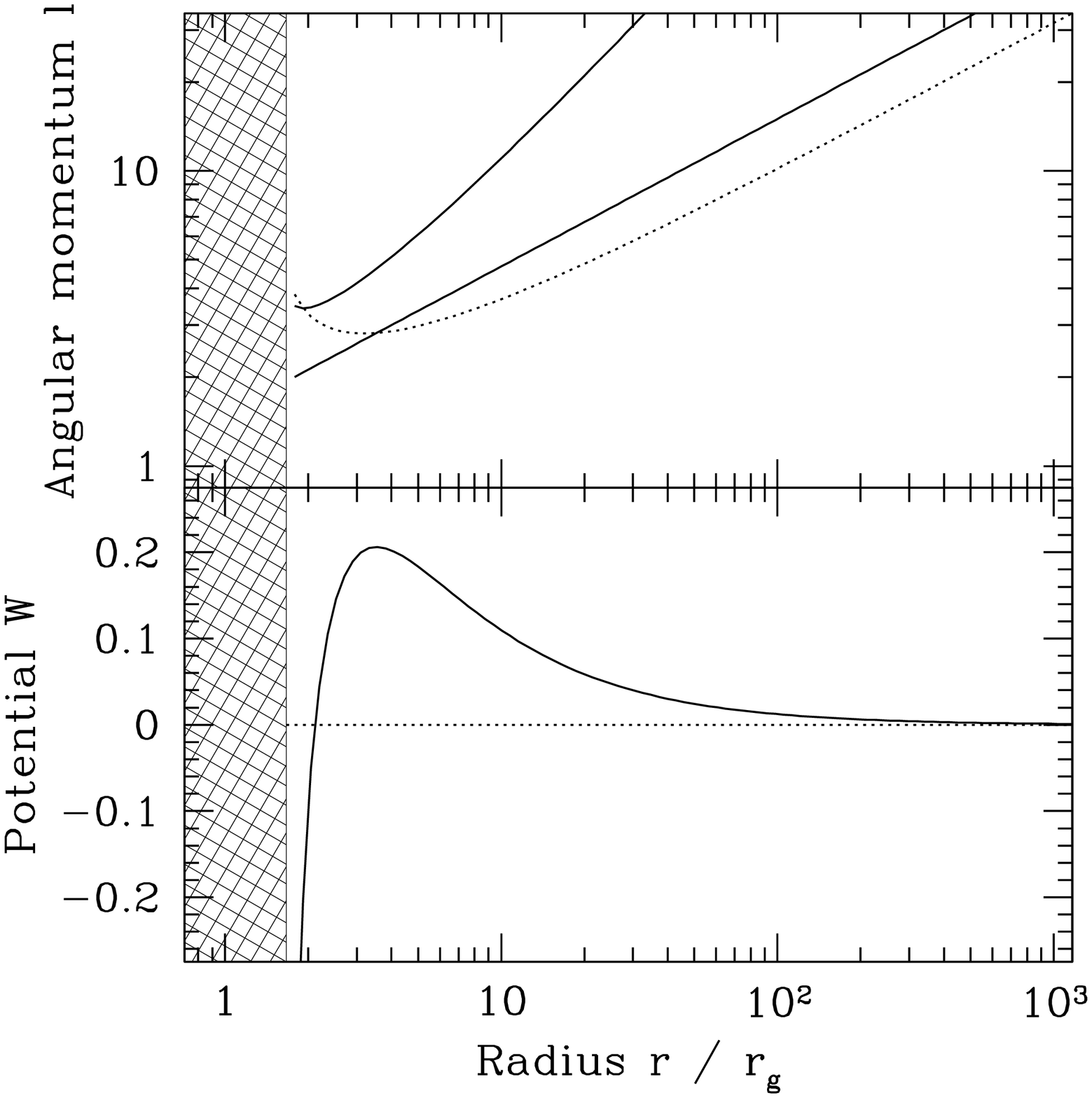,width=6.25cm} &
\psfig{file=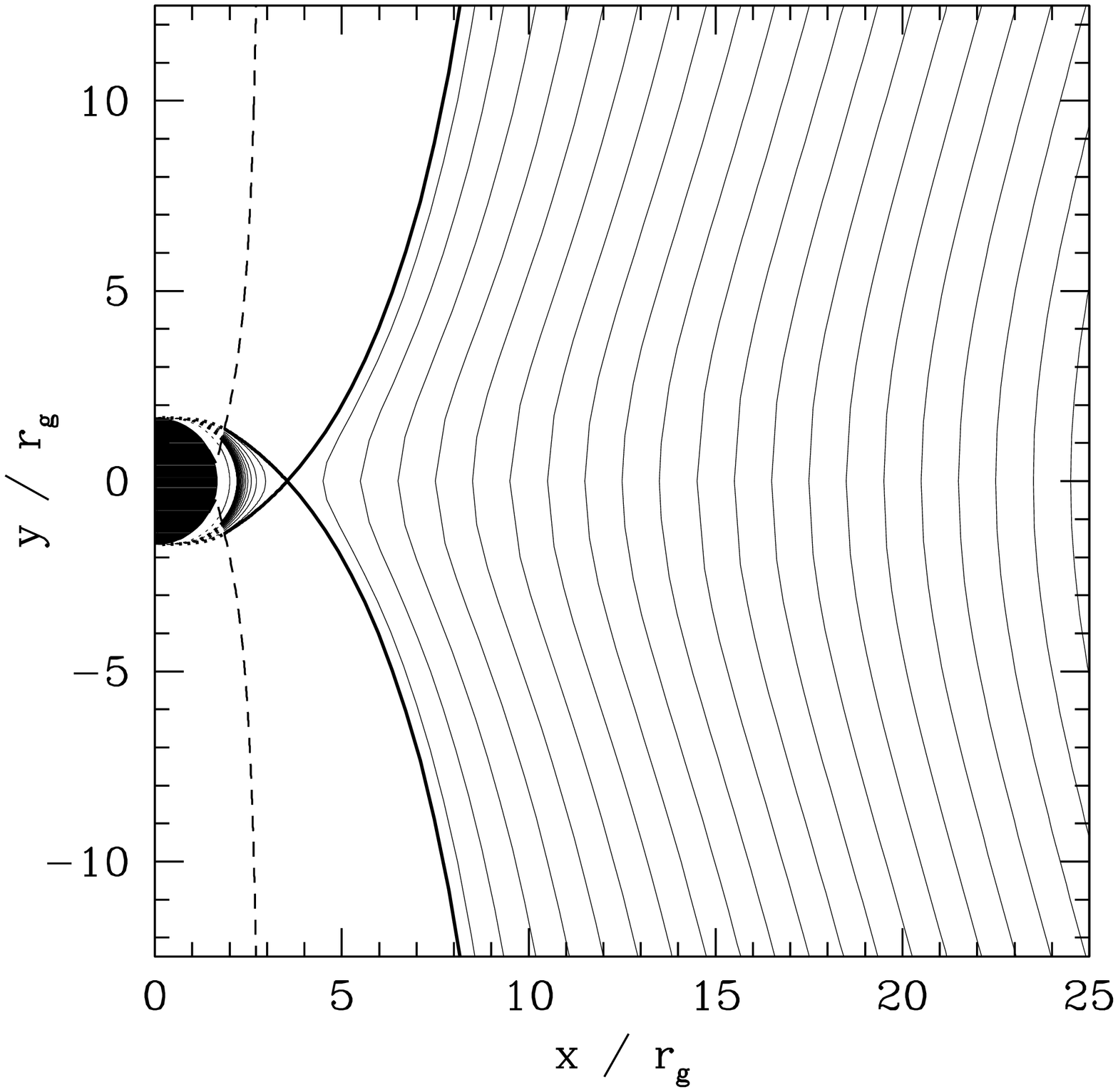,width=6.25cm} \\
\\
\multicolumn{2}{c}{\underline{\textbf{Super-Keplerian case (here with $\alpha=3/4$)}}}\\
\multicolumn{2}{c}{\textbf{Case (3): $K < K_\mathrm{max}$} (here $K= 1$; 
$r_\mathrm{cusp}\simeq 4.08$; centre at $\infty$)}\\
\psfig{file=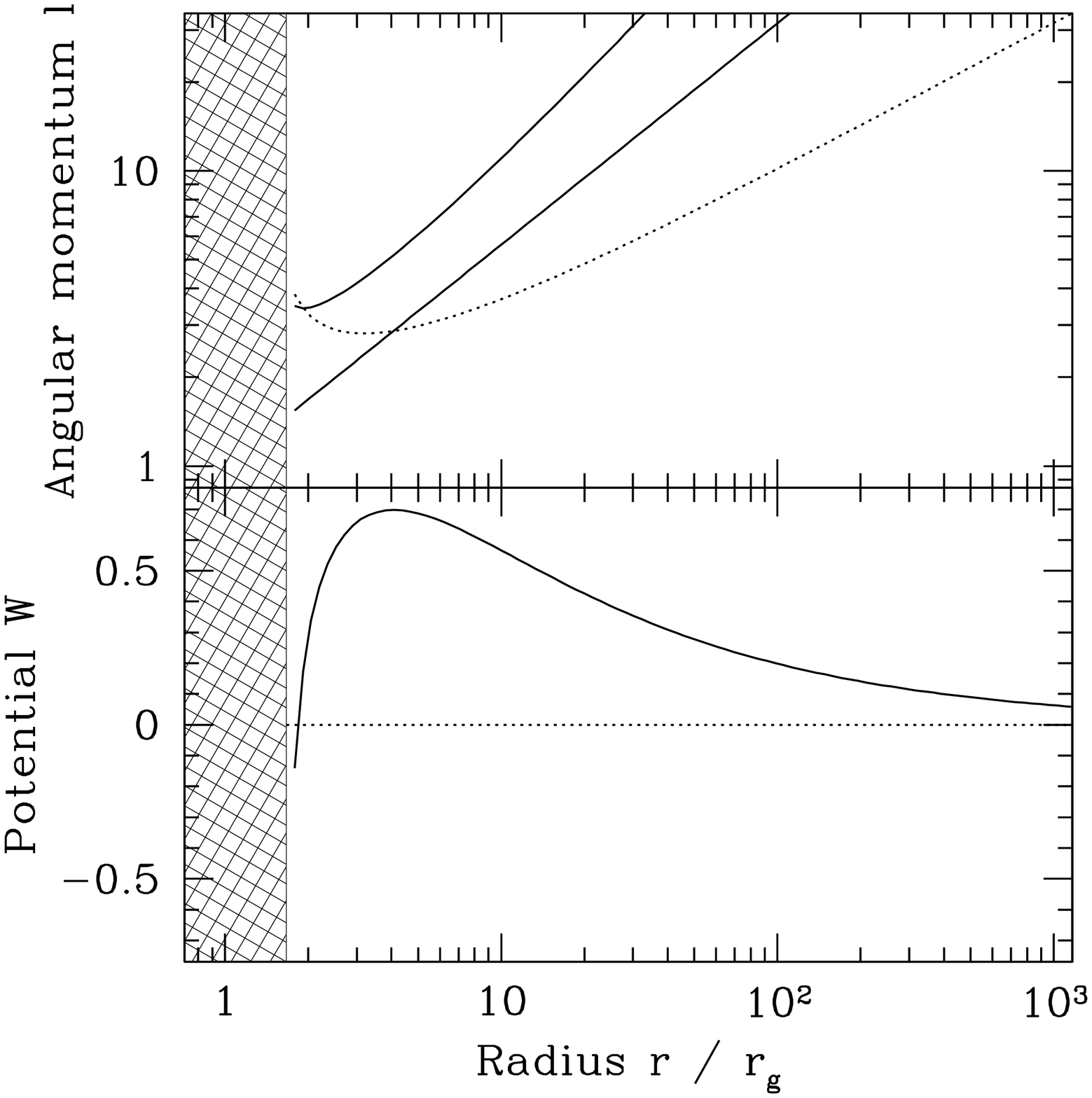,width=6.25cm} &
\psfig{file=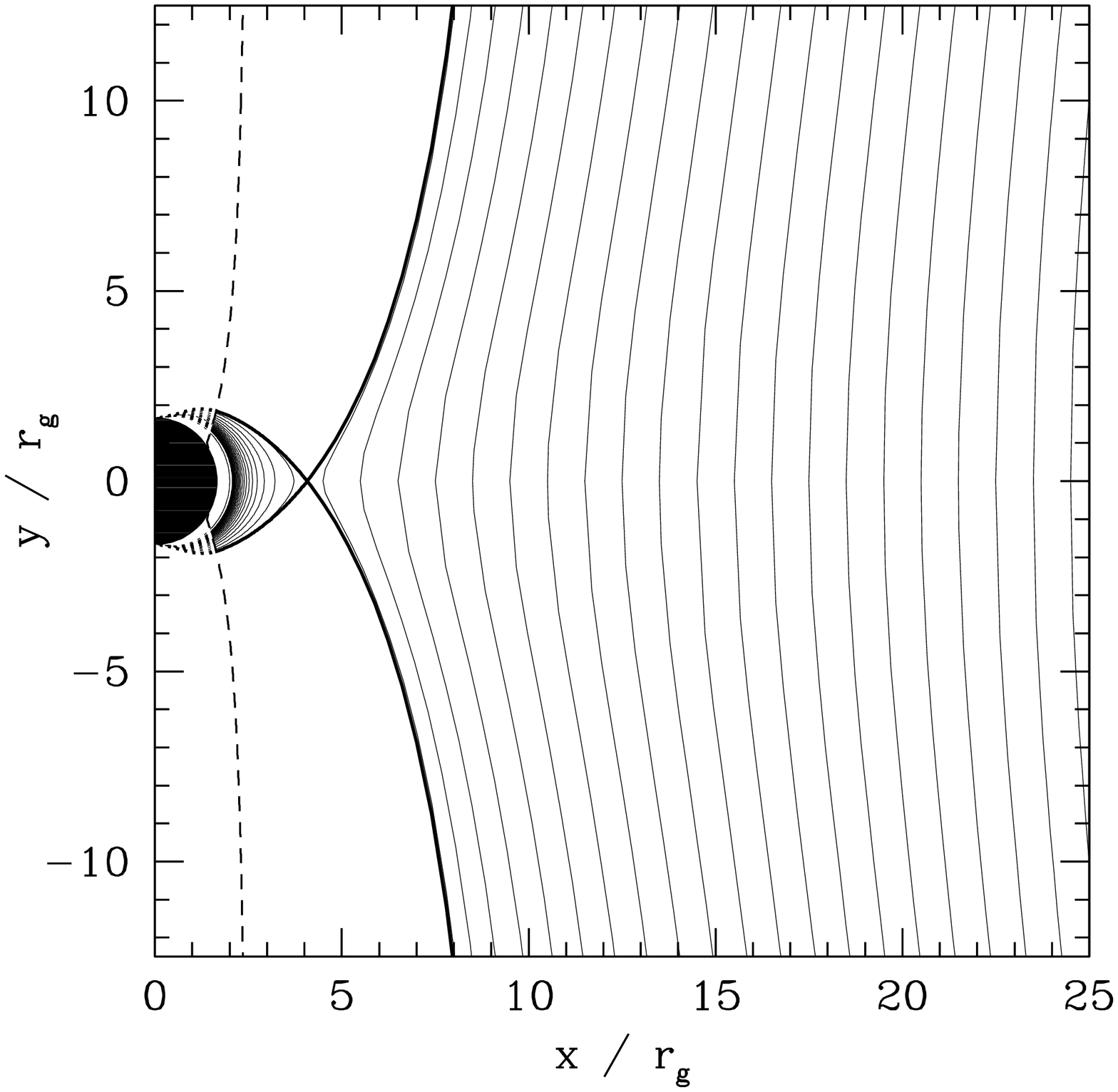,width=6.25cm}\\
\end{tabular}
\end{center}
\contcaption{\textbf{Geometry of the equipotentials: mildly rotating Kerr black hole ($a=\sqrt{5}/3$) 
and prograde disc. Keplerian and super-Keplerian cases.}}
\end{figure*}



\begin{figure*}
\begin{center}
\begin{tabular}{cc}
\multicolumn{2}{c}{\textbf{\textit{-- Extreme Kerr black hole}} ($a=1$) --}\\
\\
\multicolumn{2}{c}{\underline{\textbf{Sub-Keplerian case (here with $\alpha=1/4$)}}}\\
\multicolumn{2}{c}{\textbf{Case (1): $K<K_\mathrm{ms}$} (here $K= 1.5$; no cusp; no centre)}\\
\psfig{file=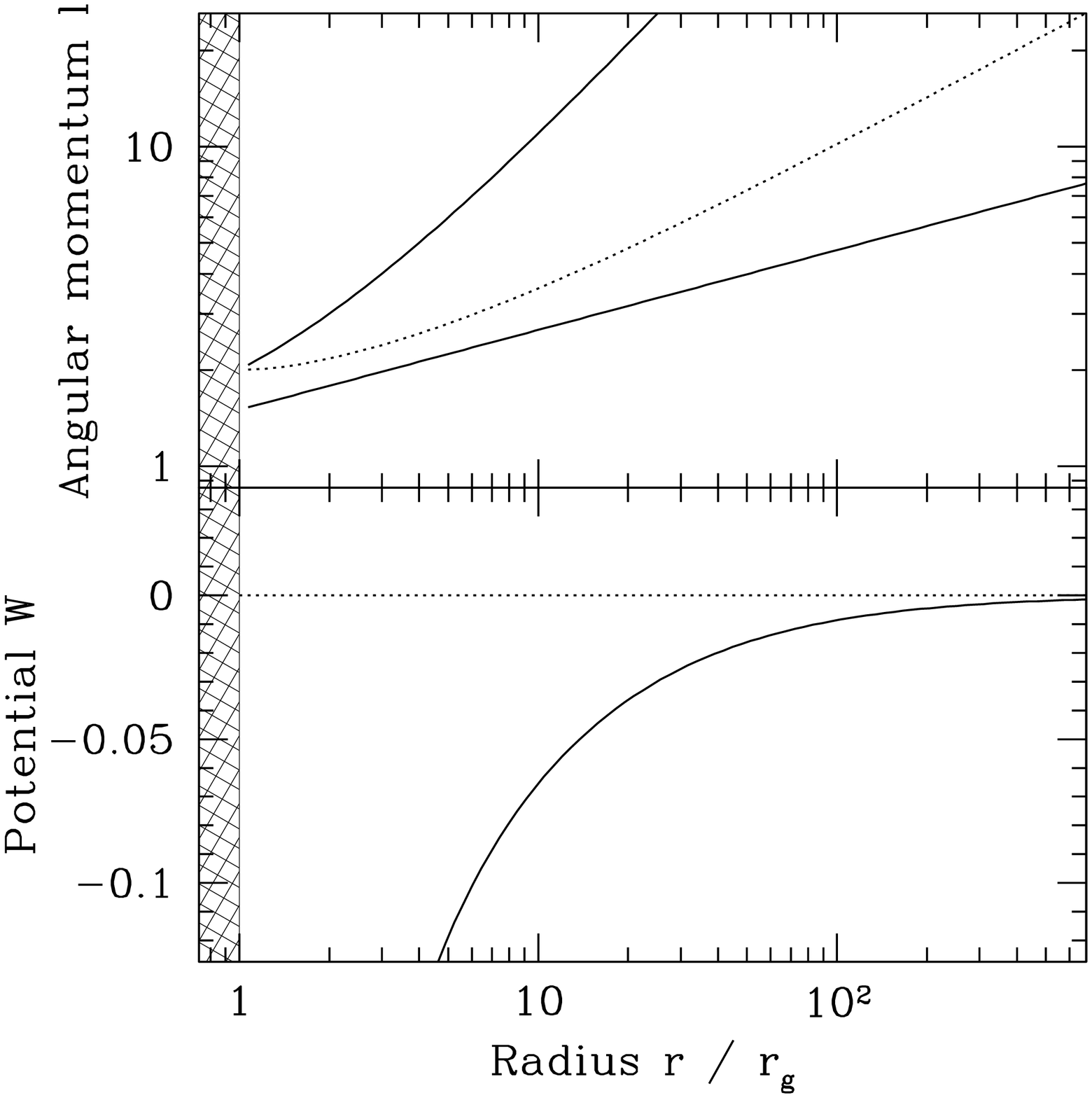,width=6.25cm} &
\psfig{file=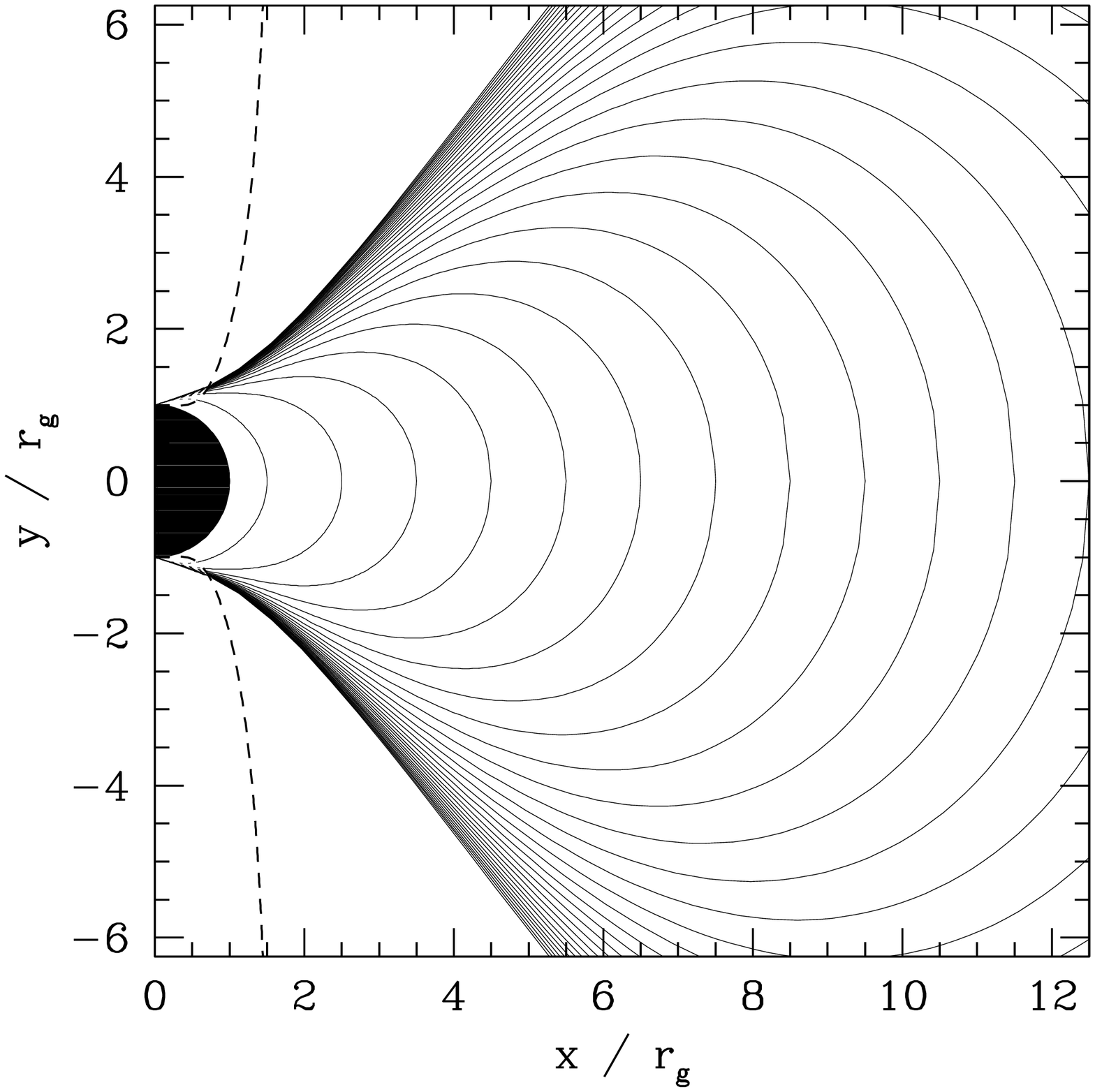,width=6.25cm}\\
\multicolumn{2}{c}{\textbf{Critical case: $K=K_\mathrm{ms}$} (here $K_\mathrm{ms}\simeq  1.82$; 
$r_\mathrm{cusp}=r_\mathrm{centre} \simeq 2.62$)}\\
\psfig{file=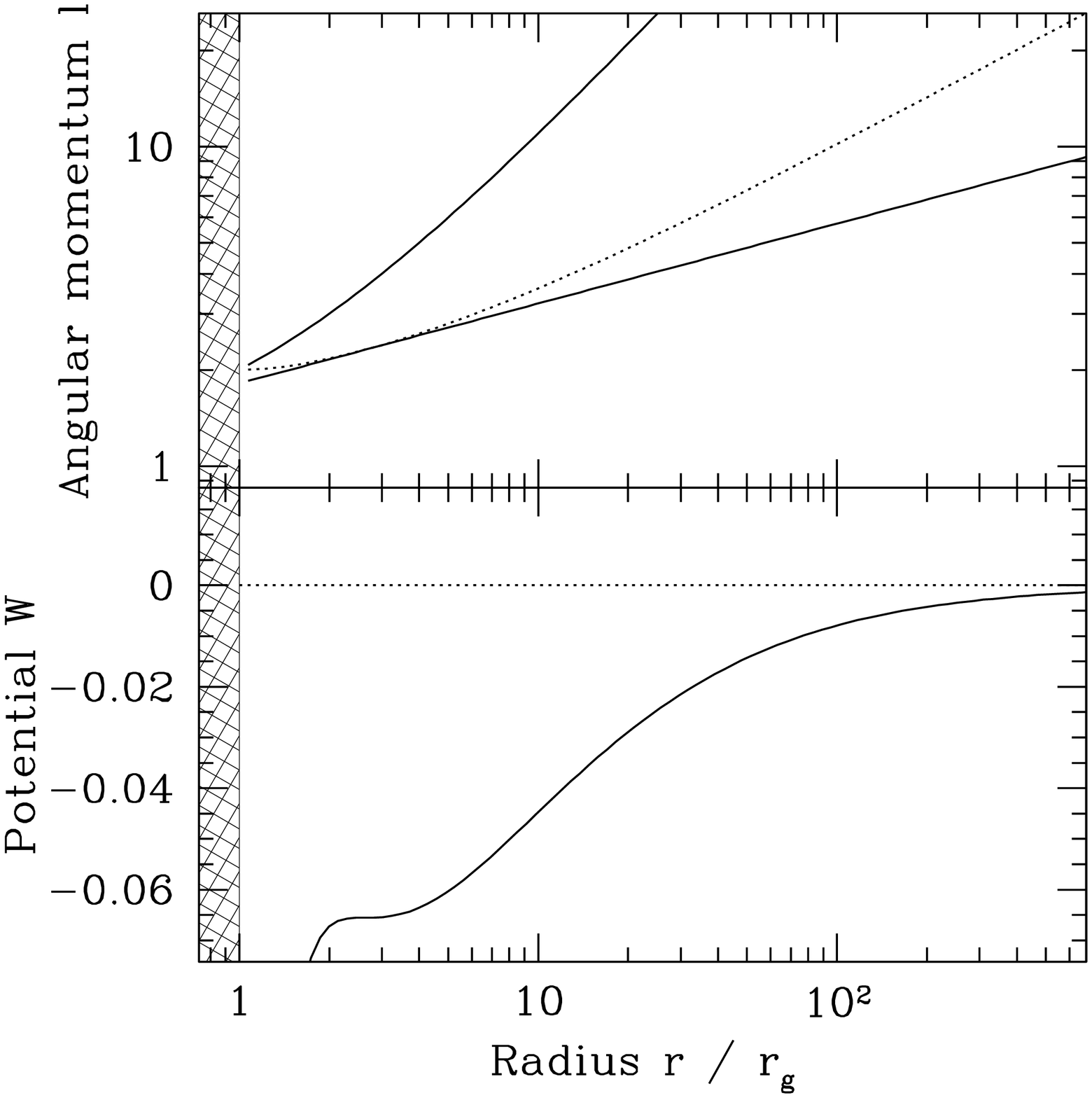,width=6.25cm} &
\psfig{file=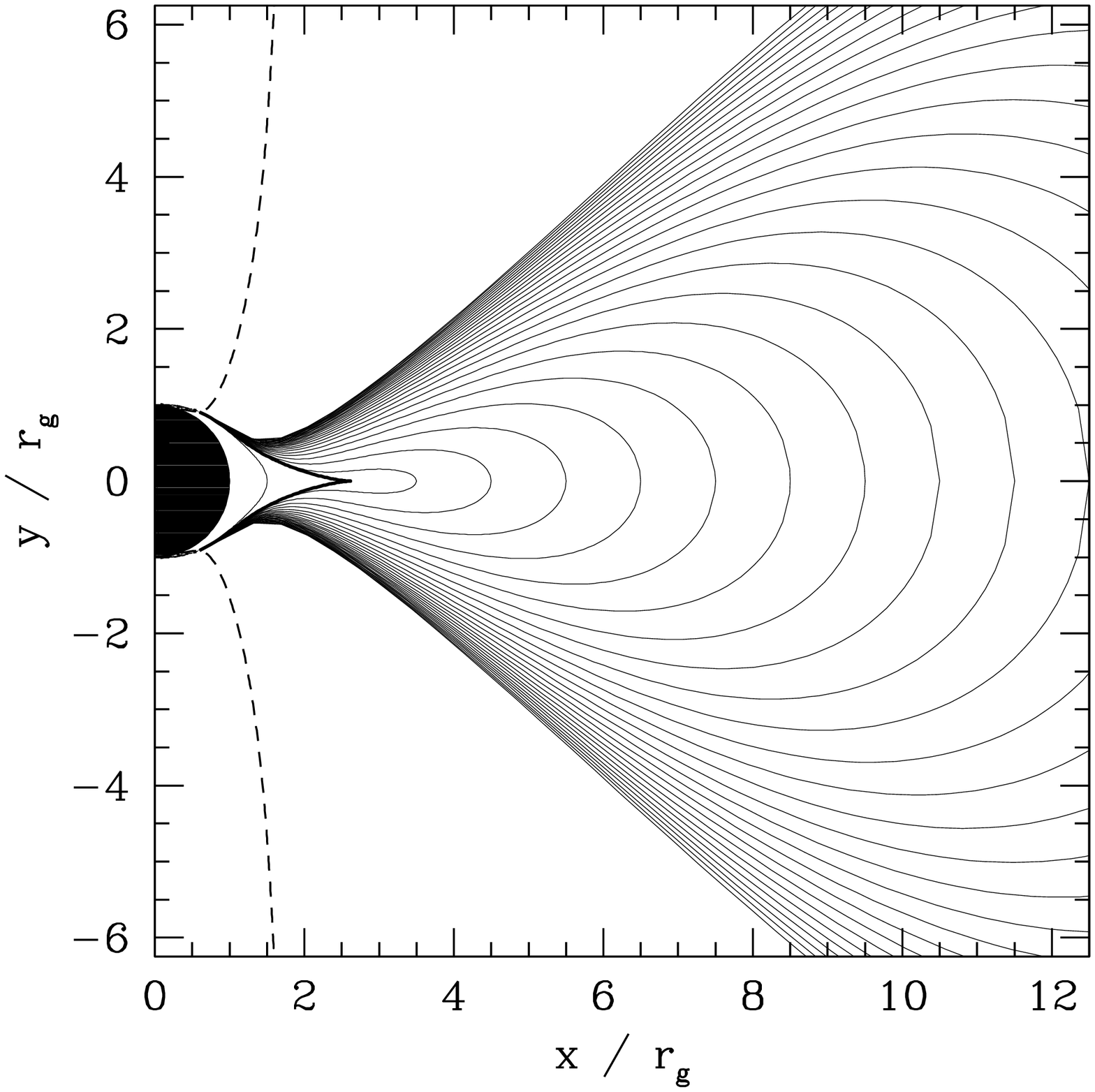,width=6.25cm}\\
\multicolumn{2}{c}{\textbf{Case (2): $K_\mathrm{ms}<K<K_\mathrm{mb}$} (here $K\simeq 1.84$; 
$r_\mathrm{cusp}\simeq 1.76$ ; $r_\mathrm{centre}\simeq 4.03$)}\\
\psfig{file=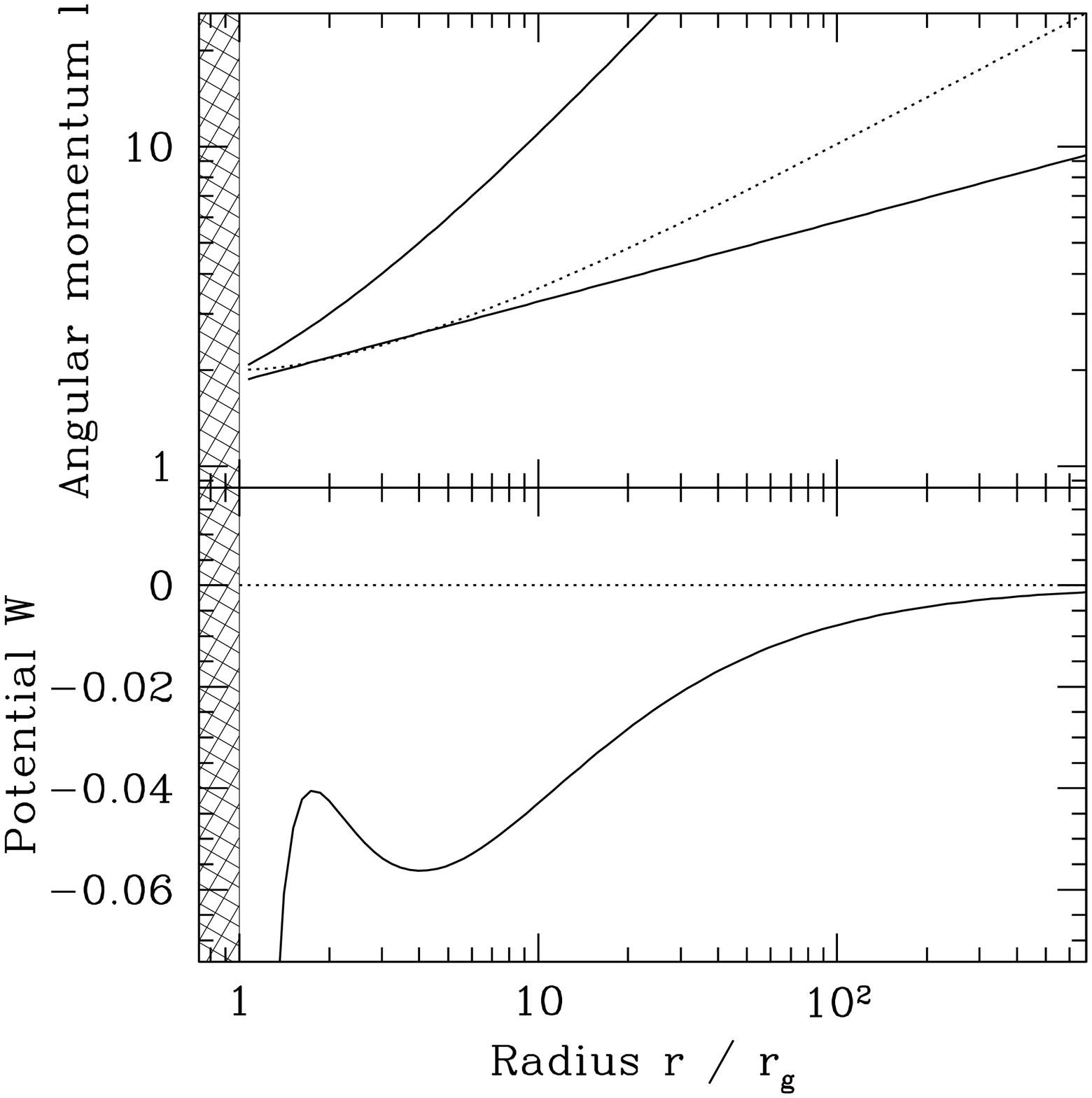,width=6.25cm} &
\psfig{file=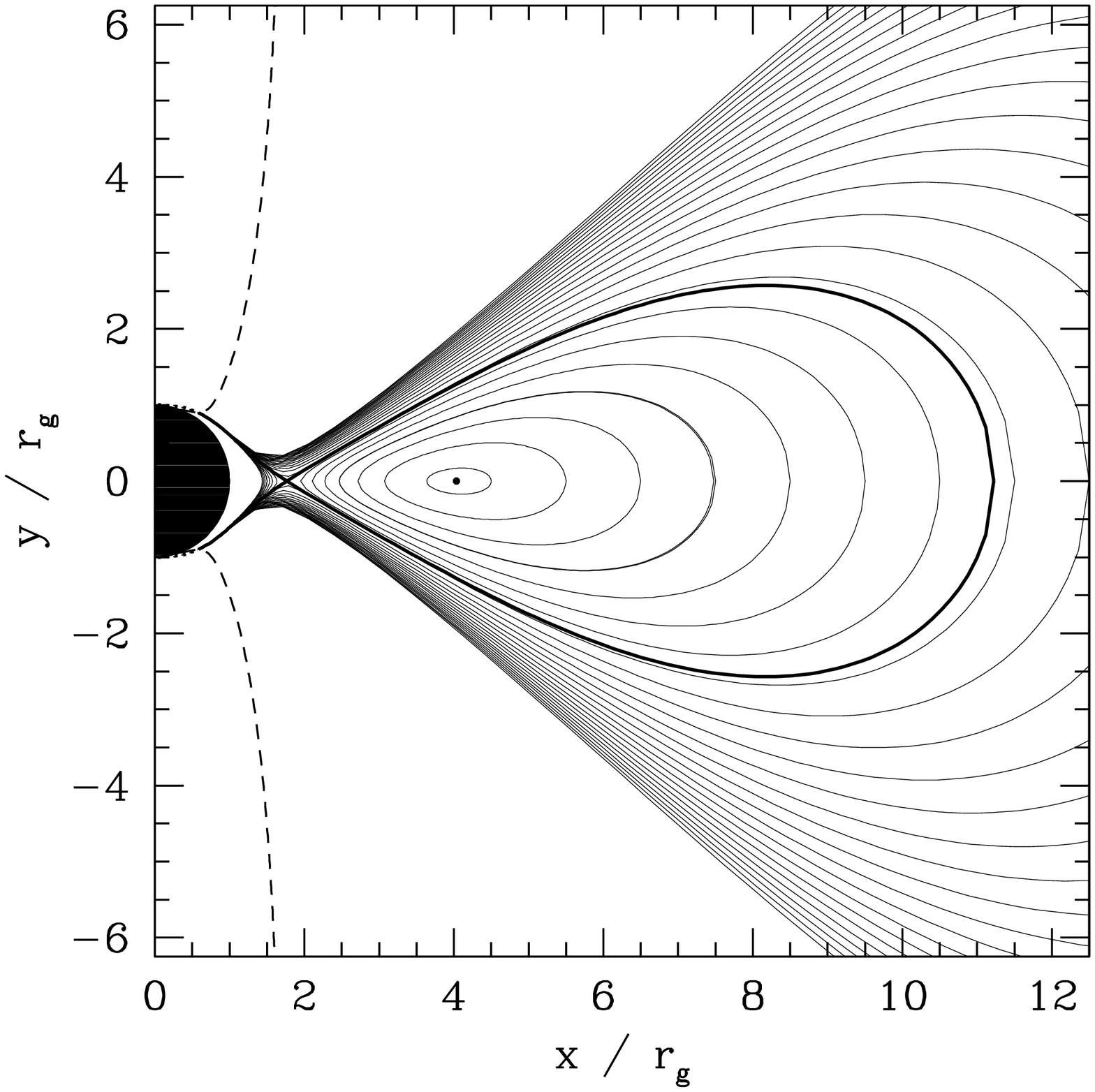,width=6.25cm}\\
\end{tabular}
\end{center}
\caption{\textbf{Geometry of the equipotentials: extreme Kerr black hole ($a=1$) and prograde disc. 
Sub-Keplerian case.} All line styles and notations are identical to those used in 
Fig.~\ref{fig:equi1pro}. Note the scale change in the plots on the right column as
compared to Figs.~\ref{fig:equi1pro} and \ref{fig:equi2pro}.}

\label{fig:equi3pro}
\end{figure*}

\begin{figure*}
\begin{center}
\begin{tabular}{cc}
\multicolumn{2}{c}{\textbf{\textit{-- Extreme Kerr black hole}} ($a=1$) --}\\
\\
\multicolumn{2}{c}{\underline{\textbf{Sub-Keplerian case (here with $\alpha=1/4$)}}}\\
\multicolumn{2}{c}{\textbf{Critical case: $K=K_\mathrm{mb}$} (here $K_\mathrm{mb}\simeq 1.86$; 
$r_\mathrm{cusp}\simeq 1.52$ ; $r_\mathrm{centre}\simeq 4.87$)}\\
\psfig{file=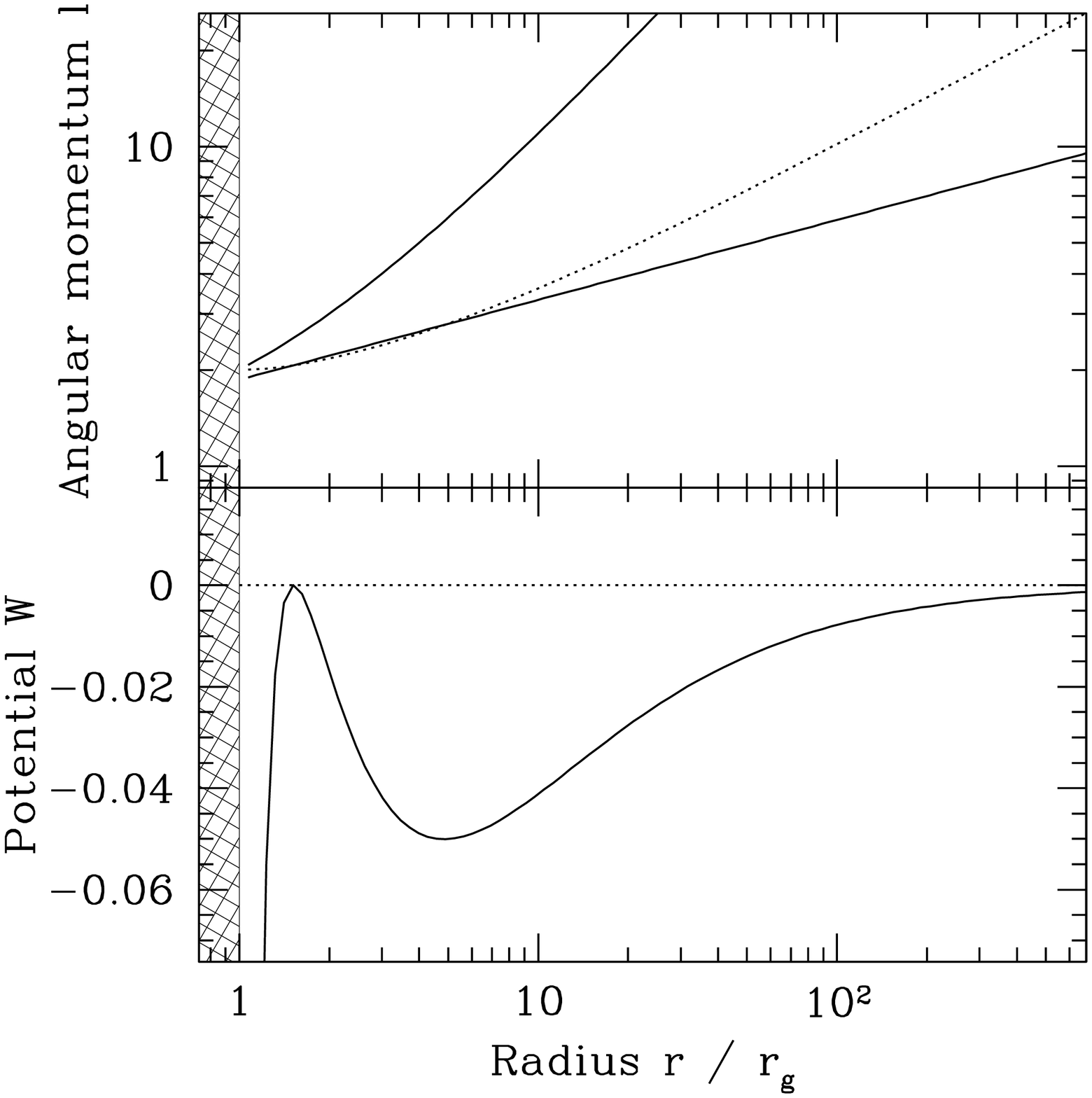,width=6.25cm} &
\psfig{file=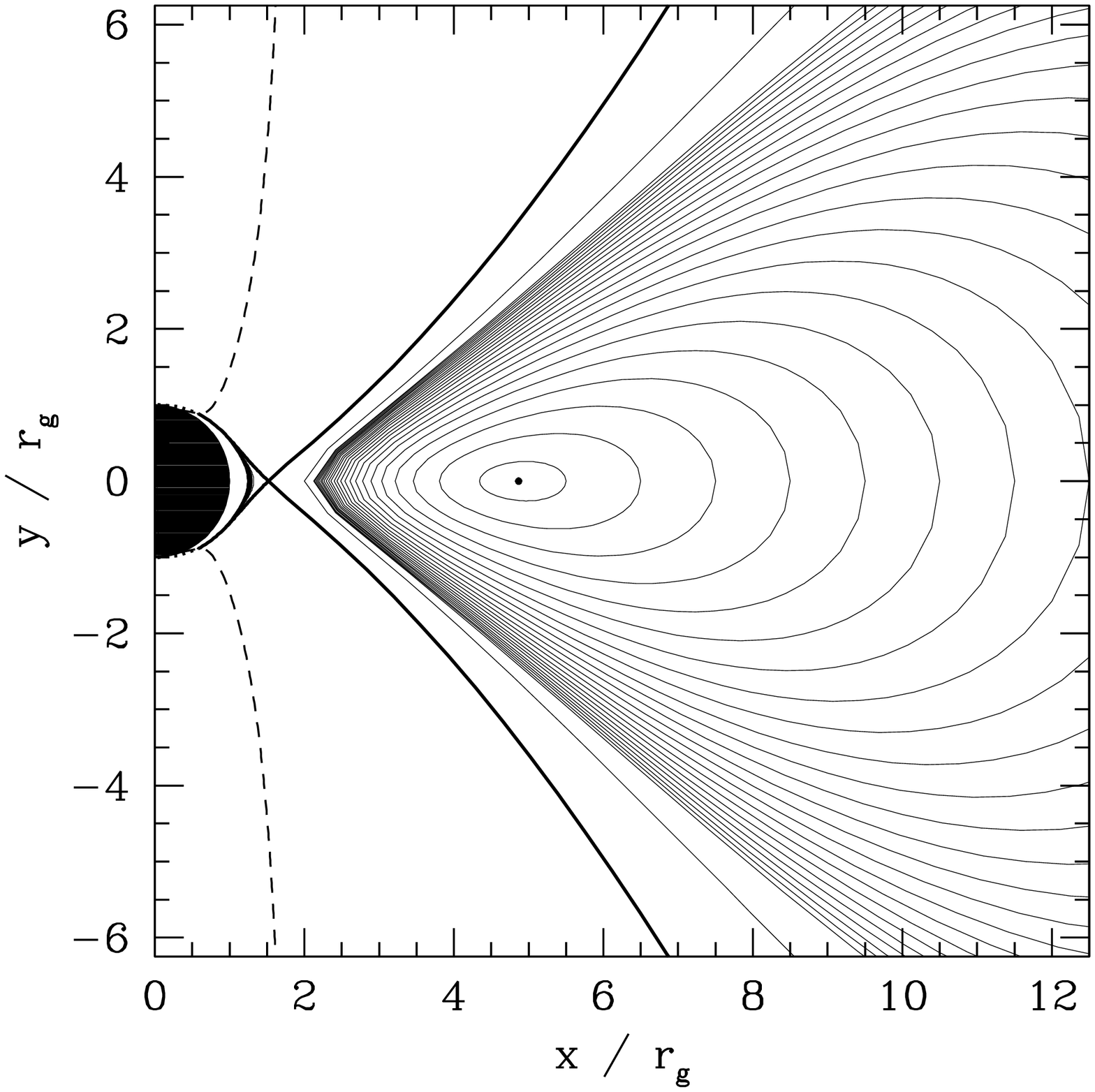,width=6.25cm}\\
\multicolumn{2}{c}{\textbf{Case (3): $K_\mathrm{mb} < K < K_\mathrm{max}$} (here $K= 1.87$; 
$r_\mathrm{cusp}\simeq 1.47$ ; $r_\mathrm{centre}\simeq 5.07$)}\\
\psfig{file=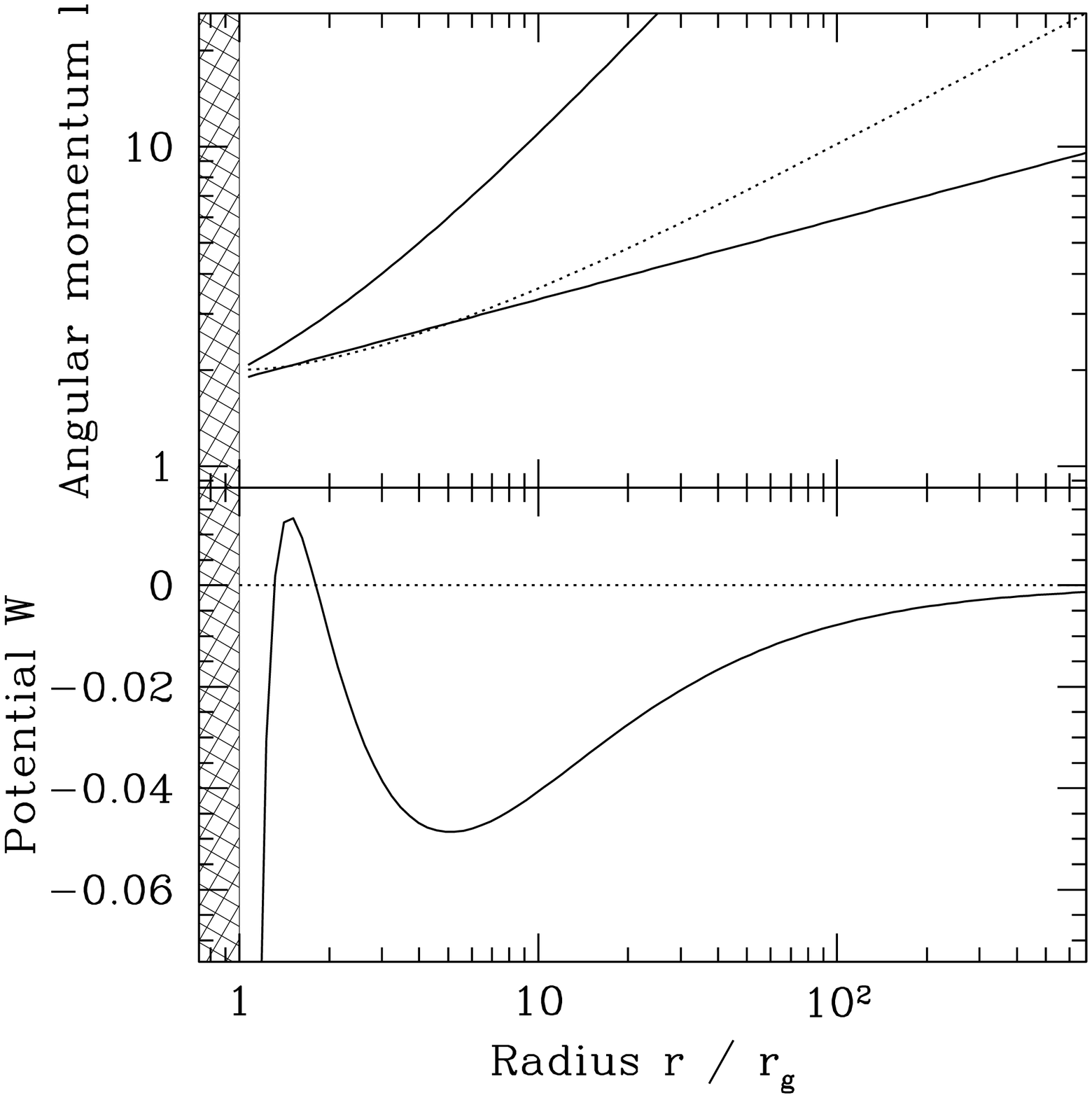,width=6.25cm} &
\psfig{file=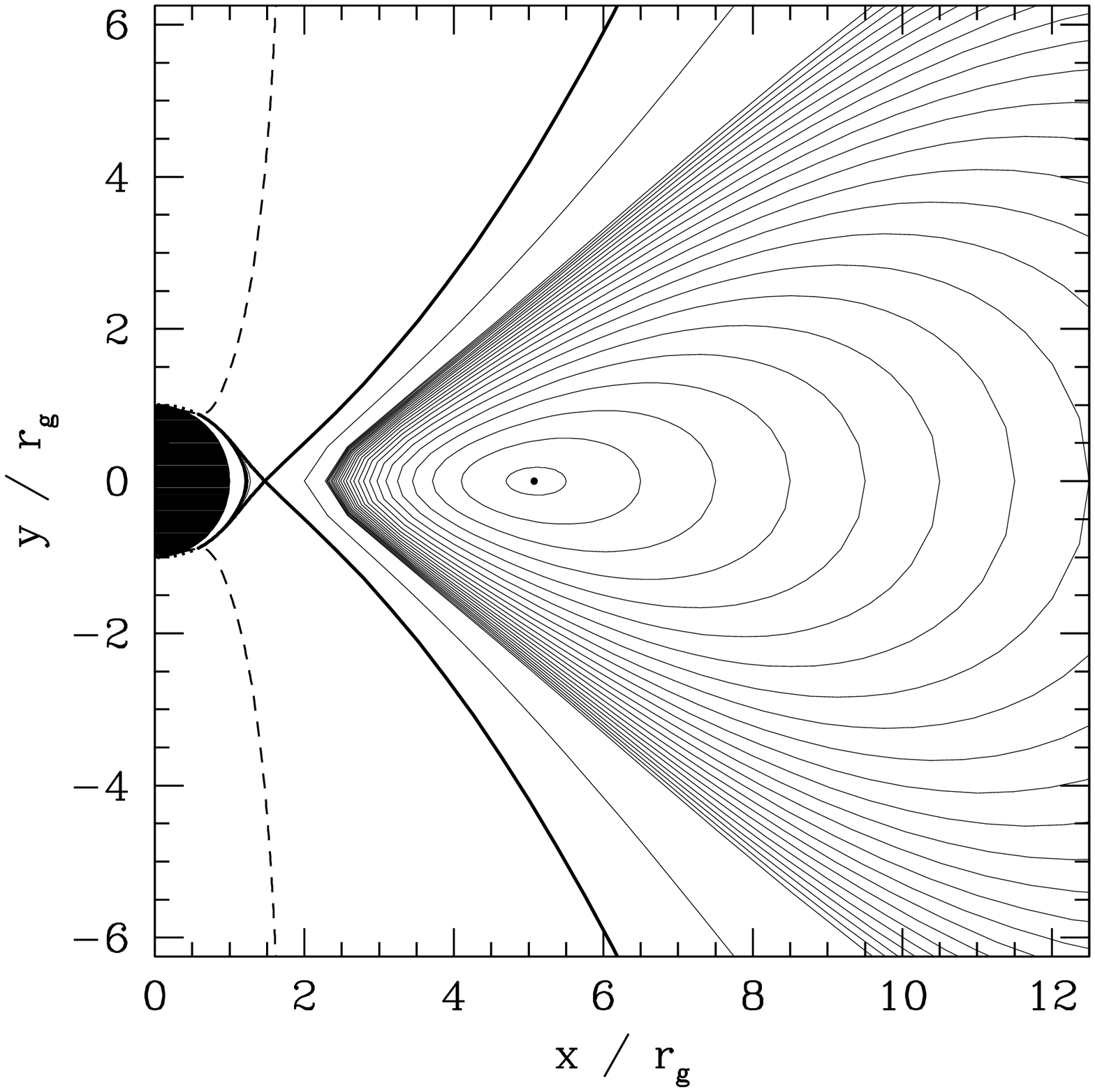,width=6.25cm}\\
\multicolumn{2}{c}{\textbf{Critical case: $K = K_\mathrm{max}$} (here $K_\mathrm{max}=2.$; 
no cusp; $r_\mathrm{centre}\simeq 9.18$)}\\
\psfig{file=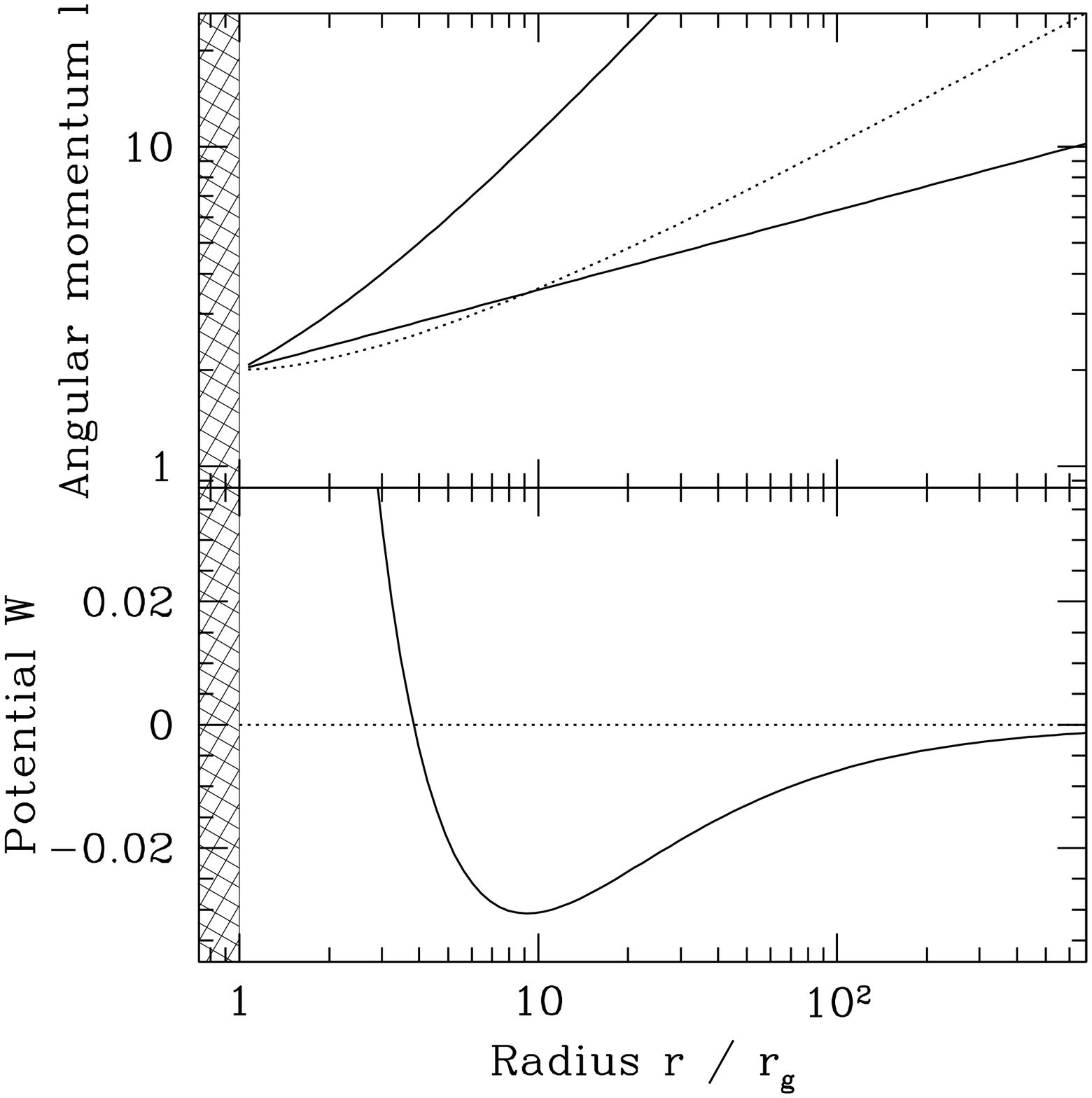,width=6.25cm} &
\psfig{file=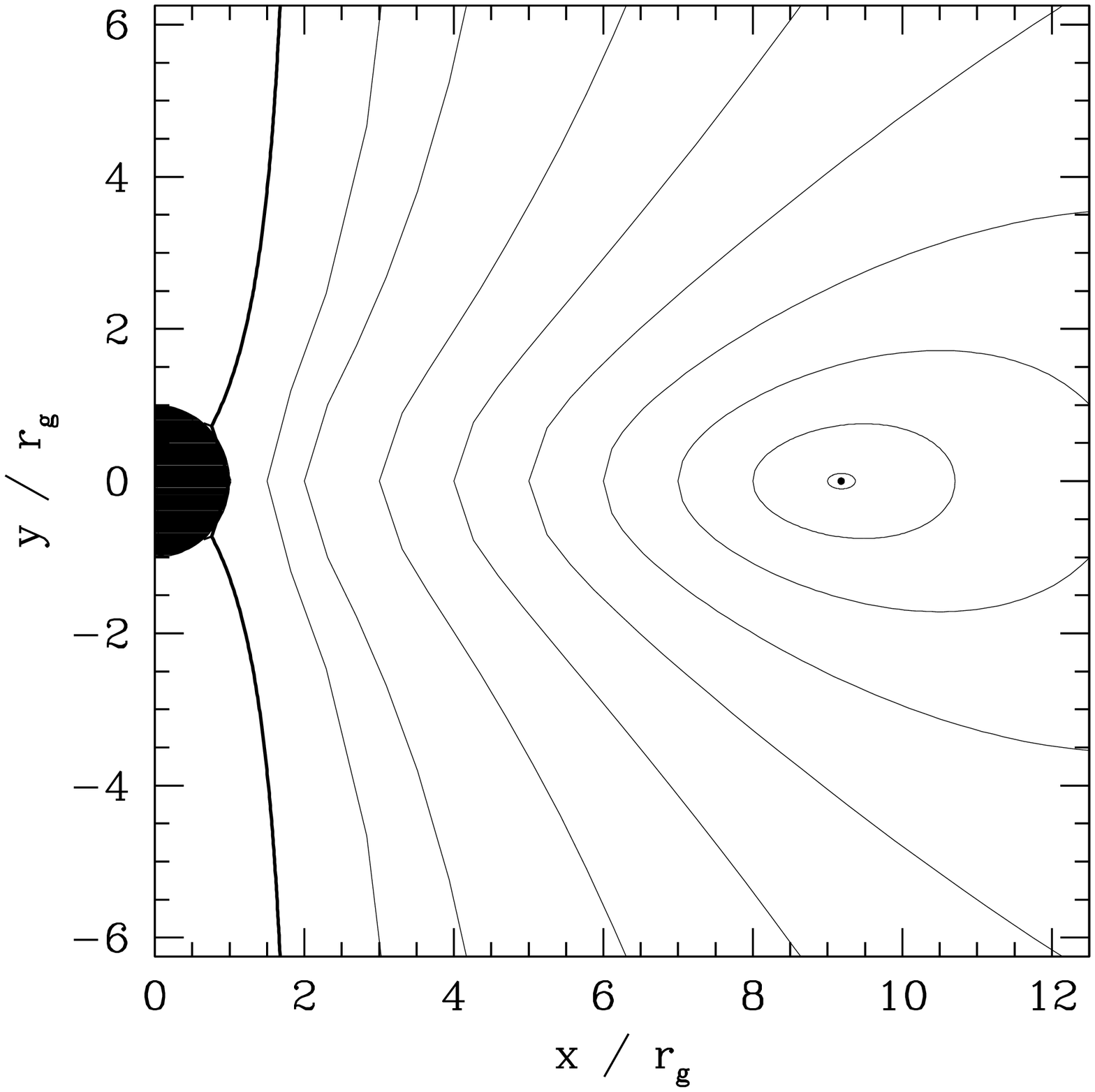,width=6.25cm}\\
\end{tabular}
\end{center}
\contcaption{\textbf{Geometry of the equipotentials: extreme Kerr black hole ($a=1$) and prograde 
disc. Sub-Keplerian case.}}
\end{figure*}

\begin{figure*}
\begin{center}
\begin{tabular}{cc}
\multicolumn{2}{c}{\textbf{\textit{-- Extreme Kerr black hole}} ($a=1$) --}\\
\\
\multicolumn{2}{c}{\underline{\textbf{Keplerian case ($\alpha=1/2$)}}}\\
\multicolumn{2}{c}{\textbf{Case (1): $K<K_\mathrm{ms}=K_\mathrm{mb}=1$} (here $K=0.5$; 
no cusp; no centre)}\\
\psfig{file=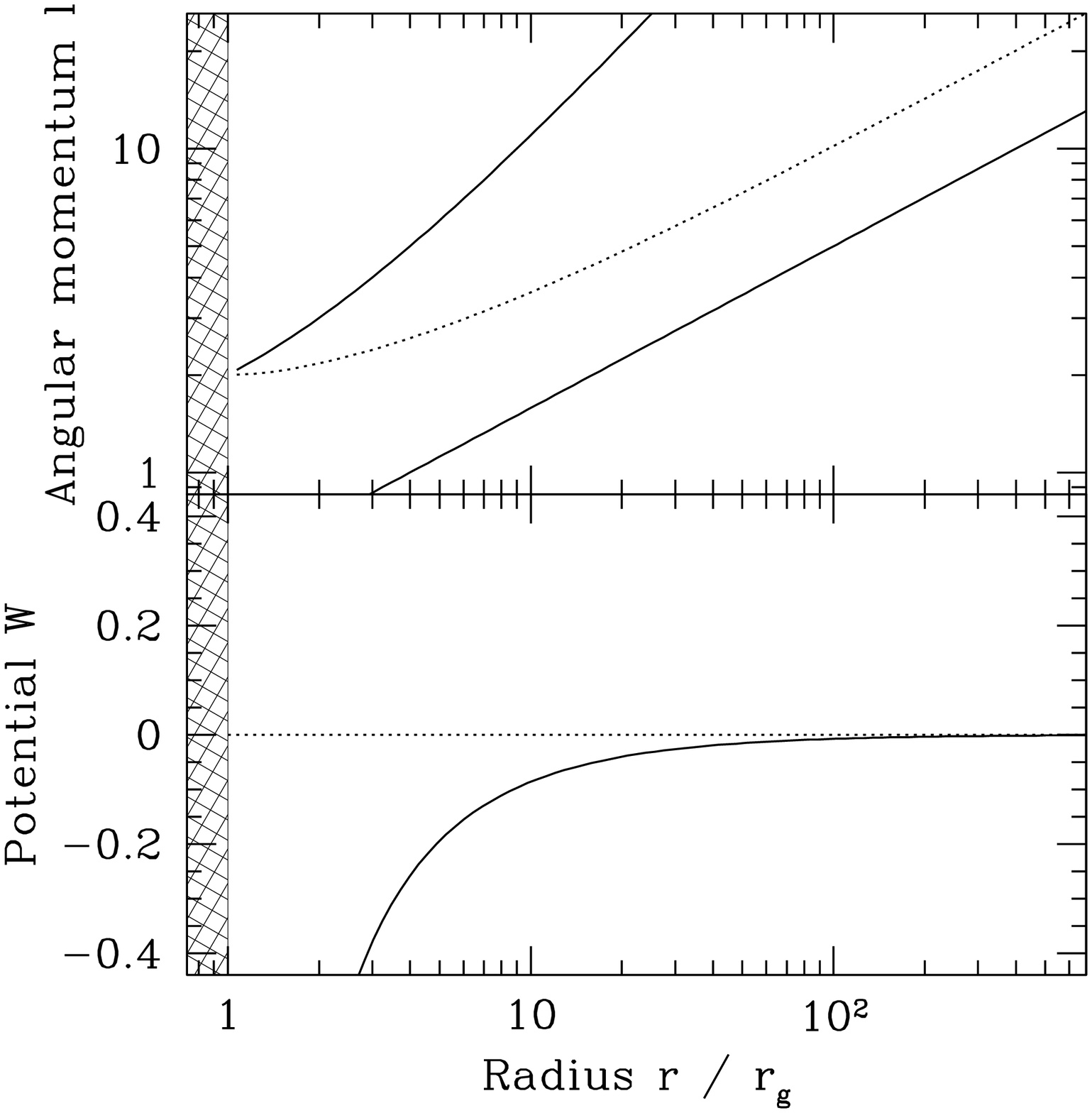,width=6.25cm} &
\psfig{file=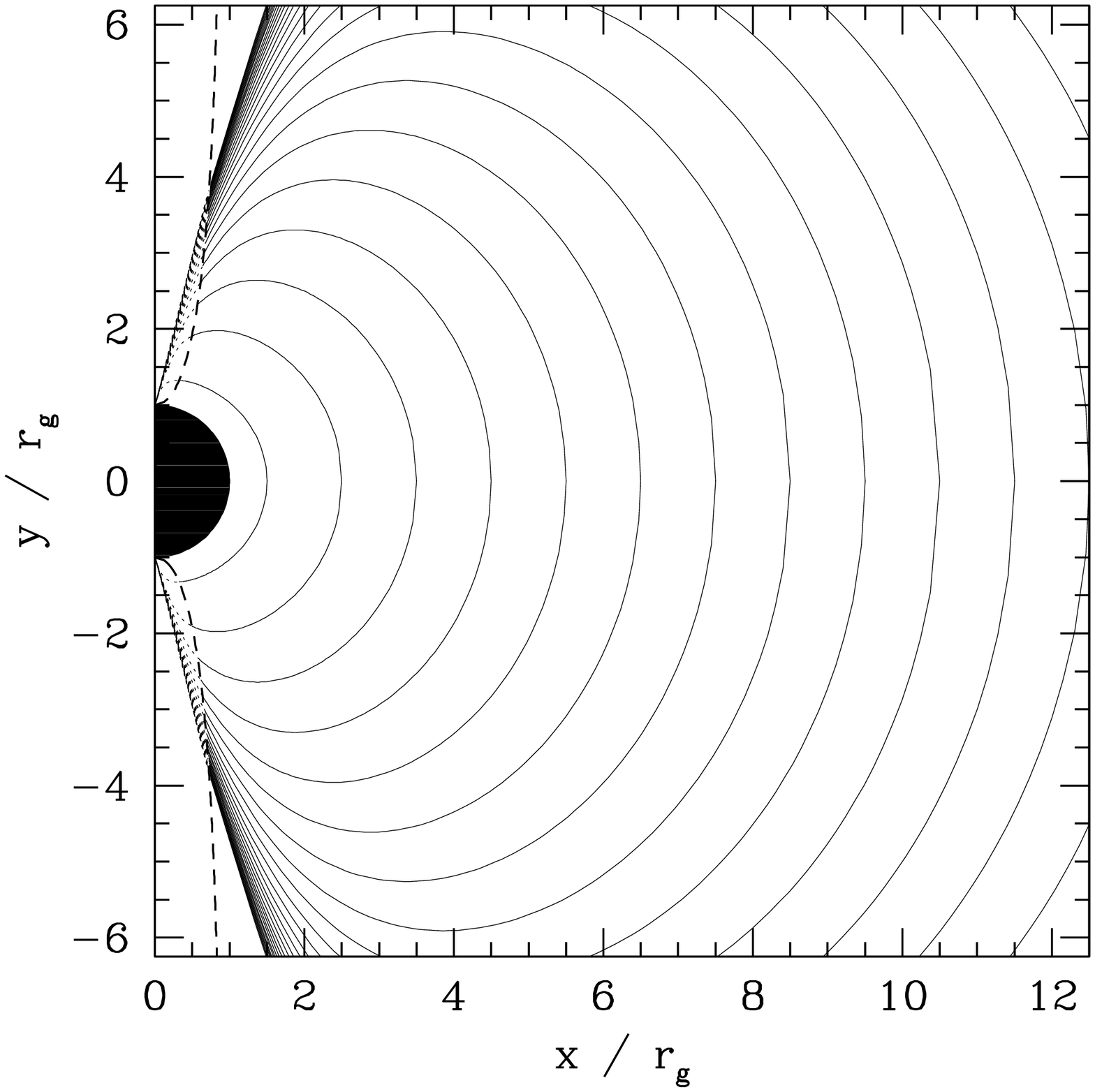,width=6.25cm}\\
\multicolumn{2}{c}{\textbf{Case (3): $1=K_\mathrm{ms}=K_\mathrm{mb} < K < K_\mathrm{max}$} 
(here $K= 3.54$; $r_\mathrm{cusp} \simeq 2.17$; centre at $\infty$)}\\
\psfig{file=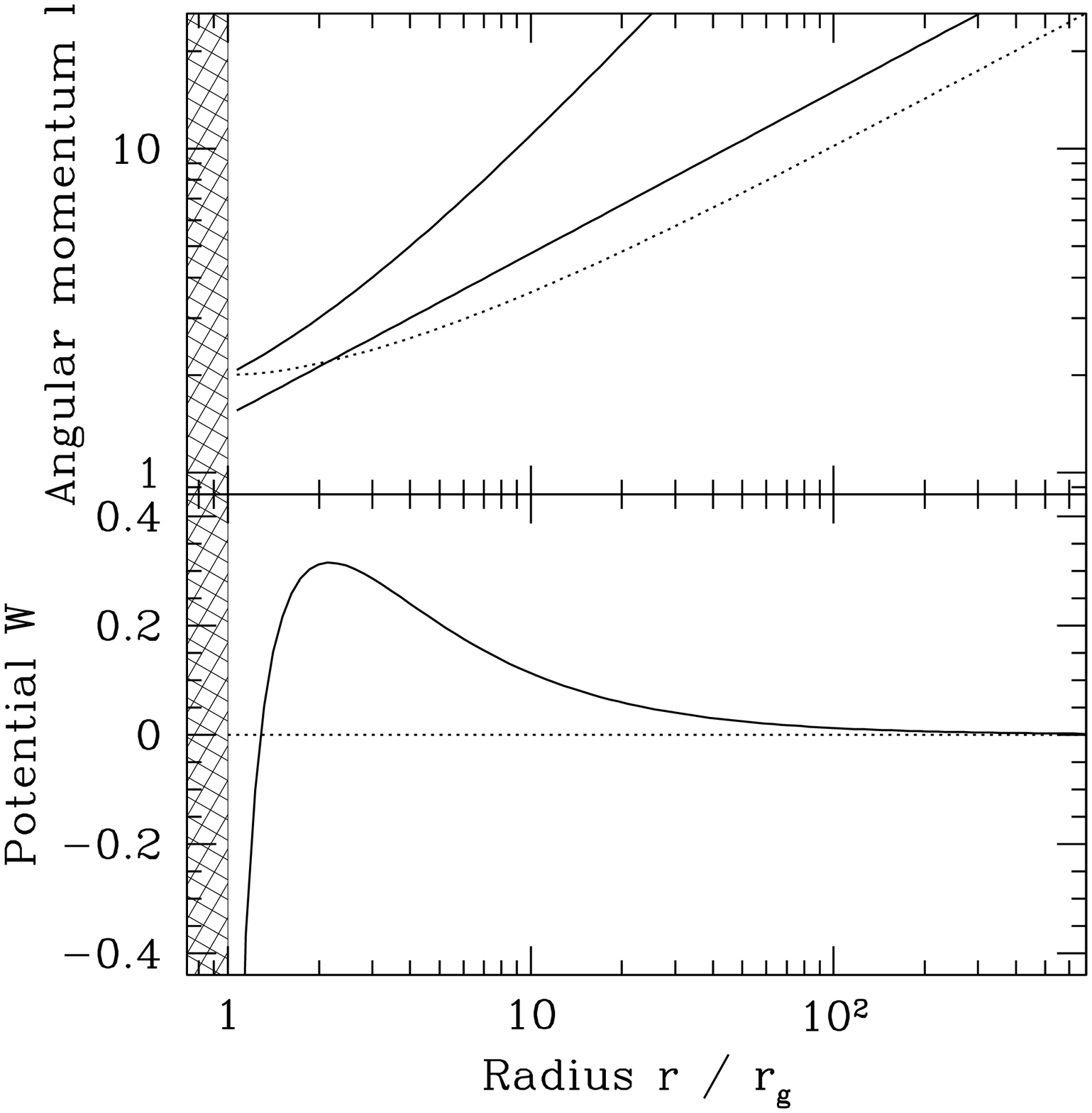,width=6.25cm} &
\psfig{file=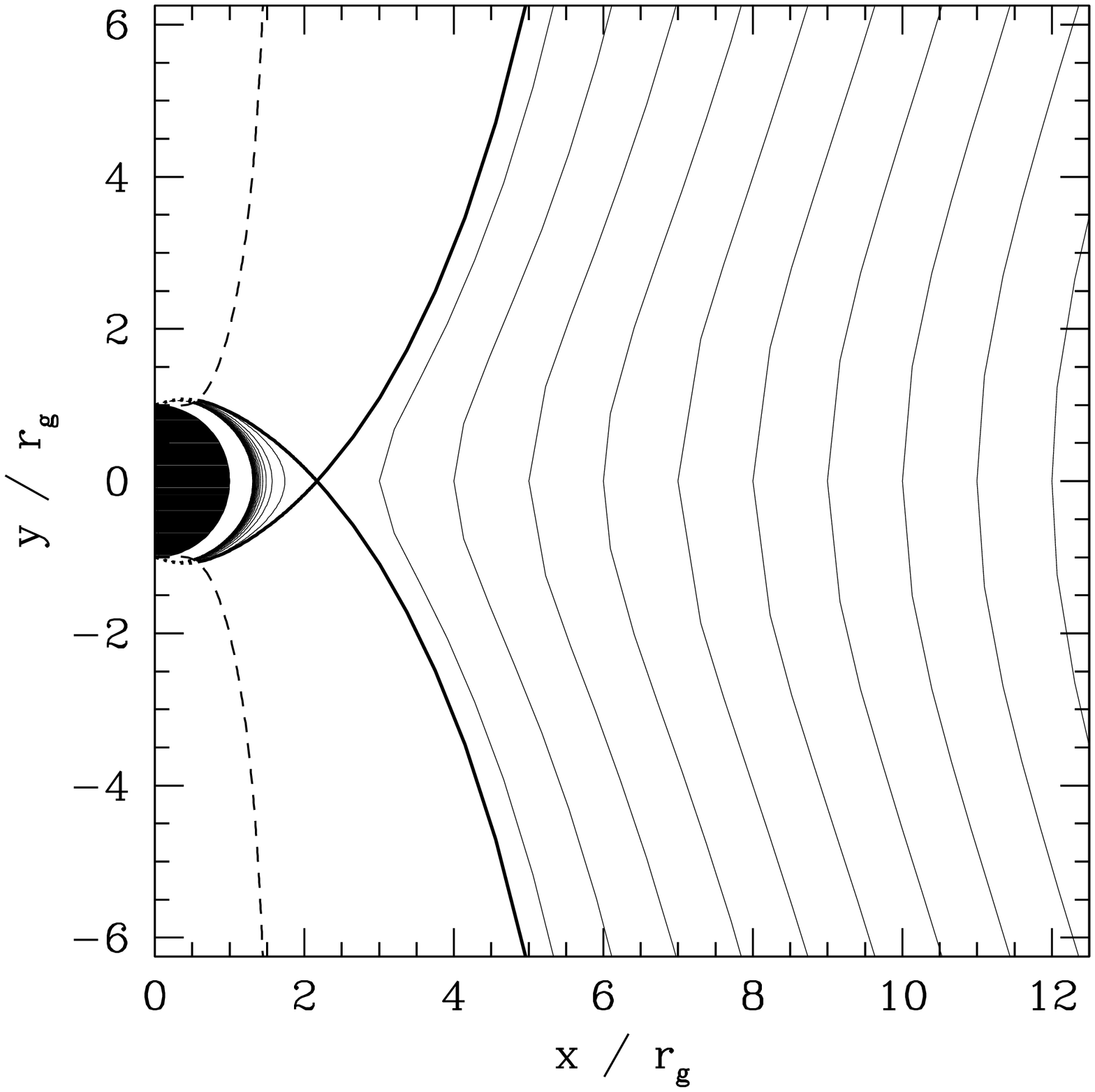,width=6.25cm} \\
\\
\multicolumn{2}{c}{\underline{\textbf{Super-Keplerian case (here with $\alpha=3/4$)}}}\\
\multicolumn{2}{c}{\textbf{Case (3): $K < K_\mathrm{max}$} (here $K= 1$; 
$r_\mathrm{cusp}\simeq 3.32$; centre at $\infty$)}\\
\psfig{file=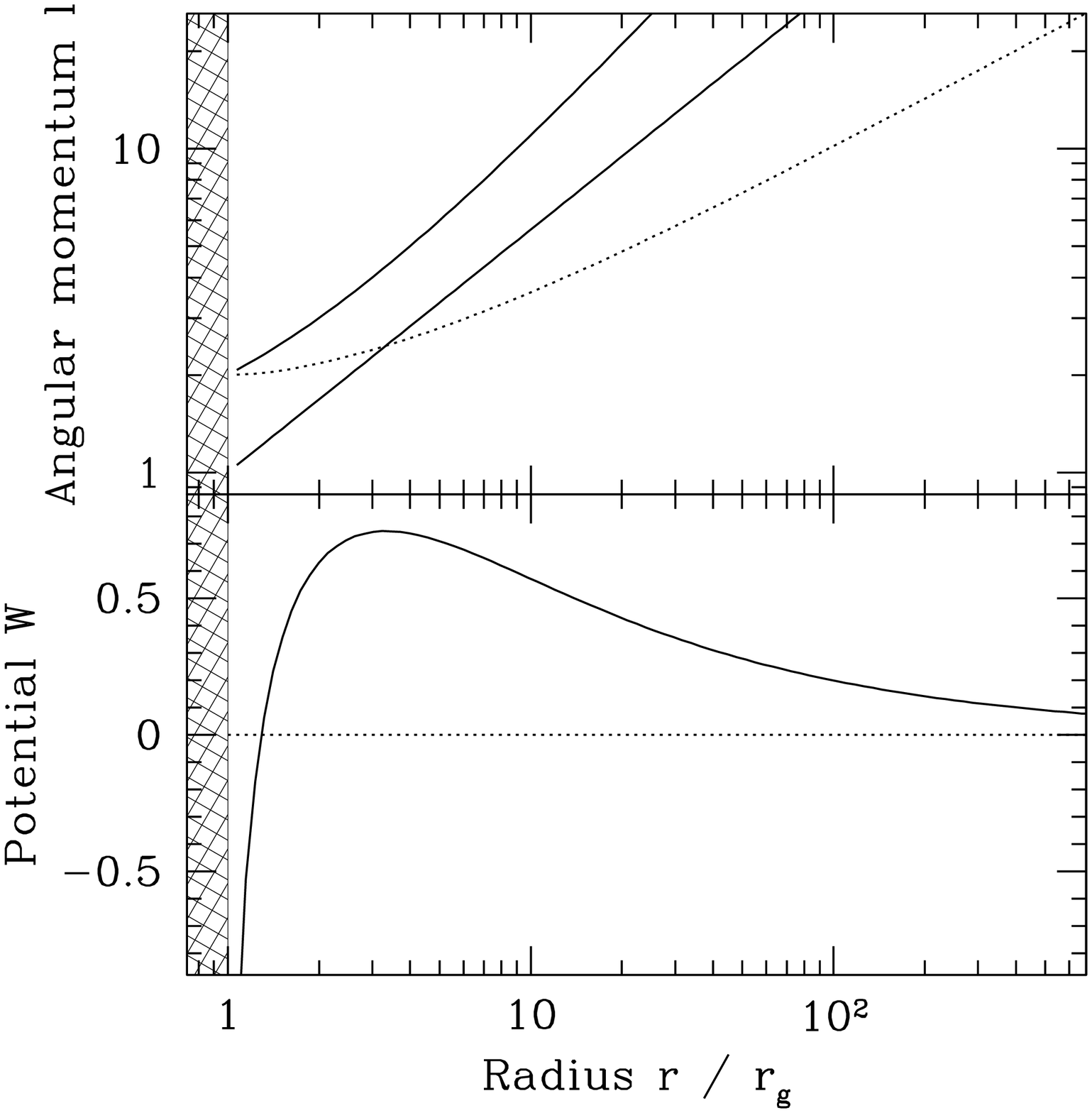,width=6.25cm} &
\psfig{file=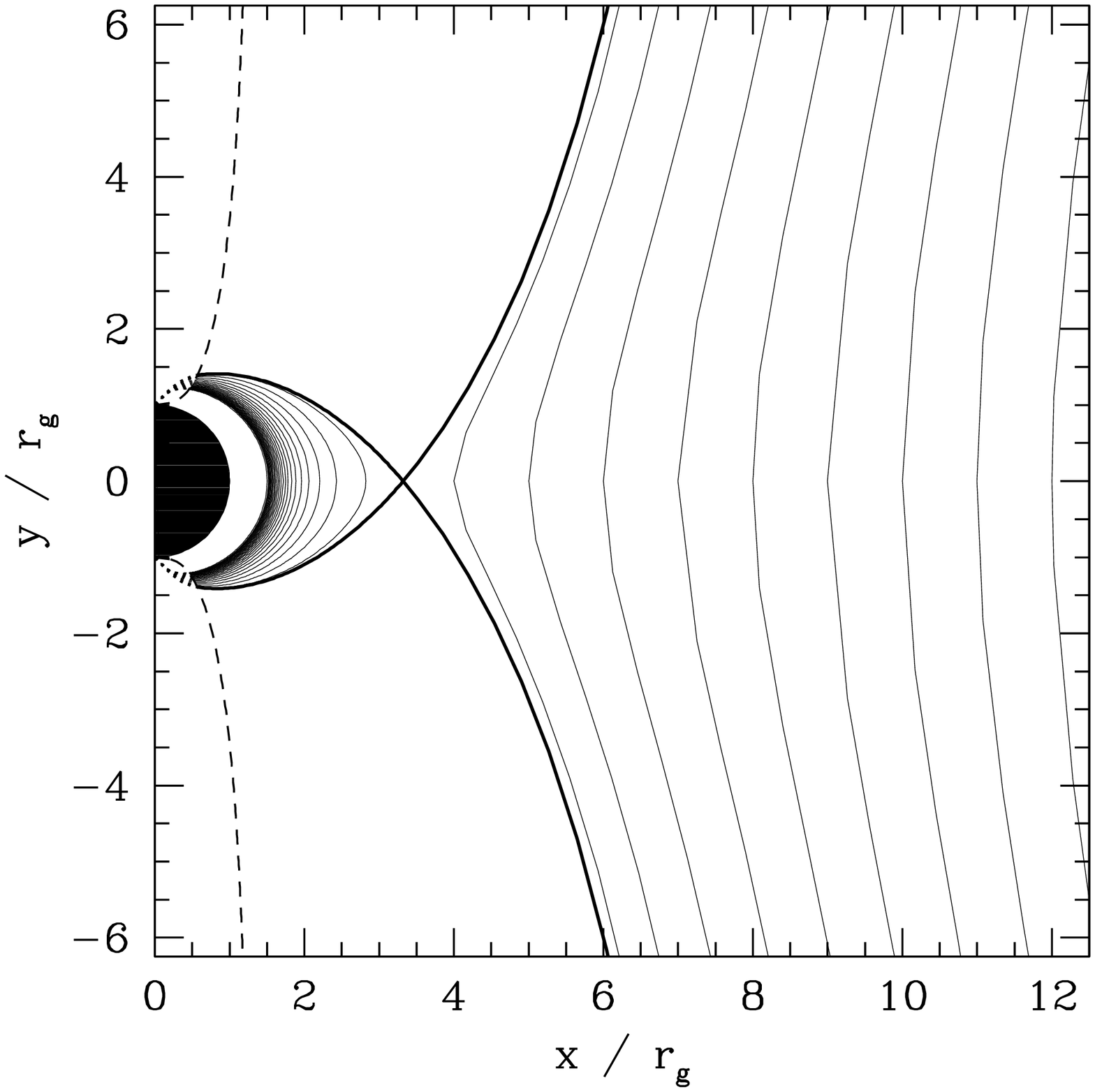,width=6.25cm}\\
\end{tabular}
\end{center}
\contcaption{\textbf{Geometry of the equipotentials: extreme Kerr black hole ($a=1$) and prograde 
disc. Keplerian and super-Keplerian cases.}}
\end{figure*}

\subsection{Method of construction}

A configuration of a thick disc orbiting a black hole is defined by 
eight parameters: (i) two parameters describing the black hole, the mass 
$M_\mathrm{BH}$ and the spin $0 \le a/M_\mathrm{BH} \le 1$; (ii) four parameters for the disc: 
the mass $M_\mathrm{D}$, the slope $0 \le \alpha < 1/2$ and the sign (prograde or 
retrograde rotation) of the angular momentum in the equatorial plane, and the 
potential barrier at the surface $\Delta W_\mathrm{in}$; (iii) two parameters 
for the EoS: the adiabatic index $\gamma$ and the polytropic constant $\kappa$. 
Once this set of parameters has been specified, we adopt the following procedure 
to build the disc:
\begin{enumerate}
\item Gravitational field: from $a$, we compute the Kerr metric coefficients 
appearing in Eq.~(\ref{blform}). 
\item Limits for the angular momentum: from $\alpha$ and the sign of $K$, we 
compute the two limits ${K}_\mathrm{ms}$ and ${K}_\mathrm{mb}$ defined by 
equations~(\ref{eq:Kms}) and (\ref{eq:Kmb}). Notice that $K_\mathrm{mb}$ is 
computed by solving Eq.~(\ref{eq:Kmb}) with a bisection procedure, where each 
iteration implies the computation of $r_\mathrm{cusp}$ and 
$W_\mathrm{eq}(r_\mathrm{cusp})$.
\item Initial value of $K$: $|{K}_\mathrm{ms}| < |K| < |{K}_\mathrm{mb}|$.
\item Potential in the equatorial plane: the function $W_\mathrm{eq}(r)$ is computed 
from Eq.~(\ref{eq:w}).
\item Characteristic radii: we compute the radius of the cusp and the centre 
$r_\mathrm{cusp}$ and $r_\mathrm{centre}$ by solving Eq.~(\ref{eq:cuspcenter}), and 
the inner radius $r_\mathrm{in}$ from $\Delta W_\mathrm{in}$.
\item Energy per unit inertial mass: the function $-u_{t}(r,\theta)$ is computed everywhere 
from Eq.~(\ref{eq:defut}), where the angular momentum $l(r,\theta)$ is computed from 
the procedure described in Subsection~\ref{sec:rotation} and Appendix~\ref{sec:vonzeipel}.
\item Potential: the function $W(r,\theta)$ is computed from $W_\mathrm{eq}(r)$ and 
$-u_{t}(r,\theta)$ using Eq.~(\ref{eq:woutside}).
\item Enthalpy, density and pressure: $h(r,\theta)$, $\rho(r,\theta)$ and $p(r,\theta)$ 
are computed from equations~(\ref{eq:h}), (\ref{eq:rho}) and (\ref{eq:p}) in the region 
where $W\le W_\mathrm{in}$ (interior of the disc) and are forced to $h=1$, $\rho=0$, $p=0$ 
elsewhere (vacuum).
\item Mass: $M_\mathrm{D}$ is computed from the integral~(\ref{eq:md}).
\item From the difference between $M_\mathrm{D}$ and its expected value, the coefficient 
$K$ is adjusted (secant method) and the iteration starts again at step (iv). This uses the 
fact that $M_\mathrm{D}/M_\mathrm{BH}$ as a function of $|K|$ is strictly increasing from 
$0$ for $K={K}_\mathrm{ms}$ to $+\infty$ for $K={K}_\mathrm{mb}$.
This procedure is stopped once the accuracy obtained for the value of $M_\mathrm{D}$ is 
good enough.
\end{enumerate}

\section{Numerical framework}
\label{sec:hydro}

The axisymmetric numerical code we use to evolve in time the equilibrium 
configurations described in the previous section is the same we used in 
Paper I. The code was originally written to account for the Kerr metric 
terms. The Schwarzschild black hole case analysed in Paper I was hence studied 
by setting to zero the black hole angular momentum parameter. In our code 
the general relativistic hydrodynamics equations (see Paper I for the 
especific expresions in Kerr spacetime) are numerically solved using an 
upwind high-resolution shock-capturing scheme based upon the HLLE Riemann 
solver. The code is second order accurate in both space and time due to 
the use of a piecewise linear, cell-reconstruction algorithm and a two-step, 
conservative Runge-Kutta time update. 

The form of the Kerr metric we employ is the one given by standard Boyer-Lindquist 
(spherical) coordinates $(t,r,\theta,\phi)$ (cf. Eq.~(\ref{blform})). Axisymmetry 
with respect to the black hole rotation axis implies that all $\phi$ derivatives 
vanish. The computational grid has $400\times 100$ zones in the radial and polar 
direction, respectively, and it is logarithmically spaced in the radial direction 
to ensure the finest resolution at smaller radii. The typical width of the innermost 
cell, where we have the highest resolution, is $\Delta r \simeq 1.99\times 10^{-2}$. 
The innermost radius is located at $r_\mathrm{min}=2.12$, sufficiently close to the 
horizon of the initial Schwarzschild black hole to avoid numerical inaccuracies 
introduced by the coordinate singularity of the coordinate system we use. It is clear 
that the use of ``horizon-adapted'' coordinate systems as the ones adopted by 
\citet{font:99}, regular at the black hole horizon, would suppress all numerical 
difficulties associated with the position of the innermost radius of the grid. This 
approach, however, has not been pursued in the present investigation but will be 
considered in the future. We postpone the explanation of the procedure we employ to 
increase the innermost radius in simulations where the size of the black hole horizon 
is growing to Section~\ref{sec:MethodBH} below. On the other hand, the location of 
the maximum radius $r_\mathrm{max}$ depends on the model. For the stationary test models
presented in Section \ref{sec:tests}, $r_\mathrm{max}=35$. Correspondingly, for the 
simulations of the runaway instability, the radial grid extends to a sufficiently large 
distance in order to ensure that the whole disc is included within the computational 
domain. The particular values of $r_\mathrm{max}$ that we choose are reported in 
Table \ref{tab:ModelParameters} below. Finally, as we did in Paper I, we use a 
finer grid in the angular direction within the torus and a much coarser grid 
outside. The angular zones are distributed according to the same law presented 
in Paper I. The computational domain in the angular direction extends from 0 to 
$\pi$.

Following the same approach than in Papers I and II we restrict here to adiabatic
hydrodynamical simulations which reduces the number of equations to solve to four, 
the relativistic counterparts of the continuity and Euler equations. This permits to 
obtain the specific internal energy through a polytropic EoS, $p=\kappa\rho^{\gamma}$, 
i.e.  $\varepsilon=\frac{\kappa}{\gamma-1}\rho^{\gamma-1}$. The boundary conditions 
for the (primitive) hydrodynamical quantities also coincide with those discussed in 
Paper I. Furthermore, the procedure to recover those variables from the evolved 
quantities (the relativistic densities of mass and momenta) is the same of Paper I. 
We refer the reader to that work for further details.

Finally, it is worth commenting on the simple yet effective procedure we adopt in order
to evolve with our hydrodynamical code the ``vacuum" zones which lie outside the disc.
Before constructing the initial torus we build a background, spherical `dust' accretion 
solution of sufficiently low density so that its presence does not affect the dynamics 
of the disc. This solution corresponds to a radially accreting flow with negligible 
pressure (geodesic motion). We even select among all possible solutions the marginally 
bound case where $u_\mathrm{t}=-1$. Our `dust' solution is then determined by the 
following equations, in which there is only one free parameter: $u_{r} = 
-\sqrt{\frac{-1-g^{tt}}{g^{rr}}}, \rho  = \frac{K_\mathrm{dust}}{\sqrt{-g}}\ 
\frac{-u_{r}}{g^{rr}}$. The pressure and the specific internal energy are computed from 
the density using a polytropic EoS. In our approach we adjust $K_\mathrm{dust}$ so that 
the maximum density in the background spherical solution is $5\times 10^{-6}$ times the 
maximum density at the centre of the disc. By doing this we have checked that the 
rest-mass present in our background solution is always negligible compared to the mass 
of the disc and that the associated mass flux corresponding to this spherical accretion 
is also negligible compared to the mass flux from the disc. 

\section{Code tests}
\label{sec:tests}

\begin{figure}
\centerline{\psfig{file=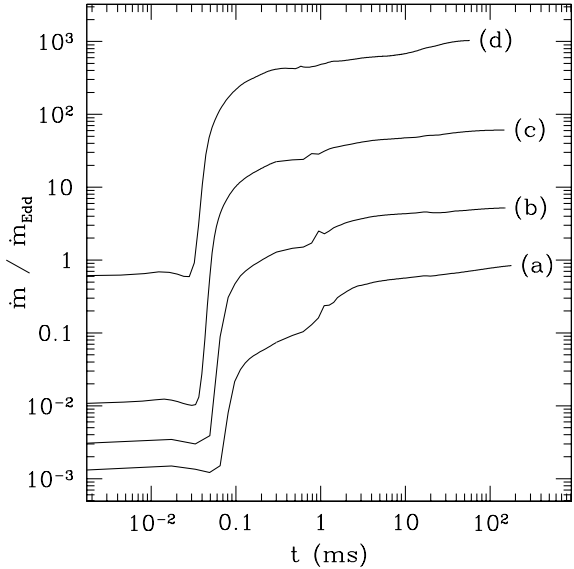,width=0.45\textwidth}}
\caption{Time evolution of the mass flux for a stationary model with a Kerr black hole 
of spin $a/M_\mathrm{BH}=0.9$ and a disc of constant angular momentum $l=2.6088$. Models (a), (b), 
(c) and (d) correspond respectively to $\Delta W_\mathrm{in}/c^{2}=0.04$, $0.08$, $0.16$ 
and $0.32$. The time is given in ms and the mass flux in units of the Eddington mass flux, 
both assuming that the mass of the black hole is $1\ M_{\odot}$. After about 1ms, the 
mass flux tends asymptotically to the stationary value. These test simulations are made 
with $\gamma=4/3$ and $\kappa = 1.5\times 10^{20}\ \mathrm{cgs}$.}
\label{fig:ib97a}
\end{figure}

\begin{figure}
\centerline{\psfig{file=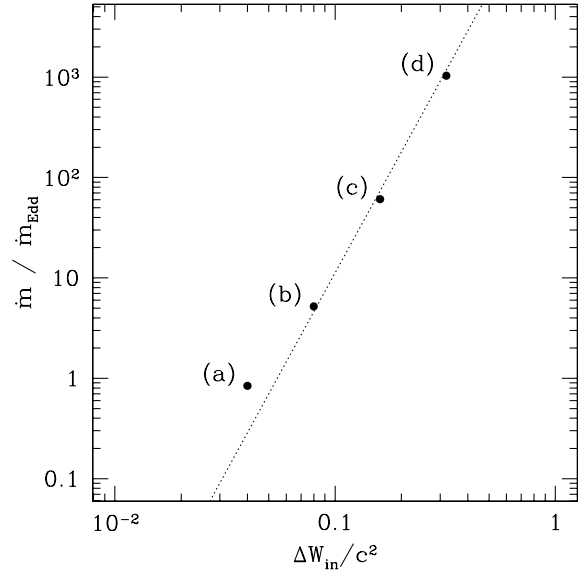,width=0.45\textwidth}}
\caption{Mass flux as a function of the potential barrier $\Delta W_\mathrm{in}/c^{2}$. 
For each of the models of Fig.~\ref{fig:ib97a}, the asymptotic stationary value of the 
mass flux is plotted as a function of the potential barrier (big dots). The analytic
dependence is $\dot{m}\propto \Delta W_\mathrm{in}^{\frac{\gamma}{\gamma-1}}$. The 
corresponding slope is 4 for $\gamma=4/3$ and is indicated with the straight line.}
\label{fig:ib97b}
\end{figure}

As we did in Paper I for the case of a Schwarzschild black hole, we have first 
tested the code by considering time-dependent simulations of stationary constant 
angular momentum discs around a rotating black hole. The aim of these simulations 
is to check whether the code is able to keep those tori in equilibrium during a 
sufficiently long period of time (much larger than the rotation period of the discs). 
In order to do so we have considered the same four stationary models analysed by
\citet{igumenshchev:97}, for a $1\ M_{\odot}$ rotating black hole with $a/M_\mathrm{BH}=0.9$. 
These four models are characterized by a constant angular momentum $l=2.6088$, a 
value in between the marginally stable and marginally bound orbits, and an increasingly 
large value of the energy gap at the cusp, $\Delta W_\mathrm{in} = 0.04$, $0.08$, 
$0.16$, and $0.32$. The adiabatic index of the EoS is $\gamma=4/3$ and the corresponding 
polytropic constant $\kappa = 1.5\times 10^{20}\ \mathrm{cgs}$. Furthermore, the mass 
and the spin of the black hole are kept constant throughout these test evolutions.

A conclusive proof of the ability of the code in keeping the stationarity of the 
various equilibrium models considered is provided by Figure~\ref{fig:ib97a}. This figure 
shows the evolution of the mass accretion rate (normalized to the Eddington value), 
$\dot{m}/\dot{m}_\mathrm{Edd}$, as a function of time (in ms). The mass flux (flux 
of rest-mass density) is computed at the innermost radial point according to 
\begin{eqnarray}
\dot{m}= 2\pi\int_0^{\pi} \sqrt{-g} D v^r d\theta,
\end{eqnarray}
where $D=\rho \Gamma$, $\Gamma$ being the Lorentz factor. The volume element for the 
Kerr metric appearing in the above equation is given by $\sqrt{-g}=\varrho^2\sin\theta$.
The Eddington mass flux $\dot{m}_\mathrm{Edd}=\frac{L_\mathrm{Edd}}{c^{2}}\simeq 1.4\times
10^{17}\ \frac{M_\mathrm{BH}} {\mathrm{M}_\odot}\ \mathrm{g/s}$ is computed for $M_\mathrm{BH}=1\
\mathrm{M}_\odot$. Rescaling of $\dot{m}$ for different polytropic constant $\kappa$ and 
black hole mass $M_\mathrm{BH}$ is given by
\begin{equation}
\frac{\left(\dot{m}/\dot{m}_\mathrm{Edd}\right)_{1}}
{\left(\dot{m}/\dot{m}_\mathrm{Edd}
\right)_{2}} = \left(\frac{\kappa_{1}}{\kappa_{2}}\right)^{-3}
\frac{M_{\mathrm{BH}\ 1}}
{M_{\mathrm{BH}\ 2}}\ ,
\end{equation}
where $\gamma=4/3$ is assumed. Fig.~\ref{fig:ib97a} shows that after a transient initial 
phase of about 1ms the mass flux rapidly tends, asymptotically, to a constant value. The 
offset observed during the initial phase corresponds to the spherical accretion mass flux 
associated with the particular background solution we use outside the torus.

Correspondingly, Figure~\ref{fig:ib97b} shows the dependence of the mass flux with the 
energy gap $\Delta W_\mathrm{in}$ for the four stationary models we consider. The values 
selected for the mass flux in each model are the asymptotic ones read off from 
Fig.~\ref{fig:ib97a}. This plot allows us to check if the code is able to reproduce 
the analytic dependence derived by \citet{abramowicz:78}
\begin{equation}
\dot{m} \propto \Delta W_\mathrm{in}^{\frac{\gamma}{\gamma-1}}.
\label{eq:mdotvsdw}
\end{equation}
For a $\gamma=4/3$ polytrope the expected slope is 4 (dashed line). As the figure shows
our results are in good agreement with this analytic prediction as well as with the 
numerical results obtained by~\cite{igumenshchev:97} for the same models.

\section{Results}
\label{sec:simulations}

We turn now to describe our simulations. The main objective of this section is to 
demonstrate the suppression of the runaway instability in thick discs around black 
holes due to the stabilizing effect of an increasing angular momentum distribution.

\subsection{The physical origin of the stabilizing effect}

As already mentioned in the introduction the runaway instability of thick discs 
around black holes was discovered by \citet{abramowicz:83} with a stationary model 
which employed Newtonian gravity and a pseudo-potential to mimic the relativistic 
gravitational field of the black hole. In Paper I we summarized all subsequent 
studies which followed this first work, as well as the current understanding of the 
physical origin of the instability. The results presented in Paper I were the first 
confirmation by hydrodynamical simulations in general relativity that the runaway 
instability is indeed occuring, at least for non self-graviting constant angular 
momentum discs. Our study also provided the first confirmation that this instability 
was extremely violent, leading to the destruction of the disc on a dynamical timescale, 
shorter than 1 s for realistic parameters. More recently these results have been 
confirmed by \citet{zanotti:02} with independent hydrodynamical relativistic simulations. 
These authors have also shown that the instability of constant angular momentum
discs is a robust feature, independent of the way accretion is induced.

The main goal of the present study is now to check -- using again hydrodynamical 
relativistic simulations -- the stabilizing effect of a \textit{non-constant} angular 
momentum distribution in the disc. In the collapsar model of GRBs, the simple constraint 
given by the \textit{Rayleigh's criterion} imposes such a distribution both in the initial 
massive star before the collapse and in the final thick disc. The constant angular 
momentum case appears as a rather unlikely limiting case. Positive slope of the angular 
momentum in the disc are indeed found in numerical simulations of the gravitational 
collapse of a massive star, as well as in simulations of the coalescence of two compact 
objects. The stability of non-constant angular momentum discs was first discovered by 
\citet{daigne:97} in pseudo-Newtonian gravity, and confirmed with a relativistic 
calculation by \citet{abramowicz:98}, both studies being based on stationary models. 
The physical origin of this stabilizing effect is simple: during the accretion process, 
the material at the inner edge of the disc is replaced by new material with a higher 
angular momentum. Therefore, the centrifugal barrier which acts against accretion 
becomes much more efficient. As shown by \citet{daigne:97} the difference between 
discs with constant and non-constant angular momentum is the location of the cusp 
(the Lagrange point $L_{1}$) with respect to the inner edge of the disc after mass 
transfer. In the first case, the inner edge of the disc moves faster than the cusp 
towards the black hole horizon (favouring the instability) whereas the cusp moves 
faster towards the horizon in the latter (regulating the accretion). 

Recently, \citet{rezzolla:03} presented a perturbative study of vertically-integrated 
discs around black holes. They found that when acoustic waves are considered, the inner 
region of the torus are of evanescent type for discs with non constant angular momentum, 
while such an evanescent region does not exist in the constant angular momentum case. 
This leads to a complementary explanation of the suppression of the runaway instability 
in non-constant angular momentum discs: the evanescent region in the inner disc prevents
the inward propagation of perturbations and acts against the accretion of mass on to the 
black hole.

\subsection{The sequence of stationary metrics approximation}
\label{sec:MethodBH}

The flux of mass and angular momentum from the disc to the black hole increases
the mass and the spin of the latter, therefore changing the metric potentials
of the underlying gravitational field where the fluid evolves. Such increase is, 
in fact, the fundamental process which triggers the runaway instability of the disc.
Thus, it needs to be included in some way in the hydrodynamical simulations.

Strictly speaking, in order to properly take into account the dynamical evolution 
of the gravitational field, one should solve the coupled system of Einstein and 
hydrodynamics equations for a self-gravitating matter source. Numerical relativity 
codes capable of evolving black hole spacetimes with perfect fluid matter are
only becoming available recently, both in axisymmetry~\citep{brandt:00,shibata:03}
as in full 3D~\citep{shibata:00,shibata2:00,alcubierre:00,font:01}. These codes,
however, have not yet been applied to investigate astrophysical aspects of accretion 
on to black holes, in particular the runaway instability. When trying to simulate
dynamical black hole spacetimes there are some issues which make this a challenging 
problem, among those the affordable resolution and the integration times required to 
study the development of the runaway instability, perhaps far too demanding for such 
relativistic codes which incorporate self-gravity. One of the most serious problem 
is, no doubt, the numerical handling of the physical singularity hidden inside the 
black hole horizon, which prevents long-term evolutions. Recently, successful, 
long-term simulations of (vacuum) black hole spacetimes have been achieved by excising 
part of the computational grid within the event horizon (see e.g. \citet{yo:03} and 
references therein). This is a promising line of research which could provide a way 
to investigate the runaway instability in full general relativity in the near future.

While these obstacles are paved away we have adopted a simplified and pragmatic
approach to the problem. In our procedure the spacetime metric is approximated at
each time step by a stationary exact black hole metric of varying mass and angular
momentum. As we mentioned in Paper I the mass $M_\mathrm{BH}$ of the black hole, 
necessary to compute the metric coefficients, is increased at each time step $\Delta t$ 
from time $t^n$ to time $t^{n+1}$ according to:
\begin{equation}
M_\mathrm{BH}^{n+1} =  M_\mathrm{BH}^{n}+\Delta t\ \dot{m}^{n}\ ,
\label{massincrease1}
\end{equation}
where the mass flux at the inner radius of the grid is evaluated by the equations
\begin{eqnarray}
\dot{m}^{n} & = & \sum_{j=1}^{j_{\mathrm{max}}} \left(\theta_{j+1}-\theta_{j}\right) 
\left.\frac{d\dot{m}}{d\theta}\right|_{i_\mathrm{min} j}\ ,
\label{massincrease2}\\
\left.\frac{d\dot{m}}{d\theta}\right|_{i_\mathrm{min} j} & = & 2\pi \sqrt{-g}_{i_\mathrm{min} j} 
D_{i_\mathrm{min} j} v^{r}_{i_\mathrm{min} j}\ .
\end{eqnarray}
Similarly, the angular momentum $J_\mathrm{BH}$ of the black hole is increased at each time step
according to:
\begin{equation}
J_\mathrm{BH}^{n+1} =  J_\mathrm{BH}^{n}+\Delta t\ \dot{j}^{n}\ ,
\label{spinincrease1}
\end{equation}
where the angular momentum flux at the inner radius is computed by the equation
\begin{equation}
\dot{j}^{n} = \eta \times \sum_{j=1}^{j_\mathrm{max}} \left(\theta_{j+1}-\theta_{j}\right) 
\left.\frac{d\dot{m}}{d\theta}\right|_{i_\mathrm{min} j}
l_{i_\mathrm{min} j}\ .
\label{spinincrease2}
\end{equation}
The angular momentum is related to the $\phi$-component of the 3-velocity used in the hydrodynamics
code by
\begin{equation}
l = \frac{v_{\phi}}{\alpha-\beta^{\phi}v_{\phi}}\ .
\end{equation}
In addition, the parameter $\eta$ appearing in Eq.~(\ref{spinincrease2}) controls the efficiency 
of the transfer of angular momentum from the disc to the black hole. Since in realistic discs angular 
momentum is transported outwards by dissipative processes or removed from the system by gravitational 
radiation, we adopt in most models a conservative value for the efficiency of the angular momentum 
transfer, assuming that only 20\% of the angular momentum of the accreted material is transferred to 
the black hole.

\begin{table*}
\begin{center}
\caption{\textbf{Initial models.} The following parameters are listed: mass of the black
hole $M_\mathrm{BH}$, disc-to-hole mass ratio $M_\mathrm{D}/M_\mathrm{BH}$, slope of the
specific angular momentum in the disc $\alpha$, ratio of the potential barrier at the inner
edge over the potential at the cusp $\Delta W_\mathrm{in}/|W_\mathrm{cusp}|$, mass flux in
the stationary regime $\dot{m}_\mathrm{stat}$, minimum and maximum radii of the grid
$r_\mathrm{min}$ and $r_\mathrm{max}$, radius of the cusp $r_\mathrm{cusp}$,
radius of the centre $r_\mathrm{centre}$ (all these radii are in unit of the gravitational
radius $r_\mathrm{g}$), and orbital period at the centre of the disc $t_\mathrm{orb}$ (in
units of $r_\mathrm{g}/c$). The last column lists the timescale associated with the runaway
instability as defined in Section \ref{sec:ResultsUnstable}. In all cases, the EoS parameters
are $\kappa = 4.76\times 10^{14}\ \mathrm{cgs}$ and $\gamma=4/3$. As mentioned in the text, 
models 3c' and 3c'' differ only from model 3c by the value of the efficiency $\eta$ of the 
angular momentum transfer.}
\label{tab:ModelParameters}
\begin{tabular}{l|cc|lccc|cc|ccc|c}
\hline
Model &
$M_\mathrm{BH}$ & $M_\mathrm{D}/M_\mathrm{BH}$ &
$\alpha$ & $\Delta W_\mathrm{in}/|W_\mathrm{cusp}|$
& $\dot{m}_\mathrm{stat}$
& $r_\mathrm{min}^{(1)}$ & $r_\mathrm{max}^{(2)}$ & $r_\mathrm{cusp}$ & $r_\mathrm{centre}$ &
$t_\mathrm{orb}$  & $t_\mathrm{run}/t_\mathrm{orb}\ ^{(3)}$\\
&  $(\mathrm{M}_\odot)$ & & & & $(\mathrm{M_{\odot}/s})$ & & & & & (geo/ms) & \\
\hline
1a & 2.5 & 1.   & 0.0   & 0.75  & 26.   & 2.12 & 20.15/242 & 4.90 & 7.59 & 131/1.61 & 4.3 \\
2a & 2.5 & 1.   & 0.05  & 0.75  & 8.1   & 2.12 & 20.15/242 & 5.02 & 8.68 & 161/1.98 & 9.4 \\
3a & 2.5 & 1.   & 0.075 & 0.75  & 4.6   & 2.12 & 20.15/242 & 5.11 & 9.29 & 178/2.19 & 28 \\
4a & 2.5 & 1.   & 0.08  & 0.75  & 4.1   & 2.12 & 20.15/242 & 5.13 & 9.42 & 182/2.24 & 41 \\
5a & 2.5 & 1.   & 0.085 & 0.75  & 3.6   & 2.12 & 20.15/242 & 5.15 & 9.56 & 186/2.29 & 73 \\
6a & 2.5 & 1.   & 0.09  & 0.75  & 3.2   & 2.12 & 20.15/242 & 5.18 & 9.70 & 190/2.34 & stable \\
7a & 2.5 & 1.   & 0.1   & 0.75  & 2.5   & 2.12 & 20.15/242 & 5.22 & 9.98 & 198/2.44 & stable \\
8a & 2.5 & 1.   & 0.15  & 0.75  & 0.69  & 2.12 & 20.15/242 & 5.53 & 11.6 & 250/3.08 & stable \\
\hline
1b & 2.5 & 1.   & 0.0   & 0.5   & 0.14  & 2.12 & 20.15/242 & 4.23 & 9.49 & 184/2.26 & 39 \\
2b & 2.5 & 1.   & 0.01  & 0.5   & 0.11  & 2.12 & 20.15/242 & 4.27 & 9.67 & 189/2.32 & 100 \\
3b & 2.5 & 1.   & 0.015 & 0.5   & 0.10  & 2.12 & 20.15/242 & 4.29 & 9.76 & 192/2.36 & $>$157 \\
4b & 2.5 & 1.   & 0.02  & 0.5   & 0.093 & 2.12 & 20.15/242 & 4.31 & 9.85 & 194/2.39 & stable \\
5b & 2.5 & 1.   & 0.025 & 0.5   & 0.085 & 2.12 & 20.15/242 & 4.33 & 9.95 & 197/2.42 & stable \\
6b & 2.5 & 1.   & 0.05  & 0.5   & 0.053 & 2.12 & 20.15/242 & 4.45 & 10.5 & 213/2.62 & stable \\
\hline
1c & 2.5 & 0.1  & 0.0   & 0.5   & 5.9   & 2.12 & 19.23/100 & 4.96 & 7.46 & 128/1.57 & 7.5 \\
2c & 2.5 & 0.1  & 0.025 & 0.5   & 3.5   & 2.12 & 19.23/100 & 5.04 & 7.91 & 140/1.72 & 17. \\
3c & 2.5 & 0.1  & 0.05  & 0.5   & 2.1   & 2.12 & 19.23/100 & 5.13 & 8.41 & 153/1.88 & 100. \\
4c & 2.5 & 0.1  & 0.055 & 0.5   & 1.9   & 2.12 & 19.23/100 & 5.15 & 8.52 & 156/1.92 & $>$ 200 \\
5c & 2.5 & 0.1  & 0.06  & 0.5   & 1.7   & 2.12 & 19.23/100 & 5.17 & 8.63 & 159/1.96 & stable \\
6c & 2.5 & 0.1  & 0.07  & 0.5   & 1.4   & 2.12 & 19.23/100 & 5.22 & 8.85 & 165/2.03 & stable \\
7c & 2.5 & 0.1  & 0.075 & 0.5   & 1.3   & 2.12 & 19.23/100 & 5.24 & 8.96 & 169/2.08 & stable \\
\hline
3c'  & 2.5 & 0.1 & 0.05 & 0.5   & 2.1   & 2.12 & 19.23/100 & 5.13 & 8.41 & 153/1.88 & stable\\
3c'' & 2.5 & 0.1 & 0.05 & 0.5   & 2.1   & 2.12 & 19.23/100 & 5.13 & 8.41 & 153/1.88 & stable\\
\hline
1d & 2.5 & 0.1  & 0.0   & 0.427 & 1.3   & 2.12 &  19.23/100 & 4.61 & 8.25 & 149/1.83 & 28 \\
2d & 2.5 & 0.1  & 0.04  & 0.466 & 1.3   & 2.12 &  19.23/100 & 4.91 & 8.62 & 159/1.96 & 180 \\
3d & 2.5 & 0.1  & 0.05  & 0.476 & 1.3   & 2.12 &  19.23/100 & 5.00 & 8.72 & 162/1.99 & $>$ 300 or stable ? \\
4d & 2.5 & 0.1  & 0.055 & 0.481 & 1.3   & 2.12 &  19.23/100 & 5.05 & 8.76 & 163/2.00 & stable \\
5d & 2.5 & 0.1  & 0.06  & 0.486 & 1.3   & 2.12 &  19.23/100 & 5.10 & 8.81 & 164/2.02 & stable \\
\hline
\end{tabular}
\end{center}
\begin{flushleft}
Notes:\\
$^{(1)}$ Initial value of the inner radius of the grid. When the spacetime is evolving, this radius
increases with time (see text).\\
$^{(2)}$ First value: inner part of the grid (high resolution); second value: outer part of the grid 
(low resolution).\\
$^{(3)}$ This value is given for the case where both the mass and the spin of the black hole increase
with time.
\end{flushleft}
\end{table*}

An important numerical aspect is the evolution of the radial grid in simulations where 
the increase of the mass and the spin of the black hole is taken into account. The radius 
of the horizon $r_\mathrm{h}=M_\mathrm{BH}+\sqrt{M_\mathrm{BH}^{2}-a^{2}}$ moves according to 
\begin{equation}
dr_\mathrm{h} = \left(1+\frac{M_\mathrm{BH}}{\sqrt{M_\mathrm{BH}^{2}-a^{2}}}\right)
d M_\mathrm{BH}-\frac{a}{\sqrt{M_\mathrm{BH}^{2}-a^{2}}}da\ .
\end{equation}
Following the procedure which has just been described, the variation of $a$ is related to the 
variation of mass via the value $l_\mathrm{in}$ of the specific angular momentum of the material 
passing through the inner limit of the grid and the efficiency $\eta$ of its transfer to the 
black hole:
\begin{equation}
\frac{da}{a}=\frac{dJ_\mathrm{BH}}{J_\mathrm{BH}}-\frac{dM_\mathrm{BH}}{M_\mathrm{BH}}=
\left(\frac{\eta l_\mathrm{in}}{a}-1\right)\frac{dM_\mathrm{BH}}{M_\mathrm{BH}}\ .
\end{equation}
By combining this expression with the previous equation and after some algebra we can express the 
variation of the radius of the horizon as
\begin{eqnarray}
\frac{d r_\mathrm{h}}{r_\mathrm{h}} & = & \left(1-\frac{1}{2}\sqrt{1-\left(\frac{a}
{M_\mathrm{BH}}\right)^{2}}- \eta \frac{l_\mathrm{in}}{l_\mathrm{cr}^{\pm}(r_\mathrm{h})}\right)\nonumber\\
& & \times\frac{2}{\sqrt{1-\left(\frac{a}{M_\mathrm{BH}}\right)^{2}}}\ \frac{dM_\mathrm{BH}}{M_\mathrm{BH}}\ ,
\end{eqnarray}
where $l_\mathrm{cr}^{\pm}(r_\mathrm{h})= 2 M_\mathrm{BH} r_\mathrm{h}/a$ is the maximum value of 
the angular momentum at the horizon. This variation is positive for $\eta l_\mathrm{in} / 
l_\mathrm{cr}^{\pm}(r_\mathrm{h}) < 1 - 1/2 \sqrt{1-(a/M_\mathrm{BH})^{2}}$, a condition which is 
always fulfilled in our simulations. Notice that for $a=0$ (Schwarzschild black hole), this 
condition reads $\eta l_\mathrm{in} < +\infty$ and for $a=M_\mathrm{BH}$ (maximally rotating 
black hole), it reads $\eta l_\mathrm{in} < l_\mathrm{cr}^{\pm}(r_\mathrm{h})$. Therefore, in all 
of our simulations we observe the expected behaviour: the horizon moves outwards. Then, as we did 
in Paper I, we follow a simple procedure to let the radial grid evolve with time. We just increase 
the index $i_\mathrm{min}$ of the innermost cell when necessary so that the condition 
$r_{i_\mathrm{min}}>r_\mathrm{h}$ is always respected. It is important to notice that for all 
disc-to-hole mass ratios considered in our simulations, the black hole mass and spin increase 
very slowly during the evolution. This implies that the metric coefficients at any time step 
differ very little from the final values which would correspond to an exact Kerr black hole 
of larger mass and spin but with no matter around.

\subsection{Initial state}

As shown in Section~\ref{sec:initialmodel}, a given initial state of the black hole plus 
non-constant angular momentum disc system is determined by eight parameters: the mass and 
the spin of the black hole, the disc-to-hole mass ratio, the slope (and sign) of the angular 
momentum in the equatorial plane of the disc, the potential barrier at the disc surface, and 
the two parameters defining the polytropic EoS. In all simulations presented in this paper, 
the black hole is initialy non-rotating ($a=0$) and therefore, the initial sign of the angular 
momentum of the disc is chosen to be positive (prograde disc) as the retrograde case is strictly 
identical. The case of initially rotating black holes and pro- or retrograde discs will be 
presented in a forthcoming paper. As already mentioned in Paper I, the computing time needed 
for one hydrodynamical simulation is too large to allow for a complete exploration of the domain 
defined by the remaining six parameters. For this reason we focus on those models which are 
expected to be found in the central engine of GRBs. 

\subsubsection{Black hole mass and disc-to-hole mass ratio}

In the two most discussed scenarios the central engine of GRBs is formed either after 
the coalescence of two compact objects or after the gravitational collapse of a massive 
star. As in Paper I, we then fix the mass of the black hole to be 
$2.5\ \mathrm{M_{\odot}}$, which is a reasonable value according to the results of various 
numerical simulations \citep{ruffert:99,kluzniak:98,macfadyen:99,shibata:00,aloy:00,shibata2:03}. 
In addition we only consider two possible values for the disc-to-hole mass ratio, either 1 or 
0.1. The second value is clearly more realistic according to the aforementioned simulations 
(even some lower values $\sim 10^{-3}-10^{-2}$ are found in some situations). This small 
mass ratio makes our initial assumption of neglecting the self-gravity of the disc less of
a concern. However, as the reservoir for the accretion process is larger in the case where 
$M_\mathrm{D}/M_\mathrm{BH}=1$, the time necessary to completely destroy the disc in the 
case of a stable accreting regime is much longer than the timescale of the runaway instability 
observed in unstable models. Therefore, our choice of this high mass ratio is justified as it  
allows a better characterization of the instability. As pointed out in Paper I, where a mass 
ratio of 0.05 was also considered, mass ratios below 0.1 are more difficult to simulate, so 
we do not consider them in this study. According to the results of Paper I, they should lead 
to very similar results to the $M_\mathrm{D}/M_\mathrm{BH}=0.1$ case.

\subsubsection{Equation of state}

We fix the adiabatic index to $\gamma=4/3$ and the polytropic constant to
$\kappa=1.2\times 10^{15}\ Y_\mathrm{e}^{4/3}$ with $Y_\mathrm{e}=0.5$. This
corresponds to an EoS dominated by the contribution of relativistic degenerate electrons
(the typical density in the disc is $\sim 10^{11}$-$10^{12}\ \mathrm{g/cm^{3}}$).
We note that such simplified EoS is nevertheless adequate for our purpose since
the work of \cite{nishida:96b} showed that, for stationary models, the effects of
realistic EoS on the stability of constant angular momentum discs is negligible,
the discs being unstable in all cases.

\subsubsection{Mass flux}

The value of the potential barrier at the disc surface is strongly related to the acrretion 
mass flux in the stationary regime, as defined in Section~\ref{sec:tests}. As we discussed 
in Paper I, this allows to limit the value of $\Delta W_\mathrm{in}$ in a domain where the 
corresponding mass flux is in the typical range expected for GRBs, $10^{-2}$--$10^{2}\ 
\mathrm{M_{\odot} s^{-1}}$, i.e. high values many orders of magnitude above the Eddington limit. 
Such very high mass fluxes are also found in the numerical simulations of compact binary coalescence 
or collapsars mentioned above. In the models listed in Table~\ref{tab:ModelParameters}, we have 
either fixed the potential barrier to a constant fraction of the potential at the cusp, 
$\Delta W_\mathrm{in}/|W_\mathrm{cusp}|=0.75$ or $0.5$, or we have adjusted it to have a constant 
mass flux in the stationary regime, $1.3\ \mathrm{M_\mathrm{\odot} s^{-1}}$. The preparation of 
this last sequence of models is much more time consuming than for the rest of models as it implies 
the following  procedure: (i) construction of the initial model for a given first guess of 
$\Delta W_\mathrm{in}$ ; (ii) evolution of this model with the hydrodynamical code with fixed 
spacetime, up to the stage where the stationary regime is reached ; (iii) from the comparison of 
the mass flux $\dot{m}_\mathrm{stat}$ with the expected value, either start again at step (i) 
with a new guess of $\Delta W_\mathrm{in}$ or stop.

All these considerations have led to the preparation of 26 initial models listed in 
Table~\ref{tab:ModelParameters}. They are splitted in four sets. Sets `a' and `b' correspond 
to a disc-to-hole mass ratio of 1, and sets `c' and `d' correspond to a mass ratio of 0.1. For 
each mass ratio, the two sets of models correspond to a different assumption for the potential 
barrier at the disc surface. The successive models within each set are obtained by increasing 
the slope $\alpha$ of the angular momentum distribution in the disc, starting from $\alpha=0$.

\subsection{Simulations}
\label{sec:ResultsUnstable}

\begin{figure*}
\begin{center}
\begin{tabular}{cc}
\hspace{1cm}
$M_\mathrm{D}/M_\mathrm{BH}=1$ and $\Delta W_\mathrm{in}=0.75 |W_\mathrm{cusp}|$ & 
\hspace{1cm}
$M_\mathrm{D}/M_\mathrm{BH}=0.1$  and $\Delta W_\mathrm{in}=0.5 |W_\mathrm{cusp}|$ \\
\psfig{file=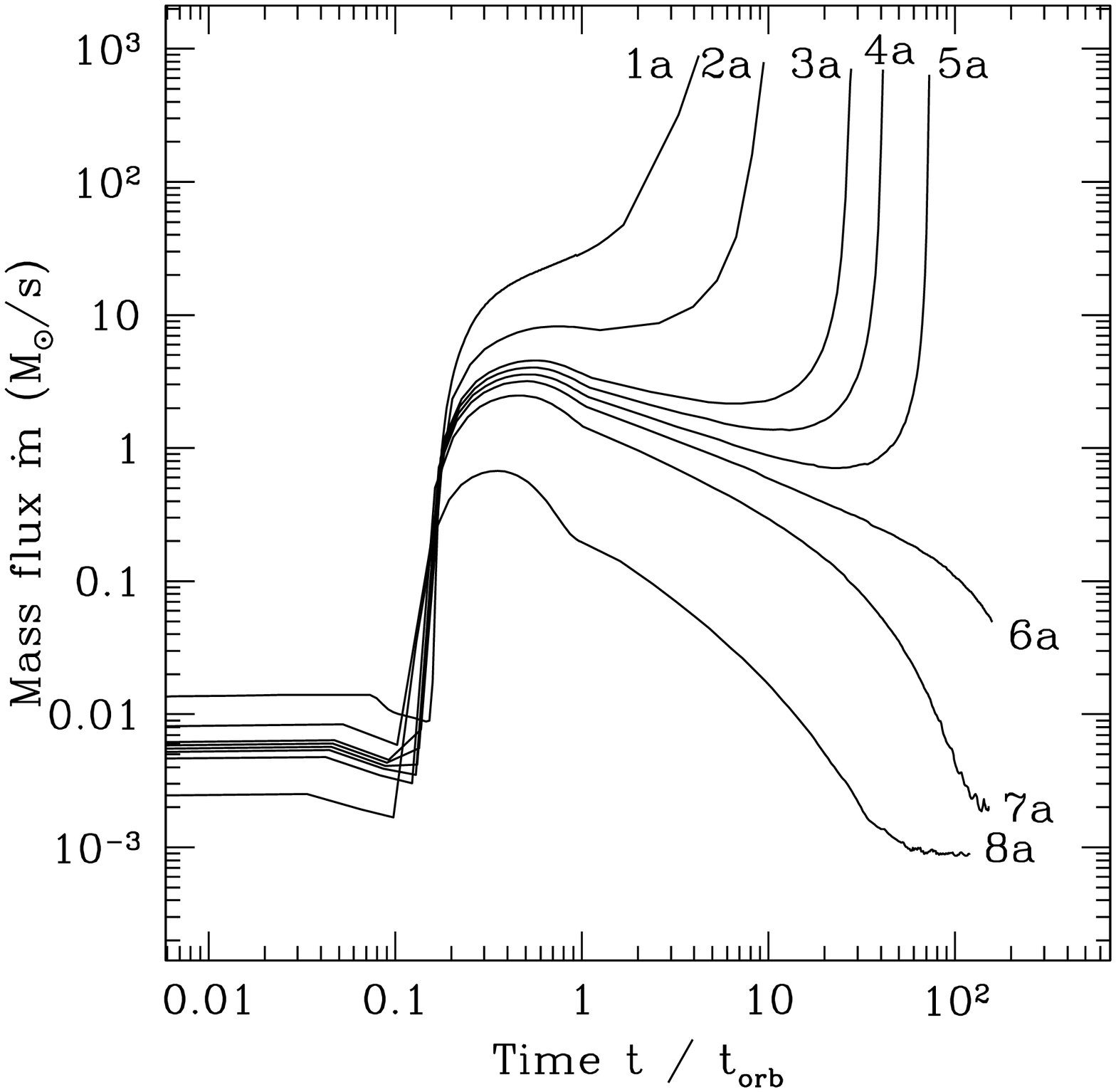,width=0.47\textwidth} & 
\psfig{file=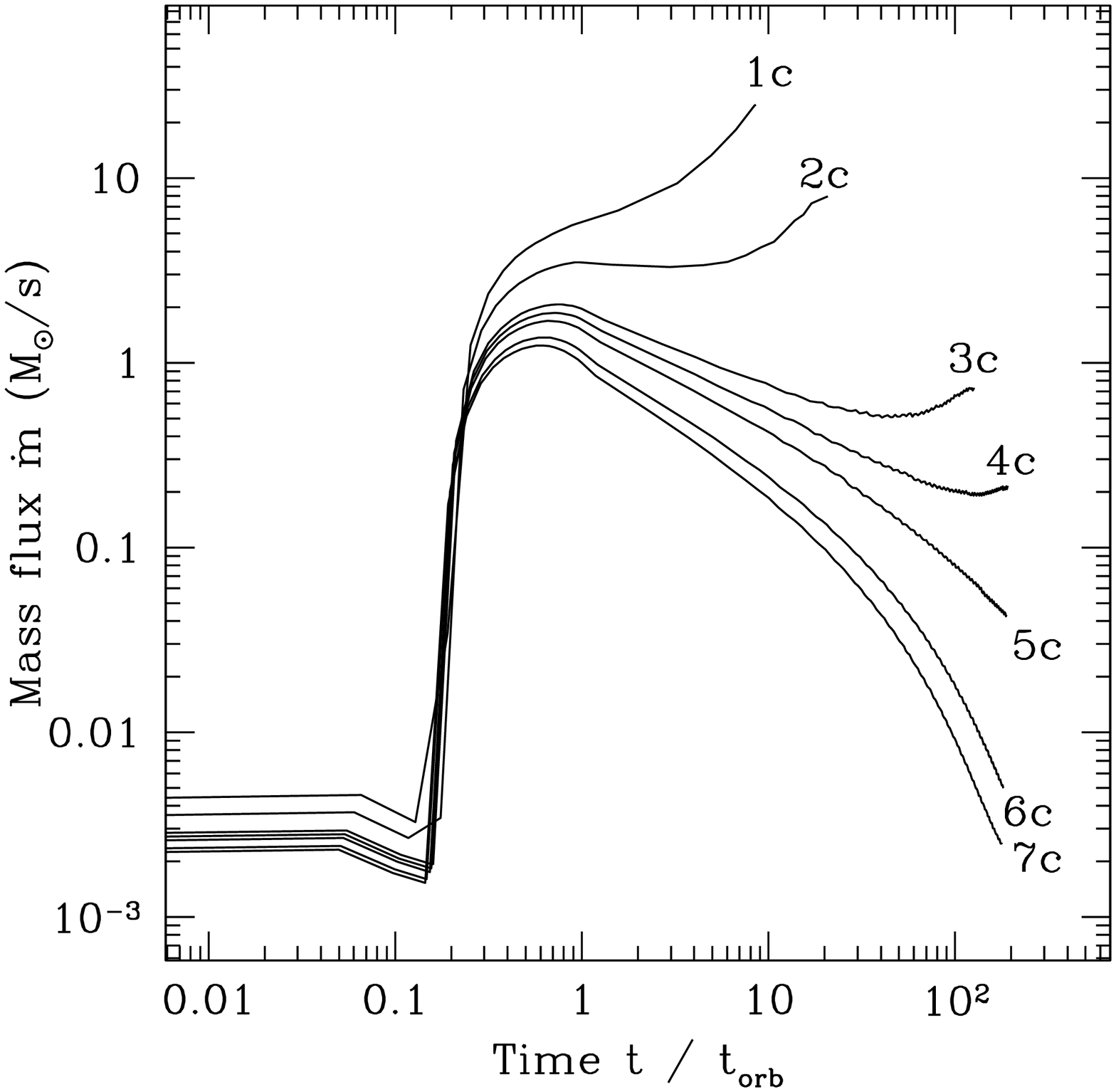,width=0.47\textwidth}\\
\hspace{1cm}
$M_\mathrm{D}/M_\mathrm{BH}=1$ and $\Delta W_\mathrm{in}=0.5 |W_\mathrm{cusp}|$ & 
\hspace{1cm}
$M_\mathrm{D}/M_\mathrm{BH}=0.1$  and $\dot{m}_\mathrm{stat}\sim 1.3\ \mathrm{M_{\odot}/s}$ \\
\psfig{file=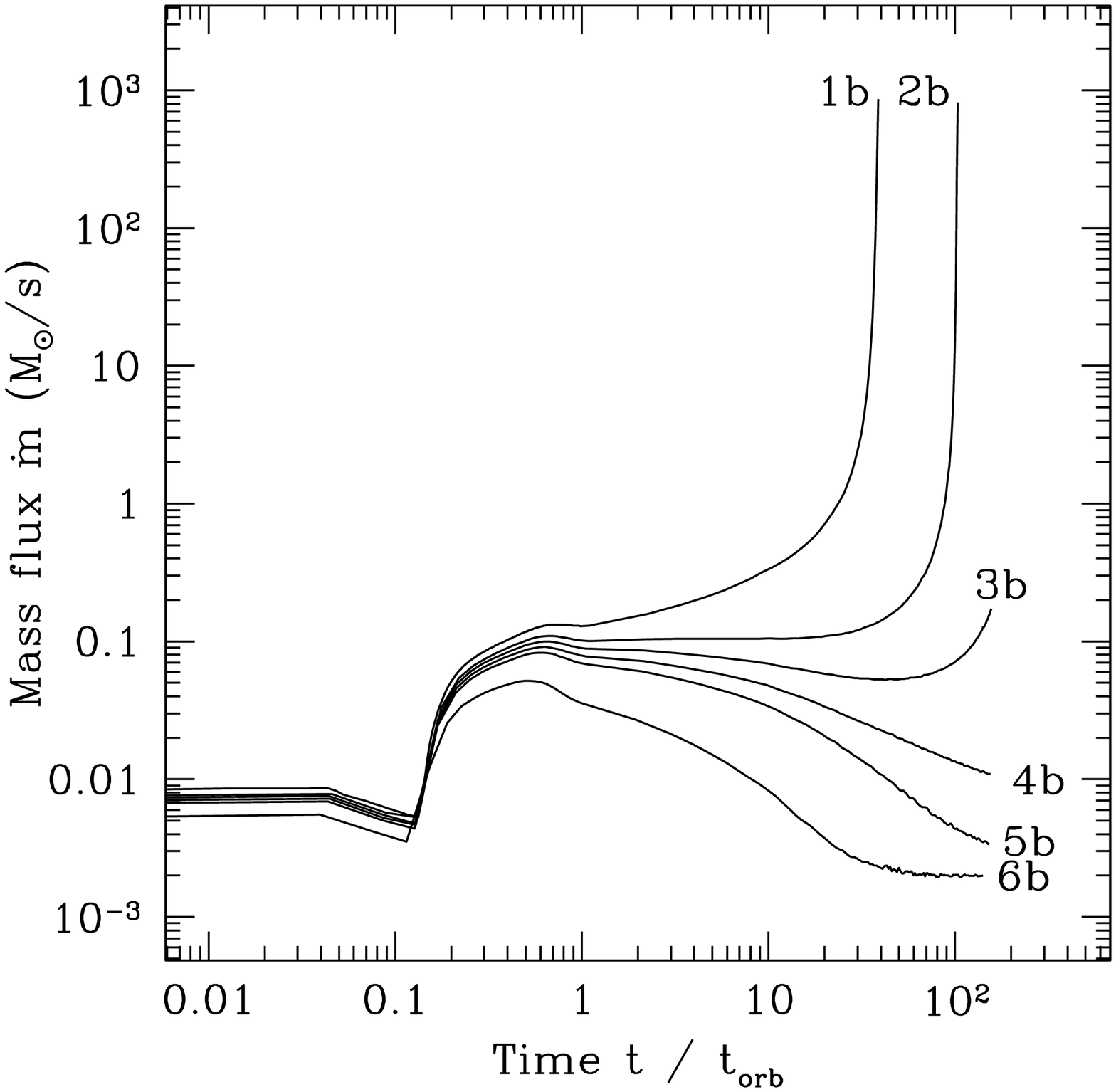,width=0.47\textwidth} & 
\psfig{file=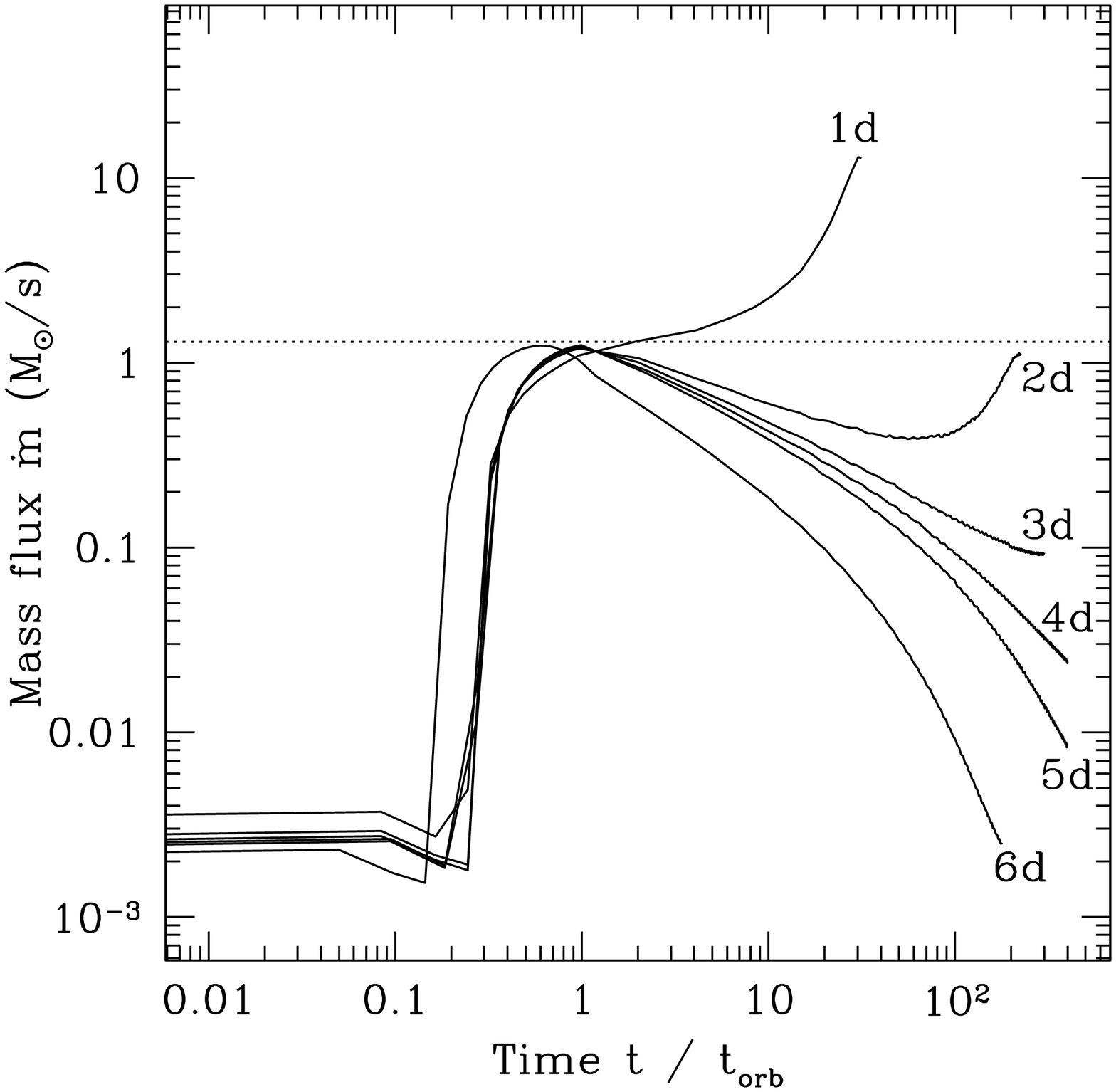,width=0.47\textwidth}\\
\end{tabular}
\end{center}
\caption{\textbf{Time evolution of the mass flux:} the mass flux is plotted as a function of time 
(normalized by the orbital time) for all 26 models listed in Table~\ref{tab:ModelParameters} in 
the case where both the mass and the spin of the black hole increase with time according to the 
procedure described in Section~\ref{sec:MethodBH}. For models 1d to 6d (bottom-right panel), a 
horizontal dotted line indicates the constant value of the mass flux in the stationary regime. 
Unstable models are characterized by the sudden divergence of the mass flux in a few orbital periods. 
The transition between stable and unstable models is clearly visible in all four panels.}
\label{massflux1}
\end{figure*}

\begin{figure*}
\begin{center}
\begin{tabular}{cc}
\hspace{1cm}
$M_\mathrm{D}/M_\mathrm{BH}=1$ and $\Delta W_\mathrm{in}=0.75 |W_\mathrm{cusp}|$ & 
\hspace{1cm}
$M_\mathrm{D}/M_\mathrm{BH}=0.1$  and $\Delta W_\mathrm{in}=0.5 |W_\mathrm{cusp}|$ \\
\psfig{file=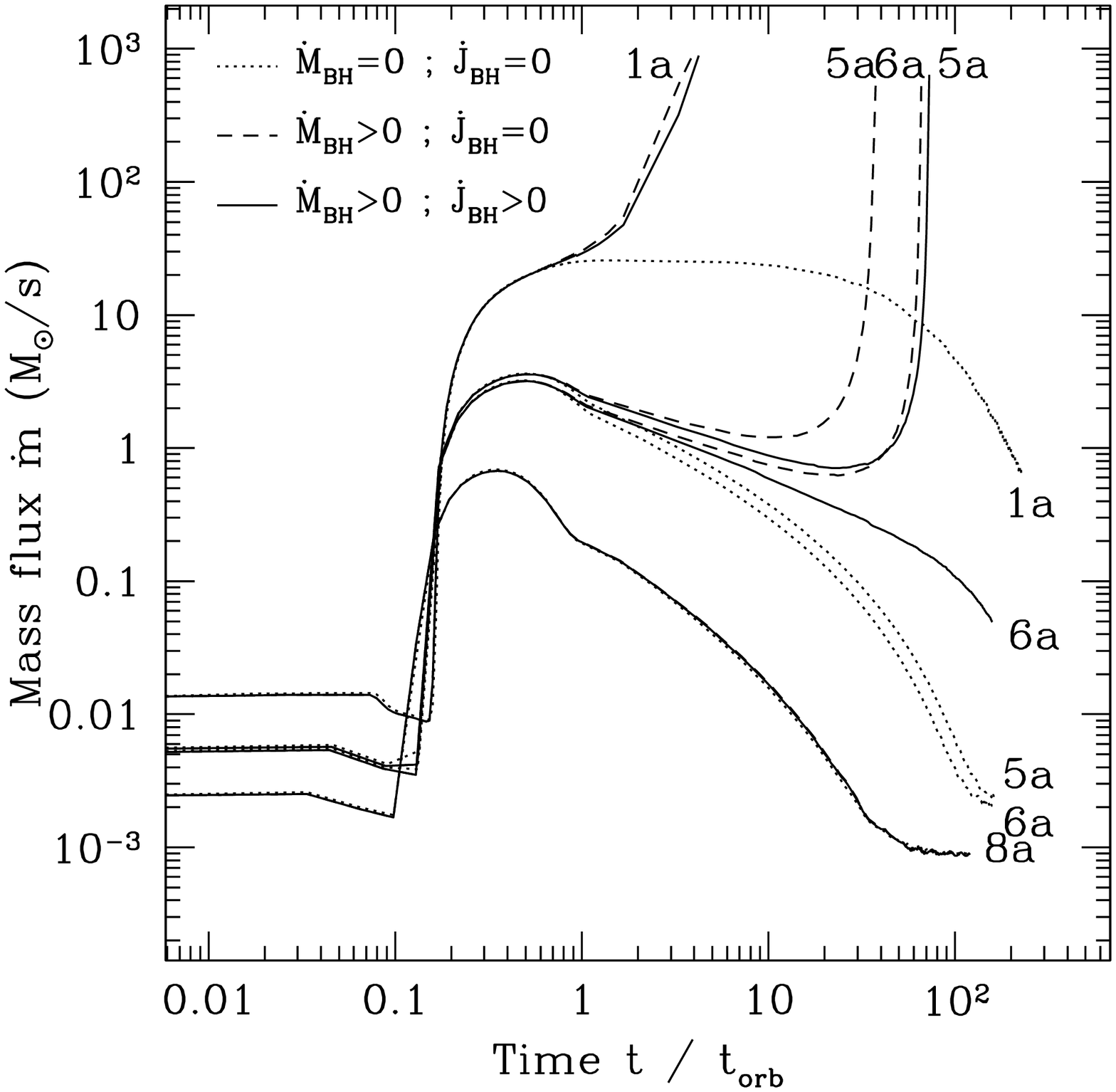,width=0.47\textwidth} & 
\psfig{file=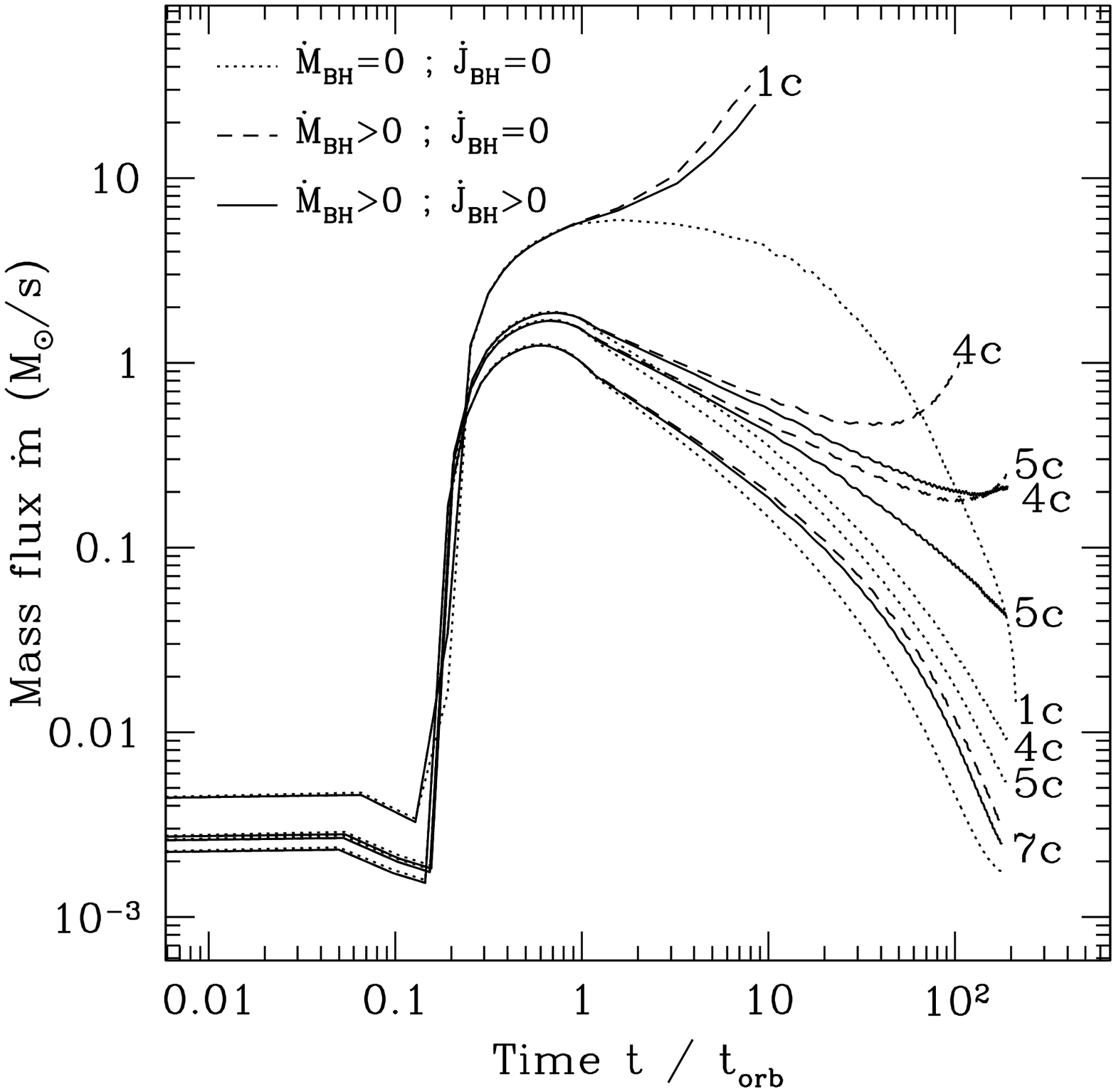,width=0.47\textwidth}\\
\hspace{1cm}
$M_\mathrm{D}/M_\mathrm{BH}=1$ and $\Delta W_\mathrm{in}=0.5 |W_\mathrm{cusp}|$ & 
\hspace{1cm}
$M_\mathrm{D}/M_\mathrm{BH}=0.1$  and $\dot{m}_\mathrm{stat}\sim 1.3\ \mathrm{M_{\odot}/s}$ \\
\psfig{file=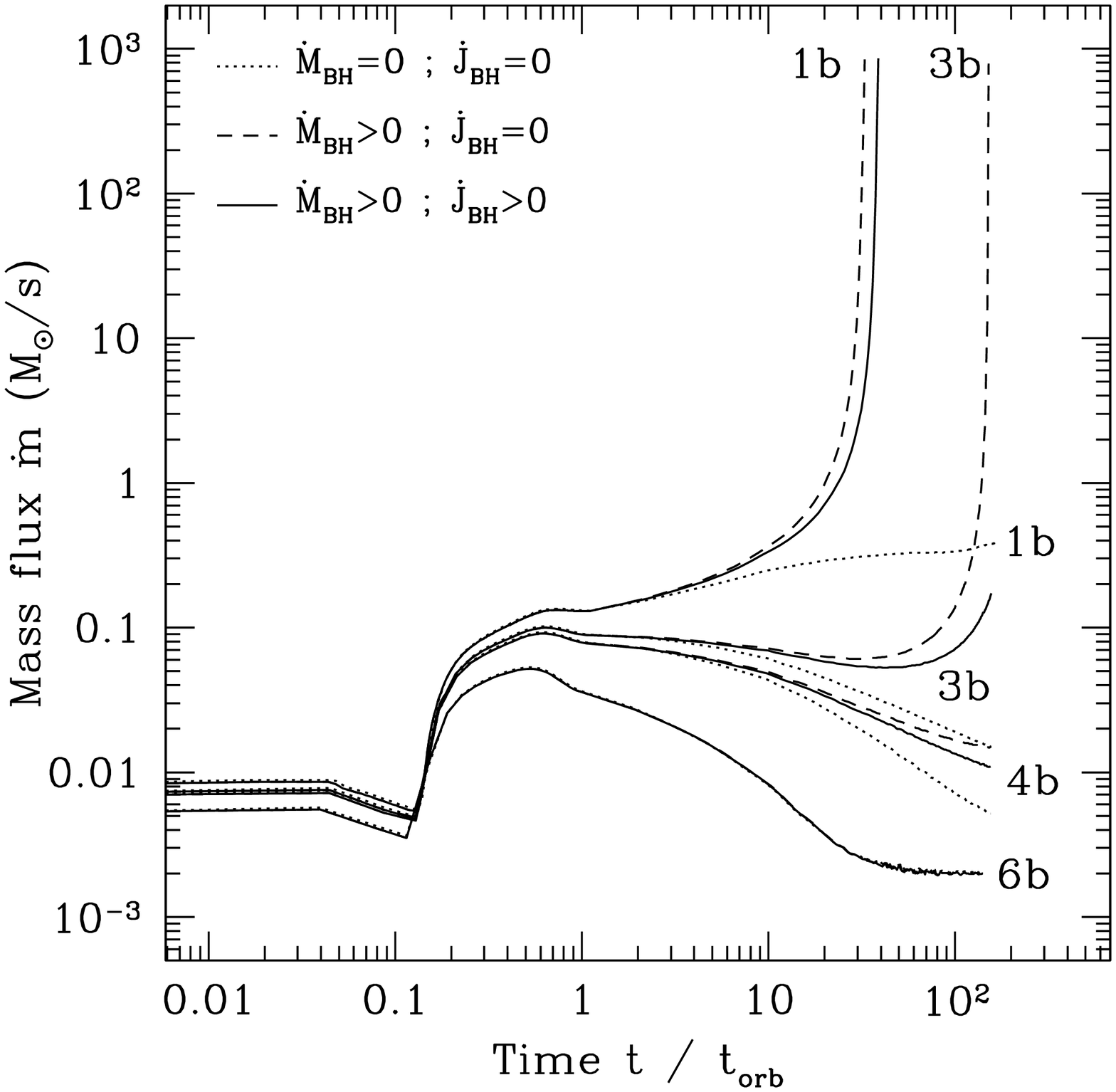,width=0.47\textwidth} & 
\psfig{file=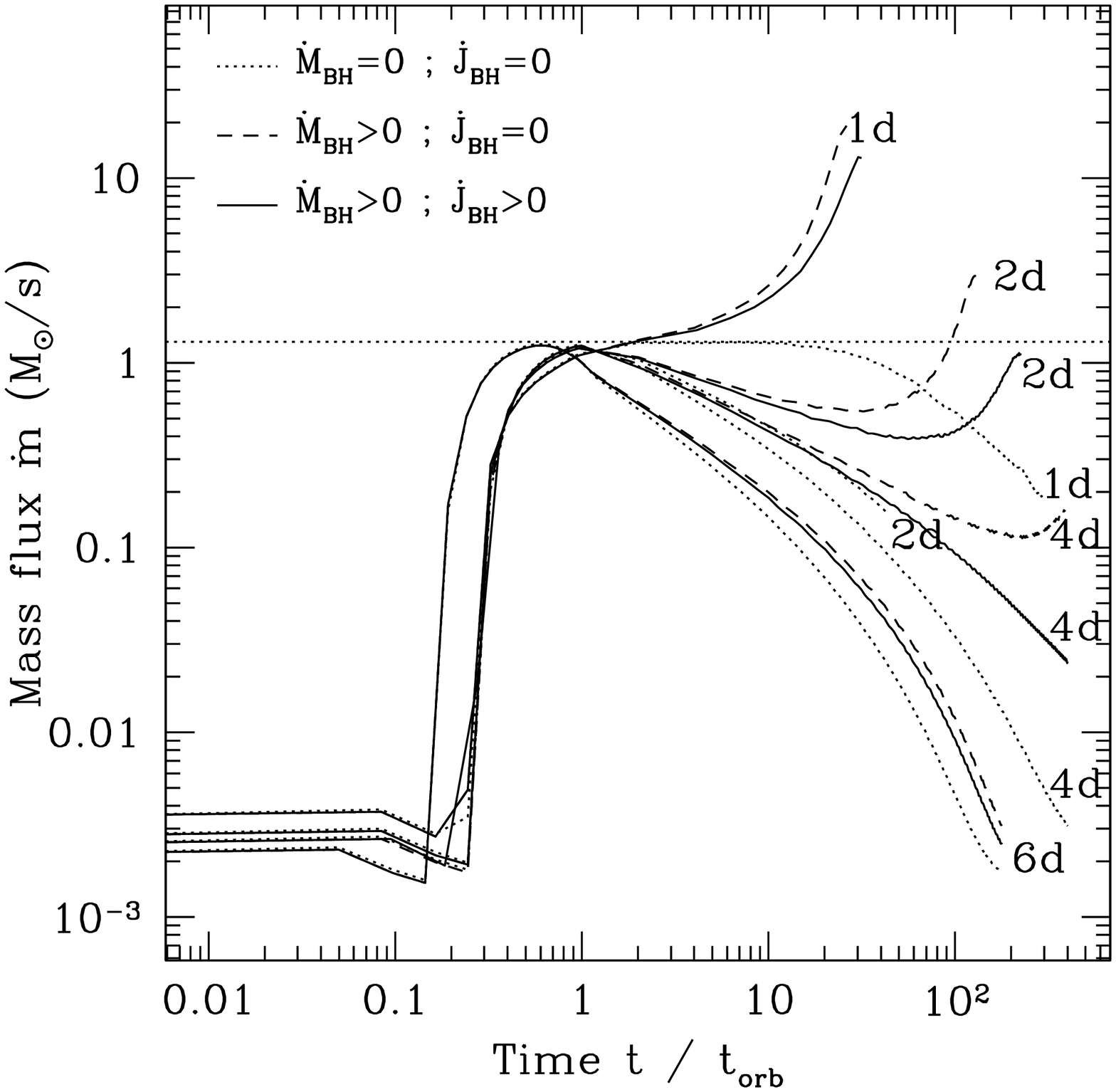,width=0.47\textwidth}\\
\end{tabular}
\end{center}
\caption{\textbf{Time evolution of the mass flux:} the mass flux is plotted as a function of time 
(normalized by the orbital time) for a representative selection of the models listed in 
Table~\ref{tab:ModelParameters}. The line styles indicate the three different cases considered: 
black hole of constant mass and angular momentum (dotted line), black hole of increasing mass 
(dashed line), and black hole of increasing mass {\it and} spin (solid line). For models 1d, 2d, 
4d and 6d, a horizontal dotted line indicates the constant value of the mass flux in the stationary 
regime. Notice that when the mass and the spin of the black hole are kept constant to their initial 
values the discs always remain stable irrespective of their angular momentum distribution.}
\label{massflux2}
\end{figure*}

\begin{figure*}
\begin{center}
\begin{tabular}{cc}
\hspace{1cm}
$M_\mathrm{D}/M_\mathrm{BH}=1$ and $\Delta W_\mathrm{in}=0.75 |W_\mathrm{cusp}|$ & 
\hspace{1cm}
$M_\mathrm{D}/M_\mathrm{BH}=0.1$  and $\Delta W_\mathrm{in}=0.5 |W_\mathrm{cusp}|$ \\
\psfig{file=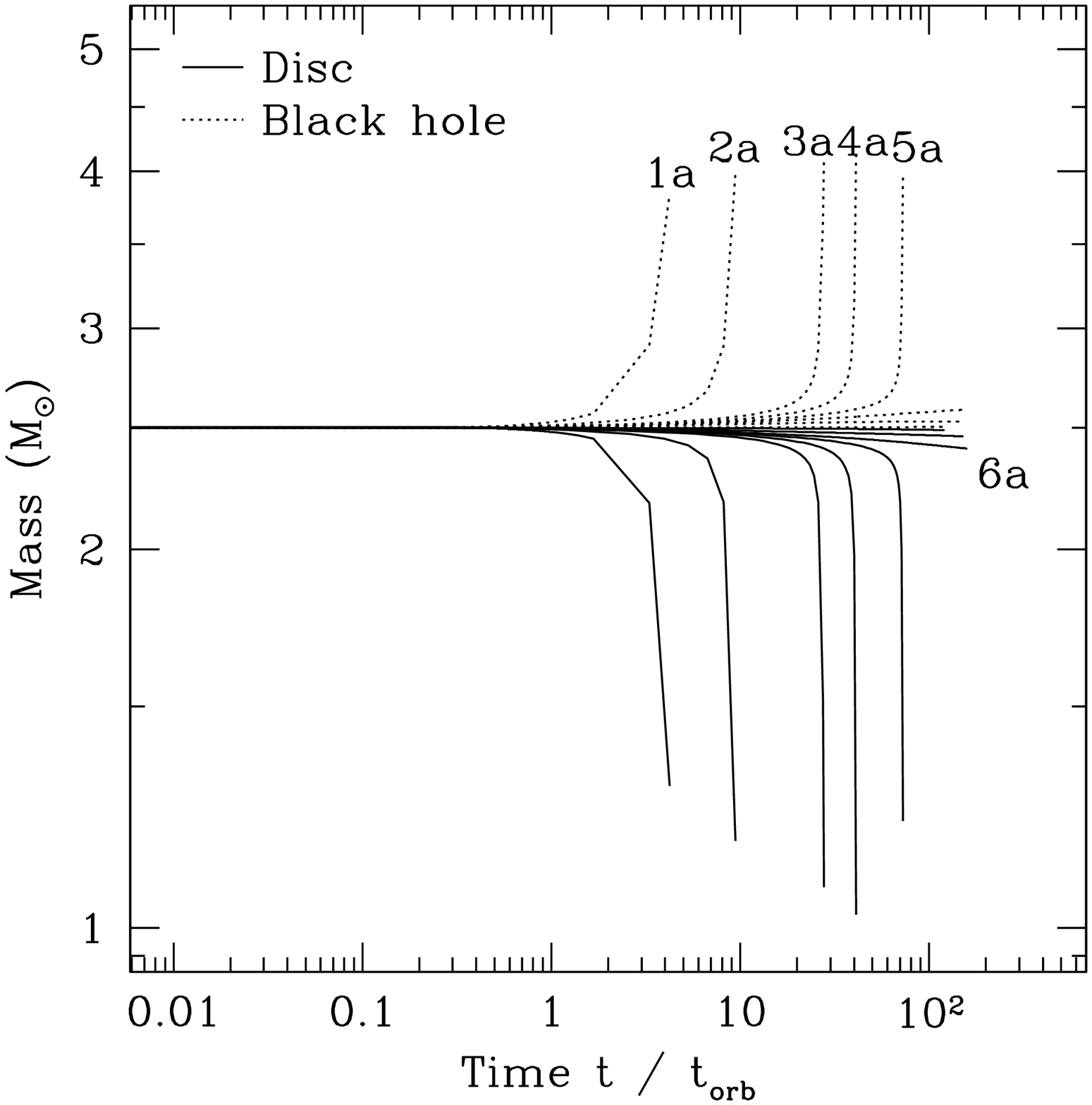,width=0.47\textwidth} & 
\psfig{file=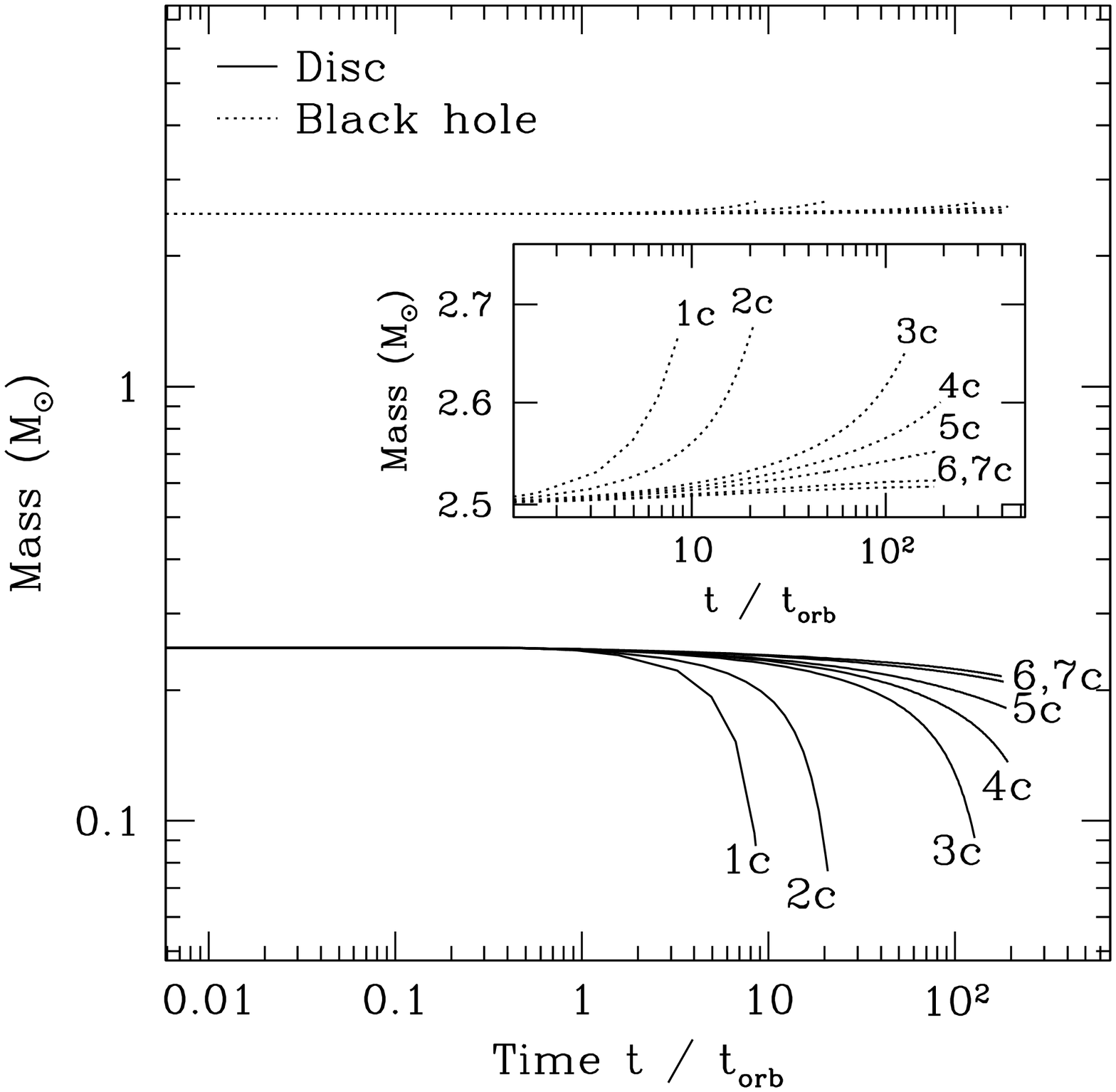,width=0.47\textwidth}\\
\hspace{1cm}
$M_\mathrm{D}/M_\mathrm{BH}=1$ and $\Delta W_\mathrm{in}=0.5 |W_\mathrm{cusp}|$ & 
\hspace{1cm}
$M_\mathrm{D}/M_\mathrm{BH}=0.1$  and $\dot{m}_\mathrm{stat}\sim 1.3\ \mathrm{M_{\odot}/s}$ \\
\psfig{file=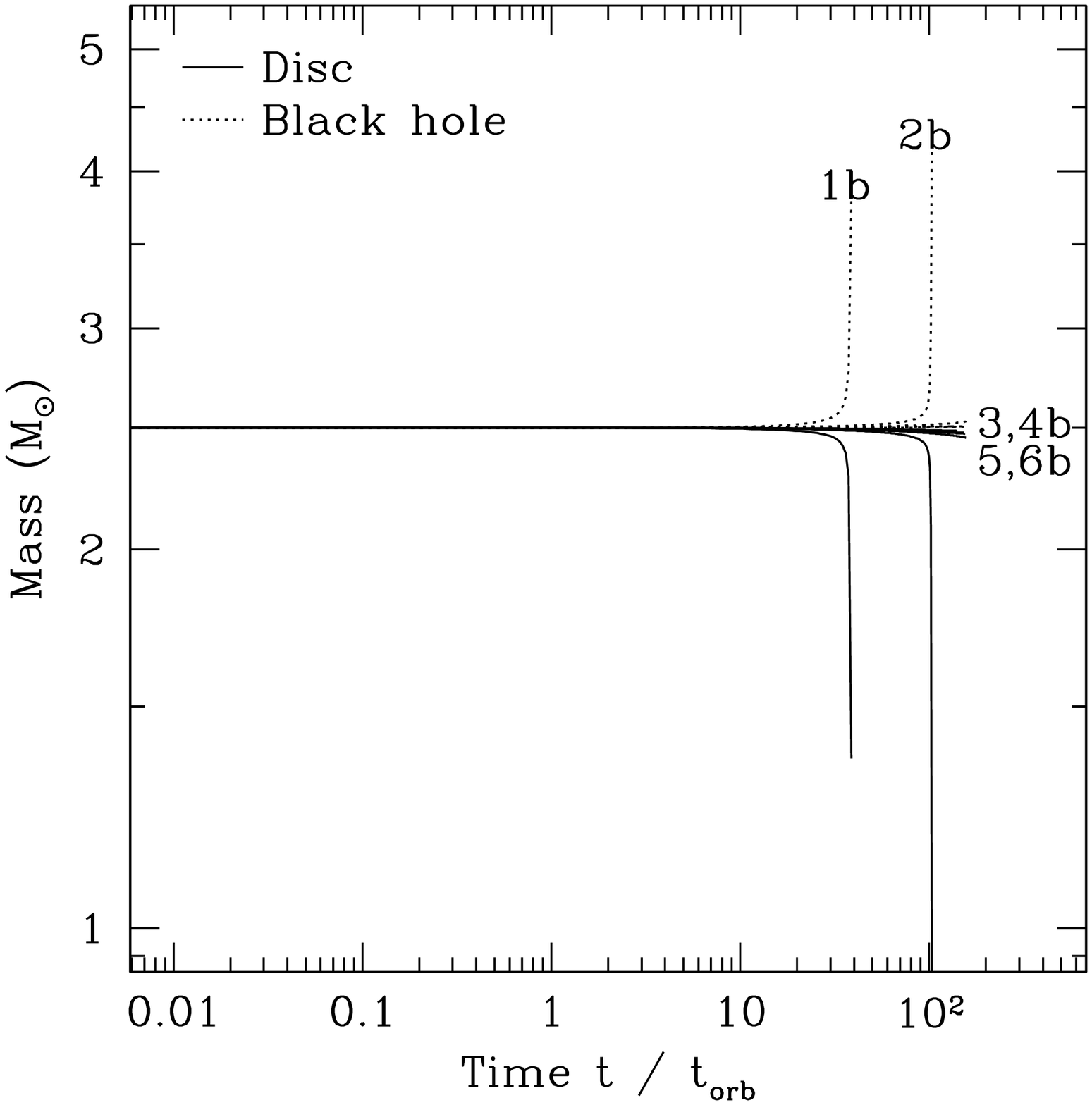,width=0.47\textwidth} & 
\psfig{file=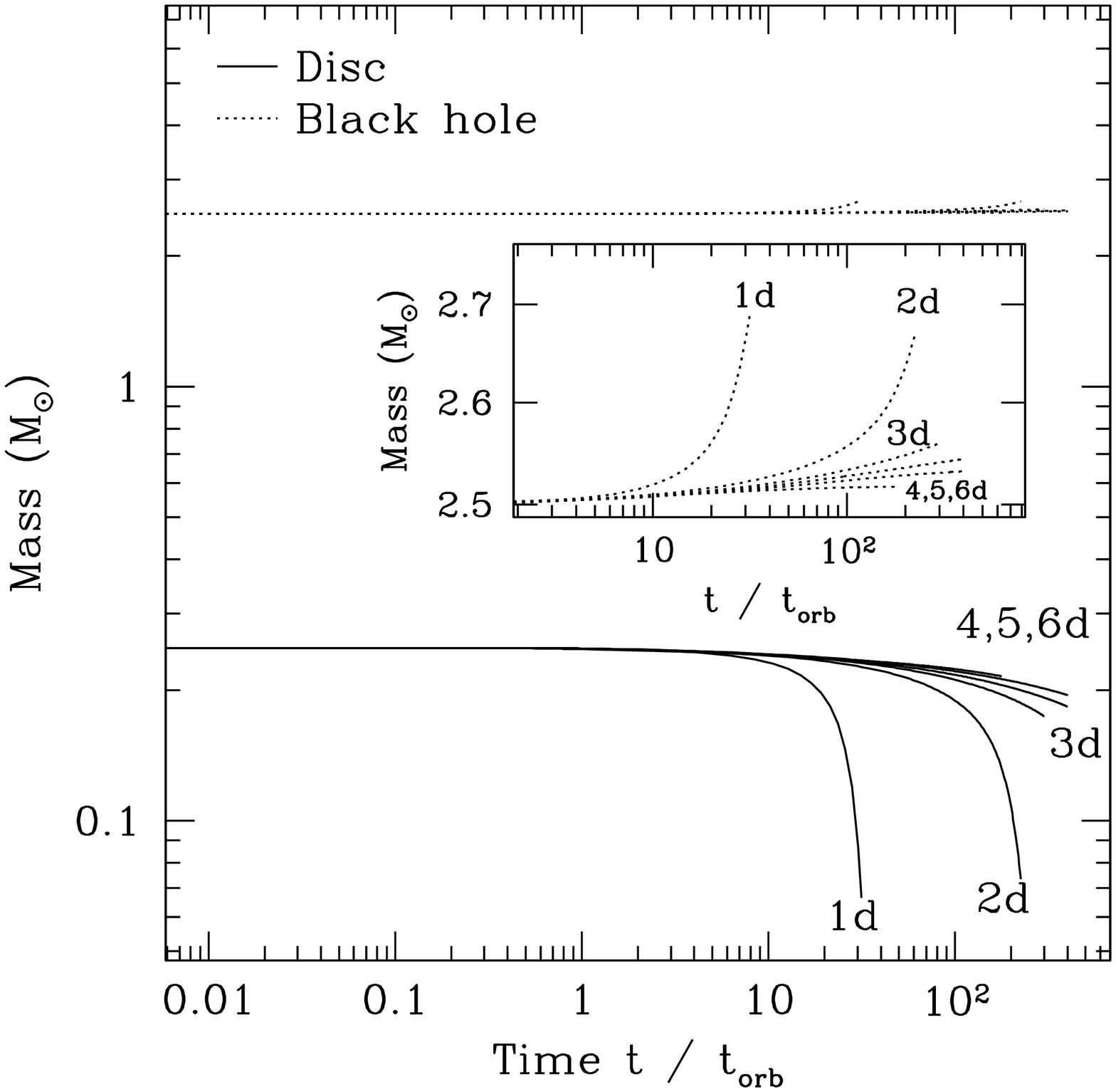,width=0.47\textwidth}\\
\end{tabular}
\end{center}
\caption{\textbf{Time evolution of the mass:} the mass of the disc (solid lines) and the mass of the 
black hole (dotted lines) are plotted as a function of time for all models listed in 
Table~\ref{tab:ModelParameters}, only in the case where both the mass and the spin of the black hole 
increase with time. For models with a disc-to-hole mass ratio $M_\mathrm{D}/M_\mathrm{BH}=0.1$
(right panels) the corresponding insets show a blow up of the evolution of the mass of the black 
hole in the latest stages of the simulation.}
\label{mass1}
\end{figure*}

\begin{figure*}
\begin{center}
\begin{tabular}{cc}
\hspace{1cm}
$M_\mathrm{D}/M_\mathrm{BH}=1$ and $\Delta W_\mathrm{in}=0.75 |W_\mathrm{cusp}|$ & 
\hspace{1cm}
$M_\mathrm{D}/M_\mathrm{BH}=0.1$  and $\Delta W_\mathrm{in}=0.5 |W_\mathrm{cusp}|$ \\
\psfig{file=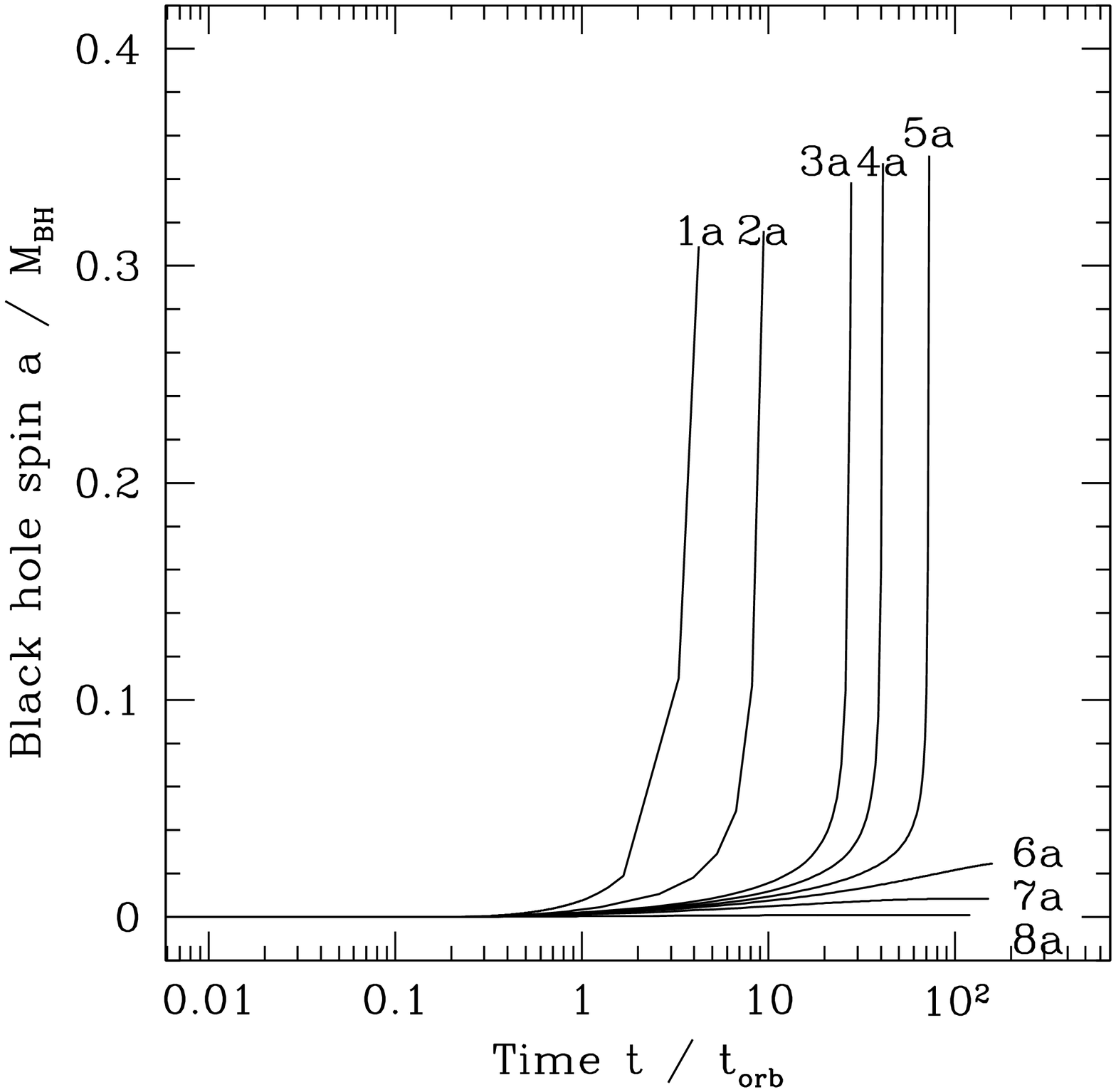,width=0.47\textwidth} & 
\psfig{file=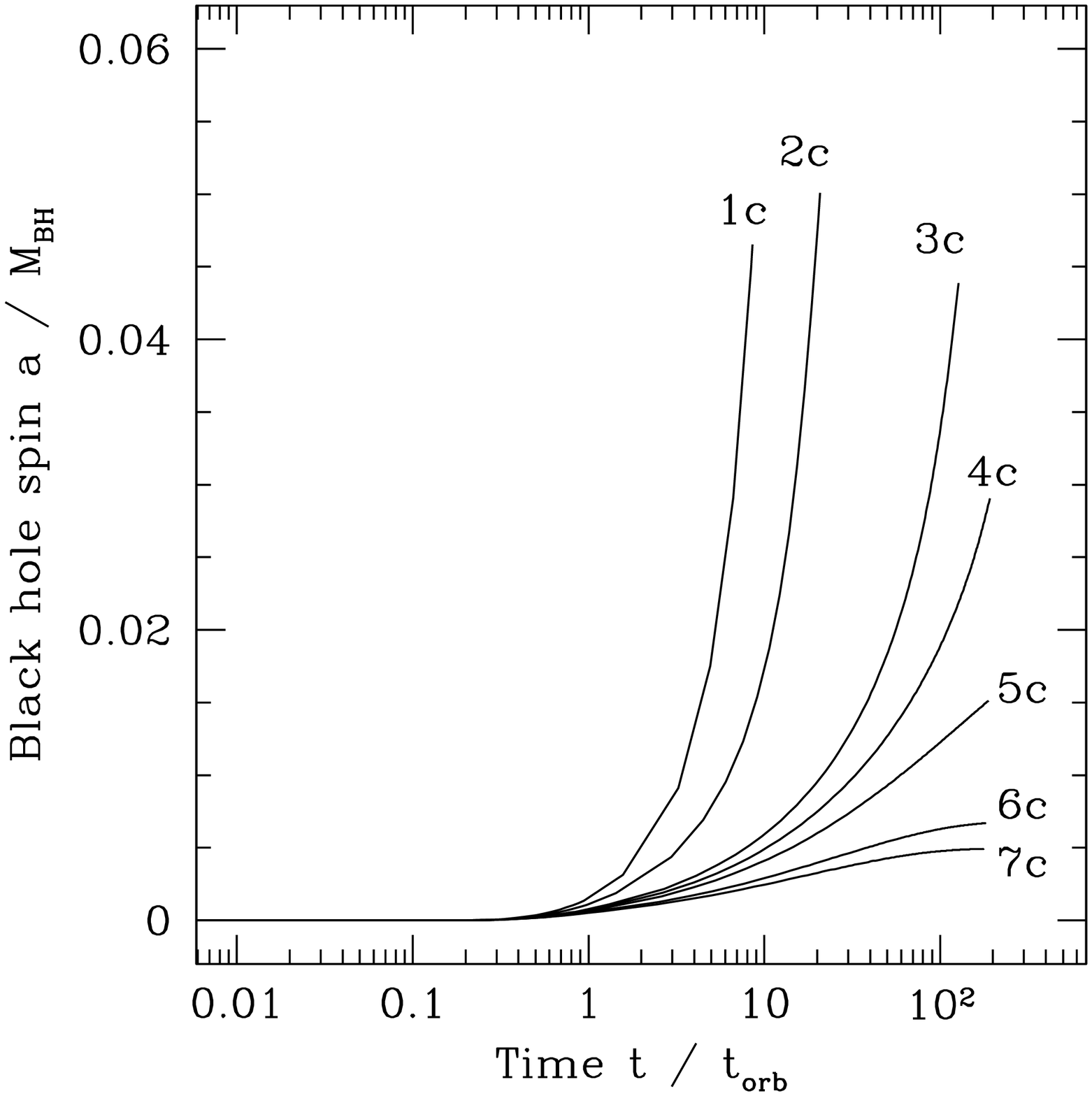,width=0.47\textwidth}\\
\hspace{1cm}
$M_\mathrm{D}/M_\mathrm{BH}=1$ and $\Delta W_\mathrm{in}=0.5 |W_\mathrm{cusp}|$ & 
\hspace{1cm}
$M_\mathrm{D}/M_\mathrm{BH}=0.1$  and $\dot{m}_\mathrm{stat}\sim 1.3\ \mathrm{M_{\odot}/s}$ \\
\psfig{file=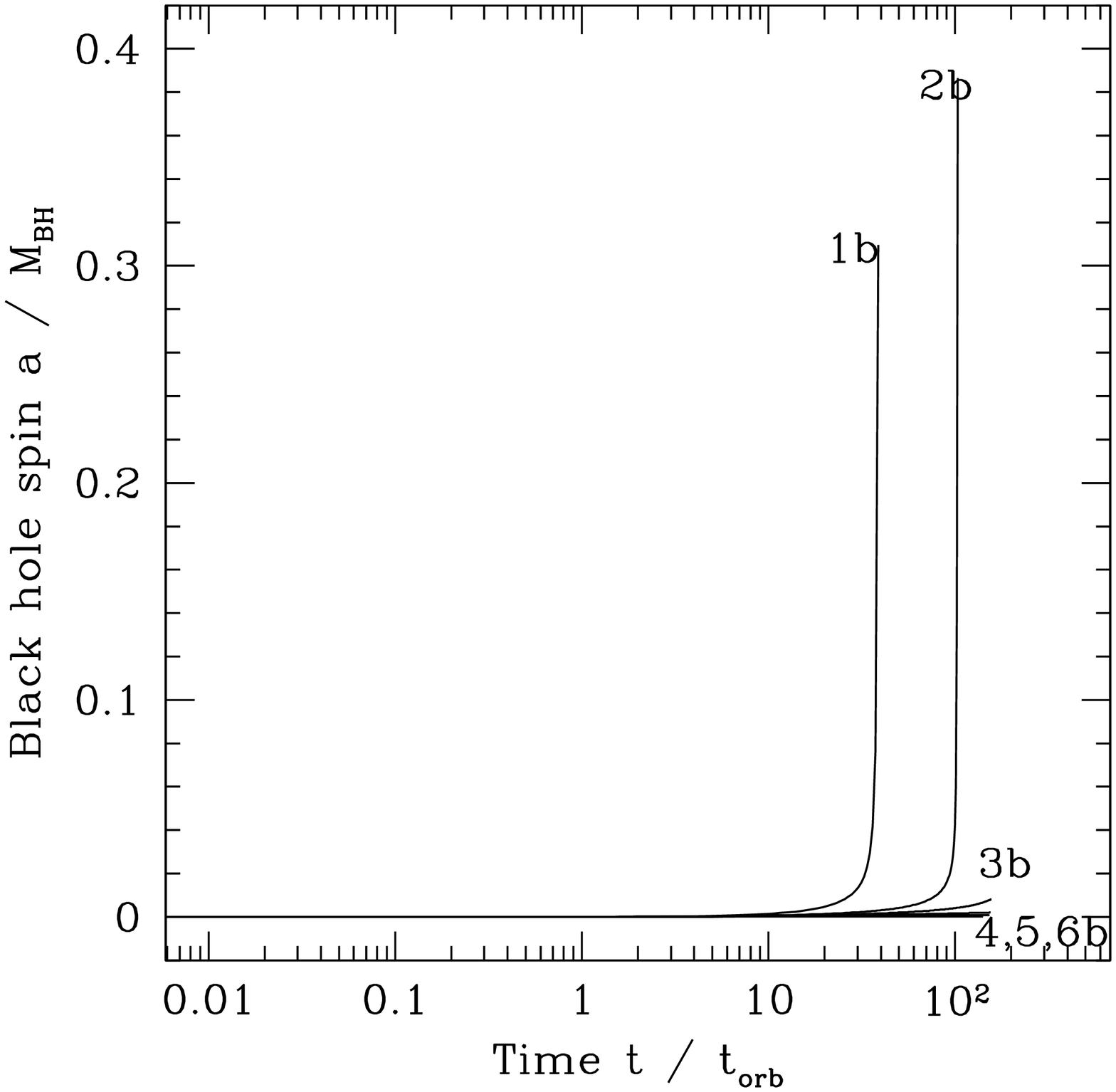,width=0.47\textwidth} &
\psfig{file=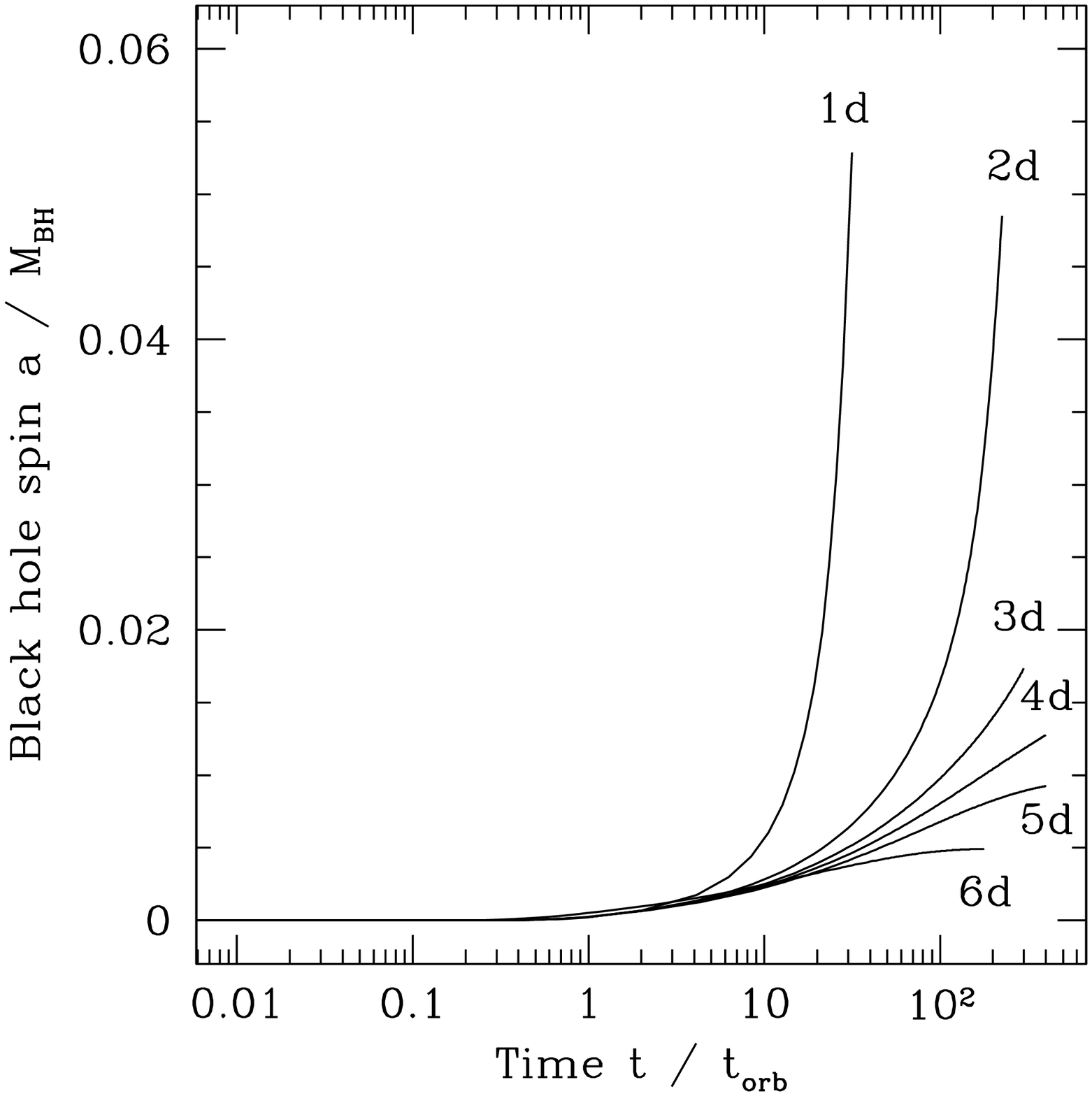,width=0.47\textwidth}\\
\end{tabular}
\end{center}
\caption{\textbf{Time evolution of the spin of the black hole:} the Kerr parameter $a$ is plotted as 
a function of time for all models listed in Table~\ref{tab:ModelParameters}. Unstable models with
disc-to-hole mass ratio $M_\mathrm{D}/M_\mathrm{BH}=1$ (left panels) spin up the initial Schwarzshild
black hole up to $a/M_\mathrm{BH}\sim 0.3-0.35$. Correspondingly, unstable models with disc-to-hole mass ratio
$M_\mathrm{D}/M_\mathrm{BH}=0.1$ only reach $a/M_\mathrm{BH}\sim 0.04-0.05$.}
\label{spin1}
\end{figure*}

Each of the 26 models listed in Table~\ref{tab:ModelParameters} is evolved three times: in 
the first series of runs (series one hereafter) both the mass and the spin of the initial 
(Schwarzschild) black hole are kept constant to their initial values. Correspondingly, in 
the second series of runs (series two hereafter) only the mass of the black hole is allowed 
to increase according to the procedure specified in the preceding section, while the black
hole angular momentum is kept fixed to zero. Finally, in the third series of runs (series 
three hereafter) both the mass and the spin of the black hole are allowed to increase. The 
growth rate is monitored through the mass and angular momentum transfer from the disc to 
the black hole across the innermost radial zone. Keeping all these various possibilities 
into account gives a total of 78 simulations. These simulations are usually stopped at 
about $t\simeq 200\ t_\mathrm{orb}$ if the runaway instability has not occured before, which 
in two cases (models 3b and 4c) does not allow for the correct determination of the timescale of 
the instability (a lower value is given in Table~\ref{tab:ModelParameters}), and in one case 
(model 3d) does not allow to determine if the model is stable or not (a lower limit on the 
timescale of the instability in the latter case is also given in Table~\ref{tab:ModelParameters}).

As mentioned in Section~\ref{sec:MethodBH} most of our initial models (of series three) are 
evolved assuming a conservative value for the efficiency of the angular momentum transfer, 
$\eta=0.2$ (cf. Eq.~(\ref{spinincrease2})). In addition to these simulations we also evolve 
model 3c with two different values for this efficiency, namely $\eta=0.5$ (model 3c') and 
$\eta=1.0$ (model 3c''). This allows us to check how sensitive our results on the runaway 
instability are on the amount of angular momentum transferred from the disc to the black hole.

The time evolution of the mass flux for the different models is plotted in Figs.~\ref{massflux1}
and~\ref{massflux2}. Additional time evolution plots for the mass of the disc and the mass of 
the black hole, as well as for the spin of the black hole, appear in Figs.~\ref{mass1} 
and~\ref{spin1}, respectively. For the sake of clarity in the presentation the time evolution 
plots of the mass flux in Fig.~\ref{massflux1} include all models listed in 
Table~\ref{tab:ModelParameters} (apart from models 3c' and 3c''), but only for series three, 
i.e. when both the mass and the spin of the black hole increase with time. Correspondingly, 
the mass flux evolution displayed in Fig.~\ref{massflux2} includes a representative selection 
of the models listed in Table~\ref{tab:ModelParameters} for all three possible combinations we 
consider for the evolution of the mass and spin of the black hole ({\it dotted line}: black hole 
of constant mass and angular momentum, series one; {\it dashed line}: black hole of increasing 
mass, series two; {\it solid line}: black hole of increasing mass {\it and} spin, series three). 
Similarly, the curves displayed in Figs.~\ref{mass1} and~\ref{spin1} correspond to all 26 initial 
models for evolutions of series three. In Figs.~\ref{massflux1}, \ref{massflux2}, \ref{mass1}, 
and \ref{spin1}, the left panels correspond to a disc-to-hole mass ratio 
$M_\mathrm{D}/M_\mathrm{BH}=1$, i.e. set `a' (top-left) and set `b' (bottom-left), and the right 
panels to $M_\mathrm{D}/M_\mathrm{BH}=0.1$, i.e. set `c' (top-right) and set `d' (bottom-right). 
The evolution of the mass flux in models 3c, 3c' and 3c'' is plotted in an additional figure 
(see Fig.~\ref{fig:massflux3} below) to illustrate the role of the efficiency $\eta$ of the
angular momentum transfer. In all these figures the time is given in units of the orbital time 
at the centre of the torus, whose precise value can be found in Table~\ref{tab:ModelParameters}. 
The results for each four classes of models listed in Table~\ref{tab:ModelParameters} are 
discussed in the following sections.

\subsubsection{Early time evolution of the mass flux} 

The time evolution of the mass flux is presented in Figs.~\ref{massflux1} and \ref{massflux2}.  
For all models and all series the early evolution is qualitatively identical: (i) initially 
the mass flux has a constant and very low value. This is due to the spherical accretion of the 
`dust' solution which we have used to fill the region outside the initial torus. This component 
of the accretion rate is negligible in the subsequent evolution of the system; (ii) the mass 
flux increases abruptly and rapidly reaches a stationary regime (in about $1-2$ orbital periods). 
This is triggered by the fact that the initial torus overfills its Roche lobe, due to the 
positive potential barrier $\Delta W_\mathrm{in}$. As shown in Paper 1 for the case of a
Schwarzschild black hole and tested again in Section~\ref{sec:tests} for an $a=0.9$ Kerr
black hole, the value $\dot{m}_\mathrm{stat}$ of the mass flux in this regime follows closely 
$\dot{m}\propto \Delta W_\mathrm{in}^{4}$, which is the theoretical expectation for $\gamma=4/3$ 
\citep{abramowicz:78}. Table~\ref{tab:ModelParameters} gives the value of the mass flux in the 
stationary regime for all models, which is independent of the considered series of runs. Notice 
that the potential barrier $\Delta W_\mathrm{in}$ of the models of class `d' have been adjusted 
to yield a constant mass flux $\dot{m}_\mathrm{stat}=1.3\ \mathrm{M_{\odot}} s^{-1}$ which is 
indicated in the bottom-right panel of Figs.~\ref{massflux1} and \ref{massflux2} by a horizontal 
dotted line. After this initial step, the time evolution of the mass flux differs strongly 
for the three series of runs.

\subsubsection{The runaway instability of constant angular momentum discs}

As we showed in Papers 1 and 2, for constant angular momentum discs (models 1a, 1b, 1c, and 1d; 
$\alpha=0$) the time evolution of the system changes dramatically when the spacetime dynamics 
is taken into account, i.e. when either $M_{\mathrm{BH}}$ only or $M_{\mathrm{BH}}$ {\it and} 
$J_{\mathrm{BH}}$ increase. In either case the runaway instability, reflected in the rapid 
growth of the mass accretion rate, appears on a dynamical timescale of $\sim 4.3\ t_{\mathrm{orb}} 
\sim 6.9\ \mathrm{ms}$ for model 1a (respectively, $\sim 39\ t_\mathrm{orb}\sim 88\ \mathrm{ms}$ 
for model 1b, $\sim 7.5\ t_\mathrm{orb}\sim 12\ \mathrm{ms}$ for model 1c, and 
$\sim 28\ t_\mathrm{orb}\sim 51\ \mathrm{ms}$ for model 1d). The time derivative of the mass flux 
also increases, which implies the rapid divergence of $\dot{m}$. As already mentioned, this effect 
is more easily observed for models with $M_\mathrm{D}/M_\mathrm{BH}=1$. By comparing the case where 
only $M_{\mathrm{BH}}$ increases (dashed line in Fig.~\ref{massflux2}) with the case where both 
$M_{\mathrm{BH}}$ and $J_{\mathrm{BH}}$ change (solid line), it can be seen that the instability 
appears slightly later in the latter case, i.e. the angular momentum transfer tends to disfavour the 
instability. This is in agreement with the result first reported by \citet{wilson:84} for constant 
angular momentum tori, who showed using stationary models that when the spin of the black hole 
increases, the cusp moves toward the black hole and the torus is stabilized. 

\subsubsection{Stabilizing effect of an increasing $\alpha$} 

The dynamical evolution of the disc, however, changes notably for non-constant angular momentum 
discs. The larger the slope $\alpha$ in the angular momentum distribution in the disc is, the 
more stable the models become, as it is visible in Figs.~\ref{massflux1} and~\ref{massflux2}, at 
least for the evolution times considered in our simulations ($\sim 200\ t_{\mathrm{orb}}$). This 
result applies to both, models of series two and series three. In the former case the instability 
is supressed for larger values of $\alpha$ than in the latter case. For instance it is possible 
to see in Fig.~\ref{massflux2} that models 1a, 2a, 3a and 4a ($\alpha \le 0.08$) undergo the 
runaway instability in series two and three, model 5a ($\alpha=0.085$) is unstable only if the 
mass of the black hole increases (series two, dashed line) and is stable when the increase of the 
black hole spin is also taken into account (series three, dotted line), and models 6a, 7a and 8a 
($\alpha\ge 0.09$) are stable in all series. The same transition appears in model 3b for set `b', 
model 5c for set `c' and model 2d in set `d'. 

\begin{figure*}
\begin{center}
\begin{tabular}{cc}
\psfig{file=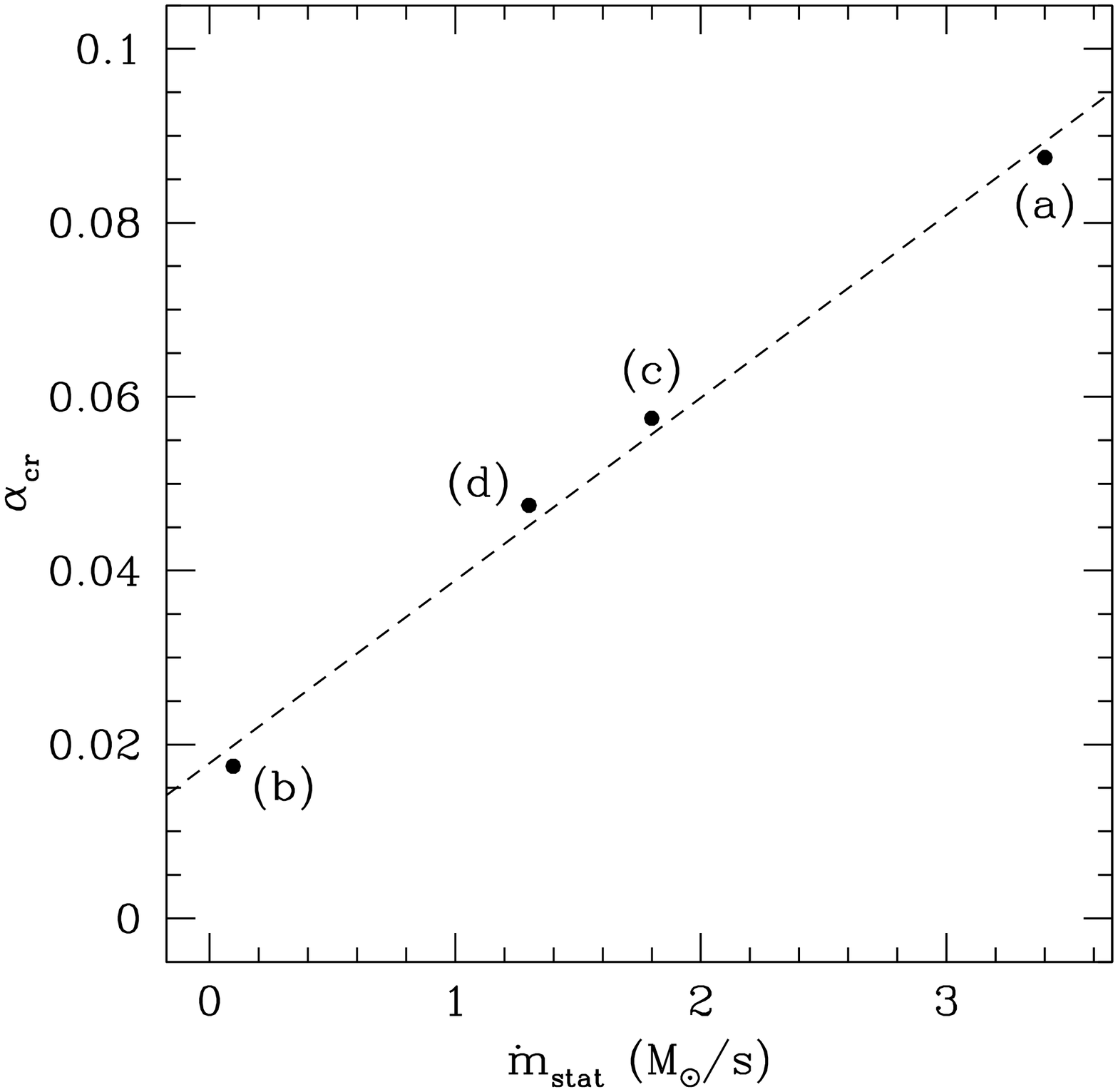,width=0.47\textwidth} &
\psfig{file=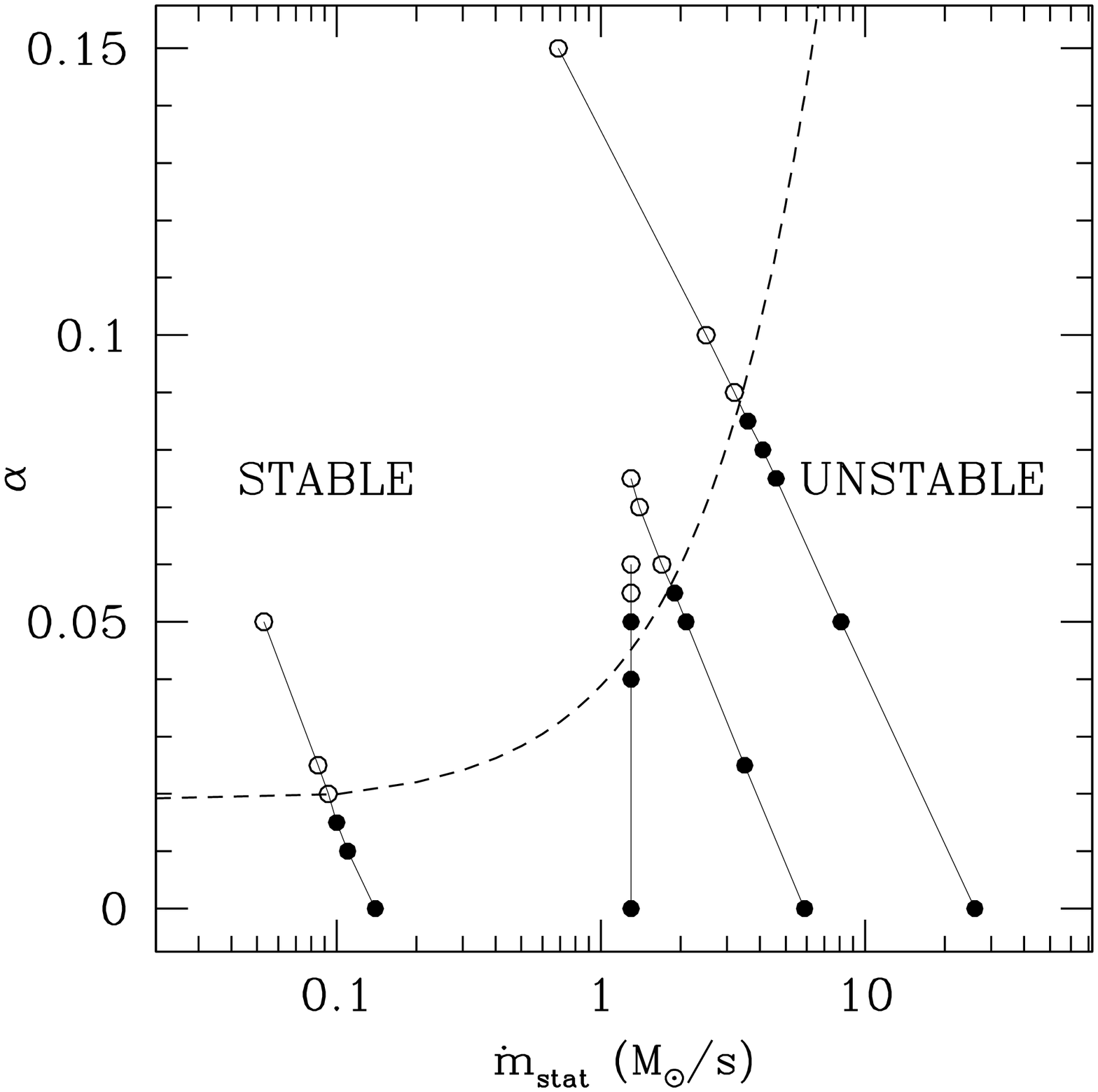,width=0.47\textwidth}\\
\end{tabular}
\end{center}
\caption{\textit{Left:} the best estimate of the critical slope $\alpha_\mathrm{cr}$ is 
plotted as a function of the initial mass flux $\dot{m}_\mathrm{stat}$. The four points 
corresponds to the mean value between models 5a and 6a (set `a'), models 3b and 4b (set `b'), 
models 4c and 5c (set `c'), and models 2d and 4d (set `d'). The dashed line indicates the 
empirical relation $\alpha_\mathrm{cr}\simeq 0.018+0.021(\dot{m}/ \mathrm{M}_{\odot} 
s^{-1})$. \textit{Right:} all models (series three) are plotted in the 
$\dot{m}_\mathrm{stat}$--$\alpha$ plane. Filled circles indicate unstable discs and empty 
circles stable discs. The dashed line separating the stable and unstable discs corresponds 
to the empirical relation shown in the left panel.}
\label{fig:timescale3}
\end{figure*}

Fig.~\ref{massflux1}, where all models of series three (both $M_\mathrm{BH}$ and $J_\mathrm{BH}$ 
increase) are plotted, shows clearly this transition between unstable and stable models. 
Correspondingly, Figure~\ref{fig:timescale3} shows the location of these models in the 
$\dot{m}_\mathrm{stat}$--$\alpha$ parameter plane. Unstable models are indicated with a black 
circle and stable models with an empty circle. The region of stability appears clearly in the 
top-left corner of the figure. The critical slope $\alpha_\mathrm{cr}$ separating the unstable 
and stable regimes depends on the initial mass flux. It is plotted in Fig.~\ref{fig:timescale3} 
and well fitted (but only for four sets of models!) by the relation $\alpha_\mathrm{cr}\simeq 
0.018+0.021 \left(\dot{m}_\mathrm{stat} / \mathrm{M_{\odot}}\right)$. The frontier between the 
two regions in the $\dot{m}_\mathrm{stat}$--$\alpha$ plane is plotted using this simple fit. 
Notice that this critical slope is always well below the Keplerian limit $\alpha=0.5$, and 
even well below the typical slope found in numerical simulations of compact binary coalescence 
and collapsars. For very high initial mass fluxes ($\dot{m}_\mathrm{stat} \gg 10\ \mathrm{M_\odot} 
s^{-1}$), the extrapolation of our estimate of the critical slope $\alpha_\mathrm{cr}$ indicates 
that models become more unstable. However, the discussion of the runaway instability is probably 
not relevant for such high mass fluxes as the stationary value of the mass flux is already large 
enough to destroy the disc in a few orbital periods.

The low value we find for the critical slope $\alpha_\mathrm{cr}$ ($\alpha_\mathrm{cr}\in 
[0.085;0.09]$ for the set of models `a'; respectively, $[0.015;0.02]$ for set `b', $[0.055;0.06]$ 
for set `c' and $[0.04;0.055]$ for set `d') indicates that a small increase of the specific 
angular momentum with the radial distance strongly stabilizes the disc, completely suppressing 
the runaway instability. This stabilizing effect that we find in our time-dependent general 
relativistic simulations is in complete agreement with previous studies based on stationary 
sequences of equilibrium configurations, either using a pseudo-Newtonian potential \citep{daigne:97} 
or in general relativity \citep{abramowicz:98}. Moreover, the precise value of the critical slope 
$\alpha_\mathrm{cr}$ is also in good accordance. \citet{daigne:97} (Newtonian) and 
\citet{abramowicz:98} (relativistic) found $\alpha_\mathrm{cr}=0.07$ and $\alpha_\mathrm{cr}=0.14$, 
respectively, for $a=0$, $M_\mathrm{BH}=2.44\ \mathrm{M_{\odot}}$ and $M_\mathrm{D}/M_\mathrm{BH}=
0.148$ ($\kappa$ and $\gamma$ being the same than the ones we use in this paper). The dependence 
of $\alpha_\mathrm{cr}$ on the mass flux is of course masked in these previous calculations as 
they assume an instantaneous transfer of mass.

\subsubsection{Time evolution of the disc and the black hole mass}

In Fig.~\ref{mass1} we plot the time evolution of the mass of the disc (solid lines) and the 
mass of the black hole (dotted lines) for all models in series three (both $M_\mathrm{BH}$ and 
$J_\mathrm{BH}$ increase). The unstable behaviour of models 1a to 5a, 1b to 3b, 1c to 3c, and 
1d to 3d is characterized by the sudden decrease of $M_{\mathrm{D}}$ and the corresponding 
increase of $M_\mathrm{BH}$. In the right panels, in which the discs are much less massive than 
in the left panels, it is seen that the high efficiency of the disc-to-hole mass transfer does 
not imply an important growth of the black hole mass. The inset of these two panels focus on 
the late time evolution when the increase is more pronounced. 

\begin{figure*}
\begin{center}
\begin{tabular}{cc}
\psfig{file=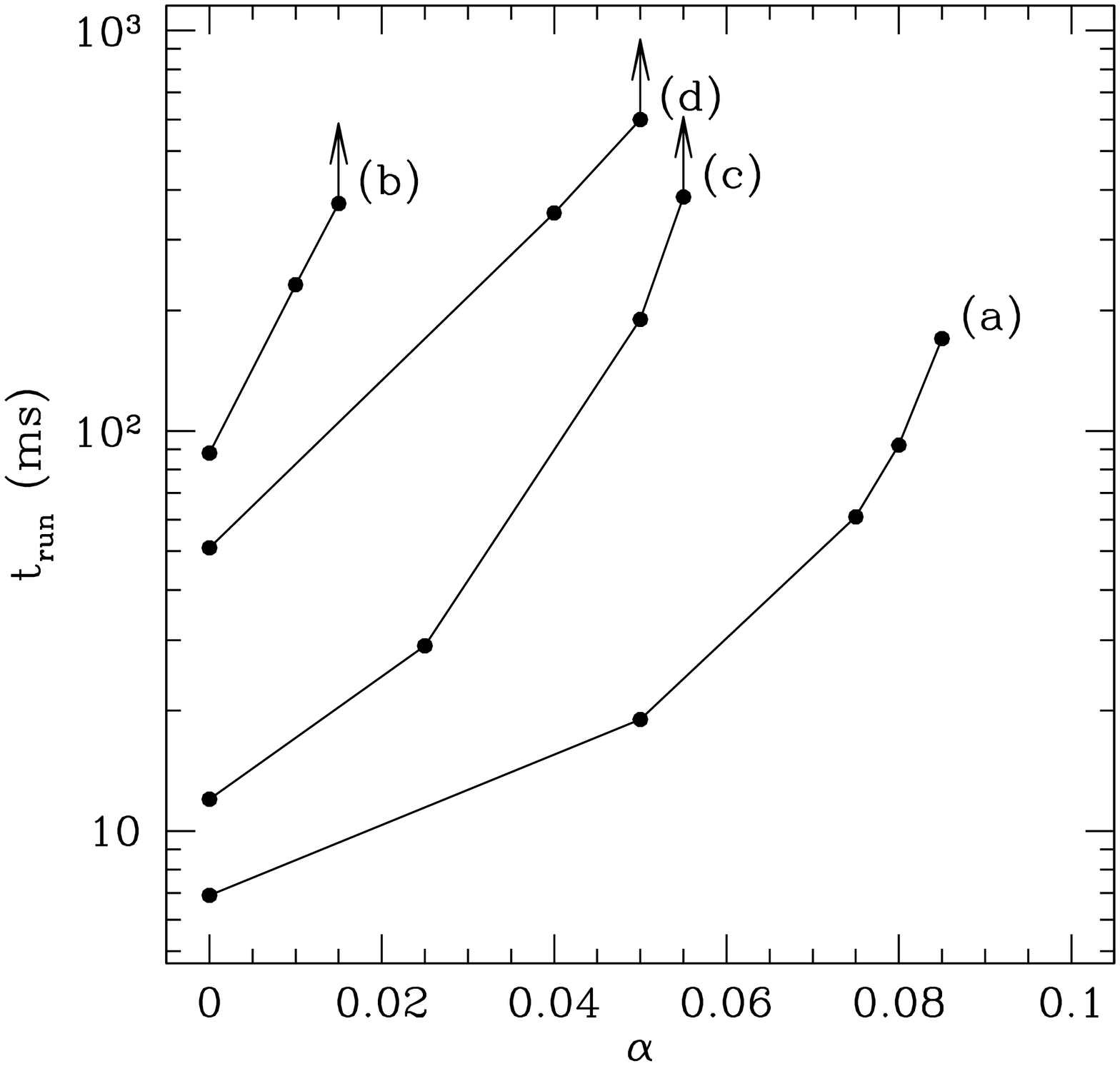,width=0.47\textwidth} &
\psfig{file=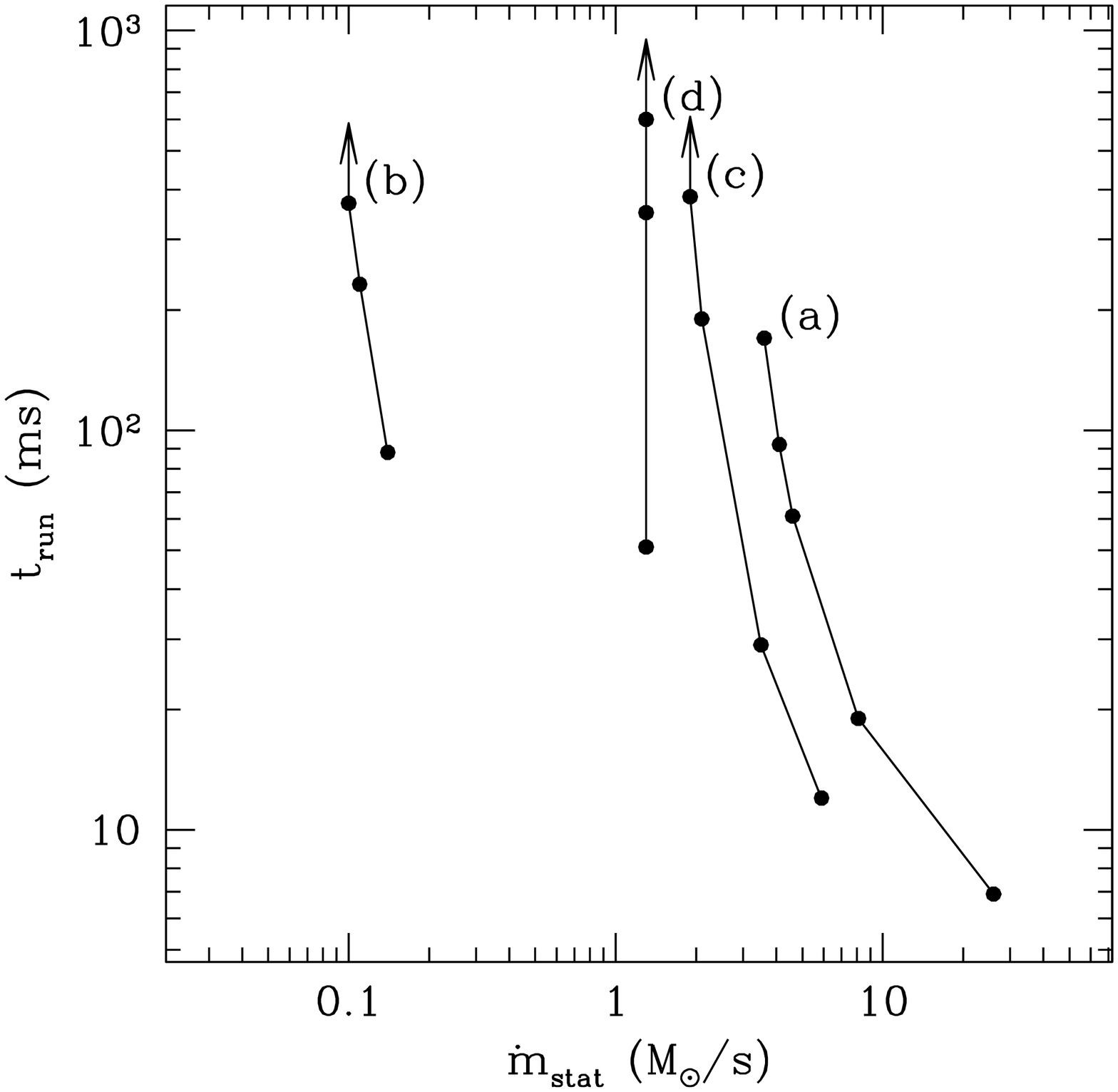,width=0.47\textwidth}\\
\end{tabular}
\end{center}
\caption{\textbf{Timescale of the runaway instability: all models.} \textit{Left:} the timescale 
$t_\mathrm{run}$ is plotted as a function of the slope of the angular momentum distribution. 
\textit{Right:} the same timescale is plotted as a function of the initial mass flux 
$\dot{m}_\mathrm{stat}$ which is induced by the potential barrier $\Delta W_\mathrm{in}$ at the 
surface of the disc.}
\label{fig:timescale1}
\end{figure*}

\begin{figure*}
\begin{center}
\begin{tabular}{cc}
\psfig{file=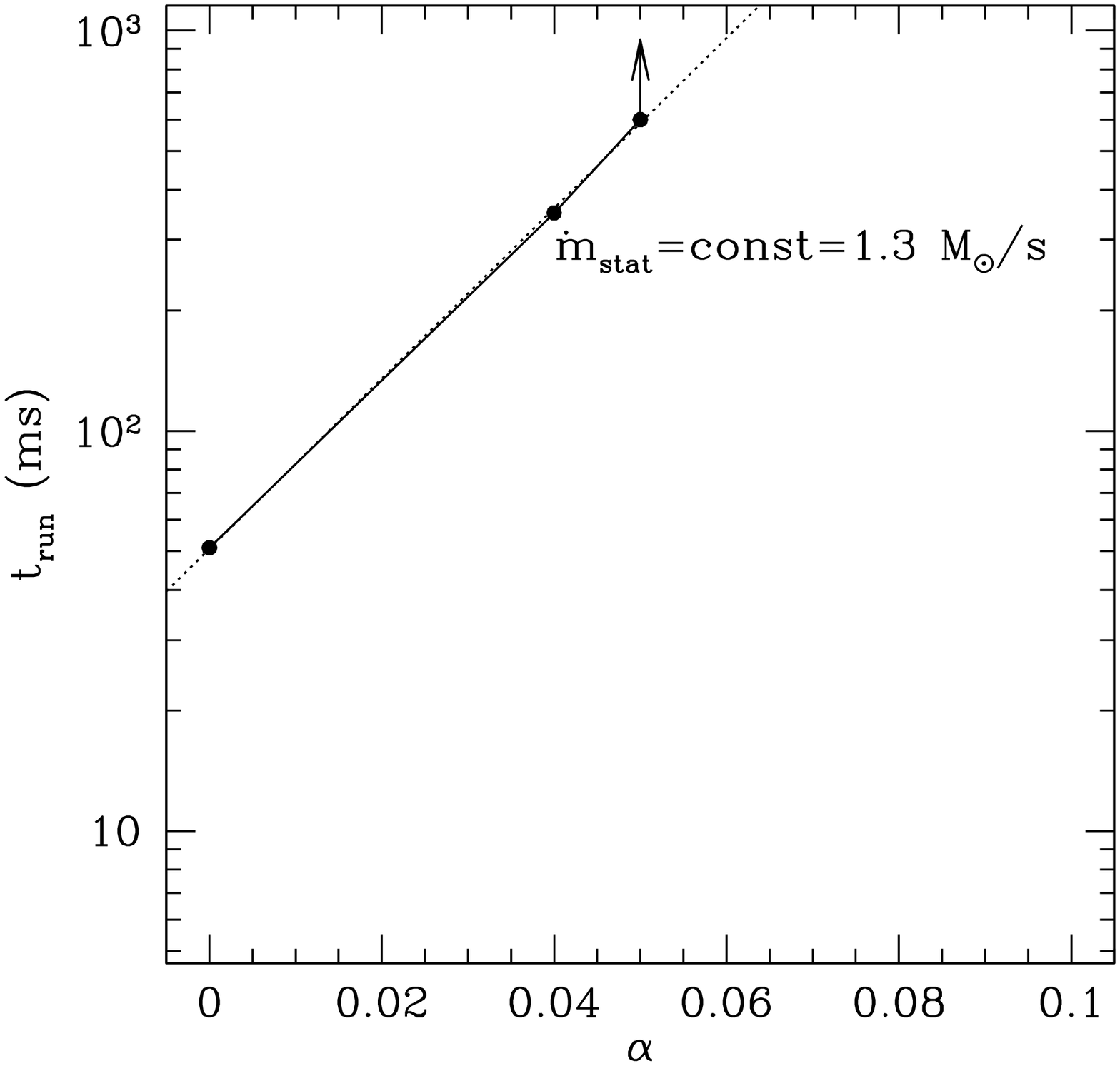,width=0.47\textwidth} &
\psfig{file=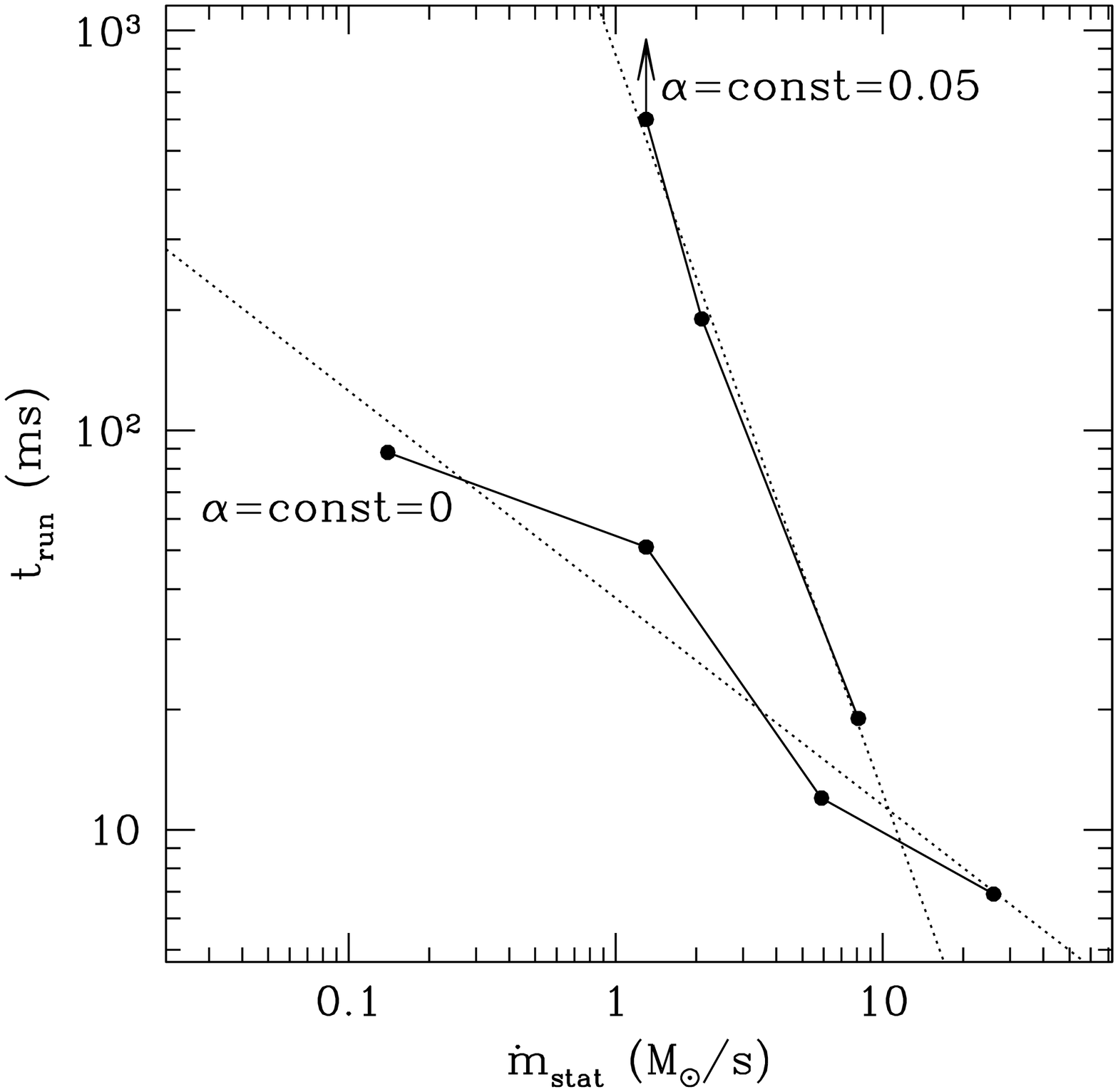,width=0.47\textwidth}\\
\end{tabular}
\end{center}
\caption{\textbf{Timescale of the runaway instability: selected models.} Compared to 
Fig.~\ref{fig:timescale1}, models with a constant mass flux have been selected and
included in the left panel, and models with constant angular momentum slope have been 
selected for the right panel. \textit{Left:} the dotted line indicates the empirical relation 
$t_\mathrm{run}=50\exp{(\alpha/0.021)}\ \mathrm{ms}$. \textit{Right:} the dotted line 
indicates the empirical relation $t_\mathrm{run}\propto \dot{m}^{-0.5}$ for $\alpha=0$ 
and $t_\mathrm{run}\propto\dot{m}^{-1.8}$ for $\alpha=0.05$.}
\label{fig:timescale2}
\end{figure*}

In all unstable models, at least 93 \% of the disc mass has been accreted by the end of the 
simulation (not displayed in the figure). The minimum value is reached for model 2d 
($M_\mathrm{BH}=2.5\ \mathrm{M_\odot}$ and $M_\mathrm{D}=0.25\ \mathrm{M_\odot}$), in which 
the final mass of the disc is only $0.016\ \mathrm{M_\odot}$. High initial mass ratios lead 
to more extreme situations: for instance in model 1a ($M_\mathrm{BH}=M_\mathrm{D}=2.5\ 
\mathrm{M_\odot}$), the final mass of the disc is only $2.8\times 10^{-4}\ \mathrm{M_\odot}$ 
and the final mass of the black hole is $5\ \mathrm{M_\odot}$. On the other hand, for models 
6a, 7a and 8a ($\alpha=0.09$, 0.1 and 0.15), 4b, 5b and 6b ($\alpha=0.02$, 0.025 and 0.05), 
5c, 6c and 7c ($\alpha=0.06$, 0.07 and 0.075), and 4d and 5d ($\alpha=0.055$ and 0.06), the 
stability properties of the non-constant distribution of angular momentum are reflected in 
the small loss of the mass of the disc throughout the evolution: less than 3.7 \% of the disc 
mass has been accreted in sets `a' and `b' (maximum reached in model 6a) and less than 27 \% 
in sets `c' and `d' (maximum reached in model 5c). The higher fraction found for the case of 
a low disc-to-hole mass ratio $M_\mathrm{D}/M_\mathrm{BH}=0.1$ is explained below in the 
subsection devoted to the long-term evolution of stable discs.

As in Paper I we define the timescale of the runaway instability $t_\mathrm{run}$ as the time 
needed for half of the disc mass to fall into the hole. An interesting outcome of the present 
study is to provide the first numerical estimation of this timescale for non-constant angular 
momentum discs. The last column of Table~\ref{tab:ModelParameters} lists $t_\mathrm{run}$ (in 
units of the orbital period) for all models (series three). It can be as short as 
$t_\mathrm{run}\sim 4.3\ t_\mathrm{orb}\sim 6.9\ \mathrm{ms}$ (model 1a) and is always shorter 
than 1 s in all models considered in our sample. The maximum values are the lower limits 
obtained for model 4c, $t_\mathrm{run}>200\ t_\mathrm{orb}\sim 0.38\ \mathrm{s}$, and model 3d, 
$t_\mathrm{run}>600\ t_\mathrm{orb}\sim 0.6\ \mathrm{s}$, the latter case being possibly stable. 

Fig.~\ref{fig:timescale1} plots the dependence of the timescale of the runaway instability with 
$\alpha$ (left panel) and with $\dot{m}_\mathrm{stat}$ (right panel) -- which is equivalent to 
$\Delta W_\mathrm{in}$ as explained above -- for the four sets of models in series three. In 
order to identify more clearly the effect of each one of these two parameters we have isolated 
in Fig.~\ref{fig:timescale2} those models that allow to study the dependence of the timescale 
with $\alpha$ for constant $\dot{m}_\mathrm{stat}$ (left panel) and with $\dot{m}_\mathrm{stat}$ 
for constant $\alpha$ (right panel). As for the case of constant angular discs (see Paper I), 
we find now that for non-constant angular momentum discs the runaway instability occurs also 
faster when the initial mass flux in the stationary regime is larger. As plotted in 
Fig.~\ref{fig:timescale2} (right panel), we get an empirical dependence $t_\mathrm{run}\propto 
\dot{m}_\mathrm{stat}^{-0.5}$ for $\alpha=0$ and an even steeper decay $t_\mathrm{run}\propto 
\dot{m}_\mathrm{stat}^{-1.8}$ for $\alpha=0.05$. Notice that the slope -0.5 found for $\alpha=0$ 
is slightly smaller than the value -0.9 found in Paper I. This is due to the stabilizing effect 
of the angular momentum transfer which was not taken into account in Paper I. On the other hand, 
the timescale of the runaway instability increases exponentially with $\alpha$ until it is 
finally suppressed when the critical slope $\alpha_\mathrm{cr}$ is reached. The left panel of 
Fig.~\ref{fig:timescale2} shows that, in the case where $\dot{m}_\mathrm{stat}=1.3\ 
\mathrm{M_\odot}$, the timescale follows the empirical dependence $t_\mathrm{run} \sim 
50\exp{(\alpha/0.021)}\ \mathrm{ms}$. This exponential increase shows again how strong is the 
stabilizing effect of a non-constant angular momentum distribution in the disc.

On the other hand Fig.~\ref{massflux1} shows (compare set `a' top-left and set `b' bottom-left) 
that a smaller overflow of the initial marginally stable potential barrier (resulting in a smaller 
initial mass flux) tends, not surprisingly, to defer the occurence of the instability. The 
timescale $t_\mathrm{run}$ is typically ten times longer in set `b' compared to set `a'. For 
the same reason Fig.~\ref{massflux1} also indicates that the smaller the initial mass flux is, 
the smaller the slope $\alpha$ necessary to suppress the instability is. The critical slope is 
$\sim 0.09$ for set of models `a' and only 0.02 for set `b'. Finally, Fig.~\ref{massflux1} 
illustrates that in comparison with the strong dependence on $\dot{m}_\mathrm{stat}$ and $\alpha$, 
the dependence of the runaway instability on the disc-to-hole mass ratio is rather weak (compare 
for instance set `a' (top-left panel) and set `c' (top-right panel) in Fig.~\ref{massflux1}).

\subsubsection{Time evolution of the black hole spin}

The transfer of mass and angular momentum from the disc to the black hole leads to the gradual 
increase of the rotation law of the initially non-rotating black hole. Fig.~\ref{spin1} shows 
the time evolution of the black hole spin $a/M_\mathrm{BH}$ for all models considered. Once 
again there is a clear distinction between stable and unstable situations. The stable discs 
only transfer a very small fraction of their angular momentum to the black hole. At the end 
of the simulation the black hole spin is less than $a/M_\mathrm{BH}=1.5\times 10^{-2}$ 
(maximum reached for model 5c) and can be as low as $1.9\times 10^{-4}$ (model 6b). On the 
contrary, for the unstable discs the transfer of angular momentum is considerably higher, 
accelerating rapidly during the runaway instability. As a result, in less than 100 orbital 
periods the initial Schwarzschild black hole turns into a mildly rotating Kerr black hole 
with $a/M_\mathrm{BH} \sim 0.3-0.35$ for sets `a' and `b' ($M_\mathrm{D}/M_\mathrm{BH}=1$)
and into a slowly rotating black hole with $a/M_\mathrm{BH}\la 0.05$ for sets `c' and `d' 
($M_\mathrm{D}/M_\mathrm{BH}=0.1$). In the second case, the small increase of the spin is 
directly related to the fact that for such disc-to-hole mass ratio the reservoir of angular 
momentum is much smaller. 

\begin{figure}
\centerline{\psfig{file=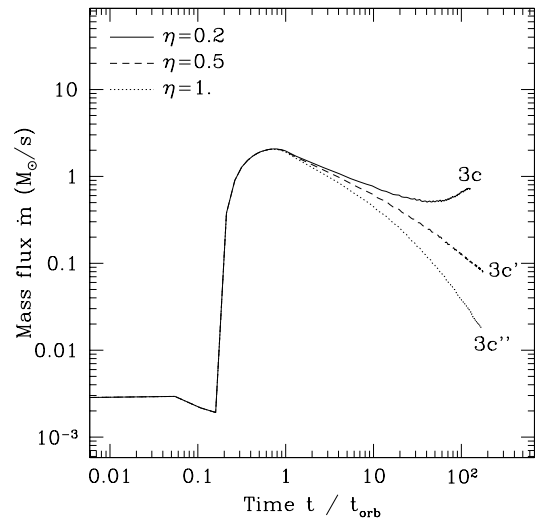,width=7cm}}
\caption{\textbf{Time evolution of the mass flux:} the mass flux is plotted as a function of time 
(normalized by the orbital time) for models 3c, 3c' and 3c'', which correspond to three different 
values of the efficiency $\eta$ of the disc-to-hole angular momentum transfer: $\eta=0.2$ (conservative 
value used in all other simulations, solid line), $\eta=0.5$ (dashed line) and $\eta=1$ (dotted line). 
Notice the stabilizing effect of an increasing $\eta$.}
\label{fig:massflux3}
\end{figure}

These results on the spin of the black hole have been obtained for a conservative value of 
the efficiency $\eta=0.2$ of the disc-to-hole angular momentum transfer. 
Figure~\ref{fig:massflux3} illustrates the dependence of our results with $\eta$. The mass 
flux is plotted as a function of time for models 3c, 3c' and 3c''. This figure clearly shows 
that increasing the efficiency of angular momentum transfer has a stabilizing effect. While 
model 3c is unstable for $\eta=0.2$, model 3c' is stable for $\eta=0.5$ and model 3c'' is 
even more stable for $\eta=1$. Therefore, except for very low values of the slope $\alpha$, 
a high efficiency $\eta$ would probably not lead to more rapidly rotating black holes but 
most likely to stable discs. On the other hand, if the initial black hole is already rotating, 
as it is inferred from the existing numerical simulations of compact binary coalescence or 
collapsars, this effect might result in black holes with high values of $a/M_\mathrm{BH}$. 
This will be studied in a forthcoming paper.

\subsubsection{Long-term evolution of stable discs}

The long-term behaviour of the mass flux evolution in stable discs simply reflects the fact 
that the initial accretion rate is far from zero but rather highly super-Eddington. For 
instance in the set of models `a', the mass flux is $\sim 3.2\ \mathrm{M_{\odot}/s}$ for 
model 6a, $\sim 2.5\ \mathrm{M_{\odot}/s}$ for model 7a, and $\sim 0.69\ \mathrm{M_{\odot}/s}$ 
for model 8a. Hence, by integrating $\dot{m}$ along the entire time of the simulation, it is 
possible to check that the total mass transferred to the black hole becomes, asymptotically, 
of the order of the initial mass of the disc, which results in the end of the accretion 
process. This effect is more pronounced for smaller disc-to-hole mass ratios 
(sets `c' and `d') as the accretion timescale $\sim\dot{m}_\mathrm{stat}/M_\mathrm{D}$ is 
shorter, or in other words, as the reservoir of mass is smaller. Furthermore, as we already 
noticed in Paper 2, a second effect is the gradual decrease of the mass flux during the 
long-term evolution of stable models (and of unstable models before the instability sets in). 
This is simply due to the fact that for non-constant angular momentum discs the accretion 
process makes the cusp move faster towards the black hole than the inner radius of the disc 
(which is precisely why the instability is physically suppressed). Therefore, the potential 
barrier at the inner edge $\Delta W_\mathrm{in}$ decreases with time and, consequently, also 
the mass flux. For this reason, as shown in Fig.~\ref{massflux1}, accretion becomes 
increasingly slower, which, in turn, makes numerically difficult to follow the evolution until 
the entire disc has been fully accreted.

\section{Conclusions}
\label{sec:conclusions}

We have presented a comprehensive numerical study aimed at clarifying the likelihood of the
so-called runaway instability in axisymmetric thick discs orbiting around (initially 
non-rotating) black holes. Our main conclusion, already pointed out in a preliminary
investigation \citep{font:02b} is that the runaway instability of (non self-gravitating) 
thick discs around (rotating) black holes \citep{abramowicz:83} is strongly suppressed when
the distribution of angular momentum in the disc is assumed to be non constant. We have 
reached this conclusion using a numerical framework based on time-dependent hydrodynamical 
simulations in general relativity. This approach departs from the most common procedures followed
in previous studies of the runaway instability in the literature, which are generally based 
on the analysis of stationary models (see. e.g. \citet{font:02a} and references there in). 
Despite the novel approach it is remarkable that our main result is in complete agreement with 
such previous studies based on stationary sequences of equilibrium configurations, either using a 
pseudo-Newtonian potential \citep{daigne:97} or in general relativity \citep{abramowicz:98}.
We further notice that our finding is also in good agreement with recent perturbative 
analysis of vertically integrated discs performed by \citet{rezzolla:03}.

The present paper extends in a number of ways the preliminary results reported in our previous
investigation \citep{font:02b}. In the first half of the article (cf. Section~\ref{sec:initialmodel}),
we have shown how to build a series of thick accretion discs in hydrostatic equilibrium around a 
Schwarzschild or Kerr black hole. All relevant aspects of the theory of stationary thick discs 
around rotating black holes have been presented in great detail. In particular our description 
has been done for discs in which the specific angular momentum (per unit mass) is assumed to 
increase outwards with the radial distance according to a power law $l=K r^{\alpha}$. All possible 
equipotential configurations for a Kerr black hole (including the limiting case of a Schwarzschild 
black hole) have been discussed in detail for both, prograde ($K>0$) and retrograde ($K<0$) rotation, 
and sub-Keplerian ($0\le\alpha < 1/2$), Keplerian ($\alpha=1/2$), and super-Keplerian ($1/2<\alpha<1$) 
discs. The various possible geometries of these configurations have been summarized in 
Table~\ref{tab:geometry} and in Figs.~\ref{fig:equi1pro} to \ref{fig:equi3pro}. These figures and 
table extend earlier results presented in \citet{font:02a} where only {\it constant} angular 
momentum discs around Schwarzschild black holes were considered.

The second half of the article has been devoted to presenting the results of a comprehensive set of
models that we have evolved with our numerical code. The main result of these simulations has been
to show, quite generically, that a small increase outwards of the specific angular momentum distribution 
in the discs, well below the Keplerian limit, results in a strong stabilizing effect. In particular
we have demonstrated that the case $\alpha=0$ (constant angular momentum disc) is {\it always} 
unstable when the black hole dynamics is taken into account, for both, $M_{\mathrm D}/M_\mathrm{BH}=1$ 
and 0.1, irrespective of the value of the potential barrier at the cusp (slightly overflowing the corresponding
Roche lobe) and of the initial mass flux in the stationary regime. We note in passing that a similar
conclusion has been recently obtained by \citet{zanotti:02} for a Schwarzschild black hole and for 
different initial data, in which models that initially are in strict stable equilibrium are conveniently 
perturbed to trigger the accretion process. However, we have shown that discs with small non-zero values 
of $\alpha$ are stable, and we have provided an estimate of the critical value of $\alpha$ separating 
stable and unstable models, for various disc-to-hole mass ratios and mass fluxes in the stationary regime. 
More precisely we have found that for $M_{\mathrm D}/M_\mathrm{BH}=1$, the critical slope is 
$\alpha_{\mathrm{cr}}\sim 0.085-0.09$ for $\dot{m}_\mathrm{stat}\sim 3.2-3.6\ \mathrm{M_{\odot} s^{-1}}$ 
and $\alpha_{\mathrm{cr}}\sim 0.015-0.02$ for $\dot{m}_\mathrm{stat}\sim 0.1\ \mathrm{M_{\odot} s^{-1}}$. 
Correspondingly, for $M_{\mathrm D}/M_\mathrm{BH}=0.1$, the critical slope is 
$\alpha_{\mathrm{cr}}\sim 0.055-0.06$ for $\dot{m}_\mathrm{stat}\sim 1.7-1.9\ \mathrm{M_{\odot} s^{-1}}$ 
and $\alpha_\mathrm{cr}\sim 0.05$ for $\dot{m}_\mathrm{stat}\sim 1.3\ \mathrm{M_\mathrm{\odot} s^{1}}$. 
These values are slightly smaller than the values found in stationary studies (0.14 in the relativistic 
case for a disc-to-hole mass ratio of $\sim 0.15$ \citep{abramowicz:98}). However these studies assume 
an instantaneous mass and angular momentum transfer from the disc to the black hole, and our study 
clearly indicates that the critical slope $\alpha_\mathrm{cr}$ increases with the mass flux 
$\dot{m}_\mathrm{stat}$. For the first time we have shown the suppression of the runaway instability 
in nonconstant angular momentum discs with a time-dependent relativistic calculation, and we have 
been able to estimate the lifetime of thick discs orbiting a black hole in both, the unstable and 
stable cases. The very strong stabilizing effect of the non-constant angular momentum distribution 
in the disc is demonstrated in our simulations by the exponential increase of the timescale for the 
instability to set in with the slope $\alpha$.

The black hole plus thick discs systems we have simulated with $M_\mathrm{D}/M_\mathrm{BH}=0.1$ 
are very close to what numerical simulations of compact binary mergers and collapsars 
predict \citep{ruffert:99,kluzniak:98,macfadyen:99,shibata:00,aloy:00}. Therefore, the results 
regarding this subset of models should be considered as, perhaps, much closer to what may be 
happening in realistic discs. However, even in this case, three important simplifications have 
been made: (i) the black hole is initially non-rotating whereas both in collapsars and 
binary mergers it is expected to form with an initial spin parameter $a/M_\mathrm{BH}\ga 0.5$; 
(ii) the self-gravity of the disc is not included; and (iii) magnetic fields, which are believed 
to play an important role on the dynamics of those objects, are also not included.

Certainly, along with analyzing the effects of varying the EoS of the matter in the disc, these 
are all issues that we plan to consider in the near future in order to extend the present
study. Among those, the inclusion of an appropriate treatment to account for the self-gravity of 
the disc is the most difficult task despite its obvious relevance. Previous studies 
\citep{khanna:92,nishida:96a,masuda:98}, based on either time-dependent simulations with 
pseudo-Newtonian potentials or stationary models in general relativity, indicate that self-gravity
seems to have a strong destabilizing effect. It has been argued \citep{nishida:96a} that in 
self-gravitating accreting tori the cusp is pushed towards the black hole by the gravitational 
force of the discs which, in principle, should act against the instability. However, during the 
accretion process, the interplay between the gravitational forces of the black hole (increasing) 
and of the disc (decreasing) make the cusp move closer to the centre of the torus than in 
non-self-gravitating models, thus favouring the instability. This important result is yet to be 
proved by means of time-dependent simulations in full general relativity. Unfortunately, successful 
attempts to long-term (numerically) stable simulations of thick discs orbiting black holes by 
solving the coupled system of Einstein and general relativistic hydrodynamic equations are still 
out of reach of present day numerical relativity codes (see, e.g. \cite{fontlr} and references 
there in for an up-to-date list of such codes). Such simulations should provide a definite answer 
to the issue of whether the runaway instability exists or not. 

We notice that neglecting the self-gravity of the disc is a more valid assumption the smaller 
the disc-to-hole mass ratio becomes. Therefore, with respect to the models analyzed in this paper 
we expect that for discs with $M_\mathrm{D}/M_\mathrm{BH} \la 0.1$ the strong stabilizing effect 
of non-constant angular momentum distribution demonstrated in our study could still overcome the 
destabilizing effect of the much less relevant disc's self-gravity.

The incorporation of magnetic fields seems to be more easily conceivable. Recently, \citet{devilliers:03} 
and \citet{gammie:03} have presented magneto-hydrodynamical simulations of constant angular momentum 
tori in general relativity. These simulations assume much smaller accretion mass fluxes than in the 
present study and do not include any time evolution of the spacetime. However, they indicate a promising 
direction to include the effect of magnetic fields in future simulations of the runaway instability and 
hence suppress the third limitation listed above. 

Nevertheless, the first aspect that we plan to consider in a forthcoming study is the time 
evolution of discs orbiting black holes which are {\it initially} rotating. This will allow 
us to pinpoint the dependence of $\alpha_{\mathrm{cr}}$ with $a$. We notice that our 
hydrodynamics code has already proved in this study its ability to deal with the Kerr 
spacetime, and the method of construction of the initial state in the case of a rotating 
black hole has been fully described in this paper. Therefore, our upcoming study does not 
involve any new numerical development.

\appendix

\section{Coefficients of the Kerr metric and their derivatives in the equatorial plane}
\label{sec:kerr}

In this appendix we provide the expressions for the Kerr metric coefficients and their derivatives
in the equatorial plane, which are needed in the construction of the equilibrium tori. In the 
following we assume that the black hole mass is $M_{\mathrm{BH}}=1$ (therefore $0\le a \le 1$) 
and we use the standard Boyer-Lindquist (spherical) coordinates $(t,r,\theta,\phi)$ to describe the 
Kerr metric (cf. Eq.~(\ref{blform}) in Section~\ref{sec:initialmodel}):
\begin{eqnarray}
g_{tt}       & = & -1+\frac{2r}{r^{2}+a^{2}\cos^{2}{\theta}}\label{eq:gtt}\\
g_{t\phi}    & = & -\frac{2 a r \sin^{2}{\theta} }{r^{2}+a^{2}\cos^{2}{\theta}}\label{eq:gtp}\\
g_{\phi\phi} & = & \left(r^{2}+\frac{2 a^{2} r \sin^{2}{\theta}}{r^{2}+a^{2}\cos^{2}{\theta}}+a^{2}\right)\sin
^{2}{\theta}\label{eq:gpp}\\
g_{rr}       & = & \frac{r^{2}+a^{2}\cos^{2}{\theta}}{r^{2}-2r+a^{2}}\\
g_{\theta\theta} & = & r^{2}+a^{2}\cos^{2}{\theta}\ .
\end{eqnarray}
The pseudo-distance $\varpi$, the radius of the horizon $r_\mathrm{h}$ and the radius of the 
ergosphere $r_\mathrm{e}$ are given by:
\begin{eqnarray}
\varpi^{2} & = & g_{t\phi}^{2}-g_{tt}g_{\phi\phi} = \left(r^{2}-2r+a^{2}\right) \sin^{2}{\theta}\ .
\label{eq:defvarpi}\\
r_\mathrm{h} & = & 1+\sqrt{1-a^{2}}\\
r_\mathrm{e}(\theta) & = & 1+\sqrt{1-a^{2}\cos^{2}{\theta}}\ .
\end{eqnarray}
The metric coefficients in the equatorial plane which are needed in Section~\ref{sec:initialmodel} 
are then given by: 
\begin{eqnarray}
\tilde{g}_{tt}       & = & -1+\frac{2}{r},\\
\tilde{g}_{t\phi}    & = & -\frac{2 a}{r},\\
\tilde{g}_{\phi\phi} & = & r^{2}+a^{2}+\frac{2 a^{2}}{r}.
\end{eqnarray}
The first and second derivatives of these coefficients are also needed:
\begin{eqnarray}
\frac{\partial\tilde{g}_{tt}      }{\partial r} & = & -\frac{2}{r^{2}}\ ,\\
\frac{\partial\tilde{g}_{t\phi}   }{\partial r} & = & \frac{2 a}{r^{2}}\ ,\\
\frac{\partial\tilde{g}_{\phi\phi}}{\partial r} & = & 2r-\frac{2 a^{2}}{r^{2}}\ ,\\
& & \nonumber\\
\frac{\partial\tilde{g}_{tt}      }{\partial \theta} & = & 
\frac{\partial\tilde{g}_{t\phi}   }{\partial \theta} =
\frac{\partial\tilde{g}_{\phi\phi}}{\partial \theta} =  0\ ,\\
& & \nonumber\\
\frac{\partial^{2}\tilde{g}_{tt}      }{\partial r^{2}} & = & \frac{4}{r^{3}}\ ,\\
\frac{\partial^{2}\tilde{g}_{t\phi}   }{\partial r^{2}} & = & -\frac{4 a}{r^{3}}\ ,\\
\frac{\partial^{2}\tilde{g}_{\phi\phi}}{\partial r^{2}} & = & 2+\frac{4 a^{2}}{r^{3}}\ ,\\
& & \nonumber\\
\frac{\partial^{2}\tilde{g}_{tt}      }{\partial r \partial \theta} & = & 
\frac{\partial^{2}\tilde{g}_{t\phi}   }{\partial r \partial \theta} = 
\frac{\partial^{2}\tilde{g}_{\phi\phi}}{\partial r \partial \theta} =  0\ ,\\
& & \nonumber\\
\frac{\partial^{2}\tilde{g}_{tt}      }{\partial\theta^{2}} & = & -\frac{4 a^{2}}{r^{3}}\ ,\\
\frac{\partial^{2}\tilde{g}_{t\phi}   }{\partial\theta^{2}} & = & \frac{4 a(r^{2}+a^{2})}{r^{3}}\ ,\\
\frac{\partial^{2}\tilde{g}_{\phi\phi}}{\partial\theta^{2}} & = & -2\left(r^{2}+a^{2}+
\frac{4 a^{2}}{r}+\frac{2 a^{4}}{r^{3}}\right)\ .
\end{eqnarray}
The pseudo-distance to the axis, Eq.~(\ref{eq:defvarpi}), is also a useful quantity. We give here 
its expression as well as its derivatives in the equatorial plane:
\begin{eqnarray}
\tilde{\varpi}^{2} & = & r^{2}-2r+a^{2}\\
& & \nonumber\\
\frac{\partial\tilde{\varpi}^{2}}{\partial r} & = & 2(r-1)\\
\frac{\partial\tilde{\varpi}^{2}}{\partial \theta} & = & 0\\
& & \nonumber\\
\frac{\partial^{2}\tilde{\varpi}^{2}}{\partial r^{2}} & = & 2\\
\frac{\partial^{2}\tilde{\varpi}^{2}}{\partial r\partial\theta} & = & 0\\
\frac{\partial^{2}\tilde{\varpi}^{2}}{\partial \theta^{2}} & = & -2\left(r^{2}-2r+a^{2}\right) 
\end{eqnarray}

\section{von Zeipel's cylinders}
\label{sec:vonzeipel}

\subsection{Equation of the cylinders}

In this appendix we describe the structure of the von Zeipel's cylinders (i.e. the surfaces of 
constant angular momentum per unit inertial mass and constant angular velocity, which coincide 
for a barotropic equation of state) in the Kerr spacetime. We use the same notations and 
conventions than in Section~\ref{sec:initialmodel} and Appendix~\ref{sec:kerr}. We limit the 
discussion to the region outside the black hole horizon and to those cylinders which have an 
intersection with the equatorial plane, where we assume that the distribution of angular 
momentum per unit inertial mass $l_\mathrm{eq}(r)$ is given. The equation of the von Zeipel 
cylinder corresponding to $l=l_{0}$ is:
\begin{equation}
g_{tt} l_{0}+g_{t\phi}\left(1+\Omega_{0}l_{0}\right)+g_{\phi\phi}\Omega_{0} = 0\ ,
\end{equation}
where $\Omega_{0}$ is the angular velocity corresponding to $l_{0}$. Both quantities are related by 
\begin{equation}
\Omega_{0} = -\frac{g_{t\phi}+g_{tt} l}{g_{\phi\phi}+g_{t\phi} l}\ .
\end{equation}
This leads to the equation of the von Zeipel's cylinder intersecting the equatorial plane at 
radius $r_{0}$:
\begin{eqnarray}
0 = \mathcal{Z}(r,\theta\ ; r_{0}) & = & \ \ \ \left[g_{tt}\tilde{g}_{t\phi}
                            (r_{0})-g_{t\phi}\tilde{g}_{tt}(r_{0})\right]\ 
                            l^{2}_\mathrm{eq}(r_{0})\nonumber\\
                      &   &+\left[g_{tt}\tilde{g}_{\phi\phi}(r_{0})-g_{\phi\phi}
                            \tilde{g}_{tt}(r_{0})\right]\
                            l_\mathrm{eq}(r_{0})\nonumber\\
                      &   &+\left[g_{t\phi}\tilde{g}_{\phi\phi}(r_{0})-g_{\phi\phi}
                            \tilde{g}_{t\phi}(r_{0})\right]\ .
\label{eq:cylinders}
\end{eqnarray}

\subsection{Intersection with the equatorial plane}

We investigate now the structure of the von Zeipel's cylinders in the vicinity of the 
equatorial plane. For $\theta=\pi/2$, $\mathcal{Z}(r_{0},\pi/2\ ; r_{0})=0$ and the 
derivatives of $\mathcal{Z}(r,\theta\ ; r_{0})$ at $(r_{0},\pi/2)$ can be computed from 
the coefficients of the metric and their derivatives (see Appendix~\ref{sec:kerr}):
\begin{eqnarray}
\frac{\partial Z}{\partial r}(r_{0},\pi/2\ ; r_{0})      & = & \frac{2}{r_{0}^{2}}\ 
\mathcal{P}\left(r_{0},l_\mathrm{eq}(r_{0})\right)\\
\frac{\partial Z}{\partial \theta}(r_{0},\pi/2\ ; r_{0}) & = & 0\\
\frac{\partial^{2} Z}{\partial r^{2}}(r_{0},\pi/2\ ; r_{0}) & = & -\frac{2}{r_{0}^{3}}\ 
\mathcal{Q}\left(r_{0},l_\mathrm{eq}(r_{0})\right)\\
\frac{\partial^{2} Z}{\partial r\partial \theta}(r_{0},\pi/2\ ; r_{0}) & = & 0\\
\frac{\partial^{2} Z}{\partial\theta^{2}}(r_{0},\pi/2\ ; r_{0}) & = & \frac{2}{r_{0}^{3}}
\left(r_{0}^{2}-2r_{0}+a^{2}\right)\ \mathcal{Q}\left(r_{0},l_\mathrm{eq}(r_{0})\right)\ ,
\nonumber\\
\end{eqnarray}
where the two polynomials $\mathcal{P}$ and $\mathcal{Q}$ are given by
\begin{eqnarray}
\mathcal{P}(r_{0},l_{0}) & = & a l_{0}^{2} + \left(r_{0}^{3} -3 r_{0}^{2}-2 a^{2}\right) 
l_{0} + a(3 r_{0}^{2} + a^{2})\\
\mathcal{Q}(r_{0},l_{0}) & = & 2 a l_{0}^{2} - \left(r_{0}^{3}+4 a^{2}\right) l_{0} + 2 a^{3}\ .
\end{eqnarray}
Then, in the vicinity of the equatorial plane, for $r=r_{0}+dr$ and $\theta=\pi/2+d\theta$, 
the equation of the cylinder is given by
\begin{eqnarray}
0 & = & \mathcal{Z}\left(r_{0}+dr,\pi/2+d\theta\ ;r_{0}\right)\nonumber\\
  & = &   \frac{2}{r_{0}^{2}}\ \mathcal{P}\left(r_{0},l_\mathrm{eq}(r_{0})\right)\  
          dr\ +\nonumber\\
  &   &   \frac{1}{r_{0}^{3}}\ \mathcal{Q}\left(r_{0},l_\mathrm{eq}(r_{0})\right)\ \left[
  dr^{2}
+ \left(r_{0}^{2}-2r_{0}+a^{2}\right) d\theta^{2}
\right]+ ...\nonumber\\
\label{eq:vicinity}
\end{eqnarray}
There are two possible cases: 

\begin{enumerate}

\item
In the normal case, $\mathcal{P}\left(r_{0},l_\mathrm{eq}
(r_{0})\right)\ne 0$ and Eq.~(\ref{eq:vicinity}) implies that the first derivative 
$dr/d\theta$ along the cylinder in the equatorial plane vanishes. In the vicinity of 
$(r_{0},\pi/2)$, the equation of the cylinder becomes 
\begin{equation}
r = r_{0}-\frac{r_{0}^{2}-2 r_{0}+a^{2}}{2 r_{0}}\ \frac{\mathcal{Q}\left(r_{0},
l_\mathrm{eq}(r_{0})\right)}{\mathcal{P}\left(r_{0},l_\mathrm{eq}(r_{0})\right)}\ 
\left(\theta-\frac{\pi}{2}\right)^{2}\ .
\end{equation} 
It means that the von Zeipel's cylinder starts tangential to the circle of radius $r_{0}$ 
and moves aside from this circle with an increasing (respectively, decreasing) radius for 
$\mathcal{Q}/\mathcal{P}<0$ (respectively $\mathcal{Q}/\mathcal{P}>0$) at $\left(r_{0},
l_\mathrm{eq}(r_{0})\right)$. 

\item
In the critical case where $\mathcal{P}\left(r_{0},
l_\mathrm{eq}(r_{0})\right)=0$, Eq.~(\ref{eq:vicinity}) implies that the cylinder 
is divided in two branches which intersect in the equatorial plane. The intersection 
$(r_{0},\pi/2)$ appears as a cusp. The equation of these two branches in the vicinity 
of the cusp is simply given by
\begin{equation}
r = r_{0} \pm \sqrt{r_{0}^{2}-2 r_{0}+a^{2}} \left(\theta-\frac{\pi}{2}\right)\ .
\end{equation}

\end{enumerate}

In the case where $a>0$ (rotating black hole), the two polynomials $\mathcal{P}$ and 
$\mathcal{Q}$ can be studied as a function of $l_{0}$ for a fixed radius $r_{0}$. 
Outside the horizon $\mathcal{P}$ has two positive roots for $r_\mathrm{h}\le r_{0} 
\le r_{0}^{(1)}$ and two negative roots for $r_{0} \ge r_{0}^{(2)}$. It is negative 
if $l_{0}$ is between the two roots when they exist and positive everywhere else. 
The radii $r_{0}^{(1)}$ and $r_{0}^{(2)}$ are the roots of $r_{0}^{3}-6 r_{0}^{2}+
9 r_{0}-4 a^{2}$ with $r_{h} \le r_{0}^{(1)}\le 3$ and $3 \le r_{0}^{(2)} \le 4$. 
For $a=0$, $r_{0}^{(1)}=r_{0}^{(2)}=3$ and for $a=1$, $r_{0}^{(1)}=r_{h}=1$ and 
$r_{0}^{(2)}=4$. The radius $r_{0}^{(1)}$ crosses the ergosphere for $a=\sqrt{2}/2$. 
The behaviour of $\mathcal{Q}$ is much simpler: it has always two positive roots 
and it is negative when $l_{0}$ is between these two roots. 

\begin{figure*}
\begin{center}
\begin{tabular}{cccc}
$a=0$ & $a=1/3$ & $a=2/3$ & $a=1$\\
\psfig{file=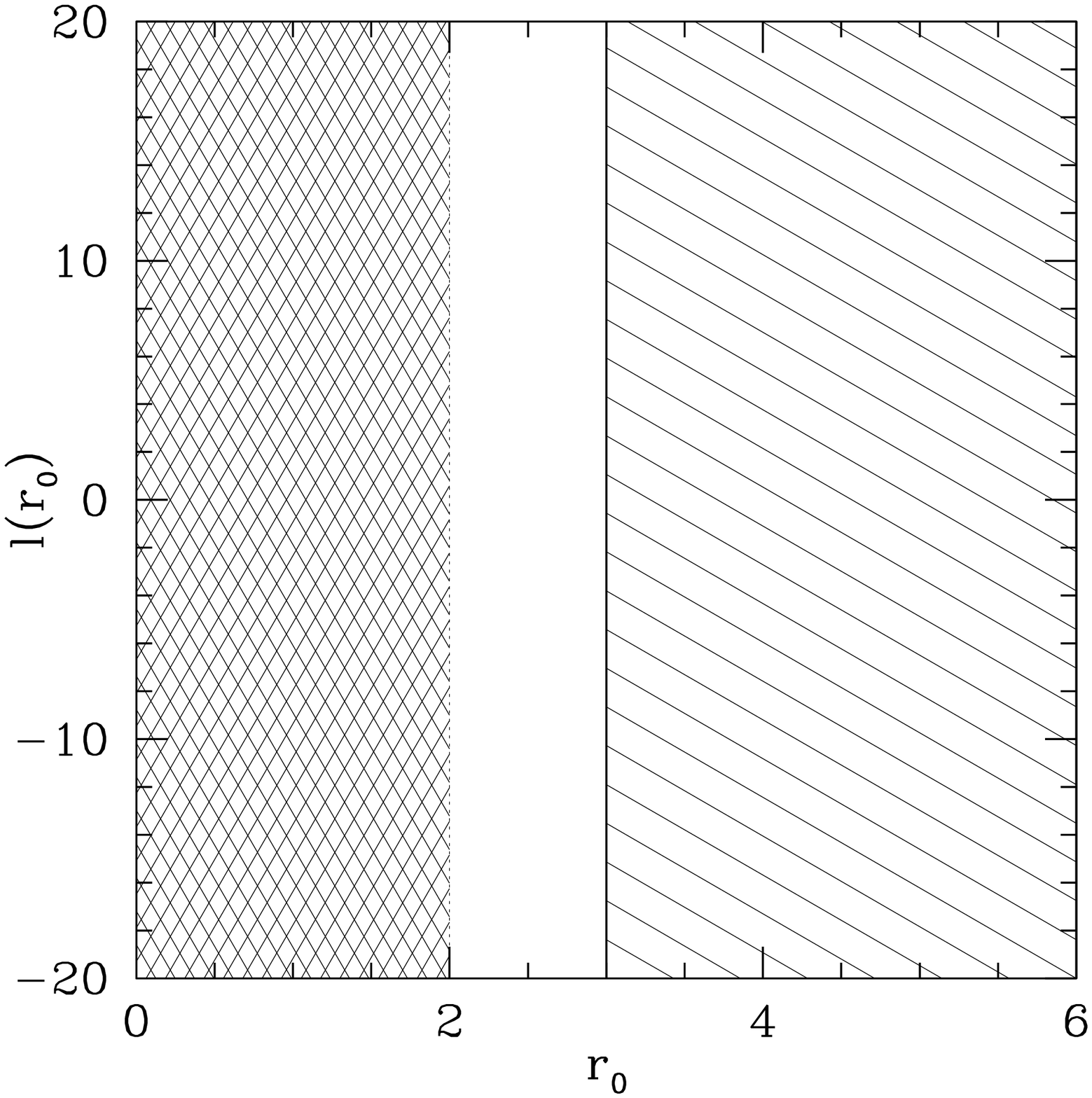,width=0.22\textwidth} &
\psfig{file=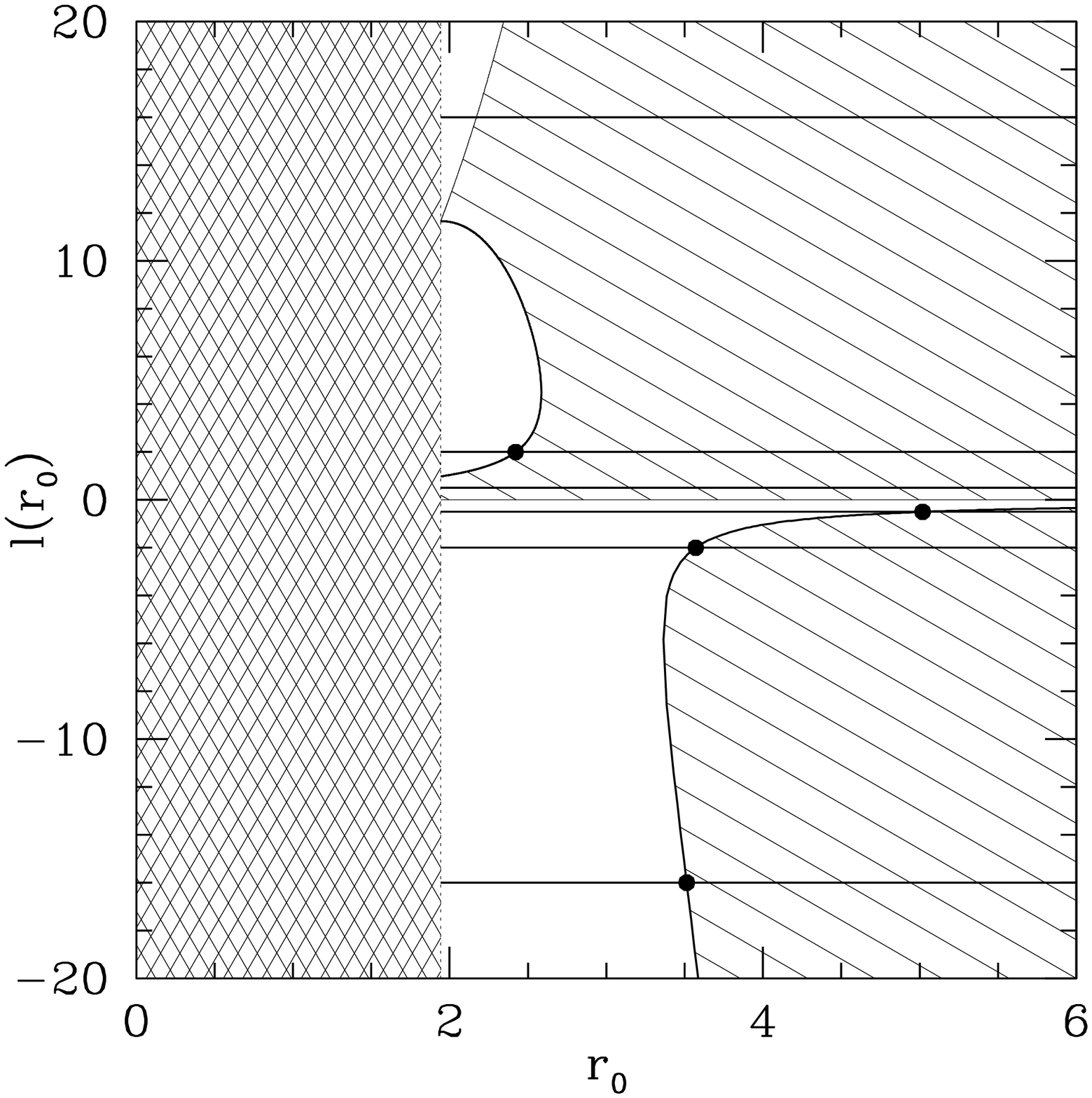,width=0.22\textwidth} &
\psfig{file=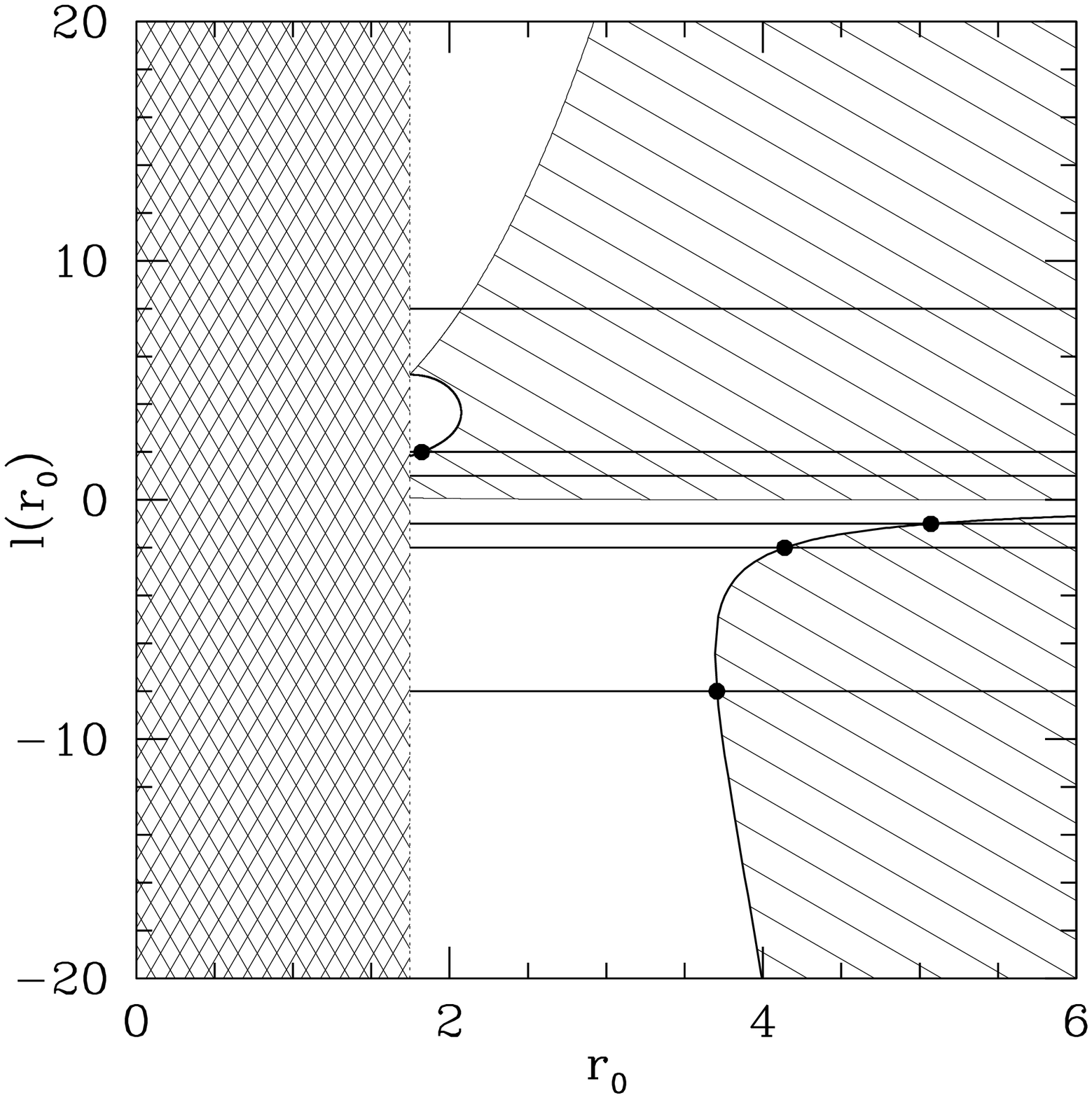,width=0.22\textwidth} &
\psfig{file=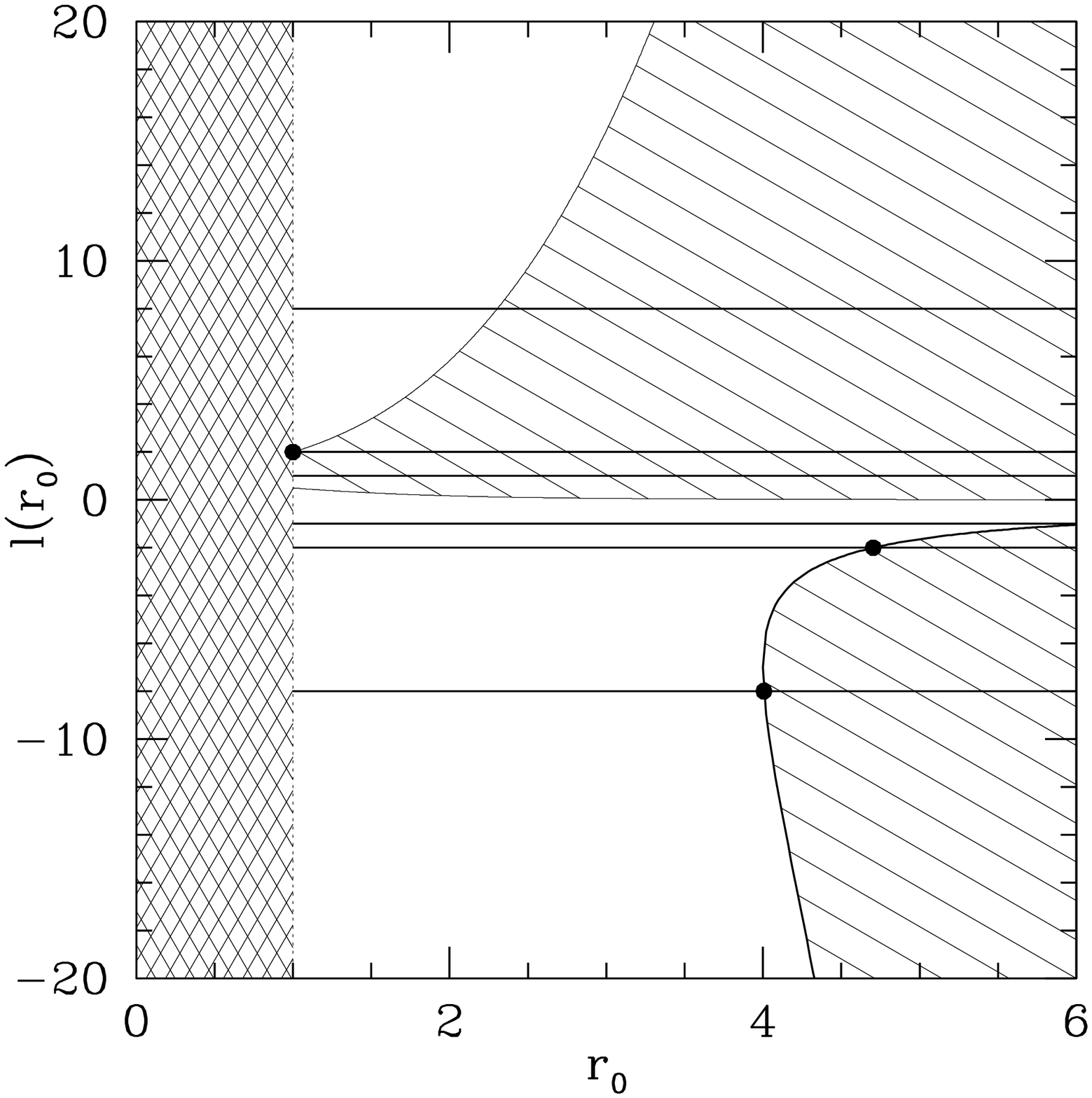,width=0.22\textwidth}\\
 &
\psfig{file=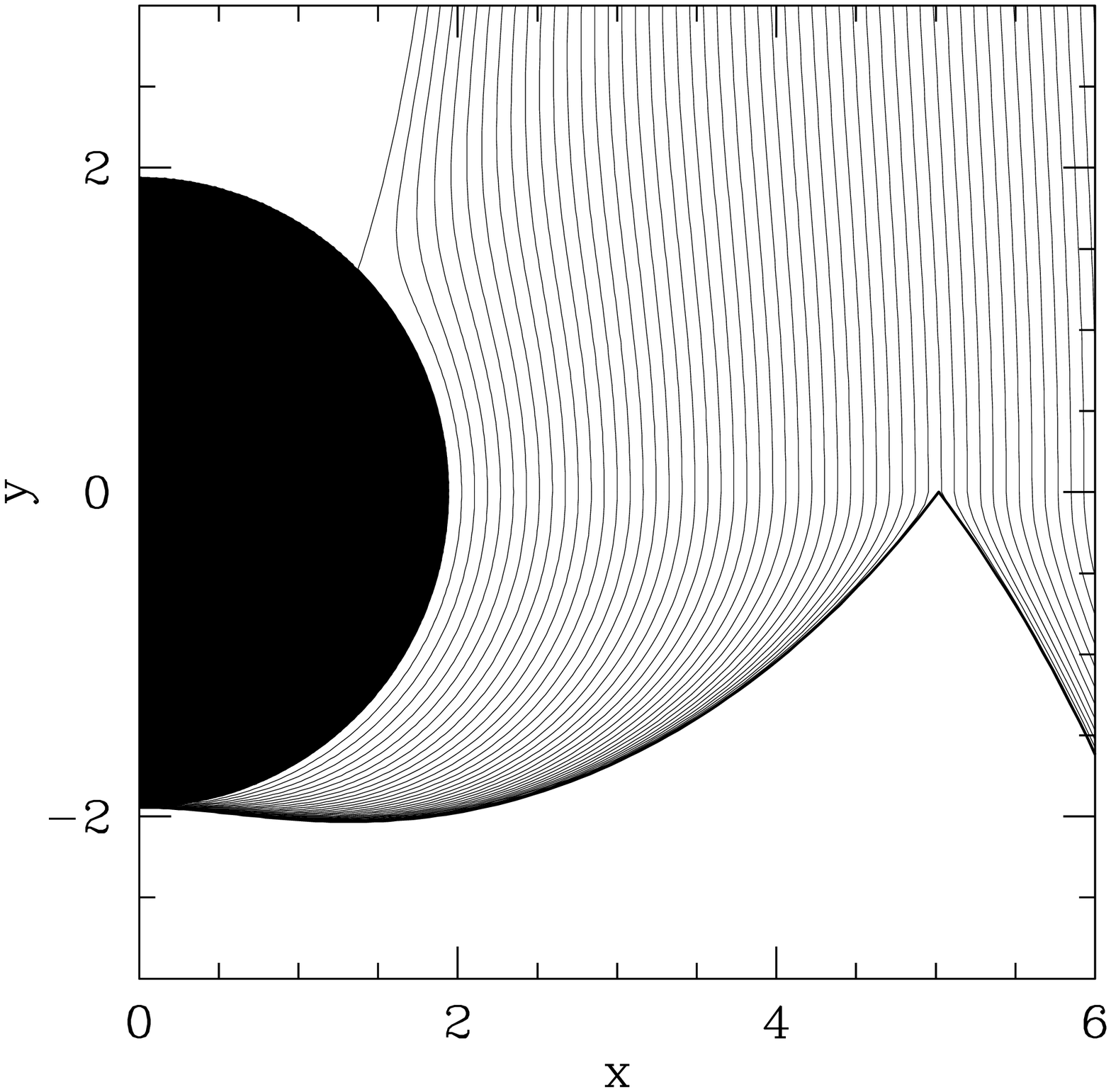,width=0.22\textwidth} &
\psfig{file=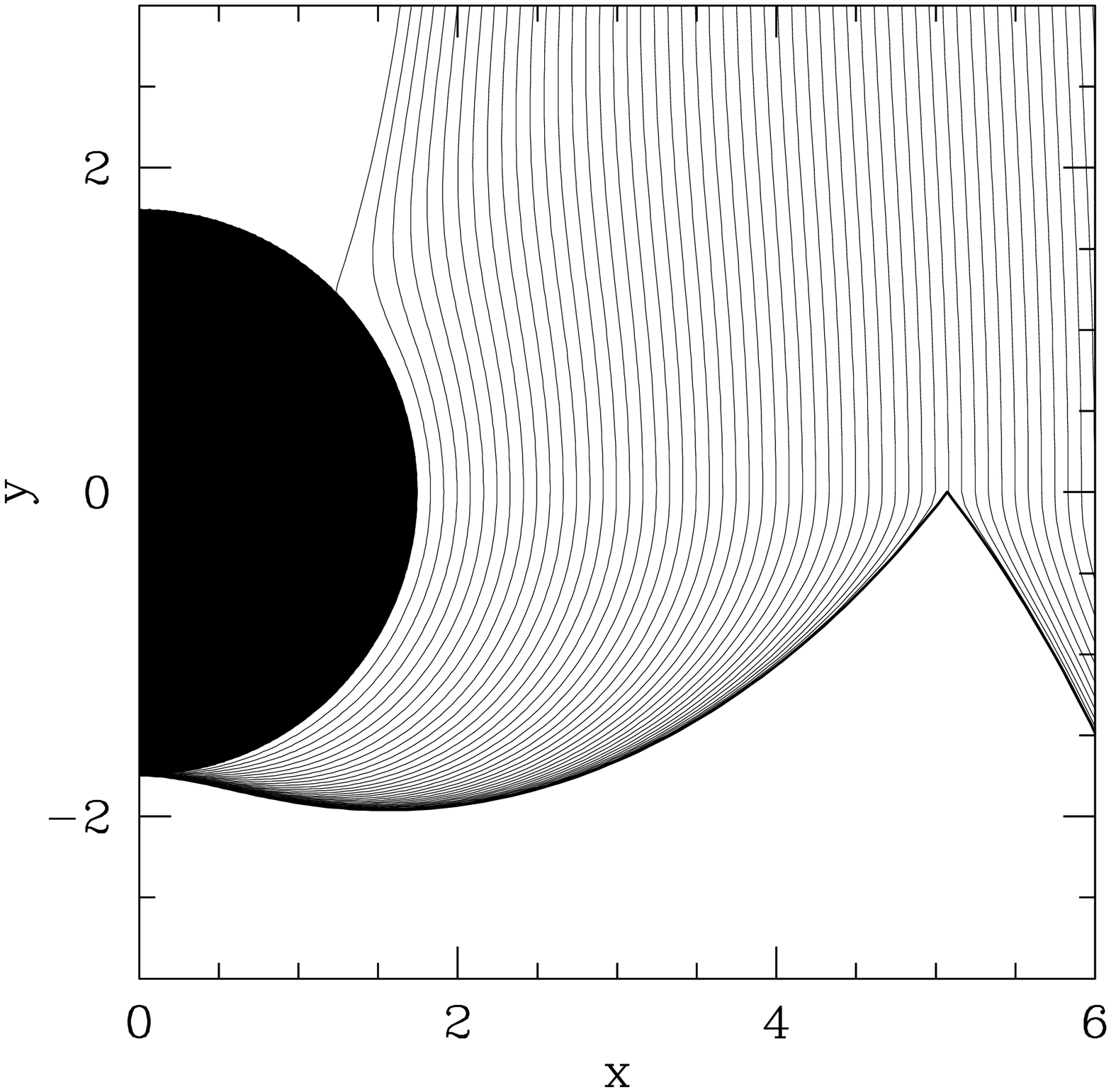,width=0.22\textwidth} &
\psfig{file=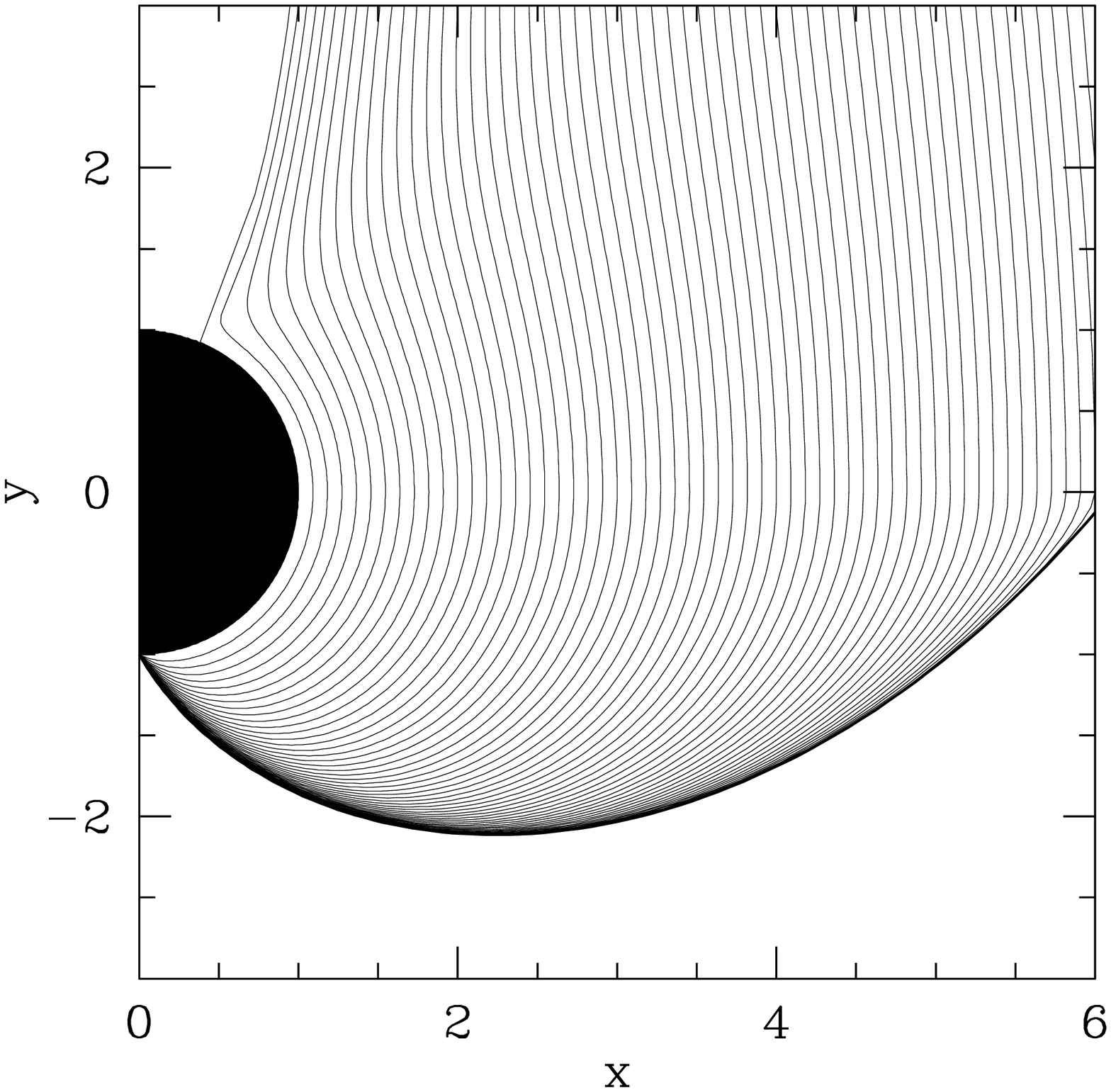,width=0.22\textwidth}\\
\psfig{file=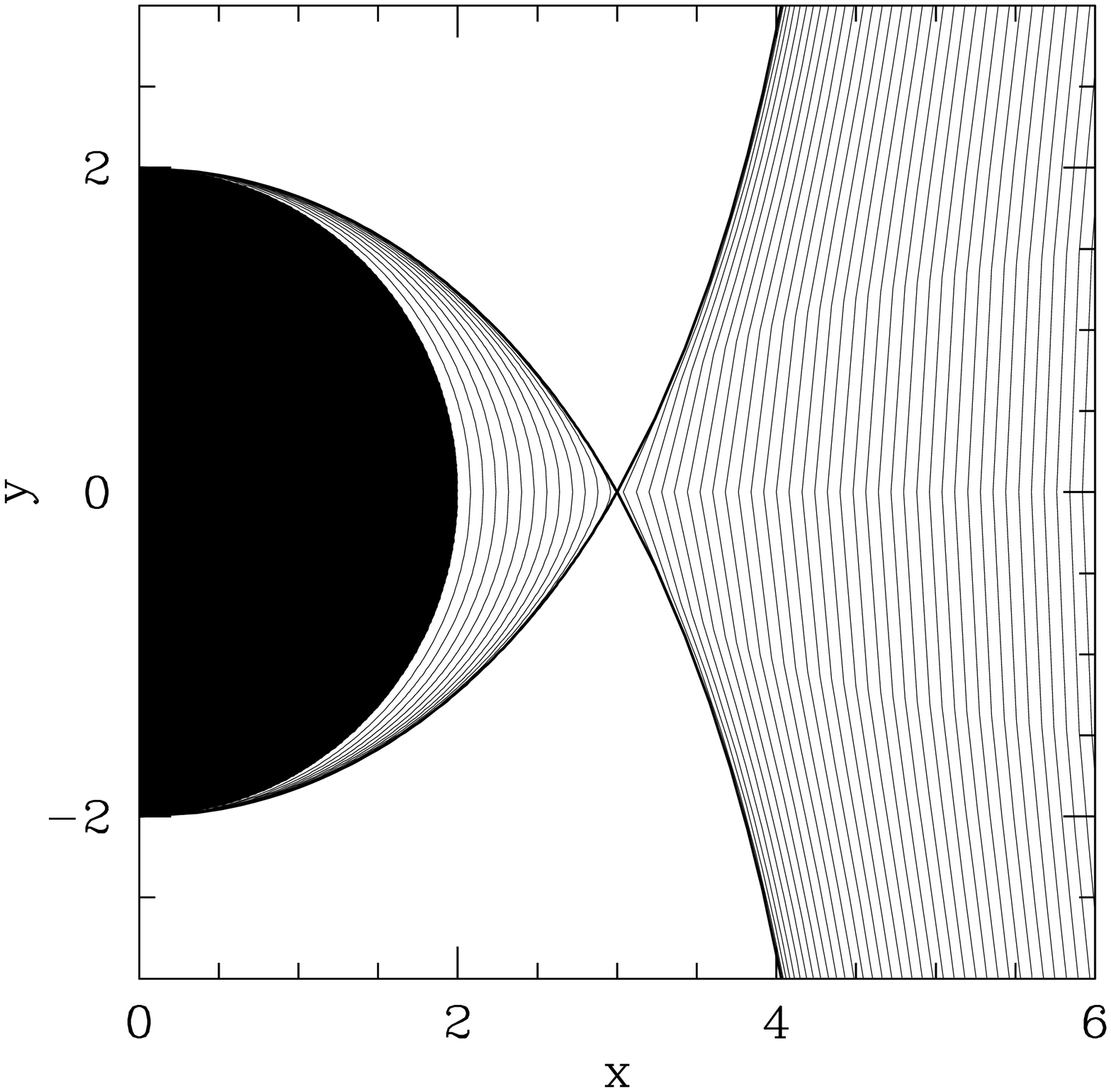,width=0.22\textwidth} &
\psfig{file=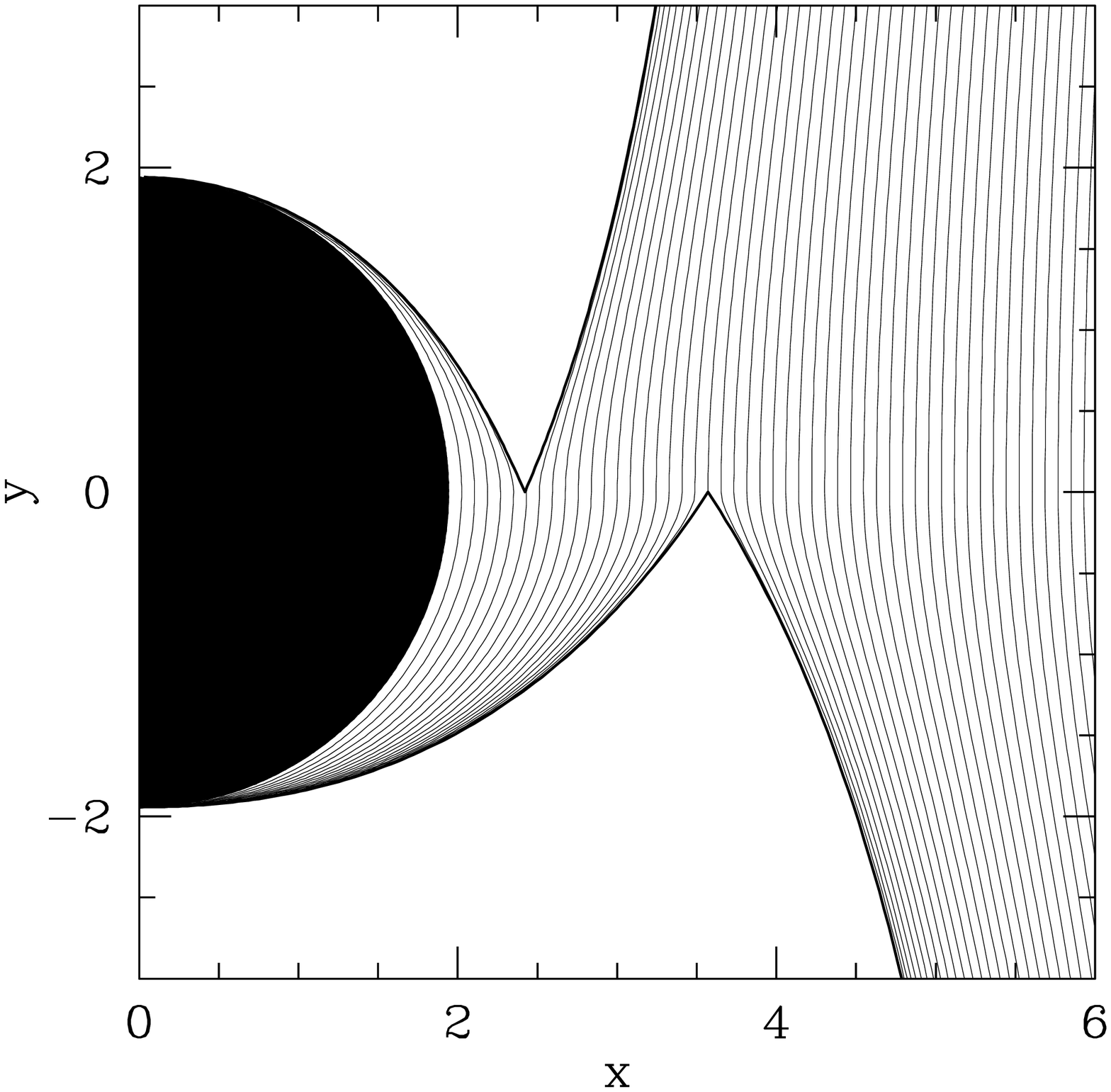,width=0.22\textwidth} &
\psfig{file=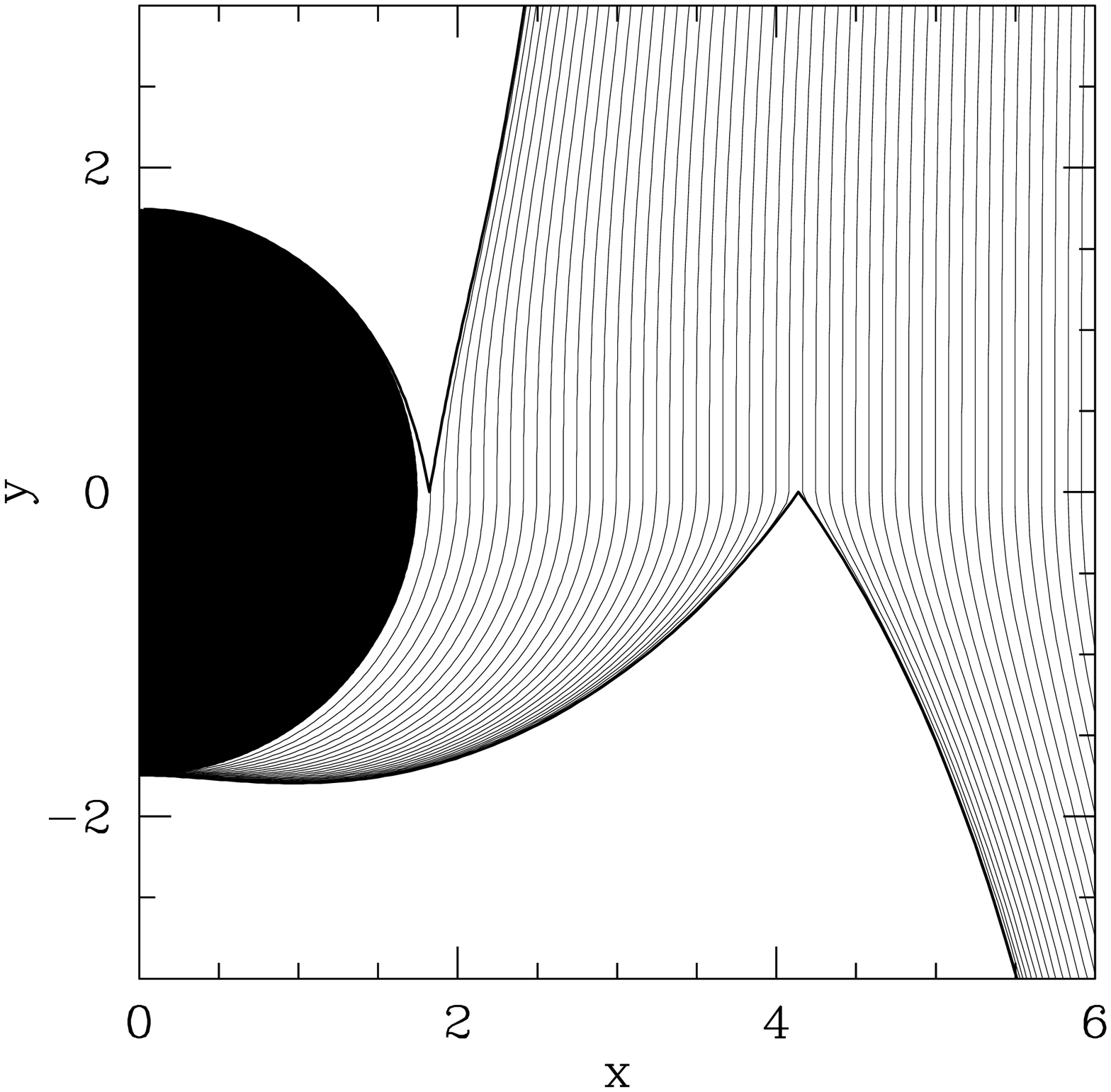,width=0.22\textwidth} &
\psfig{file=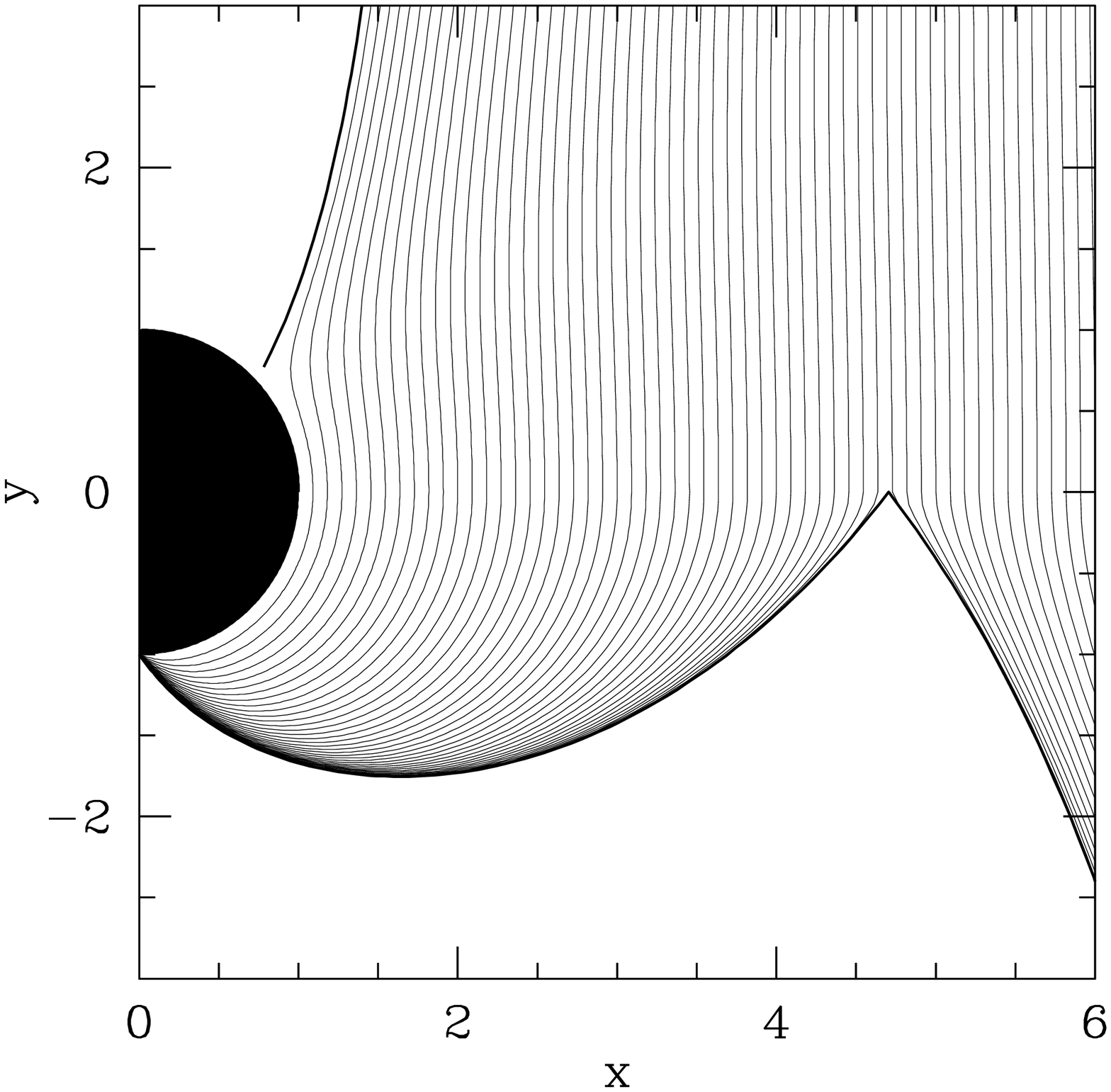,width=0.22\textwidth}\\
 &
\psfig{file=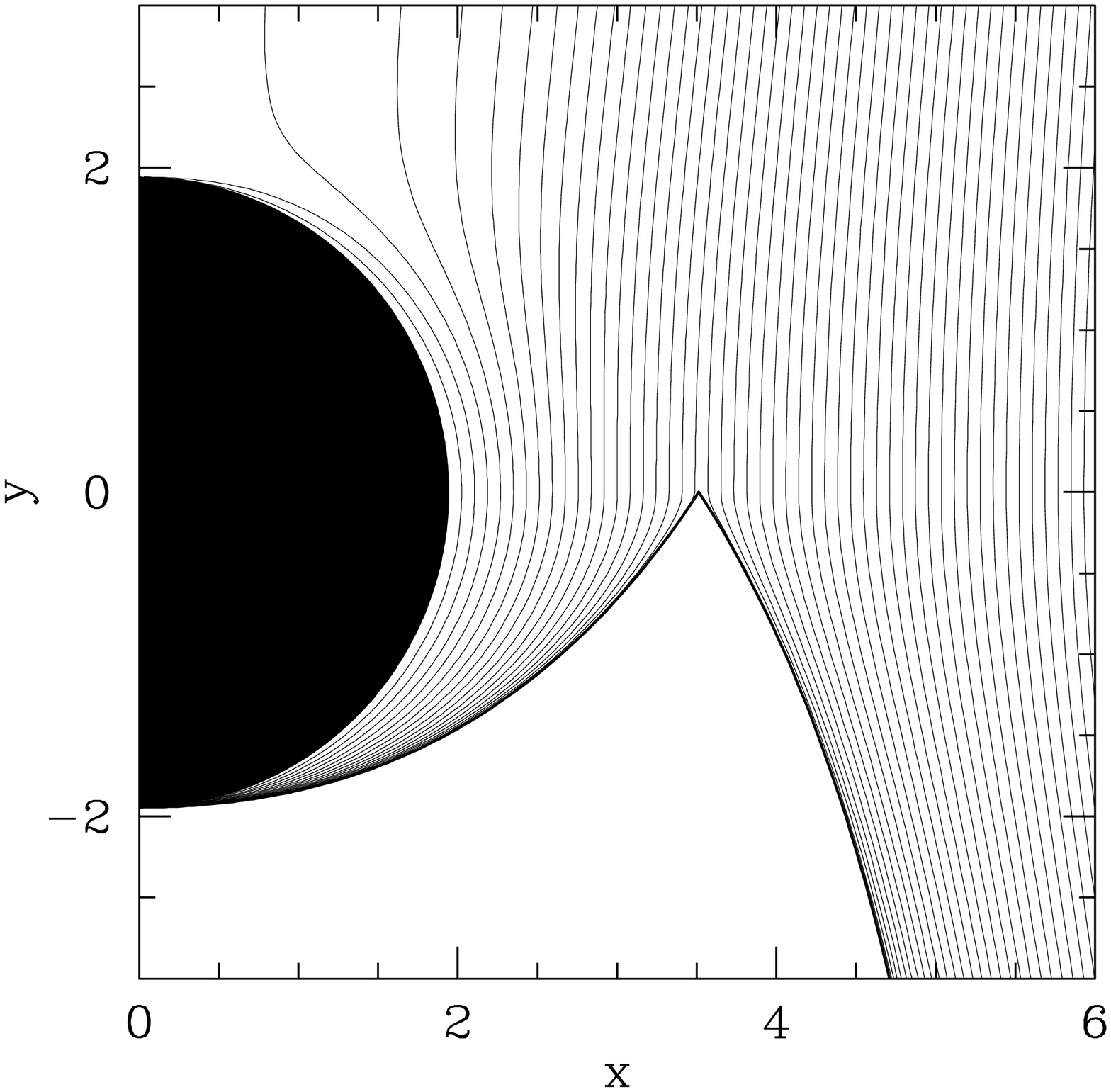,width=0.22\textwidth} &
\psfig{file=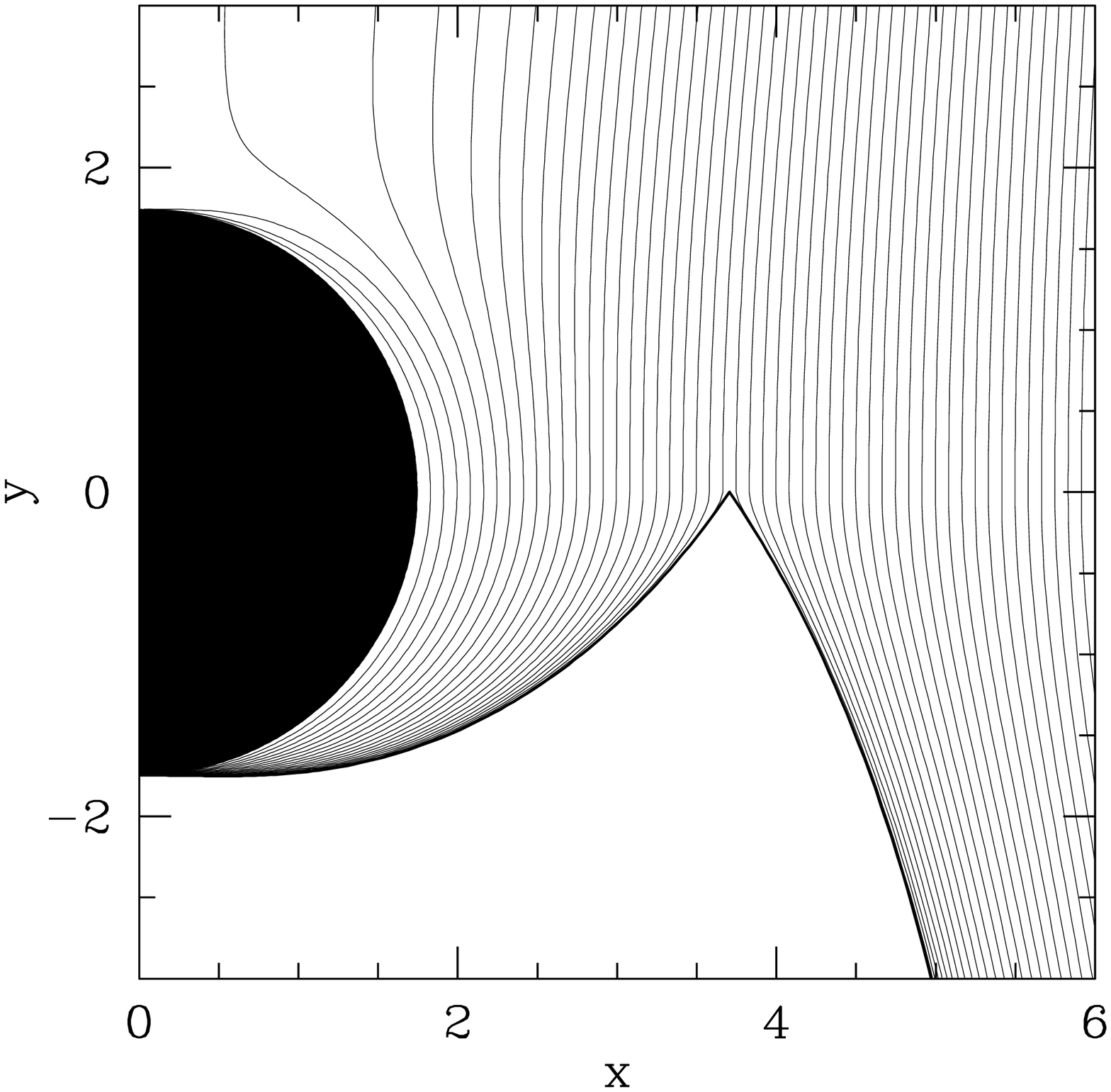,width=0.22\textwidth} &
\psfig{file=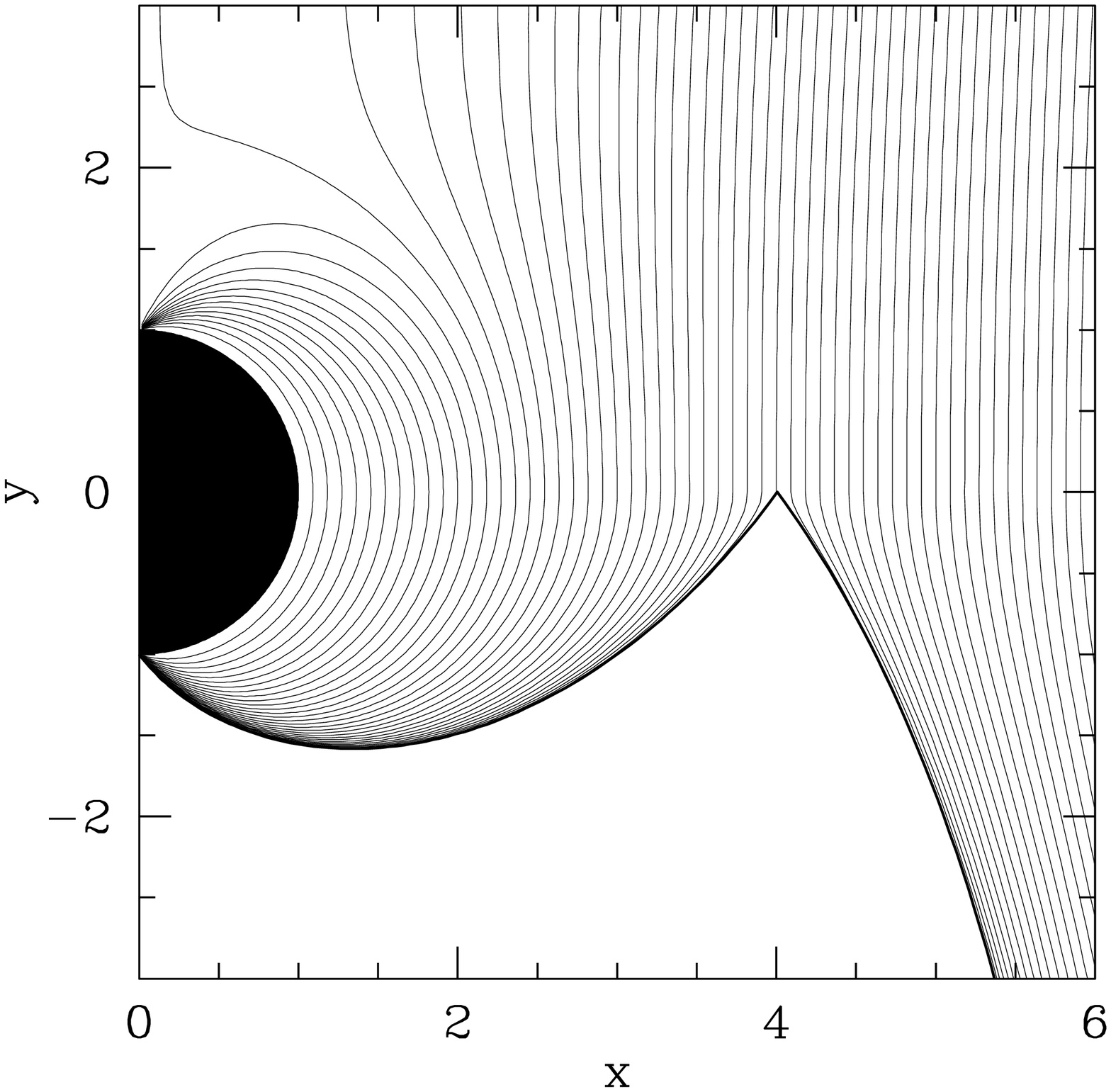,width=0.22\textwidth}\\
\end{tabular}
\end{center}
\caption{\textbf{The structure of the von Zeipel's cylinders outside the horizon:} each 
column corresponds to a different value of the spin of the black hole $a$. The first row 
shows in the $r_{0}$--$l_{0}$ plane the curve $\mathcal{P}=0$ (thick line) and the curve 
$\mathcal{Q}=0$ (thin line) as well as the region where $\mathcal{Q}/\mathcal{P}<0$ 
(shaded area), which allows to predict the behaviour of any cylinder in the vicinity of 
the equatorial plane (see text). The interior of the horizon has been shaded with crossed 
lines. For each value of $a$, the von Zeipel's cylinders are shown in the $x$-$y$ plane 
for $l_\mathrm{eq}(r_{0})=\pm 0.5$ for $a=1/3$ (resp. $\pm 1$ for $a=2/3$ and $a=1$) in 
second row, for $l_\mathrm{eq}(r_{0})=\pm 2$ in the third row and for $l_\mathrm{eq}(r_{0})=\pm 16$ 
for $a=1/3$ (resp. $\pm 8$ for $a=2/3$ and $a=1$) in the fourth row. The corresponding 
distribution $l_\mathrm{eq}(r_{0})$ are indicated in the first row as an horizontal thick line. 
In the case where there is a critical cylinder, it has been plotted with a thick line is 
the $x$-$y$ plane and the corresponding $(r_{0},l_{0})$ values have been indicated with a 
big dot in the $r_{0}$--$l_{0}$ plane. For the second, third and fourth rows, the $x$--$y$ 
plane is divided in two regions : $y\ge 0$ corresponds to prograde rotation ($l_\mathrm{eq}>0$) 
and $y \le 0$ to retrograde rotation ($l_\mathrm{eq}(r_{0})\le 0$). In the first column 
($a=0$, non-rotating black hole), the cylinders are independent of the distribution 
$l_\mathrm{eq}(r_{0})$.}
\label{fig:cylinders}
\end{figure*}

The roots of $\mathcal{P}$ (respectively, $\mathcal{Q}$) are plotted as a thick (respectively, 
thin) line in the $r_{0}$--$l_{0}$ plane in Fig.~\ref{fig:cylinders}. The region where 
$\mathcal{Q}/\mathcal{P}<0$ appears shaded in the plot (criss-crossing lines). In the same 
figure, the von Zeipel's cylinders are plotted for a constant angular momentum 
$l_\mathrm{eq}(r_{0})=l_{0}$ with different values of $|l_{0}|$. The corresponding values 
of the angular momentum are indicated as an horizontal thick line in the $r_{0}$--$l_{0}$ 
plane in each plot. In each case, the $x$--$y$ plane has been divided in two regions: for 
$y\ge 0$, the cylinders are plotted for $l_{0}=+|l_{0}|$ (prograde orbits) and for 
$y\le 0$ for $l_{0}=-|l_{0}|$ (retrograde orbits). When the line $l_\mathrm{eq}=l_{0}$ 
intersects the curve $\mathcal{P}(r_{0},l_{0})=0$, it is indicated by a thick dot in the 
$r_{0}$--$l_{0}$ plane and the corresponding two branches critical cylinder is plotted in 
the $x$--$y$ diagram with a thick line. Notice that for $a=1$ the only possible critical 
cylinder corresponding to prograde orbits is obtained with $r_{0}=r_\mathrm{h}=1$ and 
$l_{0}=2$. Its equation is given by $(r-1)^{2}\cos^{4}{\theta}-(r^{4}-4r+3)\cos^{2}{\theta}
-(r^{4}-3r^{2}+2r)=0$ (last column, third row).

In the particular case $a=0$ (non-rotating black hole), the equation of the von Zeipel's 
cylinders becomes independent of the distribution of the angular momentum per unit inertial 
mass:
\begin{equation}
0 = (r_{0}-2) r^{3}\sin^{2}{\theta}-(r-2)r_{0}^{3}.
\end{equation}
There is always one critical cylinder with a cusp located at $r_{0}=3$. This case is also 
plotted in Fig.~\ref{fig:cylinders} (first column).

\section{Derivative of the potential in the equatorial plane}
\label{sec:potential}

\subsection{Mathematical expression in Kerr spacetime}

The derivative $w_\mathrm{eq}(r)=dW_\mathrm{eq}/dr$ given in Eq.~(\ref{eq:dwdr}) 
can be expressed from Eq.~(\ref{eq:defvarpi}) and the expression of the metric 
coefficients and their derivatives in the equatorial plane given in Appendix~\ref{sec:kerr} 
as
\begin{eqnarray}
w_\mathrm{eq}(r) & = & \frac{1}{2\varpi^{2}}\times \frac{\tilde{A} 
l_\mathrm{eq}^{2}+2\tilde{B} l_\mathrm{eq} + \tilde{C}}{\tilde{g}_{tt} 
l_\mathrm{eq}^{2}+2\tilde{g}_{t\phi}l_\mathrm{eq}+\tilde{g}_{\phi\phi}}\ ,
\end{eqnarray}
where the coefficients $\tilde{A}$, $\tilde{B}$ and $\tilde{C}$ are given by
\begin{eqnarray}
\tilde{A} & = & -\left(\tilde{g}_{t\phi}^{2}\frac{\partial \tilde{g}_{tt}}
{\partial r}-2 \tilde{g}_{tt} \tilde{g}_{t\phi}\frac{\partial\tilde{g}_{t\phi}}
{\partial r}+\tilde{g}_{tt}^{2}\frac{\partial\tilde{g}_{\phi\phi}}{\partial r}\right)
\nonumber\\
& = & -\frac{2(r^{3}-4 r^{2}+4 r - a^{2})}{r^{2}}\nonumber\\
\tilde{B} & = & -\left(-\tilde{g}_{t\phi}\tilde{g}_{\phi\phi}
\frac{\partial\tilde{g}_{tt}}{\partial r}+\left(\tilde{g}_{t\phi}^{2}+
\tilde{g}_{tt}\tilde{g}_{\phi\phi}\right)\frac{\partial\tilde{g}_{t\phi}}
{\partial r}-\tilde{g}_{tt}\tilde{g}_{t\phi}\frac{\partial\tilde{g}_{\phi\phi}}
{\partial r}\right)\nonumber\\
& = & -\frac{2a\left(3 r^{2}-4 r + a^{2}\right)}{r^{2}}\nonumber\\
\tilde{C} & = & -\left(\tilde{g}_{\phi\phi}^{2}\frac{\partial\tilde{g}_{tt}}
{\partial r}-2 \tilde{g}_{t\phi}\tilde{g}_{\phi\phi}\frac{\partial\tilde{g}_{t\phi}}
{\partial r}+\tilde{g}_{t\phi}^{2}\frac{\partial\tilde{g}_{\phi\phi}}{\partial r}\right)
\nonumber\\
& = & \frac{2\left(r^{4}+2a^{2}r^{2}-4a^{2}r+a^{4}\right)}{r^{2}}\ .\nonumber
\end{eqnarray}

\begin{figure*}
\begin{center}
\begin{tabular}{ccc}
$a=0$ & $a=\sqrt{5}/3$ & $a=1$\\
\\
\psfig{file=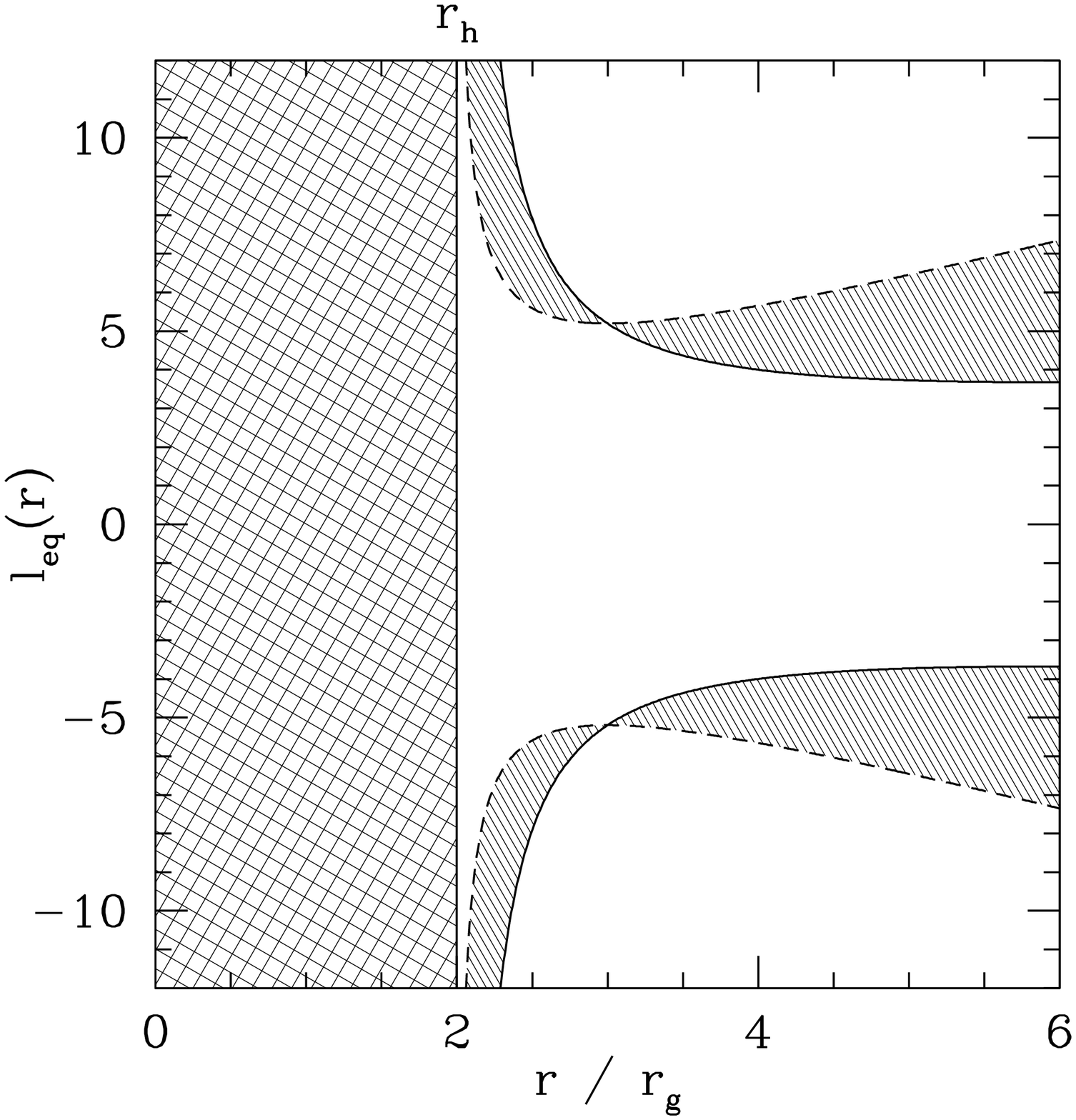,width=0.32\textwidth} &
\psfig{file=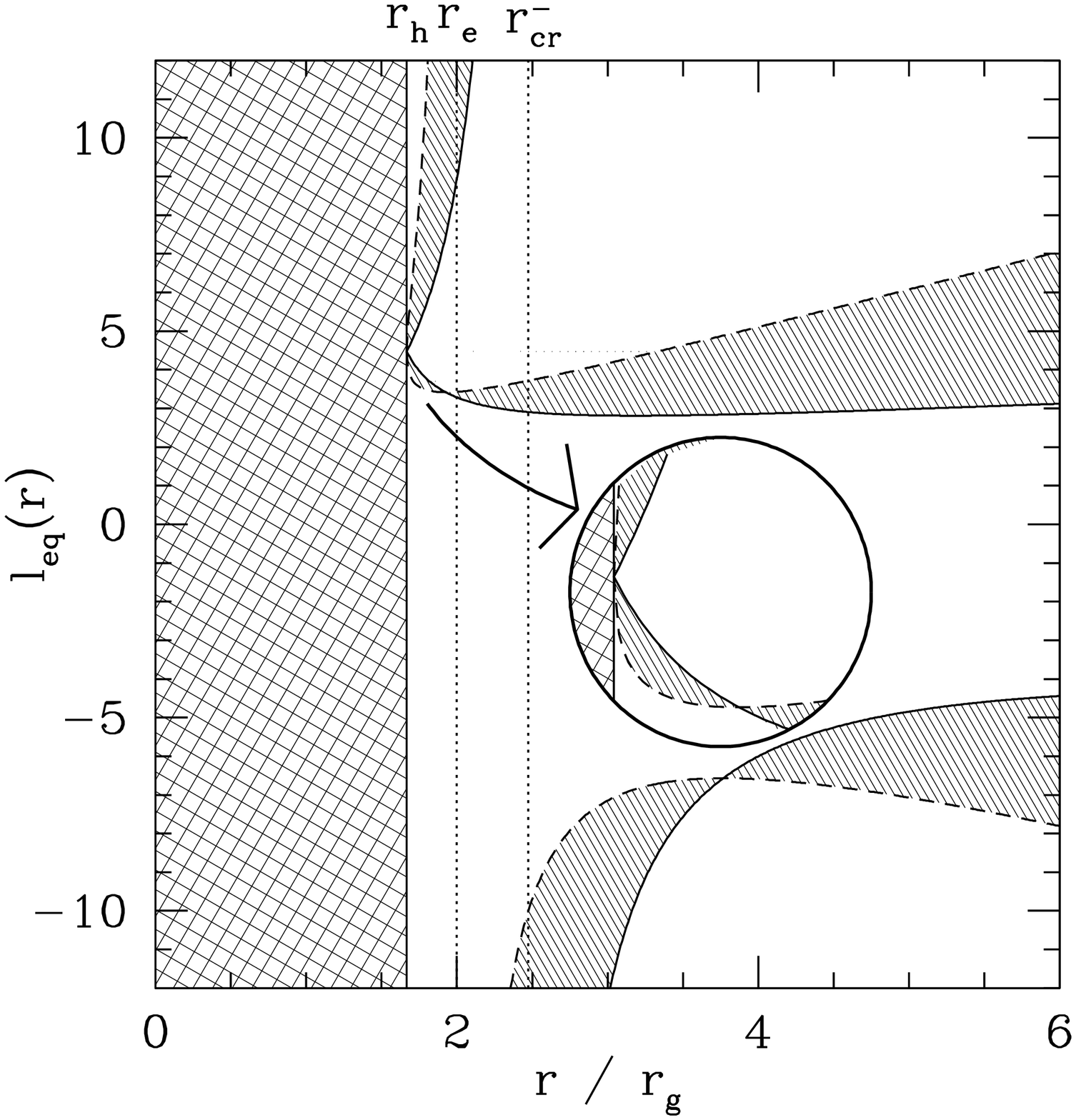,width=0.32\textwidth} &
\psfig{file=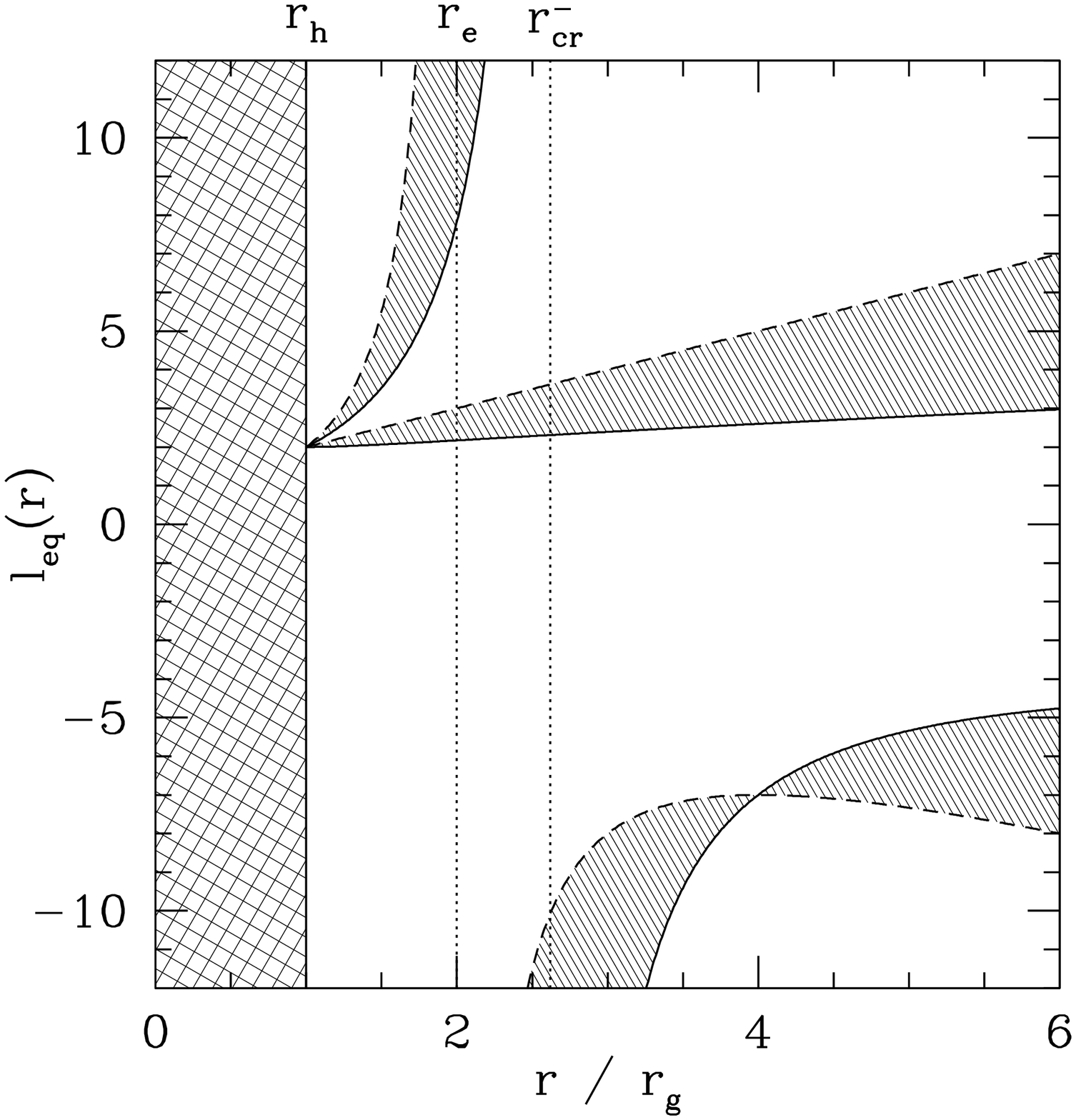,width=0.32\textwidth}
\end{tabular}
\end{center}
\caption{\textbf{The behaviour of $w_\mathrm{eq}(r)=dW_\mathrm{eq}/dr$ outside the horizon:} in the 
plane $r$--$l_\mathrm{eq}(r)$, the critical value $l_\mathrm{K}^{\pm}(r)$ (respectively, 
$l_\mathrm\mathrm{cr}^{\pm}(r)$) where the derivative of the potential $w_\mathrm{eq}(r)$ vanishes 
(respectively, becomes infinite) is plotted in solid line (respectively, dashed line). The interior 
of the horizon is the region shaded with crossed lines. The two radii $r_\mathrm{e}=2$ and 
$r_\mathrm{cr}^{-}$ are indicated by a vertical dotted line. The area shaded with oblique lines 
corresponds to the region where the derivative $w_\mathrm{eq}(r)$ is negative. The left, middle and 
right panels correspond, respectively, to three representative values of the spin of the black hole: 
$a=0$ (non rotating black hole), $a=\sqrt{5}/3$ and $a=1$ (maximally rotating black hole). Notice that 
the positions of the centre and the cusp in the disc are given by the intersection of the assumed 
rotation law $l_\mathrm{eq}(r)$ with $l_\mathrm{K}^{\pm}(r)$, if any.}
\label{fig:dwdr}
\end{figure*}

\subsection{Condition to have a finite value}

In the denominator of $w_\mathrm{eq}(r)$, the first quantity $2\varpi^{2}$ is everywhere 
positive for $r>r_\mathrm{h}$ and vanishes for $r=r_\mathrm{h}$, so that the derivative 
of the potential $W_\mathrm{eq}(r)$ is infinite at the horizon. The second term is a 
polynomial of degree 2 for the variable $l_\mathrm{eq}$. Its discriminant is simply 
$\varpi^{2}$ so that there are always two distinct real roots outside the horizon:
\begin{equation}
l_\mathrm{cr}^{\pm}(r) = \frac{-\tilde{g}_{t\phi} \mp \varpi}{\tilde{g}_{tt}} = 
\frac{-2a \pm r \sqrt{r^{2}-2 r + a^{2}}}{r-2}\ .
\end{equation}
Then, $w_\mathrm{eq}(r)$ has always a finite value in the equatorial plane outside the 
horizon except for $l_\mathrm{eq}=l_\mathrm{cr}^{\pm}(r)$. Notice that at the horizon 
$l_\mathrm{cr}^{\pm}(r_\mathrm{h})=\pm\infty$ for $a=0$ and $l_\mathrm{cr}^{\pm}(r_\mathrm{h}) 
= 2 r_\mathrm{h}/a>0$
otherwise. Concerning the sign of the denominator of $w_\mathrm{eq}(r)$, there are two cases: 
(i) inside the ergosphere ($r\le 2$) we have $l_\mathrm{cr}^{-}(r) \ge l_\mathrm{cr}^{+}(r)>0$ 
and the denominator is negative for $l_\mathrm{cr}^{+}(r) < l_\mathrm{eq}(r) < l_\mathrm{cr}^{-}(r)$; 
(ii) outside the ergosphere ($r>2$), we have $l_\mathrm{cr}^{-}(r) < 0 < l_\mathrm{cr}^{+}(r)$ 
and the denominator is negative for $l_\mathrm{eq}(r)<l_\mathrm{cr}^{-}(r)$ or $l_\mathrm{eq}(r) 
> l_\mathrm{cr}^{+}(r)$.

In Fig.~\ref{fig:dwdr}, we plot $l_\mathrm{K}^{\pm}(r)$ and $l_\mathrm{cr}^{\pm}(r)$ for 
$a=0$ (left), $a=\sqrt{5}/3$ (middle) and $a=1$ (right), indicating the region where 
$w_\mathrm{eq}(r)$ is negative.

\subsection{Condition to vanish}

The discriminant $\tilde{\Delta}=\tilde{B}^{2}-\tilde{A}\tilde{C}$ of the numerator equals
$$
\tilde{\Delta} = \frac{4(r^{2}-2r+a^{2})^{2}}{r}
$$
and is always positive so that $w_\mathrm{eq}(r)$ can vanish under the condition
$l_\mathrm{eq} = (-\tilde{B}\pm\sqrt{\tilde{\Delta}}) / \tilde{A}$. This leads to
\begin{eqnarray}
l_\mathrm{eq}(r) & = & \frac{-a(3 r^{2}-4r+a^{2})\pm (r^{2}-2r+a^{2})r\sqrt{r}}
{r^{3}-4 r^{2}+4 r - a^{2}}\nonumber\\
 & = &  \pm\frac{(r^{2}+a^{2})\mp 2a\sqrt{r}}{(r-2)\sqrt{r}\pm a}\nonumber
\end{eqnarray}
From equation~(\ref{eq:lk}), this condition is nothing else than
\begin{equation}
l_\mathrm{eq}(r)  = l_\mathrm{K}^{\pm}(r)\ .
\end{equation}
Notice that at the horizon $l_\mathrm{K}^{\pm}(r_\mathrm{h})=\pm \infty$ for $a=0$ and 
$l_\mathrm{K}^{\pm}(r_\mathrm{h})=2 r_\mathrm{h}/a=l_\mathrm{cr}^{\pm}(r_\mathrm{h})>0$ 
otherwise. The coefficient $\tilde{A}$ can be written as 
\begin{eqnarray}
\tilde{A} & = & -\frac{2}{r}\left((r-2)\sqrt{r}-a\right)\left((r-2)\sqrt{r}+a\right)\ ,
\end{eqnarray}
so that outside the horizon, $\tilde{A}$ is positive for $r < r_\mathrm{cr}^{-}$ and negative 
otherwise. The radius $r_\mathrm{cr}^{-} \ge 2$ is the unique root for $r>r_\mathrm{h}$ of 
$(r-2)\sqrt{r}=a$. Then there are two cases for the sign of the numerator of $w_\mathrm{eq}(r)$: 
(i) for $r\le r_\mathrm{cr}^{-}$, we have $l_\mathrm{K}^{-}(r) \ge l_\mathrm{K}^{+}(r)>0$ and 
the numerator is negative for $l_\mathrm{K}^{+}(r) < l_\mathrm{eq}(r) < l_\mathrm{K}^{-}(r)$; 
(ii) for $r > r_\mathrm{cr}^{-}$, we have $l_\mathrm{K}^{-}(r) < 0 < l_\mathrm{K}^{+}(r)$ and 
the numerator is negative for $l_\mathrm{eq}(r) < l_\mathrm{K}^{-}(r)$ or $l_\mathrm{eq}(r) > 
l_\mathrm{K}^{+}(r)$.

\section*{Acknowledgments}
We thank Olindo Zanotti for a careful reading of the manuscript. J.A.F. acknowledges financial 
support from the Spanish Ministerio de Ciencia y Tecnolog\'{\i}a (grant AYA 2001-3490-C02-01).
The simulations were performed on DEC-alpha workstations with generous computing
time provided by the Albert Einstein Institute in Golm (Germany).

\bibliographystyle{mn2e}
\bibliography{runaway}

\begin{thebibliography}{}

\bibitem[\protect\citeauthoryear{{Abramowicz}, {Calvani} \&
  {Nobili}}{{Abramowicz} et~al.}{1983}]{abramowicz:83}
{Abramowicz} M.~A.,  {Calvani} M.,    {Nobili} L.,  1983, Nature, 302, 597

\bibitem[\protect\citeauthoryear{{Abramowicz}, {Karas} \& {Lanza}}{{Abramowicz}
  et~al.}{1998}]{abramowicz:98}
{Abramowicz} M.~A.,  {Karas} V.,    {Lanza} A.,  1998, A\&A, 331, 1143

\bibitem[\protect\citeauthoryear{{Alcubierre}, {Br{\" u}gmann}, {Dramlitsch},
  {Font}, {Papadopoulos}, {Seidel}, {Stergioulas} \& {Takahashi}}{{Alcubierre}
  et~al.}{2000}]{alcubierre:00}
{Alcubierre} M.,  {Br{\" u}gmann} B.,  {Dramlitsch} T.,  {Font} J.~.,
  {Papadopoulos} P.,  {Seidel} E.,  {Stergioulas} N.,    {Takahashi} R.,  2000,
  Phys. Rev. D, 62, 044034

\bibitem[\protect\citeauthoryear{{Aloy}, {M{\" u}ller}, {Ib{\' a}{\~ n}ez},
  {Mart\'{\i}} \& {MacFadyen}}{{Aloy} et~al.}{2000}]{aloy:00}
{Aloy} M.~A.,  {M{\" u}ller} E.,  {Ib{\' a}{\~ n}ez} J.~M.,  {Mart\'{\i}}
  J.~M.,    {MacFadyen} A.,  2000, ApJ, 531, L119

\bibitem[\protect\citeauthoryear{{Blaes}}{{Blaes}}{1987}]{blaes:87}
{Blaes} O.~M.,  1987, MNRAS, 227, 975

\bibitem[\protect\citeauthoryear{{Brandt}, Font, Ib\'a\~nez, Mass\'o \&
  {Seidel}}{{Brandt} et~al.}{2000}]{brandt:00}
{Brandt} S.,  Font J.~A.,  Ib\'a\~nez J.~M.,  Mass\'o J.,    {Seidel} E.,
  2000, Comput. Phys. Comm., 124, 169

\bibitem[\protect\citeauthoryear{{Daigne} \& {Mochkovitch}}{{Daigne} \&
  {Mochkovitch}}{1997}]{daigne:97}
{Daigne} F.,  {Mochkovitch} R.,  1997, MNRAS, 285, L15

\bibitem[\protect\citeauthoryear{{Davies}, {Benz}, {Piran} \&
  {Thielemann}}{{Davies} et~al.}{1994}]{davies:94}
{Davies} M.~B.,  {Benz} W.,  {Piran} T.,    {Thielemann} F.~K.,  1994, ApJ,
  431, 742

\bibitem[\protect\citeauthoryear{{De Villiers} \& {Hawley}}{{De Villiers} \&
  {Hawley}}{2003}]{devilliers:03}
{De Villiers} J.,  {Hawley} J.~F.,  2003, ApJ, 592, 1060

\bibitem[\protect\citeauthoryear{{Fishbone} \& {Moncrief}}{{Fishbone} \&
  {Moncrief}}{1976}]{fishbone:76}
{Fishbone} L.~G.,  {Moncrief} V.,  1976, ApJ, 207, 962

\bibitem[\protect\citeauthoryear{{Font}}{{Font}}{2003}]{fontlr}
{Font} J.~A.,  2003, Living Reviews in Relativity, 6, 4

\bibitem[\protect\citeauthoryear{{Font} \& {Daigne}}{{Font} \&
  {Daigne}}{2002a}]{font:02b}
{Font} J.~A.,  {Daigne} F.,  2002a, ApJL, 581, L23

\bibitem[\protect\citeauthoryear{{Font} \& {Daigne}}{{Font} \&
  {Daigne}}{2002b}]{font:02a}
{Font} J.~A.,  {Daigne} F.,  2002b, MNRAS, 334, 383

\bibitem[\protect\citeauthoryear{{Font}, Goodale, Iyer, Miller, Rezzolla,
  Seidel, Stergioulas, Suen \& Tobias}{{Font} et~al.}{2001}]{font:01}
{Font} J.~A.,  Goodale T.,  Iyer S.,  Miller M.,  Rezzolla L.,  Seidel E.,
  Stergioulas N.,  Suen W.,    Tobias M.,  2001, Phys. Rev. D

\bibitem[\protect\citeauthoryear{{Font}, {Ib{\'a}{\~n}ez} \&
  {Papadopoulos}}{{Font} et~al.}{1999}]{font:99}
{Font} J.~A.,  {Ib{\'a}{\~n}ez} J.~M.,    {Papadopoulos} P.,  1999, MNRAS, 305,
  920

\bibitem[\protect\citeauthoryear{{Gammie}, {McKinney} \& {T{\' o}th}}{{Gammie}
  et~al.}{2003}]{gammie:03}
{Gammie} C.~F.,  {McKinney} J.~C.,    {T{\' o}th} G.,  2003, ApJ, 589, 444

\bibitem[\protect\citeauthoryear{{Igumenshchev} \&
  {Beloborodov}}{{Igumenshchev} \& {Beloborodov}}{1997}]{igumenshchev:97}
{Igumenshchev} I.~V.,  {Beloborodov} A.~M.,  1997, MNRAS, 284, 767

\bibitem[\protect\citeauthoryear{{Khanna} \& {Chakrabarti}}{{Khanna} \&
  {Chakrabarti}}{1992}]{khanna:92}
{Khanna} R.,  {Chakrabarti} S.~K.,  1992, MNRAS, 259, 1

\bibitem[\protect\citeauthoryear{{Kluzniak} \& {Lee}}{{Kluzniak} \&
  {Lee}}{1998}]{kluzniak:98}
{Kluzniak} W.,  {Lee} W.~H.,  1998, ApJ, 494, L53

\bibitem[\protect\citeauthoryear{{Kozlowski}, {Jaroszynski} \&
  {Abramowicz}}{{Kozlowski} et~al.}{1978}]{abramowicz:78}
{Kozlowski} M.,  {Jaroszynski} M.,    {Abramowicz} M.~A.,  1978, A\&A, 63, 209

\bibitem[\protect\citeauthoryear{{Lee}}{{Lee}}{2000}]{lee:00}
{Lee} W.~H.,  2000, MNRAS, 318, 606

\bibitem[\protect\citeauthoryear{{Lu}, {Cheng}, {Yang} \& {Zhang}}{{Lu}
  et~al.}{2000}]{lu:00}
{Lu} Y.,  {Cheng} K.~S.,  {Yang} L.~T.,    {Zhang} L.,  2000, MNRAS, 314, 453

\bibitem[\protect\citeauthoryear{{MacFadyen} \& {Woosley}}{{MacFadyen} \&
  {Woosley}}{1999}]{macfadyen:99}
{MacFadyen} A.~I.,  {Woosley} S.~E.,  1999, ApJ, 524, 262

\bibitem[\protect\citeauthoryear{{Masuda}, {Nishida} \& {Eriguchi}}{{Masuda}
  et~al.}{1998}]{masuda:98}
{Masuda} N.,  {Nishida} S.,    {Eriguchi} Y.,  1998, MNRAS, 297, 1139

\bibitem[\protect\citeauthoryear{Misner, Thorne \& Wheeler}{Misner
  et~al.}{1973}]{misner:73}
Misner C.~W.,  Thorne K.~S.,    Wheeler J.~A.,  1973, Gravitation.
W. H. Freeman, San Francisco

\bibitem[\protect\citeauthoryear{{Nishida} \& {Eriguchi}}{{Nishida} \&
  {Eriguchi}}{1996}]{nishida:96b}
{Nishida} S.,  {Eriguchi} Y.,  1996, ApJ, 461, 320

\bibitem[\protect\citeauthoryear{{Nishida}, {Lanza}, {Eriguchi} \&
  {Abramowicz}}{{Nishida} et~al.}{1996}]{nishida:96a}
{Nishida} S.,  {Lanza} A.,  {Eriguchi} Y.,    {Abramowicz} M.~A.,  1996, MNRAS,
  278, L41

\bibitem[\protect\citeauthoryear{{Rees} \& {Meszaros}}{{Rees} \&
  {Meszaros}}{1994}]{rees:94}
{Rees} M.~J.,  {Meszaros} P.,  1994, ApJL, 430, L93

\bibitem[\protect\citeauthoryear{{Rezzolla}, {Yoshida} \& {Zanotti}}{{Rezzolla}
  et~al.}{2003}]{rezzolla:03}
{Rezzolla} L.,  {Yoshida} S.,    {Zanotti} O.,  2003, ArXiv Astrophysics
  e-prints

\bibitem[\protect\citeauthoryear{{Ruffert} \& {Janka}}{{Ruffert} \&
  {Janka}}{1999}]{ruffert:99}
{Ruffert} M.,  {Janka} H.-T.,  1999, A\&A, 344, 573

\bibitem[\protect\citeauthoryear{{Ruffert}, {Janka} \& {Schaefer}}{{Ruffert}
  et~al.}{1996}]{ruffert:96}
{Ruffert} M.,  {Janka} H.-T.,    {Schaefer} G.,  1996, A\&A, 311, 532

\bibitem[\protect\citeauthoryear{{Shapiro} \& {Shibata}}{{Shapiro} \&
  {Shibata}}{2002}]{shapiro:02}
{Shapiro} S.~L.,  {Shibata} M.,  2002, ApJ, 577, 904

\bibitem[\protect\citeauthoryear{{Shibata}}{{Shibata}}{2003}]{shibata:03}
{Shibata} M.,  2003, Phys. Rev. D, 67, 24033

\bibitem[\protect\citeauthoryear{{Shibata}, {Baumgarte} \& {Shapiro}}{{Shibata}
  et~al.}{2000}]{shibata2:00}
{Shibata} M.,  {Baumgarte} T.~W.,    {Shapiro} S.~L.,  2000, Phys. Rev. D, 61,
  4012

\bibitem[\protect\citeauthoryear{{Shibata} \& {Shapiro}}{{Shibata} \&
  {Shapiro}}{2002}]{shibata:02}
{Shibata} M.,  {Shapiro} S.~L.,  2002, ApJL, 572, L39

\bibitem[\protect\citeauthoryear{{Shibata}, {Taniguchi} \& {Ury{\=
  u}}}{{Shibata} et~al.}{2003}]{shibata2:03}
{Shibata} M.,  {Taniguchi} K.,    {Ury{\= u}} K.,  2003, Phys. Rev. D (to
  appear)

\bibitem[\protect\citeauthoryear{{Shibata} \& {Ury{\= u}}}{{Shibata} \& {Ury{\=
  u}}}{2000}]{shibata:00}
{Shibata} M.,  {Ury{\= u}} K.,  2000, Phys. Rev. D, 61, 4001

\bibitem[\protect\citeauthoryear{{Wilson}}{{Wilson}}{1984}]{wilson:84}
{Wilson} D.~B.,  1984, Nature, 312, 620

\bibitem[\protect\citeauthoryear{{Woosley}}{{Woosley}}{2001}]{woosley:01}
{Woosley} S.~E.,  2001, in {Costa} E.,  {Frontera} F.,   {Hjorth} J.,  eds,
  Gamma-ray Bursts in the Afterglow Era, Proceedings of the International
  workshop held in Rome, CNR headquarters, 17-20 October, 2000 {Gamma-Ray Burst
  Models: The Central Engine}.
Berlin Heidelberg: Springer, p.~257

\bibitem[\protect\citeauthoryear{{Yo}, {Baumgarte} \& {Shapiro}}{{Yo}
  et~al.}{2002}]{yo:03}
{Yo} H.,  {Baumgarte} T.~W.,    {Shapiro} S.~L.,  2002, Phys. Rev. D, 66, 84026

\bibitem[\protect\citeauthoryear{{Zanotti}, {Rezzolla} \& {Font}}{{Zanotti}
  et~al.}{2003}]{zanotti:02}
{Zanotti} O.,  {Rezzolla} L.,    {Font} J.~A.,  2003, MNRAS, 341, 832

\end{thebibliography}
\end{document}